\documentclass[%
 aip,
floatfix,
 amsmath,amssymb,
 reprint, multicols%
]{revtex4-1}

\pdfoutput=1

\usepackage{graphicx}
\usepackage{dcolumn}
\usepackage{bm}

\usepackage[utf8]{inputenc}
\usepackage[T1]{fontenc}
\usepackage{mathptmx}
\usepackage{lipsum}
\usepackage{setspace}

\usepackage{bm}
\usepackage[mathlines]{lineno}

\usepackage{mathptmx}
\usepackage{siunitx}

\newcommand{\Bzero}{\ensuremath{\textbf{B}_0(\textbf{r})}}
\newcommand{\Brf}{\ensuremath{\textbf{B}_\text{rf}(\textbf{r},t)}}
\newcommand{\Bperp}{\ensuremath{\textbf{B}_\text{rf}^\perp(\textbf{r})}}
\newcommand{\Rzero}{\ensuremath{\Omega_0(\textbf{r})}}
\newcommand{\RzeroSq}{\ensuremath{\Omega_0^2(\textbf{r})}}

\usepackage{bm}
\usepackage{helvet}
\usepackage{amsfonts}
\usepackage{hyperref}
\usepackage{color}
\usepackage{epstopdf}
\usepackage{etex}
\usepackage{morefloats}
\usepackage{graphicx}
\usepackage{todonotes}

\newcommand{\bra}[1]{\ensuremath{\left\langle#1\right|}}
\newcommand{\mean}[1]{\ensuremath{\left\langle #1 \right\rangle}}
\newcommand{\ket}[1]{\ensuremath{\left|#1\right\rangle}}
\newcommand{\braket}[2]{\ensuremath{\left\langle #1 \middle| #2 \right\rangle}}
\newcommand{\matrixel}[3]{\ensuremath{\left\langle #1 \middle| #2 \middle| #3 \right\rangle}}

\newcommand{\Tr}{\mathrm{Tr}\,}

\newcommand{\bracket}[2]{\ensuremath{\left\langle #1 \middle| #2 \right\rangle}}
\usepackage{xcolor}

\newcommand{\pref}[1]{(\ref{#1})}
\newcommand{\eref}[1]{Eq.~\pref{#1}}
\newcommand{\fref}[1]{Fig.~\ref{#1}}

\newcommand{\figref}[1]{Fig.~\ref{#1}}

\usepackage{subfigure}

\newcommand{\subfigimg}[3][,]{%
	\setbox1=\hbox{\includegraphics[#1]{#3}}
	\leavevmode\rlap{\usebox1}
	\rlap{\hspace*{2pt}\raisebox{\dimexpr\ht1-0.5\baselineskip}{{\bfseries \large\textsf{#2}}}}
	\phantom{\usebox1}
}

\newcommand{\subfigimgraised}[3][,]{%
	\setbox1=\hbox{\includegraphics[#1]{#3}}
	\leavevmode\rlap{\usebox1}
	\rlap{\hspace*{2pt}\raisebox{\dimexpr\ht1-0.\baselineskip}{{\bfseries \large\textsf{#2}}}}
	\phantom{\usebox1}
}

\graphicspath{{./Images/}}

\newcommand{\an}[2]{\ensuremath{\hat{#1}^{\protect\phantom{\dagger}}_{#2}}}
\newcommand{\cn}[2]{\ensuremath{\hat{#1}^\dagger_{#2}}}
\newcommand{\nn}[2]{\ensuremath{\hat{n}^{#1}_{#2}}}
\newcommand{\expU}[1]{\ensuremath{e^{#1}}}
\newcommand{\abs}[1]{\left|#1\right|}

\newcommand{\densdescr}[4]{The propagation of the density excitation in time. The upper curves show the transmitted density wave into the drain lead (integrated between site \ensuremath{#1} and \ensuremath{#2}), and the lower curves the incoming and reflected wave in the source lead (\ensuremath{#3} and \ensuremath{#4}). The background density is subtracted.}
\newcommand{\idg}[1]{{\bfseries #1)}}
\newcommand{\pdif}[2]{\ensuremath{\frac{\partial#1}{\partial#2}}}

\newcommand{\avg}[1]{\ensuremath{\langle#1\rangle}}

\pdfoutput=1
\usepackage[utf8]{inputenc}
\usepackage{amsmath}
\usepackage{amsfonts}
\usepackage{braket}
\usepackage{tensor}
\usepackage{bbold}
\usepackage{graphicx}
\usepackage{empheq}
\usepackage{amsthm}

\def\be{\begin{equation}}
	\def\ee{\end{equation}}
\def\bea{\begin{eqnarray}}
	\def\eea{\end{eqnarray}}

\usepackage{hyperref}

\usepackage{xparse}
\usepackage{appendix}
\usepackage{graphicx}
\usepackage{dcolumn}
\usepackage{bm}

 \newcommand{\bray}[1]{\ensuremath{\left( #1 \right|}}
\newcommand{\kety}[1]{\ensuremath{\left| #1 \right)}}
\usepackage{bbold}

\newcommand\bes           {\begin{subequations}}
\newcommand\esu           {\end{subequations}}

\makeatletter
\newcommand{\vast}{\bBigg@{3.3}}
\newcommand{\Vast}{\bBigg@{5}}

\makeatother

\newcommand\p            {\partial}

\def\3pt#1#2#3{{\langle{#1}\vert{#2}\vert{#3}\rangle}}
\def\barray{\begin{eqnarray}}
\def\earray{\end{eqnarray}}

\usepackage{tikz}
\usetikzlibrary{calc}

\newcommand{\vc}[1]{\ensuremath{{\bf#1}}}

\newcommand{\op}[1]{\ensuremath{\hat{#1}}}

\begin{document}


\preprint{AIP/123-QED}

\begin{titlepage}

\vspace*{-1.8cm}

\title{Roadmap on Atomtronics: State of the art and perspective}
\author{L. Amico}
\altaffiliation{\scriptsize{On leave from Dipartimento di Fisica e Astronomia 'Ettore Majorana', Universit\`a di Catania, Italy}}
\affiliation{\scriptsize{Quantum Research Centre, Technology Innovation Institute, Abu Dhabi, UAE}}
\affiliation{\scriptsize{Centre for Quantum Technologies, National University of Singapore, 3 Science Drive 2, 117543 Singapore}\looseness=-1}
\affiliation{\scriptsize{LANEF \textit{'Chaire d'excellence'}, Universit\`e Grenoble-Alpes $\&$ CNRS, F-38000 Grenoble, France}\looseness=-1}
 \affiliation{\scriptsize{MajuLab, CNRS-UCA-SU-NUS-NTU International Joint Research Unit, Singapore}}
\author{M. Boshier}
\affiliation{\scriptsize{Physics Division, Los Alamos National Laboratory, Los Alamos, NM 87545, USA}}
\author{G. Birkl}
\affiliation{\scriptsize{Institut  f\"ur  Angewandte  Physik,  Technische  Universit\"at  Darmstadt,   Schlossgartenstr. 7, 64289 Darmstadt,  Germany}\looseness=-1}
\author{A. Minguzzi}
\affiliation{\scriptsize{Univ. Grenoble-Alpes,  CNRS, LPMMC, F-38000 Grenoble, France}}
 \author{C. Miniatura}
  \affiliation{\scriptsize{MajuLab, CNRS-UCA-SU-NUS-NTU International Joint Research Unit, Singapore}}
 \affiliation{\scriptsize{Centre for Quantum Technologies, National University of Singapore, 3 Science Drive 2, 117543 Singapore}\looseness=-1}
 \affiliation{\scriptsize{Department of Physics, National University of Singapore, Singapore 117542}\looseness=-1}
\affiliation{\scriptsize{Division of Physics and Applied Physics, Nanyang Technological University, 21 Nanyang Link, 637371 Singapore}\looseness=-1}
 \affiliation{\scriptsize{Universit\'{e}   C\^{o}te   d'Azur,   CNRS,   INPHYNI,   Nice,   France}}
  \author{L.-C. Kwek}
 \affiliation{\scriptsize{Centre for Quantum Technologies, National University of Singapore, 3 Science Drive 2, 117543 Singapore}\looseness=-1}
 \affiliation{\scriptsize{MajuLab, CNRS-UCA-SU-NUS-NTU International Joint Research Unit, Singapore}}
\affiliation{\scriptsize{Institute of Advanced Studies, Nanyang Technological University, 60 Nanyang View, 639673 Singapore}}
\author{D. Aghamalyan}

\affiliation{\scriptsize{School of Information Systems, Singapore Management University, 81 Victoria Street, Singapore 188065}\looseness=-1}

\affiliation{\scriptsize{Centre for Quantum Technologies, National University of Singapore,
3 Science Drive 2, 117543 Singapore}\looseness=-1}
\author{V. Ahufinger}
\affiliation{\scriptsize{Departament de F\'isica, Universitat Aut\`onoma de Barcelona, E-08193 Bellaterra, Spain}}
\author{D. Anderson}
\affiliation{\scriptsize{Department of Physics and JILA, University of Colorado, Boulder, Colorado, 80309-0440, USA}}
\affiliation{\scriptsize{ColdQuanta Inc. 3030 Sterling Circle, Boulder CO 80301, USA}}
\author{N. Andrei}
\affiliation{\scriptsize{Center for Materials Theory, Department of Physics and Astronomy, Rutgers University, Piscataway, NJ 08854}\looseness=-1}
\author{A. S. Arnold}
\affiliation{\scriptsize{Department of Physics, SUPA, University of Strathclyde, Glasgow G4 0NG, UK}}
%
\author{M. Baker} %
\affiliation{ \scriptsize{ARC Centre of Excellence for Engineered Quantum Systems, School of Mathematics and Physics, University of Queensland, Brisbane, QLD 4072, Australia
}\looseness=-1}%
\author{T.A. Bell}
\affiliation{ \scriptsize{ARC Centre of Excellence for Engineered Quantum Systems, School of Mathematics and Physics, University of Queensland, Brisbane, QLD 4072, Australia
}\looseness=-1}
\author{T. Bland}
\affiliation{\scriptsize{Joint Quantum Centre (JQC) Durham-Newcastle, School of Mathematics, Statistics and Physics, Newcastle University, Newcastle upon Tyne, NE1 7RU, U.K.}\looseness=-1}
\affiliation{\scriptsize{Institut f\"{u}r Quantenoptik und Quanteninformation, \"Osterreichische Akademie der Wissenschaften, Innsbruck, Austria}\looseness=-1}
\author{J.P. Brantut}
\affiliation{\scriptsize{Institute of Physics, EPFL, 1015 Lausanne, Switzerland}}
\author{D. Cassettari}
\affiliation{\scriptsize{SUPA, School of Physics and Astronomy, University of St. Andrews, North Haugh, St. Andrews KY16 9SS, UK}\looseness=-1}
\author{W.J. Chetcuti}
\affiliation{\scriptsize{Dipartimento di Fisica e Astronomia, Via S. Sofia 64, 95127 Catania, Italy}}
\affiliation{\scriptsize{Quantum Research Centre, Technology Innovation Institute, Abu Dhabi, UAE}}
\author{F. Chevy}
\affiliation{\scriptsize{Laboratoire Kastler Brossel, ENS-Universit\'e PSL, CNRS, Sorbonne Universit\'e, Coll\`ege de France, Paris, France}\looseness=-1}
\author{R. Citro}
\affiliation{\scriptsize{Dipartimento di Fisica "E.R. Caianiello", Universit\`a degli Studi di Salerno, Via Giovanni Paolo II, I-84084 Fisciano (SA), Italy}\looseness=-1}
\author{S. De Palo}
\affiliation{\scriptsize{CNR-IOM-Democritos National Simulation Centre, UDS Via Bonomea 265, I-34136, Trieste, Italy}\looseness=-1}
\author{R. Dumke}
\affiliation{\scriptsize{Centre for Quantum Technologies, National University of Singapore, 3 Science Drive 2, 117543  Singapore}\looseness=-1}
\affiliation{\scriptsize{Division of Physics and Applied Physics, Nanyang Technological University, 21 Nanyang Link, 637371 Singapore}\looseness=-1}
 \affiliation{\scriptsize{MajuLab, CNRS-UCA-SU-NUS-NTU International Joint Research Unit, Singapore}}
 \author{M. Edwards}
 \affiliation{\scriptsize{Department  of  Physics,  Georgia  Southern  University,  Statesboro,  GA  30460–8031  USA}\looseness=-1}
\author{R. Folman}%
\affiliation{ 
\scriptsize{Department of Physics, Ben-Gurion University of the Negev, Beer Sheva 84105, Israel.
}}%
\author{J. Fortagh}
\affiliation{%
\scriptsize{CQ Center for Collective Quantum Phenomena and their Applications, Eberhard-Karls-Universit{\"a}t T{\"u}bingen, Germany.
}\looseness=-1}
\author{S. A. Gardiner}
\affiliation{\scriptsize{Joint Quantum Centre (JQC) Durham–Newcastle, Department of Physics, Durham University, South Road, Durham DH1 3LE, UK}\looseness=-1}
\author{B.M. Garraway}
\affiliation{\scriptsize{Department  of  Physics  and  Astronomy,  University  of  Sussex,  Falmer,  Brighton,  BN1  9QH, UK}\looseness=-1}
\author{G. Gauthier}
\affiliation{ \scriptsize{ARC Centre of Excellence for Engineered Quantum Systems, School of Mathematics and Physics, University of Queensland, Brisbane, QLD 4072, Australia
}\looseness=-1}
\author{A. G{\"u}nther}
\affiliation{%
\scriptsize{CQ Center for Collective Quantum Phenomena and their Applications, Eberhard-Karls-Universit{\"a}t T{\"u}bingen, Germany.
}\looseness=-1}
\author{T. Haug}
\affiliation{\scriptsize{Centre for Quantum Technologies, National University of Singapore,
3 Science Drive 2, 117543 Singapore}\looseness=-1}
\author{C. Hufnagel}
\affiliation{\scriptsize{Centre for Quantum Technologies, National University of Singapore, 3 Science Drive 2, 117543 Singapore}\looseness=-1}
 \author{M. Keil}%
\affiliation{ 
\scriptsize{Department of Physics, Ben-Gurion University of the Negev, Beer Sheva 84105, Israel.
}}
\author{P. Ireland}
\affiliation{\scriptsize{SUPA, School of Physics and Astronomy, University of St. Andrews, North Haugh, St. Andrews KY16 9SS, UK}\looseness=-1}
\author{M. Lebrat}
\affiliation{\scriptsize{Institute for Quantum Electronics, ETH Zürich, 8093 Zürich, Switzerland and Department of Physics, Harvard University, Cambridge, MA 02138, USA}\looseness=-1}
\author{W. Li}
\affiliation{\scriptsize{Centre for Quantum Technologies, National University of Singapore,
3 Science Drive 2, 117543 Singapore}\looseness=-1}
 \affiliation{\scriptsize{Department of Physics, National University of Singapore, Singapore 117542}\looseness=-1}

\author{L. Longchambon}
\affiliation{\scriptsize{Laboratoire de physique des lasers, CNRS UMR 7538, Universit\'e Paris 13 Sorbonne Paris Cit\'e, 99 avenue J.-B. Cl\'ement, F-93430 Villetaneuse}\looseness=-1}
\author{J. Mompart}
\affiliation{\scriptsize{Departament de F\'isica, Universitat Aut\`onoma de Barcelona, E-08193 Bellaterra, Spain}}

\author{O. Morsch}
\affiliation{\scriptsize{CNR-INO and Dipartimento di Fisica 'Enrico Fermi', Largo Bruno Pontecorvo 3, 56127 Pisa, Italy}\looseness=-1}
\author{P. Naldesi}
\affiliation{\scriptsize{Universit\'{e} Grenoble-Alpes, LPMMC $\&$ CNRS, LPMMC, F-38000 Grenoble, France}}
\affiliation{\scriptsize{LANEF \textit{'Chaire d'excellence'}, Universit\`e Grenoble-Alpes $\&$ CNRS, F-38000 Grenoble, France}\looseness=-1}
\author{T.W. Neely}
\affiliation{ \scriptsize{ARC Centre of Excellence for Engineered Quantum Systems, School of Mathematics and Physics, University of Queensland, Brisbane, QLD 4072, Australia
}\looseness=-1}
\author{M. Olshanii}
\affiliation{\scriptsize{Department of Physics, University of Massachusetts Boston, Boston, MA 02125, USA}}
\author{E. Orignac}
\affiliation{\scriptsize{Univ Lyon, Ens de Lyon, Univ Claude Bernard, CNRS, Laboratoire de Physique, F-69342 Lyon, France}\looseness=-1}
\author{S. Pandey}
\affiliation{{\scriptsize{Institute of Electronic Structure and Laser, Foundation for Research and Technology - Hellas, Heraklion, 70013, Greece}}\looseness=-1}

\author{A.~P\'erez-Obiol}
\affiliation{\scriptsize{Laboratory of Physics, Kochi University of Technology, Tosa Yamada, Kochi 782-8502, Japan}\looseness=-1}
\author{H. Perrin}
\affiliation{\scriptsize{Laboratoire de physique des lasers, CNRS UMR 7538, Universit\'e Paris 13 Sorbonne Paris Cit\'e, 99 avenue J.-B. Cl\'ement, F-93430 Villetaneuse}\looseness=-1}
\author{L. Piroli}
\affiliation{\scriptsize{Max-Planck-Institut f\"ur Quantenoptik,
  Hans-Kopfermann-Str. 1, 85748 Garching, Germany}\looseness=-1}
%
\author{J.~Polo}
\affiliation{\scriptsize{Quantum Systems Unit, Okinawa Institute of Science and Technology Graduate University, Onna, Okinawa 904-0495, Japan}\looseness=-1}
\author{A.L. Pritchard}
\affiliation{ \scriptsize{ARC Centre of Excellence for Engineered Quantum Systems, School of Mathematics and Physics, University of Queensland, Brisbane, QLD 4072, Australia
}\looseness=-1}
\author{N.~P.~Proukakis}
\affiliation{\scriptsize{Joint Quantum Centre (JQC) Durham-Newcastle, School of Mathematics, Statistics and Physics, Newcastle University, Newcastle upon Tyne, NE1 7RU, U.K.}\looseness=-1}
\author{C. Rylands}
\affiliation{\scriptsize{Joint Quantum Institute and Condensed Matter Theory Center, Department of Physics,
University of Maryland, College Park, Maryland 20742-4111}\looseness=-1}
\author{H. Rubinsztein-Dunlop}
\affiliation{ \scriptsize{ARC Centre of Excellence for Engineered Quantum Systems, School of Mathematics and Physics, University of Queensland, Brisbane, QLD 4072, Australia
}\looseness=-1}
\author{F. Scazza}
\affiliation{\scriptsize{CNR-INO and LENS, Universit\`a di Firenze, 50019 Sesto Fiorentino, Italy}}
\author{S. Stringari}
\affiliation{\scriptsize{CNR-INO BEC Center and Dipartimento di Fisica, Universit\`a di Trento, 38123 Povo, Italy}\looseness=-1}
\author{F. Tosto}
\affiliation{\scriptsize{Centre for Quantum Technologies, National University of Singapore, 3 Science Drive 2, 117543 Singapore}\looseness=-1}
\author{A. Trombettoni}
\affiliation{\scriptsize{Department of Physics, University of Trieste, Strada Costiera 11, I-34151, Trieste,Italy}\looseness=-1}
\affiliation{\scriptsize{CNR-IOM DEMOCRITOS Simulation Center and SISSA, Via Bonomea 265, I-34136 Trieste, Italy}\looseness=-1}

\author{N. Victorin}
\affiliation{\scriptsize{Universit\'{e} Grenoble-Alpes, LPMMC $\&$ CNRS, LPMMC, F-38000 Grenoble, France}}
\author{W. von Klitzing}
\affiliation{{\scriptsize{Institute of Electronic Structure and Laser, Foundation for Research and Technology - Hellas, Heraklion, 70013, Greece}}\looseness=-1}
\author{D. Wilkowski}
\affiliation{\scriptsize{MajuLab, CNRS-UCA-SU-NUS-NTU International Joint Research Unit, Singapore}}
\affiliation{\scriptsize{Centre for Quantum Technologies, National University of Singapore, 3 Science Drive 2, 117543 Singapore}\looseness=-1}
\affiliation{\scriptsize{Centre for Disruptive Photonic Technologies, The Photonics Institute, Nanyang Technological University, 637371 Singapore}\looseness=-1}
\affiliation{\scriptsize{Division of Physics and Applied Physics, Nanyang Technological University, 21 Nanyang Link, 637371 Singapore}\looseness=-1}
\author{K.~Xhani}
\affiliation{\scriptsize{Joint Quantum Centre (JQC) Durham-Newcastle, School of Mathematics, Statistics and Physics, Newcastle University, Newcastle upon Tyne, NE1 7RU, U.K.}\looseness=-1}
\affiliation{\scriptsize{European Laboratory for Non-Linear Spectroscopy (LENS), Universit\`{a} di Firenze, 50019 Sesto Fiorentino, Italy}\looseness=-1}
\author{A. Yakimenko}
\affiliation{\scriptsize{Department of Physics, Taras Shevchenko National University of Kyiv, Ukraine}\looseness=-1}

             

\begin{small}
\begin{abstract}

\vspace{-0.4cm}

{\bf Abstract.}
Atomtronics deals  with  matter-wave circuits of ultra-cold atoms manipulated through magnetic or laser-generated guides with different shapes and intensities.  
In this way,  new types of quantum networks can be constructed, in which coherent fluids are controlled with the know-how developed in the atomic and molecular physics community.
 In particular,  quantum devices with enhanced precision, control and flexibility of their operating conditions can be accessed.  Concomitantly,  new  quantum simulators and emulators harnessing on the coherent current flows can  also be developed.  Here, we survey  the landscape of atomtronics-enabled quantum technology  and draw  a roadmap  for the  field in the near future.  We review some of the latest progresses achieved in matter-wave circuits design and atom-chips. Atomtronic networks  are deployed  as promising platforms  for probing many-body physics with a new angle and a new twist. The latter can be done both at the level of  equilibrium and non-equilibrium situations.  Numerous relevant  problems in mesoscopic physics, like persistent currents and quantum transport in circuits of fermionic or bosonic atoms, are studied  through a new lens.  We summarize some of the atomtronics quantum devices and sensors. Finally, we discuss  alkali-earth  and  Rydberg atoms as  potential platforms for the realization of atomtronic circuits with special features.   
 \end{abstract}
\end{small}
\maketitle

\end{titlepage}



\tableofcontents

 \setcounter{tocdepth}{1}
%

\section{INTRODUCTION}

Quantum technologies are enabling important innovations in the 21st century, with applications in areas as diverse as computation, simulation, sensing, and communication. At the core of this new technological development is the ability to control quantum systems all the way from the macroscopic scale down to the single quantum level.  The latter has been achieved in physical systems ranging from atomic and spin systems to artificial atoms in the form of superconducting circuits\cite{dowling2003quantum,acin2018quantum}.    

{In this article, we mostly focus on cold atom systems, where recent technological developments have delivered a collection of magnetic or laser-generated networks and guides in which atomic matter-waves can be controlled and manipulated coherently\cite{rubinsztein2016roadmap,dumke2016roadmap}.} \emph{Atomtronics} exploits the state-of-the-art in this field to realize matter-wave circuits of ultra-cold atoms\cite{dumke2016roadmap,amico2017focus}. Some key aspects of this emerging field give atomtronic circuits great promise as a quantum technology. {First, since atomtronic circuits employ matter-waves of neutral atoms, spurious  circuit-environment interactions, that might, e.g., lead to decoherence, are expected to be less serious than in networks employing  electrically charged fluids sensitive to Coulomb forces. }  
Second,  atomtronic networks can realize new types of circuits with current carriers having bosonic and/or fermionic quantum statistics, along with tunable particle-particle interactions ranging from short-range to long-distance, and from attractive to repulsive. Third, recent progress in the manipulation of optical guiding potentials enables engineering of time-dependent circuits whose topology can be reconfigured while they operate\cite{schnelle2008versatile,henderson2009experimental,haase2017versatile,gauthier2016direct,rubinsztein-dunlop2016roadmap}. 

The name \emph{Atomtronics} is inspired by the analogy between circuits with ultracold atomic currents and those formed by electron-based networks of conductors, semiconductors or superconductors. {For example, a Bose-Einstein Condensate (BEC) confined in a linear optical lattice with a suitable abrupt variation of the particle density can exhibit behaviour very similar to that of an electronic diode\cite{seaman2005effect,pepino2010open}.}  
As another example,  a BEC in suitable optical ring trap is the atomic counterpart of the superconducting SQUID of quantum electronics \cite{ryu_experimental_2013,aghamalyan_atomtronic_2016,aghamalyan_coherent_2015}, displaying the SQUID's defining characteristics of quantum interference\cite{ryu2020quantum} and hysteresis\cite{eckel2014hysteresis}. {It is important to note that because atomtronics is based entirely on flexible potential landscapes and not limited to material properties, it is expected to be possible to create quantum devices and simulators with new architectures and functionalities that have no analog in conventional electronics. }

The quantum nature of ultracold atoms as coherent matter waves enables interferometric precision measurements and new platforms for quantum information processing with applications in fundamental science and technology \cite{birkl2007micro,schaff2014interferometry}. {At the same time, atomtronic circuits can serve as powerful probes of many-body quantum regimes: analogous to solid state I-V characteristics, many-body cold atom systems can be probed by monitoring the current flowing in them while changes are made to external parameters and applied (effective) fields. In this way, atomtronic platforms can be thought of as extensions to the scope of conventional quantum simulators, revisiting textbook scenarios in many-body physics such as frustration effects, topological constraints, edge state formation etc., with the advantages of tunable boundary conditions and minimal finite size effects. }
Another interesting domain in which atomtronics can play an important role is mesoscopic physics\cite{kulik2012quantum,fazio2003new,nazarov2012quantum}. Important themes in the field of mesoscopic physics such as persistent currents in ring-shaped structures, problems of quantum coherent transport, etc. can be explored with a new twist.  

{For the implementation of the program sketched above, an important challenge to face in the years to come is to optimize the control of the matter wave currents in complex networks as, for example, optical lattices, guiding circuits for matter waves based on optical or magnetic fields, or cold atoms-solid state hybrid circuits. On one hand, such a step would be instrumental to harness current and transport for investigations on quantum many-body physics and artificial matter, both in static and dynamic conditions. In particular, Rydberg atoms and ultracold fermionic systems with SU($N$) symmetry provide novel interesting directions to go to. Experimental challenges for this goal are to design improved schemes for controlling the resulting matter-wave interactions and for including advanced schemes for their detection. On the other hand, the control of complex quantum networks would be opening the way to work out new types of devices based on integrated atomtronic circuits. In particular, new chips integrating different technologies, for example silicon-based electronics and the various atomtronics approaches, would provide a milestone in quantum technology.  Concerning potential applications, a certainly important direction pursued in the current research in \emph{Atomtronics} is devoted to interferometry and inertial sensing with enhanced performance, but quantum simulation and computation, as well as all other aspects of quantum technology are accessible.  In this context, stabilizing the atomic coherence on small-to-intermediate spatial scales, for example by smoothing the wave guides are important challenges to be solved in order to harness the full power of cold-matter-wave quantum technology.}      

In this review, we summarise recent activities in \emph{Atomtronics} and discuss the future of the field. In the first three chapters, we review fabrication principles for atomtronic platforms, ranging from reconfigurable optical potentials employing  acousto-optic deflectors, digital micro-mirror devices, and liquid-crystal spatial light modulators to micro-optical systems and hybrid solid state  - cold atom systems circuits where a scanning focused laser beam modifies the current density of a superconducting chip to create the desired trapping potential. These new capabilities open the way to  addressing the dynamics of many-body systems, as described in chapters \ref{NonEqDyn} and \ref{DynProt}. Chapters \ref{Persistent_toroid} and  \ref{PhaseSlips} deal with persistent currents in toroidal and ring-shaped condensates. {These systems, the simplest atomtronic circuits with a closed architecture, enable the study of basic questions in many-body physics in a variety of new and different conditions.} Atomtronic quantum sensors and devices are discussed in chapter \ref{DeviceSensors}. Ring-shaped bosonic circuits are investigated as ideal platforms for matter-wave SQUIDs (the Atomtronic QUantum Interference Device - AQUID) and flux qubits in chapter \ref{MQD}.  These studies have also touched upon a number of fundamental questions, such as macroscopic quantum coherence, the nature of superfluidity in restricted geometries, vortex dynamics, etc.  Transport in fermionic and bosonic circuits are discussed in chapters \ref{TransportFermi} and \ref{TransportBose} respectively. Chapter \ref{Ladder}  deals with bosonic ladders. In addition to their potential relevance to basic research in many-body physics, we envisage that they will be instrumental to the fabrication of coupled atomtronic circuits. In Chapter \ref{Solitons}, we discuss atomtronic circuits that exploit bright solitons both for studying fundamental questions in many-body quantum dynamics and for realizing quantum devices with enhanced performances. The final two chapters, \ref{alkaline}, \ref{Rydberg} deal with  alkali-earth atoms with $SU(N)$ symmetry, and  Rydberg atoms. To date, the latter have received little attention, but we believe that they offer great promise as an atomtronic quantum technology. 
 
The present article was inspired by the Atomtronics@Benasque conference series. The Benasque staff is warmly acknowledged for their invaluable help in the organization of these workshops and we thank the Benasque director Jose-Ignacio Latorre for his constant support of this line of research.

\section{DYNAMICALLY SCULPTED LIGHT}
\label{sculpted_light}
\vspace*{-0.5cm}
\par\noindent\rule{\columnwidth}{0.4pt}
{\bf{\small{ M. Baker, G. Gauthier, T.W. Neely, H. Rubinsztein-Dunlop, F. Tosto,  R. Dumke, P. Ireland, D. Cassettari}}}
\par\noindent\rule{\columnwidth}{0.4pt}
%
%
%
%
%
%
%
%

In recent years, many experiments have been carried out with cold neutral atoms in arbitrary, reconfigurable optical potentials. Single atoms have been trapped in arbitrarily-shaped arrays, \cite{barredo2016atom,barredo2018synthetic,demello2019defect,kim2016in,barredo2020three} which have subsequently led to the demonstration of topological phases of interacting bosons in one-dimensional lattices. \cite{deleseleuc2019observation} Various configurations of atomtronic circuits have been demonstrated, namely closed waveguides and Y-junctions, \cite{ryu2015integrated} oscillator circuits, \cite{gauthier2019quantitative} atomtronic transistors, \cite{caliga2016transport} rings and atomtronic SQUIDs (AQUIDs).\cite{ryu2013experimental,eckel2018rapidly,bell2016bose} Reconfigurable optical potentials have also been used to realise Josephson junctions in rubidium condensates \cite{ryu2013experimental} and in fermionic lithium superfluids in the BCS-BEC crossover.\cite{kwon2020strongly} {They have even been used for the optimization of the rapid cooling to quantum degeneracy.~\cite{PhysRevA.93.043403}} Finally, another area of interest is the realisation and study of quantum gases in uniform potentials. \cite{gaunt2013bose,saint-jalm2019dynamical} Some of these experiments are described in detail in the subsequent chapters of this review. 

{Static holographic potentials, as opposed to reconfigurable, also play an important role in atomtronics and have been implemented  with  great  success.~\cite{boiron1998cold, newell2003dense, bakr2009quantum, tempone2017high,eckel2016contact, scherer2007vortex} In particular, static holograms can provide substantial advantages for the generation of Laguerre-Gaussian and higher order Hermite-Gaussian modes.~\cite{meyrath2005high, campbell2012generation, tempone2017high} Static hologram techniques such as optical nano-fiber evanescent wave trapping,~\cite{PhysRevLett.104.203603} structured nano-surfaces to create trapping potentials,~\cite{Mildner_2018, PhysRevA.98.023429} and the use of engineered quantum forces~\cite{chang_trapping_2014} (a.k.a. London-van der Waals or Casimir forces) are promising emerging technologies that will benefit the field of atomtronics. However this section focuses on recently-adopted \emph{dynamic} technologies which have opened new avenues of research.}

More generally, we note that sculpted light has many more applications beyond cold atom physics, e.g. to microscopy, optical tweezers and quantum information processing with photonic systems.\cite{rubinsztein2016roadmap} In this chapter, we {review} the tools and techniques that underpin all these experiments: scanning acousto-optic deflectors (AODs), digital micromirror devices (DMDs), and liquid-crystal spatial light modulators (SLMs).

\subsection{\label{AOD} Fast-scanning AOD\lowercase{s} }
By rapidly scanning a trapping laser beam much faster than the trapping frequencies for the atoms, the atoms experience the time-average of the optical potential. Under these conditions, despite the modulated scanning action of the beam, the density of the atom cloud remains constant in time. The spatial location of the beam can be scanned in arbitrary 2D patterns, ``painting” the potential landscape, simply by modulating the RF frequencies driving the crystal. \cite{schnelle08versatile,henderson2009experimental} Control over the RF power at each scan location allows local control over the potential depth. This feature can be used to error-correct, ensuring smooth homogeneous potentials, or can be deliberately engineered to implement barriers, wells, or gradients in the trap. The trapping geometry can be dynamically changed, with the use of deep-memory arbitrary waveform generators, {or field programmable gate array (FPGA) technology which combined with non-destructive measurement allows for real-time correction of the potential}. Given the weak axial confinement provided by the scanned beam, this is best used in conjunction with an orthogonal light sheet, which provides tight confinement along the axis of the scanned beam, and ensuring  excitation and phase fluctuations in the axial dimension are minimised.\cite{henderson2009experimental}

\subsubsection{Feed-forward control}

The diffraction efficiency of AODs can change with the drive frequency. In order to correct for this, it is generally necessary to use feed-forward to compensate by adjusting the RF power of the AOD crystal, and hence beam intensity, for each (x,y) location. To correct for imperfections in other elements of the trapping potential, one can measure the atomic density distribution in the trap using absorption imaging and apply iterative correction to the RF power at each (x,y) location.\cite{bell2016bose}

\subsubsection{Phase evolution in time-averaged potentials}

A full treatment time-averaged potentials needs to include the phase evolution of the condensate under the effect of the scanning beam. The time-varying potential $V(x,y,t)$ acts to imprint a phase $\phi$, with the evolution $\hbar  {\partial \phi(x,y,t)}/{\partial t}=V(x,y,t)$. For sufficiently fast scan rates, {the imprinted phase effect is negligible}, but at slower scan rates, this phase imprinting action can accumulate local phase, leading to residual micromotion in the condensate, the signatures of which have been observed.\cite{bell2018phase} This is an important consideration for atomtronic applications where the phase is an observable of interest, such as {for} guided Sagnac interferometry.\cite{lenef1997rotation}

\subsubsection{Atomtronics with time-averaged optical traps}

\begin{figure}[htbp]
\includegraphics[width=0.4\textwidth]{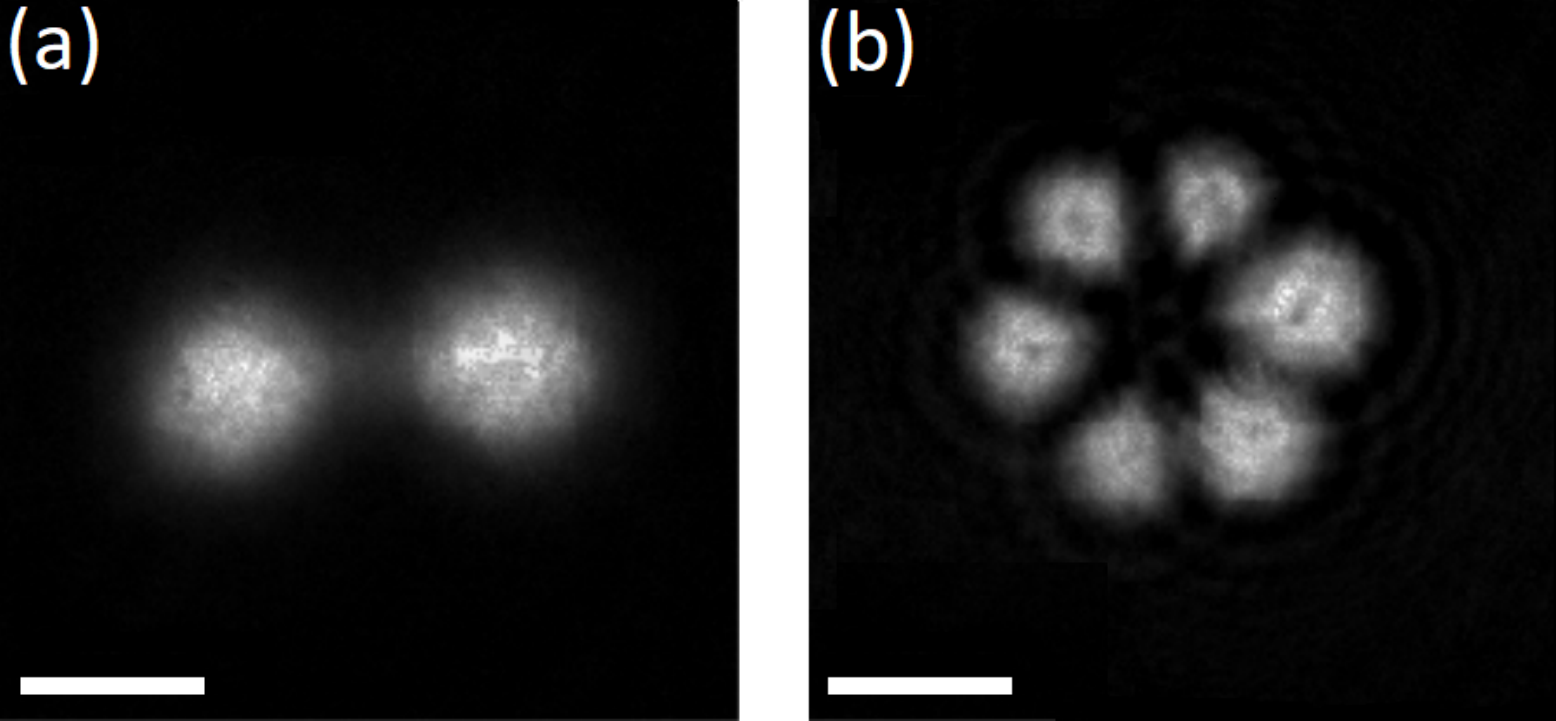}
\caption{Example geometries of time averaged optical dipole potentials -  (a) BEC trapped in dumbbell potential, with two reservoirs connected through a channel of tunable length and width (b) Ring lattice of BECs. The scale bar on each image indicates 50 $\mu$m.}
\label{fig:TAOP_1}
\end{figure}

The time-averaged optical dipole traps are extremely versatile, allowing a variety of geometries to be generated, and dynamically changed in structure by real-time adjustment of the scanning pattern. In the context of atomtronic geometries, BECs have been trapped into flat bottom line-traps, rings,\cite{aycock2017brownian,bell2016bose} lattices, \cite{chisholm2018three} and dumbbell reservoirs (Fig.~\ref{fig:TAOP_1}). Additionally, single mode matter-wave propagation and coherent phase splitting has been demonstrated in circuit elements such as waveguides and beamsplitters.\cite{ryu2015integrated} The time-averaged optical beams can be used to introduce multiple repulsive barriers and stirring elements to study persistent currents and superfluid transport in atomtronic circuits.\cite{ryu2013experimental}

\subsection{\label{SLM}Optical potentials with liquid-crystal SLM\lowercase{s}}

A liquid-crystal SLM spatially modulates the phase of the light. The phase pattern on the SLM acts as a generalised diffraction grating, so that in the far field an intensity pattern is formed, which is used to trap atoms. In practice, the far field is obtained by focusing the light with a lens, so that the intensity pattern that traps the atoms is created in a well-defined "output plane" coinciding with the lens focal plane. The SLM acts effectively as a computer-generated hologram, and the light field in the output plane is the Fourier transform of the light field in the SLM plane.

The first experiments with these holographic traps go back over ten years ago. \cite{bergamini2004holographic,boyer2006dynamic} A reason for the use of phase-only SLMs, rather than amplitude modulators, is that the former do not remove light from the incident beam. This is advantageous from the point of view of light-utilisation efficiency. Moreover, as is shown below, a phase-only SLM allows the control of both the amplitude and phase on the output plane.

The calculation of the appropriate phase modulation to give the required output field is a well-known inverse problem which, in general, requires numerical solution. Iterative Fourier Transform Algorithms (IFTAs) are commonly used, and variants which control both phase and amplitude have been recently demonstrated.~\cite{tao2015beam,wu2015simultaneous} {The  removal  of  the  singularities (e.g. vortices)  that  particular pattern optimization techniques  can  introduce  is  widely  researched  due  to  their importance for controlled beam shaping~\cite{aagedal1996theory, senthilkumaran2005vortex, gaunt2012robust, Bowman2017May} and in particular to confine BEC  in  uniform  potentials.~\cite{gaunt_bose_2013, navon_emergence_2016} One such example} is a conjugate gradient minimisation technique which efficiently minimises a specified cost function.\cite{harte2014conjugate,bowman2017high} The cost function can be defined to reflect the requirements of the chosen light pattern, such as removing optical vortices from the region of interest. 

\begin{figure}[htbp]
\includegraphics[width=0.48\textwidth]{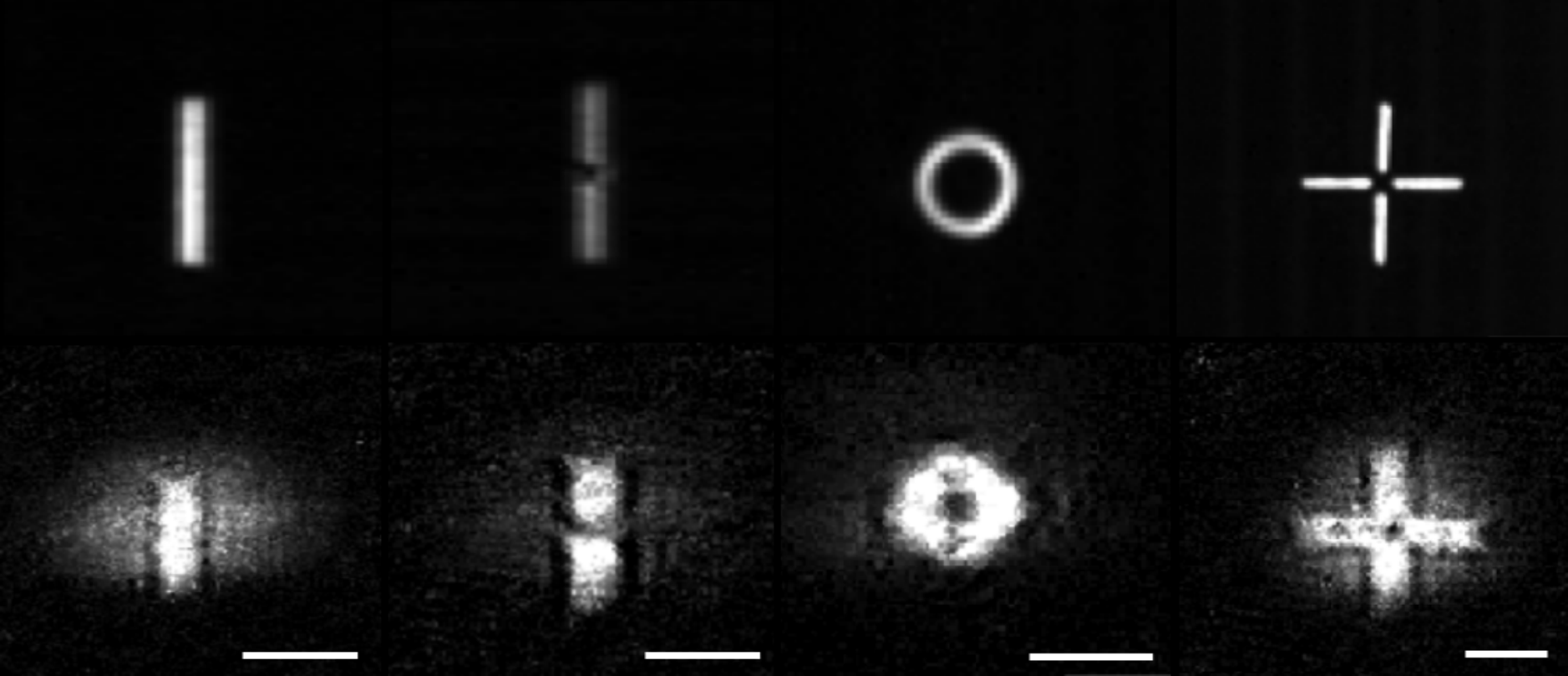}
\caption{\label{fig:SLM}$^{87}$Rb condensates in SLM-generated optical potentials. The top row shows the intensity patterns used for trapping and the bottom row shows the condensates after a 2 ms time-of-flight. The ring trap contains $\sim 10^6$ atoms, while the other traps contain $\sim 5\times 10^5$ atoms. The scale bar on the images indicates 100 $\mu$m.}
\end{figure}

The intensity patterns obtained with this method are shown in the first row of Fig. \ref{fig:SLM}. They are taken at a wavelength of 1064~nm, i.e. red-detuned relative to the rubidium transition, causing rubidium atoms to be trapped in the regions of high intensity. The SLM light is focused on the atoms by a f~=~40~mm lens, giving a diffraction limit of the optical system of 6~$\mu$m at 1064~nm.

Going from left to right in Fig. \ref{fig:SLM}, {shown are} a simple waveguide, a waveguide with a potential barrier halfway across, a ring trap, and a cross-like pattern. The latter has been proposed for the study of the topological Kondo effect. \cite{buccheri2016holographic} In all these light patterns, the phase is constrained by the algorithm. For the simple waveguide, the ring and the cross, a flat phase {is programmed} across the whole pattern. {Controlling the phase this way leads to a well maintained intensity profile shape as it propagates} out of the focal plane for up to $\sim10$ times the Rayleigh range. By comparison, a pattern with random phase loses its shape much sooner. 

Differently from the other three patterns, for the waveguide with the barrier a sharp $\pi$ phase change halfway across  the  line {was programmed}.  In  the  resulting  intensity  profile,  this  phase discontinuity causes the intensity to vanish hence creating the potential barrier whose width is close to the diffraction limit. 

The second row of Fig. \ref{fig:SLM} shows Rb BECs trapped in the potential created by the SLM light patterns, combined with an orthogonal light sheet that provides tight confinement along the axis of propagation of the SLM light. \cite{PhilPhD,DavePhD} The clouds are imaged after a 2 ms time-of-flight and undergo mean-field expansion during this time, leading to a final density distribution that is more spread out compared to the transverse size of the SLM traps.

Controlling  the  phase  of  the  light  pattern opens new possibilities for the trapping and manipulation of ultracold atoms. Here we have shown that phase control gives an alternative way to create barriers close to the diffraction limit by using discrete phase jumps. {Liquid-crystal SLMs were also used to
transfer phase structure in a four-wave-mixing process in rubidium vapour, in particular trans-spectral orbital angular momentum transfer from near-infrared pump light to blue light.~\cite{Walker2012} Additionally, they have enabled research into uniform 3D condensates.~\cite{gaunt_bose_2013} More recently, they were used in the realization of bottle  beams which have  been used  to  create  3D  optical  trapping  potentials  for  confining Rydberg atoms.~\cite{barredo_three-dimensional_2020}} In addition to this, phase control can also be useful for many atomtronics applications, for instance phase imprint via a Raman transition, \cite{ramanathan2011superflow} and the realisation of artificial gauge fields.\cite{huo2015solenoidal,lembessis2015graphene}

\subsection{\label{DMD}Direct imaged DMD optical potentials}

\begin{figure}[htbp]
\includegraphics[width=0.4\textwidth]{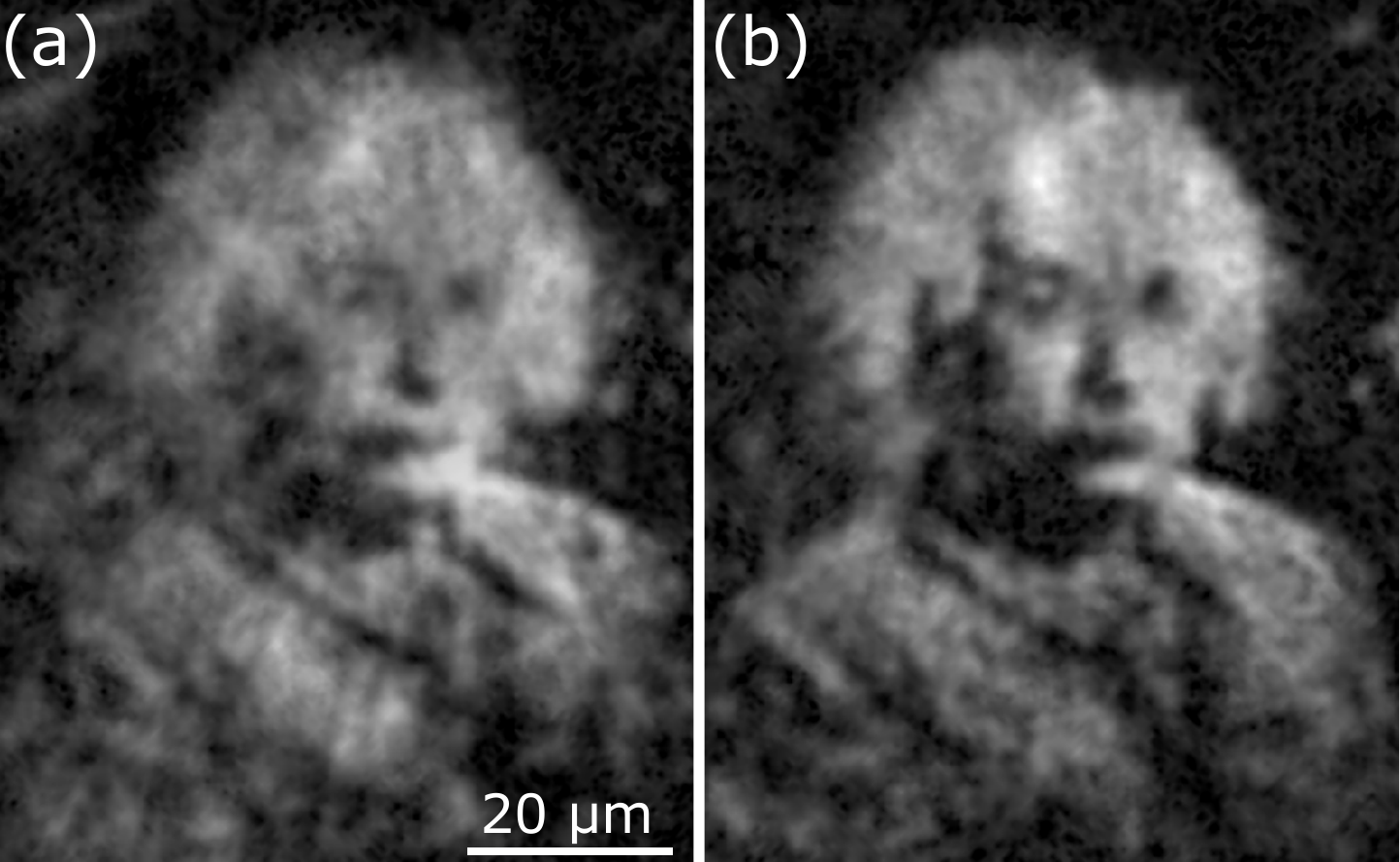}
\caption[Einstein]{Generating complex potentials using half-toning. (a) Initial in-situ image of the BEC density in a half-toned potential of Einstein, calculated using the optical system parameters. (b) Final converged BEC image after 11 feedback iterations where the atomic density is used to iteratively correct for imperfection in the density, {generated using the method described by \citet{tajik2019designing} and \citet{gauthier2019transport}}.}
\label{fig:DMD_Density_Images}
\end{figure}

A recent addition to the spatial light modulator family is the digital micromirror device (DMD). Developed for digital light processing (DLP) applications, DMDs consist of millions of individually addressable, highly reflective mirrors. Each hinged mirror, {of typical size $7.56$~-~$10.8$~$\mu$m}, is mounted on a silicon substrate on top of control electrodes. The application of a control voltage tilts the mirrors between two `on' or `off' angles, typically $\pm 12^{\circ}$. The mirror array acts as a dynamical configurable amplitude mask for light reflected from their surface.  The DMDs can be placed in the Fourier plane of the imaging/project system, similar to typical phase-based SLMs, where it can modulate both the phase and the amplitude of the  light.~\cite{zupancic2016ultra} If phase modulation is not required, the DMD can be used as a binary amplitude mask in the object plane, similar to its DLP applications.~\cite{gauthier2016direct,ville2017loading} In `DC' mode, the mirrors are fixed to the `on' angle and a static pattern can be projected. The true versatility of the device, however, lies in its dynamical (`AC') capability, with full frame refresh rates exceeding 20 kHz. 

\subsubsection{Half-toning and time-averaging}

The projected image from the DMD is binary in nature. Although this would appear as a significant limitation in producing arbitrary optical potentials, a number of techniques exist to overcome this issue. The first of these is half-toning, or error-diffusion, which takes advantage of the finite optical resolution of the projection optical system to increase the amplitude control. With suitably high magnification, such that the projected mirror size is smaller than the resolution, multiple mirrors contribute to each resolution spot in the projected plane.~\cite{liang20091} In this way, half-toning can be used to create intensity gradients in the light field, as shown in Fig.~\ref{fig:DMD_Density_Images}(a). Same as in the case of time-averaged AOD traps, feed-forward using the atomic density~\cite{bell2016bose,tajik2019designing,gauthier2019transport} can be performed to correct for imperfections in the projection potential, as shown in Fig.~\ref{fig:DMD_Density_Images}(b).\cite{gauthier2019transport}

One can also make use of the high-speed modulation of the mirrors to further improve the intensity control. The mirror array of the DMD is capable of switching speeds from DC to 20 kHz. By varying the on/off time of individual mirrors (pulse-width modulation), the time-average of the resulting light field can be utilised to improve the smoothness of the projected potentials.\cite{gauthier2019transport}  

\begin{figure}[htbp]
\includegraphics[width=0.45\textwidth]{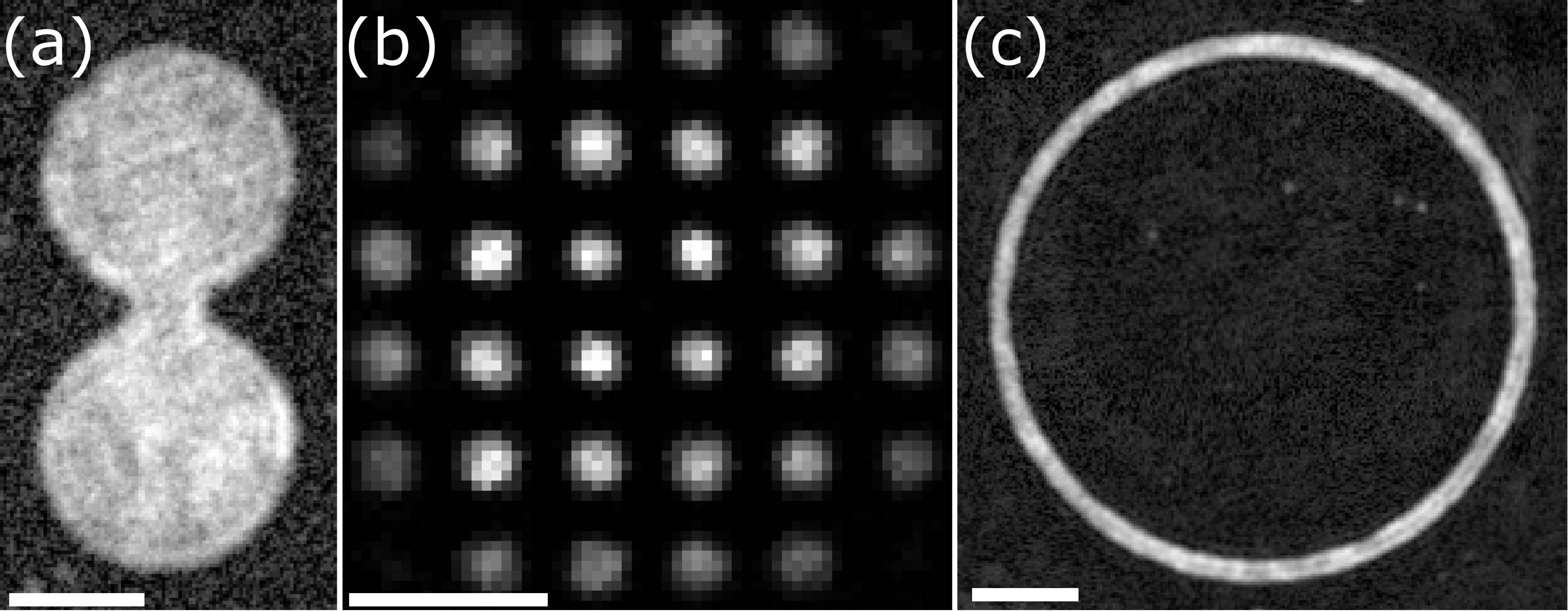}
\caption{Useful atomtronics geometries created with directly-imaged DMD trapping. (a) A dumbbell geometry of two reservoirs connected by a channel where the reservoir size, channel length and channel width can be varied to study superfluid transport. (b) Square lattice of BECs with 10~$\mu$m period formed using the projection of a half-toned DMD pattern. The lattice period can be dynamically increased or decreased. (c) A ring-shaped BEC with a 110~$\mu$m diameter and 10~$\mu$m radial width  useful for interferometry and transport experiments. The scale bar on each image indicates 20~$\mu$m.}
\label{fig:DMD_Atomtronics_Geometries}
\end{figure}

\subsubsection{Atomtronics with DMDs}

Atomtronics studies how to use neutral atom currents to create circuits that have properties similar to existing electrical devices. The advances in control and increased resolution of trapping potentials have been instrumental in the development of this field. The dynamic control over the potential given by DMDs have allowed time dependent implementations. Combined with other techniques such as the optical accordion lattice,\cite{ville2017loading} which allows smooth transitions between quasi-2D and 3D systems, they open up further avenues of control for future studies. The high resolution projection of DMD optical potentials enables the creation of complex masks. These have facilitated the study of superfluid transport in a variety of traps. {Figure~\ref{fig:DMD_Atomtronics_Geometries} shows three relevant geometries for superfluid transport studies.}

\subsubsection{Turbulence with DMDs}

\begin{figure}[htbp]
\includegraphics[width=0.45\textwidth]{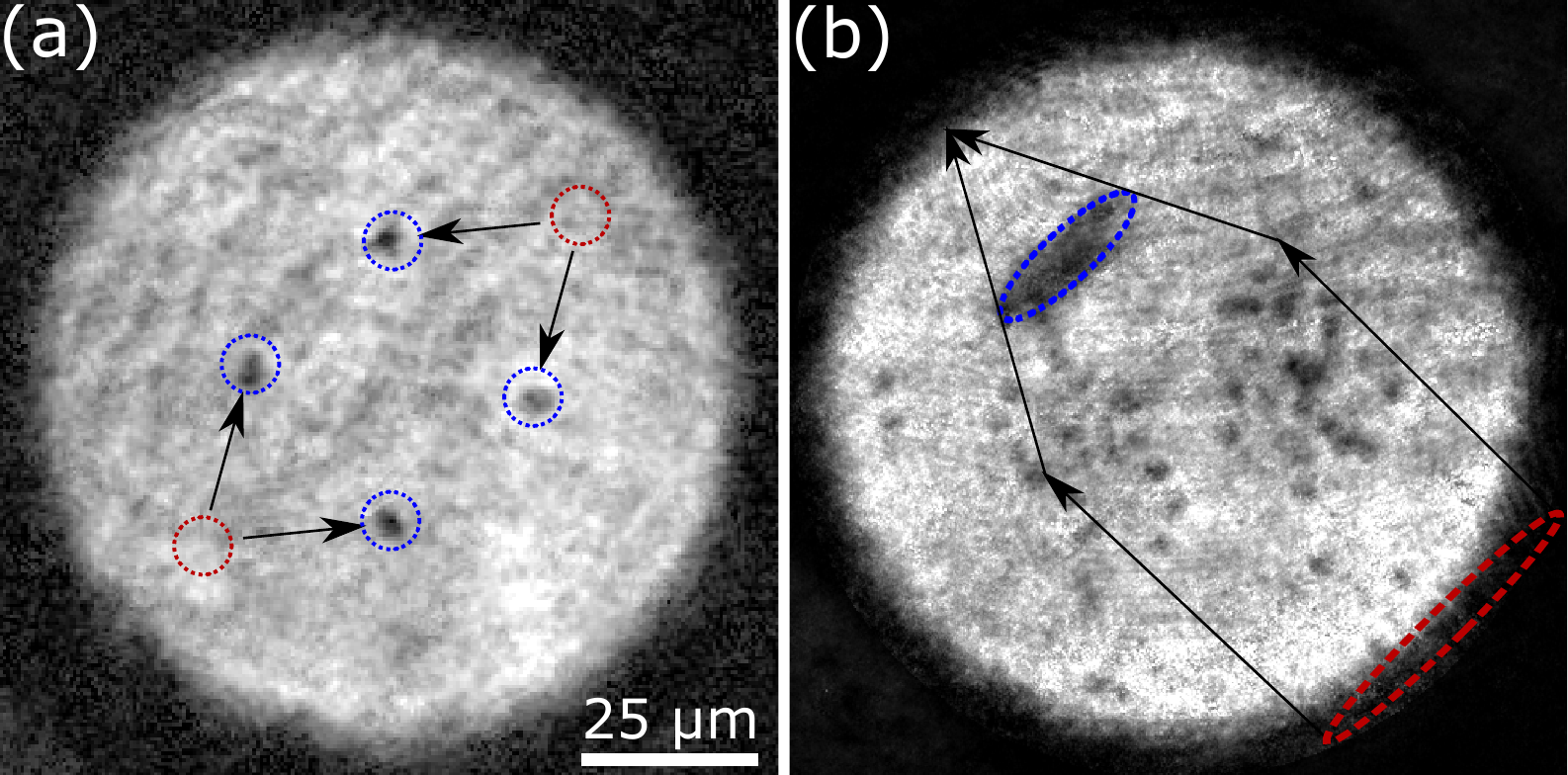}
\caption{Creating vortices using the dynamical capabilities of DMDs. (a) Deterministic creation of vortices using the `chopstick' method outlined in~\protect\citet{samson2016deterministic}. The dashed red (blue) circle represent the initial (final) position of the optical barriers, with arrows indicating their trajectories. After a short 3~ms time-of-flight, the vortices are visible as density dips (black dots).  (b) Creation of a vortex cluster, {similar to procedure used in~\protect\citet{gauthier2019giant} and~\protect\citet{johnstone2019evolution},} by sweeping a paddle-shaped optical barrier through a condensate. The red (blue) dashed ovals represent the paddle position at initial (picture) time with the arrows indicating the trajectory of the edges of the paddle.}
\label{fig:DMD_Vortex_Generation}
\end{figure}

The dynamic properties of DMDs can be used for the creation of turbulence. As shown in Fig.~\ref{fig:DMD_Vortex_Generation}, these techniques have been used to study Onsager vortices and their emergence in superfluids,\cite{gauthier2019giant,johnstone2019evolution} the creation of tunable velocity solitons,\cite{fritsch2020creating} and equilibration of chiral vortex clusters.\cite{stockdale_universal_2020}

\subsection{\label{hybridA}Hybrid atomic-superconducting quantum systems}

Superconducting (SC) atom chips have significant advantages in realizing trapping structures for ultracold atoms compared to conventional atom chips. \cite{dumke2016roadmap,skagerstam2006spin,hohenester2007spin,fermani2010heating,scheel2005atomic,dikovsky2009superconducting,emmert2009measurement} {These advantages have been extended further by the development of} the ability to dynamically tailor the superconducting trap architecture. {This is done by modifying the current density distribution in SC film through the local heating of the film using dynamically shaped optical fields. This allows for the creation of desired magnetic trapping potentials without having to change the chip or the applied electrical field.}

\begin{figure}[htbp]
\includegraphics[width=0.45\textwidth]{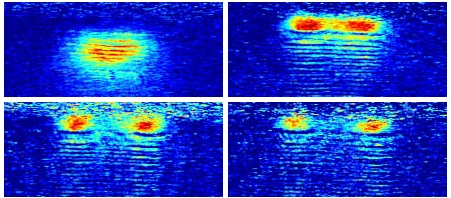}
\caption{Absorption images of the atomic cloud splitting taken after $0$ (top-left), $20$ (top-right), $30$ (bottom-left) and $40$ (bottom-right) ms of illumination time respectively.}
\label{fig:splitting}
\end{figure}

\begin{figure}[htbp]
\includegraphics[width=0.23\textwidth]{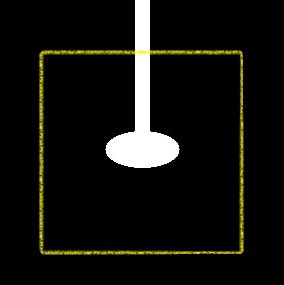}
\includegraphics[width=0.23\textwidth]{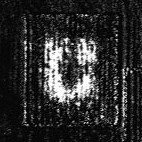}
\caption{DMD image on the left, where the real dimension of the SC film is highlighted in yellow and absorption image of the atomic cloud on the right.}
\label{fig:ring}
\end{figure}

{Typically,} a high-power laser and a DMD {are used to create and shape the light field used} to destroy the superconductivity and influence the shape and structure of a trap. Various trapping potentials {have been realized using this technique}, in particular, to split a single trap (see Fig. \ref{fig:splitting}) or to transform it into a crescent or a ring-like trap (see Fig. \ref{fig:ring}). Since the atomic cloud evolves with the trapping potential, cold atoms can be used as a sensitive probe to examine the real-time magnetic field and vortex distribution. Simulations of the film heating, the corresponding redistribution of sheet current density, and the induced trapping potentials {have been found to agree closely with} experiments. Such simulations help to better understand the process and can be used to design {traps} with the needed properties.

More complex structures can be achieved by increasing the heating pattern resolution. This method can be used to create magnetic trap lattices for ultracold atoms in quantum computing applications and, in particular, optically manipulated SC chips open new possibilities for ultracold atoms trapping and design of compact on-chip devices for investigation of quantum processes and applications in atomtronics. \cite{bensky2011controlling,thiele2014manipulating,hermann2014long,sarkany2015long,yu2016superconducting,yu2018stabilizing,petrosyan2019microwave}

\subsection{Concluding remarks and outlook}
{In this section we have described the suite of technologies available to the experimenter for creating configurable optical potentials for ultracold atoms, primarily discussing AOMs, SLMs and DMDs. Hybrid technique utilizing optical configurable potentials to shape magnetic potentials through superconducting quantum chips intermediary were also discussed. These technologies have drastically improved the control and manipulation of ultracold neutral atoms. 

Although previously available static holograms technologies provided great control for the creation of optical potentials and are still usually better for 3D trapping potentials, the dynamic manipulation capabilities presented here have enabled new classes of experiments with ultracold atoms. For example, dynamically modulated DMDs have facilitated new studies of two-dimensional-quantum turbulence~\cite{gauthier2019giant,johnstone2019evolution,stockdale_universal_2020} and condensate evolution in response to rapidly quenched trapping potentials.~\cite{eckel2018rapidly,aidelsburger_relaxation_2017} AOMs have enabled steerable arrays of single atoms,~\cite{doi:10.1063/1.5041481,endres2016atom,barredo_synthetic_2018} facilitating quantum simulation experiments. Furthermore, the rapid reconfigurability of DMD traps has enabled groundbreaking studies in the emerging field of atomtronics, where the system parameters can be easily tuned.~\cite{hausler2017scanning,fritsch2020creating,gauthier2019quantitative,kwon2020strongly,luick_ideal_2020}

As the technology behind optical manipulation continues to mature and evolve through increases in SLMs pixel array size and switching frequencies, these sculpted light and hybrid techniques are sure to have an even bigger impact on the development of atomtronics.}

{\it Acknowledgments} The UQ group has been funded by the ARC Centre of Excellence for Engineered Quantum Systems (project number CE1101013), and ARC Discovery Projects grant DP160102085. G. G. acknowledges support of ARC Discovery Project number DP200102239 and T.W.N. acknowledges the support of ARC Future Fellowship FT190100306. The St Andrews group acknowledges funding from the Leverhulme Trust (RPG-2013-074) and from EPSRC (EP/G03673X/1; EP/L015110/1). 


%
%

\section{IMPLEMENTING  RING CONDENSATES}
\label{ringCondensates}
\vspace*{-0.5cm}
\par\noindent\rule{\columnwidth}{0.4pt}
{\bf{\small{M. Baker, T.A. Bell, T.W. Neely, A.L. Pritchard, G. Birkl, H. Perrin, L. Longchambon, M.G. Boshier,
B.M. Garraway, S. Pandey, and W. von Klitzing}}}
\par\noindent\rule{\columnwidth}{0.4pt}
%


The many interesting properties of degenerate quantum gases, such as
phase coherence, superfluidity, and vortices naturally make the geometry of
these systems of great interest.
Ring systems are of particular interest, as the simplest multiply
connected geometry for coherent matter-wave guiding and as a potential
building block for circuital atomtronic devices.
In addition, ring systems have interesting properties such as persistent flow, quantum hall states and the potential for Sagnac interferometry.

Advances in the control of quantum gases have seen the development of atom 
waveguides formed from both magnetic trapping and magnetic resonance, and optical dipole trapping, and more recent implementations using hybrids of both. 
These approaches satisfy the criteria needed for coherent quantum matter-wave flow: i.e.\ the waveguides are \emph{smooth} and can form \emph{loops} that are \emph{dynamically} controllable.

{
\subsection{General features of ring traps}

Irrespective of the mechanism of trapping, magnetic or optical, some common parameters for ring traps can be described. We will restrict our discussion to ring traps that can be considered approximately harmonic; in cylindrical co-ordinates, the ring potential with radius $R$ is expressed in terms of radial and vertical trapping frequencies $\omega_\rho$ and $\omega_z$ respectively:

\begin{equation}\label{eq:potential-ring}
V(\rho,z)=\frac{1}{2}m\omega_\rho^2(\rho-R)^2+\frac{1}{2}m{\omega_z}^2z^2.
\end{equation}

Considering now a trapped gas within this potential, the connected geometry of the ring trap results in modifications to the usual derivation for the condensate critical temperature $T_c$ for a 3D harmonically trapped gas, yielding \cite{mathey2010phase}

\begin{equation}\label{eq:ring-Tc}
T_c=\left(\frac{\sqrt{2}N_0\hbar^3\omega_{\rho}\omega_{z}}{1.514 {k_b}^{5/2}m^{1/2}\pi R}\right)^{2/5}
\end{equation}

\noindent where $N_0$ is the atom number. For sufficiently elongated geometries, such as cigar traps, or ring traps with long azimuthal length, a regime of thermally driven phase-fluctuations in the condensate can exist \cite{ohberg1997low,petrov2001phase}, even at temperatures below $T_c$.  These phase-fluctuations are suppressed when the correlation length is larger than the system size, which for a ring geometry is half the azimuthal circumference, or $\pi R$. As we are typically interested in fully phase coherent ring traps, we can define this transition temperature $T_{\phi}$:~\cite{mathey2010phase}

\begin{equation}\label{eq:ring_Tphi}
T_{\phi}=\frac{\hbar^2N_0}{k_b m \pi R^2}.
\end{equation}

Finally, in the Thomas-Fermi approximation, where the interaction energy dominates,  the chemical potential in the ring trap can be expressed in terms of the trapping parameters\cite{Morizot2006}:

\begin{equation}\label{eq:chempot-ring}
\mu=\hbar\sqrt{\frac{2{N_0}\omega_{\rho}\omega_{z}a_s} {\pi R}},
\end{equation}

\noindent and $a_s$ the s-wave scattering length.\\}

In this chapter we will discuss the experimental and theoretical developments in all three types of waveguide approach. In what follows, in Section \ref{sec:magnetic-ring-traps} we discuss approaches primarily involving magnetic
and radio-frequency fields, and in Section \ref{sec:Optical-ring} we
we will discuss optical and hybrid approaches to implementing ultra-cold atoms and condensates in rings before concluding in Section \ref{sec:outlook-ring}.\\

\subsection{\label{sec:magnetic-ring-traps}Techniques based on magnetic traps}

Experimental techniques for trapping atoms in magnetic fields are well developed since the first BECs, and it is natural to consider such an approach, and build on that approach, to make ring waveguides.   
Nevertheless, this brings particular challenges because of the need to satisfy Maxwell's equations for fields trapping in a ring geometry, the need to avoid the loss of atoms from Majorana spin flips, occurring in the vicinity of field zeros, and the desire, for some experiments, to have trapping systems with high symmetry.

The earliest examples of waveguides for ultra-cold atoms were produced using static magnetic fields, where DC current carrying wires were used to create large area ring \cite{sauer2001storage} and stadium \cite{wu2004bidirectional} geometries which initially trapped thermal atoms. 
With Ref.\ \onlinecite{arnold2006large} we had the first demonstrations of a ring waveguide with a Bose-condensed gas.
Subsequent experimental developments can be divided into systems which principally use macroscopic coils for generating the magnetic trap, and those systems which employ microfabricated structures in an \emph{atom chip} to generate the spatially varying potentials. We will briefly discuss the latter next and the former in Sections \ref{sec:RF-Dressing}--\ref{sec:Dynring}.

 %
 The appeal of atom-chip traps is their compact footprint, potential portability, and the ability to fabricate quite complex geometries, switches, and antenna components into a compact package \cite{fortagh2007magnetic,keil2016fifteen}. Additionally, the close proximity of the wires allows high trapping frequencies to be achieved, even for modest currents. However, trapping in close proximity to a surface brings with it its own challenges. Foremost of these are the corrugations in the magnetic guiding potential that arise from imperfectly directed currents in the conducting material. An additional challenge is the perturbing effect of the end connections, to supply current in and out of the conducting ring. Although these problems can be alleviated to some degree by the use of AC fields \cite{trebbia2007roughness}, which provides a smooth time-averaged current in the wire, as well as switching elements at the end connections to minimise the perturbative phase effects on the ring condensate \cite{baker2009adjustable}, they cannot be removed completely. A comprehensive survey on the implementation of ring traps based on atom-chips, and their applications, is covered in detail in the recent review Ref.\  \onlinecite{garridoalzar2019compact}.

{Here we will focus our attention on ring traps derived from a combination of static magnetic traps, with RF and modulated fields. Using macroscopically large  conducting elements requires the use of high currents and occupies a greater size, but there are significant gains in the resulting trap smoothness, as the conducting elements are far from the trapping region. This makes such magnetic traps ideal for producing corrugation free toroidal waveguides for coherent matter, detailed in this section.}

However, the complexity of the fields requires an atom-chip approach to a pure magnetic waveguide system \cite{fortagh2007magnetic,keil2016fifteen} and this brings a difficult problem for the perfect ring waveguide because of the need to get the currents into, and out of, the wires that define the waveguide. We can try to live with this \cite{garridoalzar2019compact}, but asymmetry seems inevitable. We can think of tricks, for example, as the atoms go around the ring, we can switch the current between different sets of 130
 wires as in Ref. \onlinecite{baker2009adjustable}. This would avoid the bumps and humps in the waveguide which occur in the places where current enters and leaves the defining structures at the expense of potential losses and heating as the guides are switched over.




\subsubsection{\label{sec:RF-Dressing}RF dressing and bubbles}



%
It is not obvious that micron-scale trapping structures for ultra-cold atoms can be created using macroscopic scale magnetic coils.
However, by means of the addition of radio-frequency coils, magnetic traps with a simple trapping geometry can be transformed into ring traps and other topologies.
The theoretical basis is to treat the atom and 
radio-frequency field with adiabatic following and the dressed-atom theory
\cite{cohen1977dressed}. Originally introduced in the optical domain by Cohen-Tannoudji and
Reynaud in 1977 we adapt it here in the radio-frequency domain where it has found several
applications (see also sections \ref{sec:TAAP}, \ref{sec:Dynring} and \ref{sec:BubbleAndSheet}). 
The approach is suitable for ultra-cold atoms in magnetic traps where the trap potential is governed by the spatially varying Zeeman energy and the spatially varying energy difference between Zeeman levels can be in the radio-frequency range \cite{garraway2016recent,perrin2017trapping}.
The method relies on the adiabatic following of local eigenstates and it is notable that the superpositions of Zeeman states can provide some resilience to temporal noise and surface roughness \cite{trebbia2007roughness}.
The combination of static magnetic fields and radio-frequency fields with their different
spatial and vector variation allows a flexibility in the resulting potentials for the
creation of shell potentials, rings, tubes, and toroidal surfaces amongst
others\cite{garraway2016recent,perrin2017trapping}. 

As a simple example we can consider a simple spatially varying static field and a uniform radio-frequency field. 
A simple spatially varying magnetic field (obeying Maxwell's equations) is the quadrupole field
\begin{equation}
  \label{eq:B-Quad}
  \Bzero = b'( x   \, \hat{\mathbf{e}}_x 
             + y   \, \hat{\mathbf{e}}_y 
             - 2 z \, \hat{\mathbf{e}}_z     )\,,
\end{equation}
where $b'$ represents the gradient of the field in the $x$-$y$ plane. This field is often generated by a pair of coils with current circulating in opposite directions. When an atom interacts with this static field via its magnetic dipole moment $\boldsymbol{\mu}$ we obtain the ubiquitous interaction energy 
\begin{equation}
 U(\mathbf{r}) = - \bm{\mu}\cdot\mathbf{B}_0(\mathbf{r}) 
 \longrightarrow  m_F g_F \mu_B | \Bzero |\label{eq:B-Int}
 \,,
\end{equation}
responsible for magnetic potentials and the Zeeman energy splitting. The second form for $U(\mathbf{r})$ has the integer, or half integer $ m_F = -F \ldots F$ which arise from the quantisation of the energy, along with the
Land\'e $g$-factor $g_F$ and Bohr magneton $\mu_B$. For our example static field (\ref{eq:B-Quad}) the resulting potential is 
$
  U(\mathbf{r}) = m_F\hbar \alpha \sqrt{ x^2 + y^2 + 4 z^2} 
$
where $\alpha = g_F \mu_B b' / \hbar$. 

In the next step towards radio-frequency dressed potentials we add the RF field. The interaction is still given by 
Eq.~(\ref{eq:B-Int}), but with the replacement $ \Bzero \rightarrow \Bzero + \Brf$. The oscillating radio-frequency field $\Brf$ is, in general, off-resonant to the local Larmor frequency, or local Zeeman energy spacing  $  | g_F| \mu_B | \Bzero | $
and we define a spatially varying detuning of the RF field as
\begin{equation}
  \label{eq:detuning-delta}
  \delta(\mathbf{r}) 
      =  \omega_\text{rf} - \omega_L(\mathbf{r})
  \,.
\end{equation}
Those locations defined by $\delta(\mathbf{r}) \rightarrow 0 $ typically define a surface in space where RF resonance is found, and correspondingly there is a minimum in the interaction energy overall \cite{garraway2016recent,perrin2017trapping}. In the linear Zeeman regime the local Larmor frequency is given by 
\begin{equation}
  \label{eq:B-Larmor}
\omega_L(\mathbf{r}) = \frac{ | g_F| \mu_B | \Bzero | }{\hbar}  
\,,
\end{equation}
which is derived from the static potential $ U(\mathbf{r})$. The oscillating field $\Brf$ yields an interaction energy \cite{garraway2016recent,perrin2017trapping} in terms of a Rabi frequency $\Rzero$
\begin{equation}
  \label{eq:Rabi-def}
 \hbar \Rzero = \frac{g_F \mu_B}{2}  |\Bperp| 
 \,
\end{equation}
where the factor of two arises from the rotating wave approximation in the case of linear polarisation (more general polarisations are discussed in Ref.\ \onlinecite{perrin2017trapping}), and $\Bperp$ is the component of $\Brf$ perpendicular to the local static field $\Bzero$. Finally, by combining the energies (\ref{eq:B-Int}) and (\ref{eq:Rabi-def}) through diagonalisation of the Hamiltonian in a full treatment\cite{garraway2016recent,perrin2017trapping}, we obtain the local eigenenergies, or dressed potentials,
\begin{align}
  \label{eq:dressed-potentials}
  U(\mathbf{r}) &  
                = m_F' \hbar  \sqrt{  \delta^2(\mathbf{r}) + \RzeroSq } \nonumber\\
                &= m_F' \sqrt{ 
            \big[ \hbar  \omega_\text{rf} - \hbar\omega_L(\mathbf{r})   \big]^2
        +  \big[ g_F \mu_B  |\Bperp|  / 2 \big]^2
                            }
  \,,
\end{align}
where the $m_F'$ are a set of integers, or half-integers, similar to the $m_F$ described above.

The result of this is that slow atoms are confined by the potential
(\ref{eq:dressed-potentials}), which in a typical configuration, and to a first
approximation, confines atoms to an iso-$B$ surface defined by
$ \hbar \omega_\text{rf} - \hbar\omega_L(\mathbf{r}) = 0$, which approximately reduces the
value of $U(\mathbf{r})$ in Eq. (\ref{eq:dressed-potentials}).
The term $g_F \mu_B |\Bperp| / 2 $ also plays a role, and in particular it can be zero at
certain locations on the trapping surface allowing the escape of atoms. This latter effect
prevents the trapping of atoms in a shell potential by using the static quadrupole field
(\ref{eq:B-Quad}). However, shell potentials are possible with different field
arrangements such as those arising from the Ioffe-Pritchard trap and variations
\cite{zobay2004atom,zobay2001two,perrin2017trapping,garraway2016recent,lundblad2019shell,sinucoleon2020optimised} which have become candidates for experiments on the International Space Station \cite{elliott2018nasa}. 
The requirement is simply for a local extremum in the \emph{magnitude} of the field $\Bzero$ together with a non-zero $\Bperp$.
The reason for the interest in shell potentials in earth orbit is that on the earth's surface a gravitational term $mgz$ should be added to Eq. (\ref{eq:dressed-potentials}), which plays an important role for larger and interesting shells (e.g.\ see section \ref{sec:Dynring}).

Although the matter-wave \emph{bubbles} produced by shell potentials have become an object
of great interest, the shell potentials themselves are the building blocks for other
potentials of interest such as ring traps: we will see an example in section \ref{sec:BubbleAndSheet}.
Another example is in the next section \ref{sec:TAAP} where a modulated bias field is used to
make a ring trap: then $ \Bzero \rightarrow \Bzero +\mathbf{B}_\text{m}(\mathbf{r},t) $
and $\mathbf{B}_\text{m}(\mathbf{r},t) $ is a field varying in space, and time, but
typically at a frequency rather lower than the radio-frequency case.

\subsubsection{\label{sec:TAAP}Waveguides formed from Time-Averaged Adiabatic Potential (TAAP)}
\begin{figure}[hbt]
  \includegraphics[width=1\columnwidth]{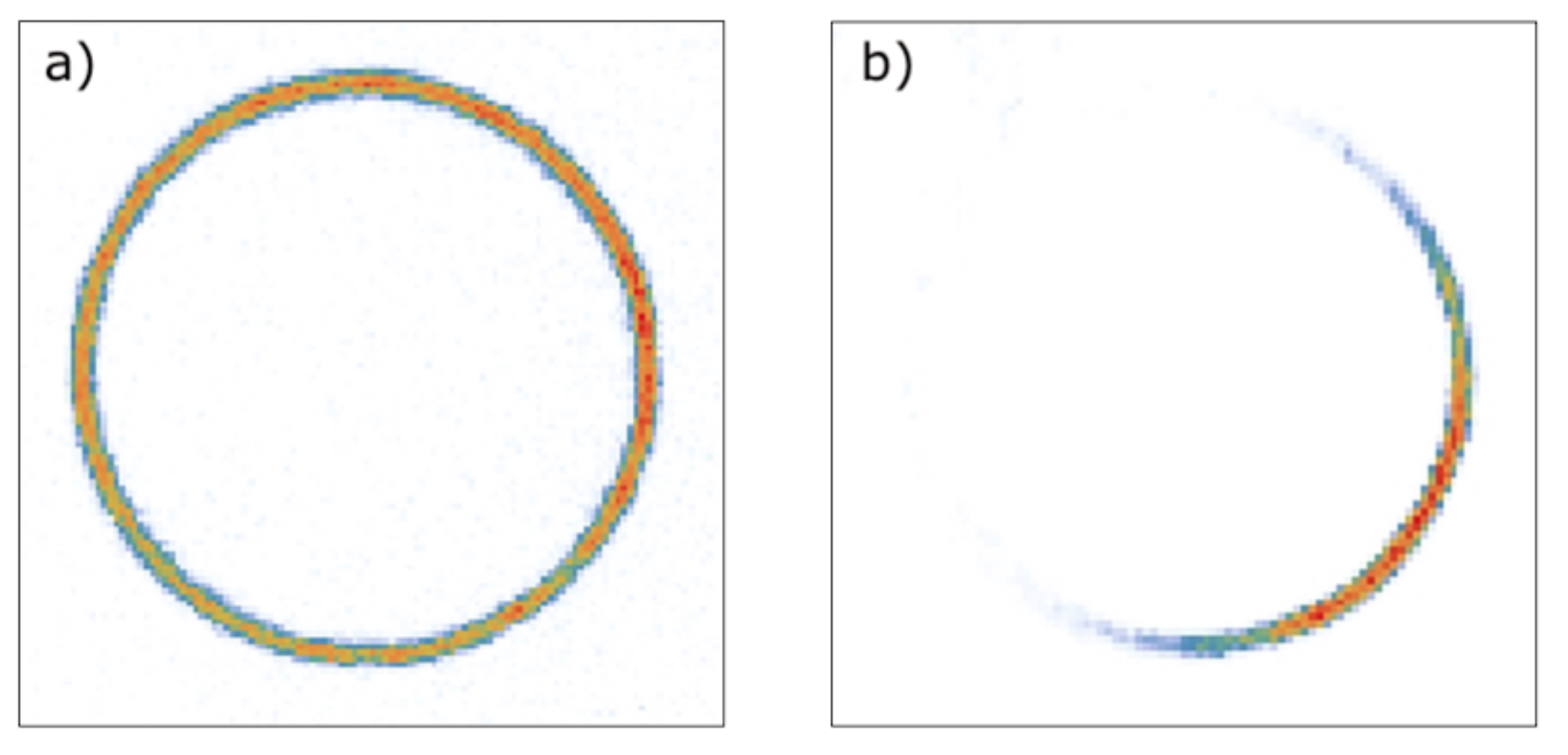}
  \caption{\label{fig:two-rings}a) BEC in a symmetric, ring-shaped TAAP ring with a diameter of $470\,\mu$m. b) BEC in a circular TAAP matterwave guide with superimposed gravito-magnetic modulation in the azimuthal direction.}
\end{figure}

%
%
%
%
Time averaged adiabatic potentials (TAAPs) allow the generation of extremely smooth matterwave guides \cite{pandey2019hypersonic}, which can be shaped into a half-moon or ring (see Fig.\,\ref{fig:two-rings}). 
They are an excellent candidate for matterwave optics, long-distance transport experiments, and interferometry in an atomtronic circuit \cite{navez2016matter,Pandey2021PRL,pandey2019hypersonic}. 
TAAPs are formed by applying an oscillating homogeneous potential to the adiabatic bubble traps described in Sec.\,\ref{sec:RF-Dressing}. 
If the modulation frequency $(\omega_\text{m}=2\pi f_\text{m})$ is small compared to the Larmor frequency, but fast compared to the trapping frequency of the bubble trap, then the effective potential for the atoms is the bubble potential time-averaged over one oscillation period \footnote{A small micro-motion of the atom cloud remains present but can be safely ignored in most circumstances since it is very small for large modulation frequencies \cite{minogin1998dynamics,cleary2010manipulation}.}.
%
%
Let us consider TAAP potentials formed from a quadrupole bubble trap and  an oscillating homogeneous field of the form $\mathbf{B}_\text{m}=\left\{0,0,B_\text{m} \sin \omega_\text{m} t\right\}$.
The modulation field simply displaces the quadrupole (and thus the bubble trap) by $z_\text{m}= \alpha^{-1} B_\text{m}\sin \omega_\text{m} t$
at an instant in time.
In order to find the \emph{effective} potential that the atoms are subjected to by this method, one calculates the time-average.
Time-averaging of a concave potential increases the energy of the bottom of the trap, as is readily illustrated by taking the time average of a harmonic potential jumping between two positions: the curvature does not change since it is everywhere the same; however, the energy of the trap bottom increases since it is at exactly the crossing point between the two harmonic potentials.
Returning to the modulated bubble trap, one notices that the modulation is orthogonal to the shell
at the poles of the shell  $(x=y=0)$, but tangential to the shell on the equator $(z=0)$.
Therefore, the time averaging causes a larger increase in the trapping potential at the poles rather than the equator---and therefore creates a ring-like structure. 

Assuming that $\omega_\text{RF}$ is modulated such as to stay resonant on the ring and  to keep $\Omega_\text{RF}$ constant, the vertical and radial trapping frequencies can be controlled via the relative amplitude of the modulation $\beta=  g_{\mathrm{F}} \mu_{\mathrm{B}} B_\text{m}/\hbar \omega_\text{RF}$ as
$\omega_{\rho}= \omega_{0}\,\left( {1 + \beta^2 } \right)^{ - 1/4}$ and
$\omega_{\mathrm{z}}=2\omega_{0}\sqrt{1-\left(1+\beta^2\right)^{-1/2}}$,
where the radial trapping frequency of the bare bubble trap is $\omega_{0}=
m_{\mathrm{F}} g_{\mathrm{F}} \mu_{\mathrm{B}}\,\alpha\, \left(m\,\hbar \Omega_\text{RF}\right)^{-1/2}$  with the mass of the atom $m$, and the $g_\text{F}$ is the Land\'e $g$-factor of the considered hyperfine manifold,  $\mu_\text{B}$ is the Bohr magneton, and $\Omega_\text{RF}$ the Rabi frequency of the dressing RF. 
In order to achieve large RF field strengths ($\approx 0.3$--1\,G) and  Rabi Frequencies, $(\Omega_\text{RF})$, one usually has to use RF-resonators, which make it very difficult to tune the RF frequency, and which results in a somewhat weaker confinement in the axial (i.e.\ vertical) direction.
Trapping frequencies of the order of a hundred Hz are readily achieved. 

In many cases it is also desirable to confine the atoms azimuthally.  
This is readily achieved either by tilting the ring away from being perfectly horizontal or by modifying the polarization of the rf-field.
The half-moon shaped BEC in Fig.\,\ref{fig:two-rings}b) was formed this way.
A gravito-magnetic trap results from tilting  the direction of the $B_\text{m}$ and thus tilting the ring against gravity \cite{navez2016matter}.
The gravito-magnetic potential forms a single minimum much like a tilted rigid pendulum.
One can also create a trap by changing the polarization of the dressing RF: tilting a linear polarization from the z-axes will cause, due to its projection on the local B-field, a sinusoidal modulation of the Rabi frequency along the ring resulting in a two minima on opposite sides of the ring.
Alternatively an  elliptical  RF polarization creates a single minimum.
Combining these modulation techniques permits the creation of two arbitrarily placed traps along the ring, or more generally  any longitudinal confinement of the form $a_1 \sin (\phi +\phi_1)+a_2 \sin (2\phi +\phi_2)$, where $\phi$ is the azimuthal angle and $\phi_1$and $\phi_2$ are phase offsets.  
Note that there are no angular spatial Fourier components higher than $2 \phi$ present in the system.

Thermal atoms and BECs are readily loaded into the gravito-magnetic TAAPs from a trapping-frequency-matched dipole trap.
This can be done fully adiabatically  by ramping down the dipole confinement and at the same time ramping up the TAAP trap.
With a sufficiently high level of control on the rf-fields, one can also load them from a TOP trap via a tilted dumbbell-shaped trap \cite{navez2016matter}.
Once in the ring, one can then manipulate the atoms with a simple manipulation of the time-averaging fields: The depth of the azimuthal trap can be changed by modifying the degree of tilt applied to the modulation field $(\mathbf{B}_\text{m})$.  
By changing the direction of the tilt (i.e. the phase between the modulation fields in the x and y directions) one can move the trap along the ring.
This can be used, e.g.\ to  accelerate the atoms along the ring, with angular momenta of $40\,000\, \hbar$ per atom being readily achieved  \cite{pandey2019hypersonic}.
They can then travel in the waveguides over distances of tens of centimeters without any additional heating associated with the propagation.
One can also remove the azimuthal confinement and allow the condensates to expand around the ring.
Viewed in the co-rotating frame at high angular momenta, the atoms see an exceptionally flat potential with the largest resulting density fluctuations corresponding to an energy difference of a few hundred picokelvin: this is equivalent to a few nanometers in height   \cite{pandey2019hypersonic}.
Current experiments have been performed  with BECs in the Thomas-Fermi Regime with about 20 transverse vibration modes occupied. The 1D regime is readily accessible simply by reducing the atom number and increasing the radius of the ring.

The complete lack of any roughness combined with a picokelvin level control of the trapping parameters make  the TAAP waveguides a very good candidate for guided matterwave interferometry and  the study of ultra-low energy phenomena such as long-distance quantum tunneling.
A remaining challenge is to completely fill the ring with a phase coherent condensate. 
Current experiments allowed a condensate to expand along the ring, which converts the chemical potential of the BEC into kinetic energy. 
When the condensate touches itself at the opposite side of the ring, the two ends have a finite velocity in opposite directions, resulting in a spiral BEC, i.e.\ a BEC wrapped around itself. 
Using atom-optical manipulation of the expansion process kinetic energies in the pico-kelvin range (a few hundred micrometers per second) can readily be achieved.
It will be interesting to study the very low energy collisions that will lead to a thermalisation of this system.
A promising approach for a fully phase-coherent ring-shaped condensate is to first fill a small ring  and then increase its radius.
This should not induce any additional phase fluctuations, despite the fact that the lowest excitation has an energy of $E=\hbar^2/(2 m r^2)$, which for a ring of 1\,mm radius is 3 femtokelvin.

\subsubsection{\label{sec:Dynring}Dynamical ring in an rf-dressed adiabatic bubble potential}

%
There is a formal analogy between the Hamiltonian of a neutral gas in rotation and the one of a quantum system of charged particles in a magnetic field. This makes rotating superfluids natural candidates to simulate condensed matter problems such as type II superconductors or the quantum Hall effect~\cite{cooper2008rapidly,bloch2012quantum}.
For a quantum gas confined in a harmonic trap of radial frequency $\omega_r$ and rotating at angular frequency $\Omega$ approaching $\omega_r$, the ground state of the system reaches the atomic analog of the lowest Landau level (LLL) relevant in the quantum Hall regime~\cite{ho2001bose,aftalion2005vortex,bloch2008many}.
Reaching these fast rotation rates is experimentally challenging in a harmonic trap because the radial effective trapping potential in the rotating frame vanishes due to the centrifugal force. To circumvent this limit, higher-order confining potentials have been developed \cite{bretin2004fast}, which allow to access the regime where $\Omega$ even exceeds $\omega_r$.

The adiabatic bubble trap  has many features that make it a very good candidate to explore this regime. Indeed, it is very smooth and easy control of its anisotropy is possible through the dressing field polarization \cite{merloti2013two}. This allows us to deform the bubble and rotate the deformation around the vertical axis in a very controlled way, allowing us to inject angular momentum into the cloud.  The curved geometry of the bubble provides naturally the anharmonicity required to rotate the atoms faster than the trapping frequency $\omega_r$ at the bottom.

In the experiment at LPL~\cite{guo2020supersonic}, the atoms are placed in a quadrupole magnetic field of symmetry axis $z$ dressed by a radio-frequency (rf) field of maximum coupling $\Omega_{0}$ at the bottom of the shell. Here, the equilibrium properties in the absence of rotation ($\Omega=0$) are well known~\cite{merloti2013two}: the minimum of the trapping potential is located at $r=0$ and $z=z_0$ and around this equilibrium position the potential is locally harmonic with vertical and radial frequencies $\omega_z=2\pi\times\SI{356}{\hertz}$ and $\omega_r=2\pi\times\SI{34}{\hertz}$, without measurable in-plane anisotropy. 
This trap is loaded with a pure BEC of $\num{2.5e5}$ ${}^{87}$Rb atoms with no discernible thermal fraction.
This atomic cloud has a chemical potential of $\mu/\hbar=2\pi\times\SI{1.8}{\kilo\hertz}$ which is much greater than $\omega_r$ and $\omega_z$, and well in the three-dimensional Thomas-Fermi (TF) regime.
In addition to the dressing field, a radio-frequency knife with frequency $\omega_{\rm kn}$ is used to set the trap depth to approximately $\omega_{\rm kn}-\Omega_{0}$ by outcoupling the most energetic atoms in the direction transverse to the ellipsoid~\cite{garridoalzar2006evaporative,kollengodeeaswaran2010rf}.\\
In a frame rotating at frequency $\Omega$, the  effective dressed trap potential is the usual trap described above with the addition of a $-\frac{1}{2}M\Omega^2r^2$ term taking into account the centrifugal potential.
\begin{figure}[t]
\includegraphics[height=3.5cm]{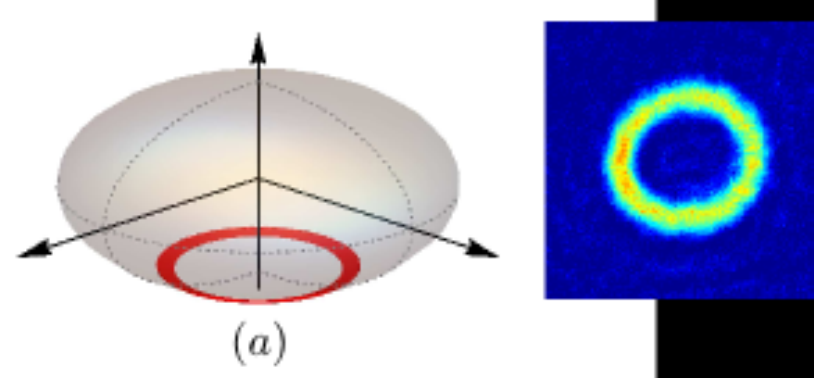}
\caption {\label{ring358}(a): Density contour (red annulus) for a BEC rotating at $1.06\,\omega_r$ in the shell trap (gray ellipsoid). (b): \emph{in situ} integrated 2D density of a dynamical ring. Picture size: $130\times 130~\mu$m$^2$. Reprinted with permission from Y. Guo, R. Dubessy, M. d. G. de Herve, A. Kumar, T. Badr, A. Perrin, L. Longchambon, and H. Perrin, Phys. Rev. Lett. 124, 025301 (2020). Copyright 2020 American Physical Society.
 }
\end{figure}
In this frame, the atomic ground state consists of an array of vortices of quantized circulation, each vortex accounting for $\hbar$ of angular momentum per atom. When only a few vortices are present, the velocity field differs strongly from the one of a classical fluid, but for a sufficiently large number of vortices the superfluid rotates as a solid body with a rotation rate $\Omega$. 
When $\Omega<\omega_r$ the equilibrium position remains on the axis $r=0$ at $z=z_0$, and the only difference is a renormalization of the radial trapping frequency: $\omega_r^{\rm eff}=\sqrt{\omega_r^2-\Omega^2}$. Of course, as this frequency decreases, the trap anharmonicity becomes more important in the determination of the cloud shape.

For  $\Omega>\omega_r$ the trap minimum is located at a non-zero radius.
In this situation, a hole grows at the trap center above a critical rotation frequency $\Omega_h$ \cite{cozzini2005oscillations}, leading to an annular two-dimensional density profile (Fig.~\ref{ring358}(a)) which we will refer to as a ``dynamical ring''~\cite{guo2020supersonic}. Moreover, the velocity of the atomic flow is expected to be supersonic~\cite{kasamatsu2002giant} \emph{i.e.}\ exceeding by far the speed of sound. For increasing $\Omega$, one expects the annular gas to sustain vortices in its bulk up to a point where the annulus width is too small to host them. The gas should then enter the so-called ``giant vortex'' regime~\cite{kasamatsu2002giant,fetter2005rapid} where all the vorticity gathers close to the center of the annulus.

The experimental sequence is the following: angular momentum is injected into the cloud by rotating the trap with an ellipsoidal anisotropy at a frequency $\Omega=\SI{31}{\hertz}$. The trap rotation is then stopped and isotropy is restored. At this moment,  which we take as $t=0$, the cloud shape goes back to circular with an increased radius due to its higher angular momentum. An additional evaporation process, selective in angular momentum, continuously accelerates the superfluid and increases its radius\cite{guo2020supersonic}.
Due to this size increase, the chemical potential is reduced and the gas enters the quasi-2D regime $\mu\leq\hbar\omega_z$.
After a few seconds a density depletion is established at the center of the cloud which is a signature of $\Omega$ now exceeding $\omega_r$.
After a boost in selective evaporation due to a lowering of the frequency of the rf knife, a macroscopic hole appears in the profile, indicating that $\Omega$ is now above $\Omega_h$ and that a fast rotating dynamical ring with a typical radius of $\sim\SI{30}{\micro\meter}$ has formed as can be seen in Fig.~\ref{ring358}(b). The rotation keeps increasing and a ring is still observable after $t=\SI{80}{\second}$. Rotational invariance is critical in that regard, and is ensured at the $10^{-3}$ level by a fine tuning of the dressing field polarization and of the static magnetic field gradients \cite{perrin2017trapping}.\\ A Thomas-Fermi profile convoluted with the imaging resolution is much better at reproducing the experimental density profile than a semi-classical Hartree-Fock profile, demonstrating that the samples are well below the degeneracy temperature.
Using the Thomas-Fermi model we can estimate the properties of the cloud. For example the ring obtained at $t=\SI{35}{\second}$ has a chemical potential of $\mu/\hbar\simeq2\pi\times\SI{84}{\hertz}$ and an averaged angular momentum per particle $\langle \hat L_z\rangle/N\simeq\hbar\times317$. Interestingly the estimated peak speed of sound $c=\sqrt{\mu/M}\simeq\SI{0.62}{mm/s}$ at the peak radius $r_{\rm peak}$ is much smaller than the local fluid velocity $v=\Omega r_{\rm peak}\simeq\SI{6.9}{mm/s}$: the superfluid is therefore rotating at a supersonic velocity corresponding to a Mach number of 11.
Moreover, due to the continuous acceleration of the rotation, the dynamical ring radius grows gradually with time which results in a decrease of the chemical potential and an increase of the Mach number.
For $t>\SI{45}{s}$ the chemical potential is below $2\hbar\omega_r$ and the highest measured Mach number is above 18.

\begin{figure}[t]
\includegraphics[height=2.5cm]{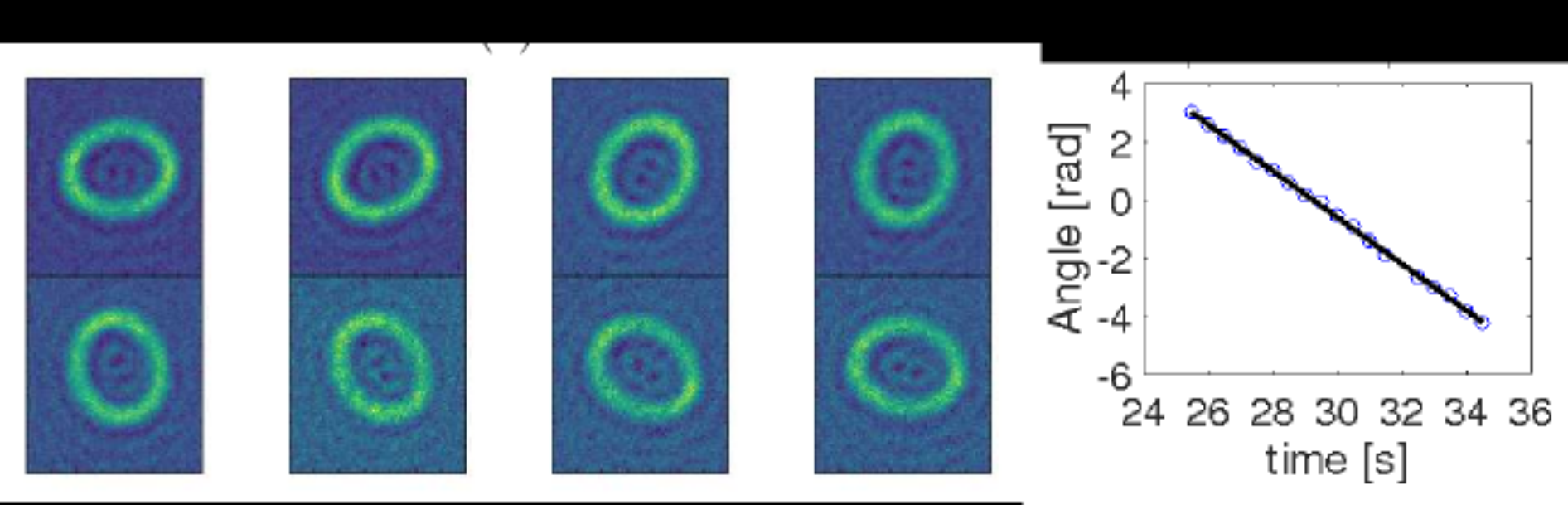}
\caption {\label{quadmode}(a): \textit{in situ} evolution of a quadrupole deformation in the dynamical ring. The elliptical deformation is rotating against the supersonic flow. (b): Time evolution of the orientation of the ellipse major axe.}
\end{figure}

Superfluidity in the dynamical ring has also been evidenced by the observation of quadrupole-like collective modes. After the ring formation, the rotation rate, while accelerating, crosses a value where the quadrupole collective mode is at zero frequency, such that any elliptical static anisotropy can excite it resonantly. A very small bubble anisotropy is enough to excite this mode characterized by an elliptic ring shape rotating with a period of approximately $\SI{10}{\second}$ in the direction opposite to the superfluid flow (Fig.~\ref{quadmode}). This counterpropagating effect is not predicted by a mean-field theory, and has been confirmed by resonant spectroscopy of the quadrupole mode during the ring acceleration~\cite{guo2020supersonic}.

The persistence of superfluidity at such hypersonic velocity raises fundamental questions about the decay of superfluidity in the presence of obstacles, and how superfluidity can be preserved at such speeds: nonlinear effects, the presence of vortices and the dependence on temperature would be particularly interesting to study experimentally and compare with theoretical predictions~\cite{law2000motional,pavloff2002breakdown,paris-mandoki2017superfluid,bradley2016breaking,dries2010dissipative}.
This hypersonic superflow is not yet a giant vortex, but it is an important step towards this long-sought regime whose transition rotation frequency is not theoretically clearly identified. Moreover, the well-known elementary excitation spectrum for a connected rotating superfluid is strongly modified when the ring appears, and the important discrepancies observed between the experimental results and a mean-field theoretical approach for a quadrupole-like collective mode highlight the need to refine the description of fast rotating superfluids in anharmonic traps.

An alternative way of generating large angular momentum states in rf-dressed adiabatic bubble potentials is to first generate them in a TAAP ring and then reduce the vertical modulation, thus adiabatically transferring the atoms into the bubble.

\subsection{\label{sec:Optical-ring}Trapping in rings with optical potentials}

Potentials for ultracold atoms can be formed through the use of focused far-detuned optical beams~\cite{grimm2000optical}. Since the potential is directly proportional to the intensity of the optical field, ring-shaped condensates may be created through the implementation of ring-shaped optical patterns. The most significant advantage in optical dipole ring traps is the insensitivity to the hyperfine state, allowing multi-component and spinor BECs to be trapped. {Additional advantages include the imprinting of superfluid flow, either through phase imprinting or through Raman transitions that can directly transfer angular momentum to the cloud.} The advent of spatial light modulator technologies means the optical ring trap has become highly configurable, allowing more complex geometries to be generated.







{
\subsubsection{Optical trapping} 
The light-matter interaction can be parameterized through the complex polarizability, where the real part is associated with the dipole trapping potential and the imaginary component results in the absorptive scattering of photons. Trapping cold atoms requires that absorption is minimised to avoid scattering loss of atoms from the trap. Defining $\Delta = \omega - \omega_0$, the detuning of the trapping laser from resonance, the scattering loss rate reduces as $\Delta^{-2}$ while the trapping potential reduces as $\Delta^{-1}$. Thus, sufficient detuning of the optical field will result in an optical potential that is approximately conservative. The potential arising for far-detuned dipole  trapping light is given by 

\begin{equation}\label{eq:dipole-ring}
U_\textrm{dip}(\mathbf{r})=\frac{\pi c^2}{2\omega_0^3}\frac{\Gamma}{\Delta}I(\mathbf{r})
\end{equation}

\noindent where $I(r)$ is the intensity profile of the light, and $\Gamma$ is the transition linewidth. Since the trapping force is determined through the gradient of Eq.~\ref{eq:dipole-ring}, a trapping potential requires a non-uniform optical intensity, obtained by shaping and focusing the intensity profile $I(\mathbf{r})$. Ring traps,  can either be created from attractive (red-detuned) or repulsive (blue-detuned) light, usually by combining the ring shaped intensity profile with a perpendicular light sheet that provides confinement along the propagation direction of the projected ring pattern.}

\subsubsection{\label{sec:pure-optical}Optical ring traps} 

{We begin by looking at some of the optical beam techniques for ring traps that are in use, and outline their potential for atomtronic applications.}

{
\paragraph{Laguerre-Gauss beams:}
One of the first proposed methods for a ring optical dipole trap was the use of Laguerre-Gaussian (LG) modes having circular symmetry \cite{wright2000toroidal}. For far-off-resonance light, these provide the spatial structure for a toroidal trap. An additional advantage  of such LG modes is that they also carry orbital angular momentum. With pulses of near-resonant light, the LG modes can be tailored to provide two-photon Raman transitions that transfer exact quanta of circulation to the condensate. \\

The LG$_{0N}$ modes are typically generated by phase transformation of a Gaussian TEM$_{00}$ mode, which transforms the spatial profile of the beam into a doughnut mode carrying $N\hbar$ units of orbital angular momentum. A number of methods exist, including spiral phase plates,  computer generated holograms, or through the use of phase based spatial light modulators. The toroidal intensity profile of the LG$_{01}$ mode is given by \cite{RamanathanThesis2011}

\begin{equation}\label{eq:LG-ring}
I_{LG_{01}}(r)=\frac{4P_{LG_{01}}}{\pi{r_0}^2}\left(\frac{r}{r_0}\right)^2 e^{\frac{-2r^2}{r_0^2}},
\end{equation}

\noindent where $P_{LG}$ is the total laser power in the LG beam, and $r_0$ is the radius at the peak intensity of the LG mode.  Correction for imperfections in the spatial structure, and obtaining sufficient power in higher order modes is typically a challenge. Ring traps and circulating currents using LG modes have been demonstrated in both single state and multi-component spinor gases, and were early demonstrations of all-optical trapping of BEC in a ring geometry \cite{ramanathan2011superflow,moulder2012quantized,beattie2013persistent}. To date, they have been used to realise small optical rings, for the study of quantized superfluid flows.}





\paragraph{Painted optical traps:}
%
%

\begin{figure}[t]
\includegraphics[width=\columnwidth]{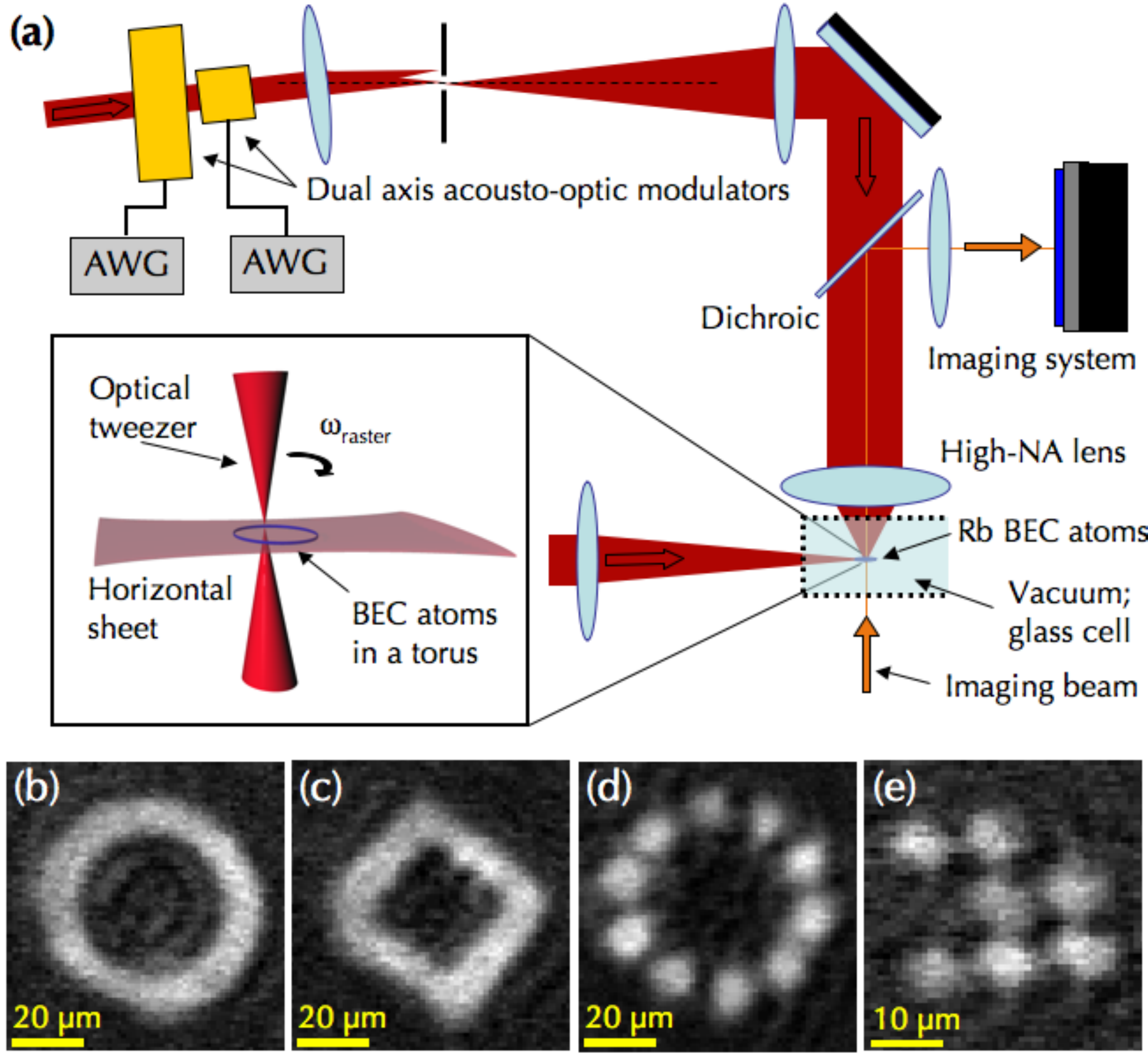}
\caption{(a) A painted potential system in which a tightly-focused rapidly-moving red-detuned laser beam paints the desired potential on a horizontal light sheet providing vertical confinement. (b)~In-trap absorption images of BECs formed in painted potentials.  The technique can create a BEC in any shape that can be drawn on a sheet of paper. Reprinted with permission from K. Henderson, C. Ryu, C. MacCormick, and M. G. Boshier, New Journal of Physics 11, 043030 (2009), under a Creative Commons Attribution License. }
\label{fig:PaintedPotential}
\end{figure}

An alternative to projecting a ring shaped beam is to build a time-averaged potential with a moving, red-detuned, focused laser beam. By rapidly steering a Gaussian beam in a circular orbit, a ring trap can be generated. This is achieved through the use of two acousto-optical deflectors (AOD) controlling the two axes of the painting beam {by driving the deflectors with lists of frequency points that are repeatably iterated at high speed.}\cite{henderson2009experimental, bell2016bose}.  This approach was used to create the first ring BEC \cite{henderson2009experimental}, as shown in Fig.~\ref{fig:PaintedPotential}. 

The advantages of this technique is that it allows adapting the intensity locally to create desired features in the potential landscape and to flatten imperfections due to possible laser inhomogeneity \cite{ryu2015integrated}; the available laser power is used in an efficient way as only the relevant trapping locations are illuminated; the painting laser itself can be used as a stirrer to set the quantum fluid into rotation and demonstrate quantized superfluid flows \cite{ryu2014creation}; the technique also enables more complex geometries. As an example, the atomtronic analogue of a Josephson junction has been demonstrated and used to realize a DC atomtronic SQUID~\cite{ryu2013experimental}.  More recently, the dynamic potentials possible with painting were used to show that the atomtronic SQUID exhibits quantum interference \cite{ryu2020quantum}. 

The painting approach also comes with specific technical constraints that may need to be addressed. The phase of the time-averaged beam loop plays a role on the fine details of the potential, {which results in imprinting of the condensate phase}, and has to be compensated for \cite{bell2018phase}. This is particularly relevant for the application of such traps in atom interferometry schemes.


\paragraph{Conical refraction:}
A novel approach to generating ring traps has been demonstrated with the use of conical refraction occurring in biaxial crystals.  A focused Gaussian beam passing along the optical axis of the crystal transforms, at the focal plane, into one or more concentric rings of light. In the case of a double-ring, the light field encloses a ring of null intensity, called the {Poggendorff dark ring} \cite{turpin2015blue}. For a blue-detuned laser field the atoms are trapped between the bright rings. The advantage of this configuration is that it minimizes spontaneous scattering of photons responsible for heating when the laser beam is not very far detuned from resonance. Further advantages include the high conversion efficiency of the incoming Gaussian beam to the ring-trap light field and the access to different ring configurations. The ring diameter is defined by the refractive indices of the the biaxial crystal and its length. The width of each ring is given by the focal waist of the focused Gaussian beam. A variation of the ratio of these numbers (e.g. by changing the focal waist) allows for a variation of the resulting light field topology from a single bright ring to a bright ring with a central bright spot and further to bright double rings of increasing diameter. {As with LG modes, there are challenges in alignment of the optical beams through the biaxial crystal. On the other hand, the conversion efficiency from a Gaussian TEM$_{00}$ mode to the ring pattern can be close to unity.} The first results on BECs transferred into a Poggendorff ring have been reported \cite{turpin2015blue}. Ongoing work is directed towards implementing quantum sensors (e.g Sagnac interferometers) for rings with large diameter and atomtronic SQUIDs for small rings. 

\begin{figure}[t]
\includegraphics[width=\columnwidth]{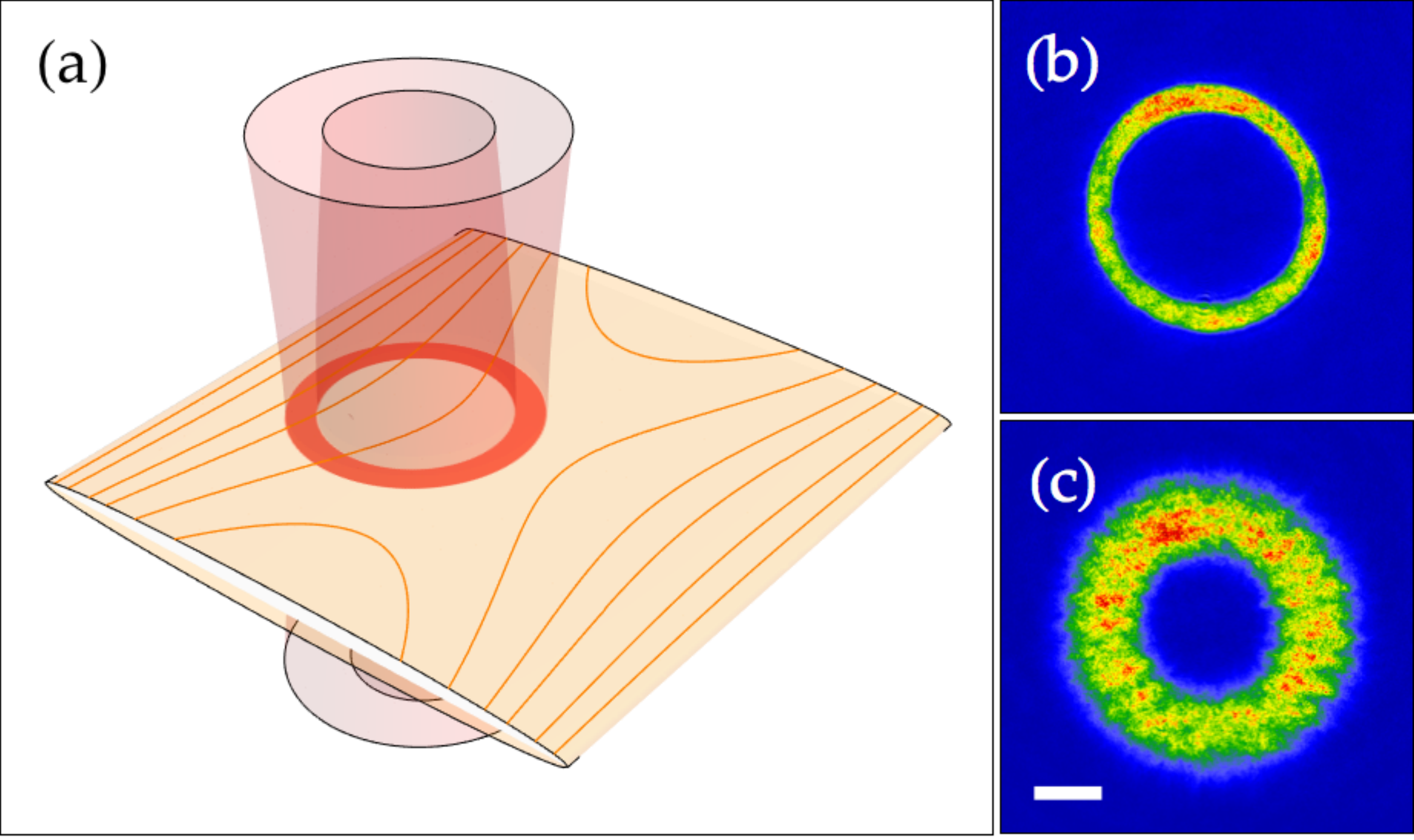}
\caption{(a)~A typical optical ring trap configuration; the potential is formed at the intersection of the vertically focused ring pattern, and horizontal sheet beam. 
(b)~In-trap absorption image of a ring BEC formed in a 164~$\mu$m diameter time-averaged optical potential with $(\omega_\rho,\omega_z)=2\pi\,(50,140)\,\text{Hz}$ trapping frequencies. (c)~Expanded ring after 20~ms time-of-flight. The scale bar is 50~$\mu$m.}
\label{fig:OpticalRingBEC}
\end{figure}

\paragraph{Digital micromirror direct projection:}

Direct imaging of digital micromirror devices (DMDs) has recently emerged as a powerful tool for the all-optical configuration of BECs~\cite{gauthier2016direct,ville2017loading,kumar2016minimally, Gauthier2019}. Ring traps can be created by {directly} projecting the DMD-patterned light onto a vertically confining attractive {light-sheet} potential ~\cite{gauthier2016direct,kumar2016minimally}, similarly to Fig.~\ref{fig:OpticalRingBEC}, or onto a vertically oriented accordion lattice~\cite{ville2017loading}. {This can be accomplished using a relatively simple optical system, usually consisting of an infinite conjugate pair. Due to the the large magnification factors required to reduce the DMD image to the typical $100~\mu$m scale of the BEC, the final element in the imaging system is typically an infinity corrected microscope objective~\cite{gauthier2016direct,kumar2016minimally}. DMDs may also be used in the Fourier plane of the imaging system~\cite{zupancic2016ultra}, where the DMD implements an amplitude-only hologram. A detailed discussion of holographic techniques is beyond the scope of this chapter, and the reader is referred to more complete reviews of the subject~\cite{dufresne2001computer}.}

\begin{figure}[t]
\includegraphics[width=\columnwidth]{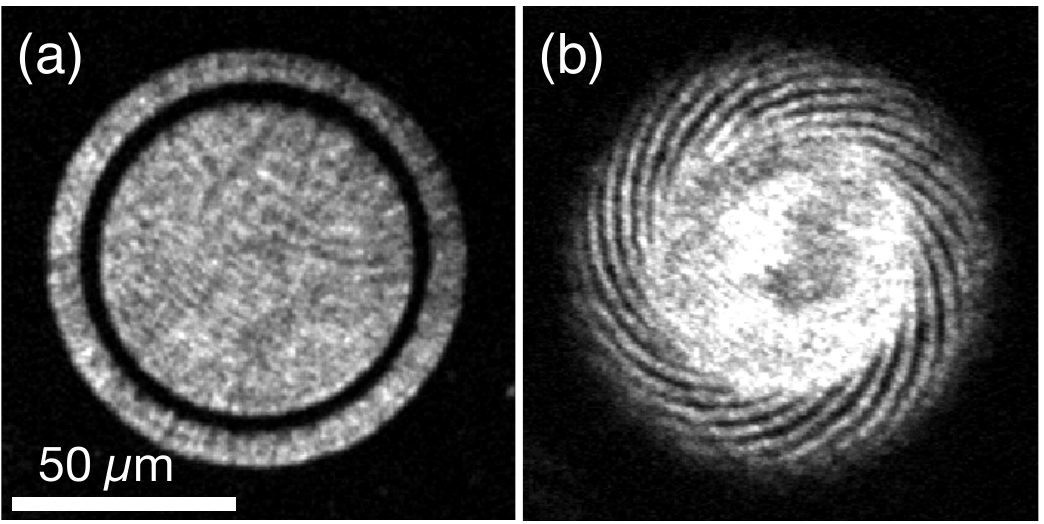}
\caption{(a) A ring trap with a diameter of 100$\mu$m is created using a repulsive DMD-patterned potential combined with an attractive horizontal sheet beam. By using the dynamic control of the DMD, a stirring barrier is introduced into the ring and then accelerated through 90$^{\circ}$, before being removed from the ring. The resulting persistent current is imaged through interference with the central BEC in a short time-of-flight, determining a net circulation quanta of $N=21$.}
\label{fig:PC_Interference}
\end{figure}

In Fig.~\ref{fig:PC_Interference}, direct imaging of a DMD is used to create a ring trap, along with a central phase-uncorrelated reference BEC. By introducing a stirring barrier with the DMD, and circulating the barrier around the ring, a 21-quanta persistent current results, corresponding to an angular momentum of $\sim132\,\hbar$ per atom. The winding number of the current is visualised through interference with the reference central BEC after a short 5~ms time of flight~\cite{mathew2015self}. The DMD technology can also be used to phase imprint an azimuthal light gradient such that angular momentum can be imparted to the atoms \cite{kumar2018producing} and a circulating current created \cite{bell2020engineering}.

%

\paragraph{Micro-fabricated optical elements:}
An approach combining flexibility, integrability, and scalability can be based on the application of micro-fabricated optical elements for the generation of complex architectures of dipole traps and guides \cite{birkl2001atom}. It draws its potential from the significant advancement in producing diffraction-limited optical elements with high quality on the micro- and nanometer scale. Lithographic manufacturing techniques can be used to produce many identical systems on one subtrate for a scalable configuration \cite{dumke2002micro}. On the other hand, state-of-the art direct laser writing gives high flexibility in producing unique integrated systems and allows for fast prototyping \cite{schaeffner2020arrays}. Applications range from integrated waveguides and interferometer-type structures \cite{dumke2002interferometer} to arrays of dipole-traps for quantum information processing \cite{schlosser2011scalable} and single-atom atomtronics implementations \cite{benseny2010atomtronics}. In combination with DMD-based control of the light field (see previous section), access to dynamic reconfiguration becomes possbile. Integrability is not limited to the generation of light fields for dipole potentials but can be extended to the integration of light sources and detectors or even complex quantum-optical systems such as an entire magneto-optical trap \cite{birkl2001atom}.

{
\subsubsection{Imperfections in optical traps}

Defects in the optical potential will influence the ability to sustain superfluid flow without dissipation, or may introduce unwanted phase perturbations on the condensate if the optical potential is time-varying. We can gain some measure of the significance, and the level of control required for optical traps useful in atomtronics, by considering the density of the BEC in a ring potential. In the Thomas-Fermi limit, with sufficient atom number  in the trap, the interaction energies dominate over kinetic energy terms, leading to a simplified GPE equation $\left[V(\mathbf{r})+g |\Psi(\mathbf{r})|^2 \right]\Psi(\mathbf{r}) = \mu \Psi(\mathbf{r})$, giving the density $n(\mathbf{r}) = |\Psi(\mathbf{r})|^2 = [\mu - V(\mathbf{r})]/g$, where $\mu$ is defined by Eq.~\ref{eq:chempot-ring}. The density occupies the spatial profile of the ring trap. In the context of the intensity of the optical potential, assuming a fixed light sheet, the trap depth scales directly scales with the ring optical intensity $I_0$, while the chemical potential of the BEC more weakly follows as $\mu \propto I_0^{1/4}$. This means that for a typical condensate, the chemical potential is on the order of tens of nK, and is only weakly effected by the trapping intensity, while large optical trap depths on the order of $1~\mu$K or larger may be easily achieved and utilised. 
Since the density of the condensate closely follows the optical potential, small perturbations in the optical field can result in significant fluctuations on at the energy scale of the condensate, and thus significant density fluctuations; variations in the optical intensity must typically be limited to less than 1\% in order to avoid unwanted perturbations. The precision of the optical projection is thus a key consideration when implementing configurable optical potentials. These aspects however also mean that the condensate density provides a very sensitive probe of the optical potential, and the atom density can be used to feedforward corrections to the optical potential~\cite{bell2016bose}.}

\subsection{\label{sec:BubbleAndSheet}Hybrid traps: RF bubble plus light sheet(s)}

\begin{figure}[b]
\includegraphics[height=2.5cm]{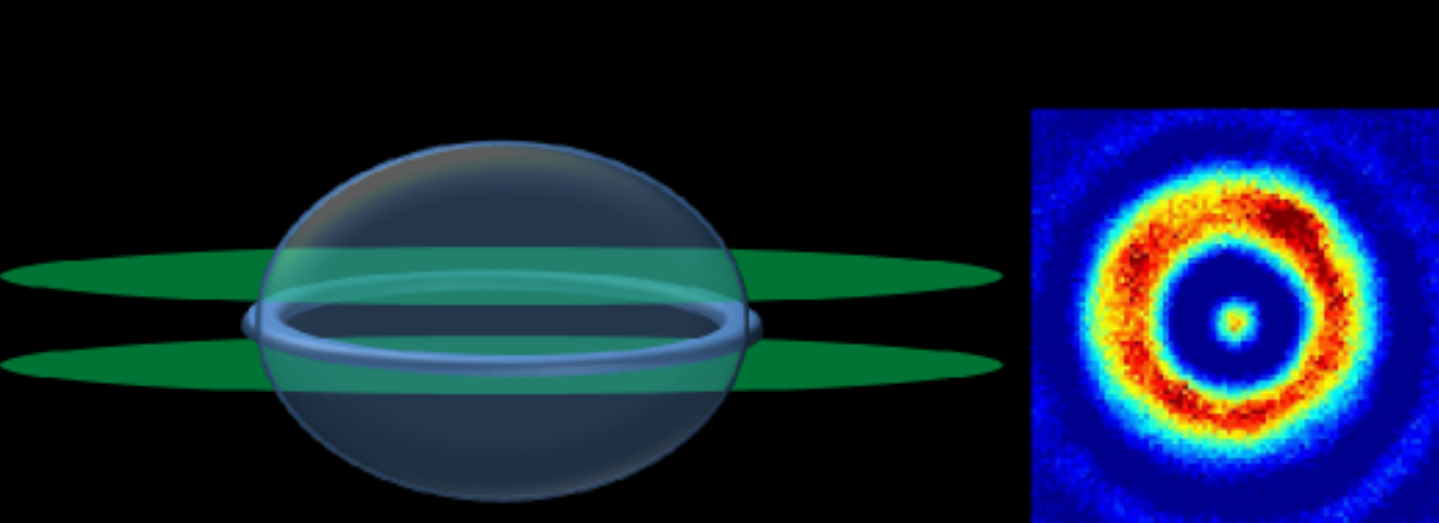}
\caption {\label{principle}(a): Principle of the ring trap {based on the combination of a magnetic bubble trap and two blue-detuned light sheets.} (b): \textit{in situ} image of the ring.The spot at the center is not due to the presence of atoms inside the ring but only to optical diffraction from the ring.}
\end{figure}

{One can also combine optical potentials and  magnetic trapping to produce a hybrid trap and exploit the advantages of each technique for ring trap generation. As mentioned above, optical potentials can achieve large trapping frequencies, while magnetic traps are very smooth due to the macroscopic size of the coils generating them. 
The bubble geometry described in Sections \ref{sec:RF-Dressing} and \ref{sec:Dynring} is particularly suited to create a ring trap: by combining the rf-dressed bubble trap and an optical light sheet as in Section \ref{sec:pure-optical}, one can create a toroidal trap.}
%
%
The principle is depicted in Fig.~\ref{principle}: a horizontal light sheet is superimposed with a bubble trap which is rotationally invariant around the vertical direction~\cite{morizot2006ring,heathcote2008ring}. The light sheet is designed to achieve {a strong optical confinement} in the vertical direction, and the radial confinement is ensured by the bubble trap {itself, made with the same rf-dressed quadrupole trap as in Section \ref{sec:Dynring}}. Maximum radial trapping and maximum radius will be attained if the light sheet is located at the equator of the bubble, a situation which also ensures maximum decoupling between the vertical trapping frequency $\omega_z$ and the radial trapping frequency $\omega_r$.

%
%

Experimentally, the optical trap is formed between two horizontal light sheets which are made repulsive by their large blue detuning from the atomic transition. {The bubble radius is significantly smaller than the light sheets width and also the vertical Rayleigh length to minimize the azimuthal potential variations. The choice of a small radius also comes with a higher critical temperature and a larger chemical potential, which reduces the relative density fluctuations around the ring due to optical imperfections from residual light scattering of the vacuum glass cell (see Section~\ref{sec:pure-optical})}. One then creates a trapped toroidal degenerate gas of approximately $\SI{e5}{}$ atoms (Fig.~\ref{principle}(b)). {With further reduction of optical imperfections in the light sheets, one could enter with $\num{e4}$ atoms the quasi-1D condensate regime\cite{gerbier2004quasi}, where large-scale correlations and solitons play an essential role in the dynamics.}
 
\begin{figure}[h]
\includegraphics[height=3cm]{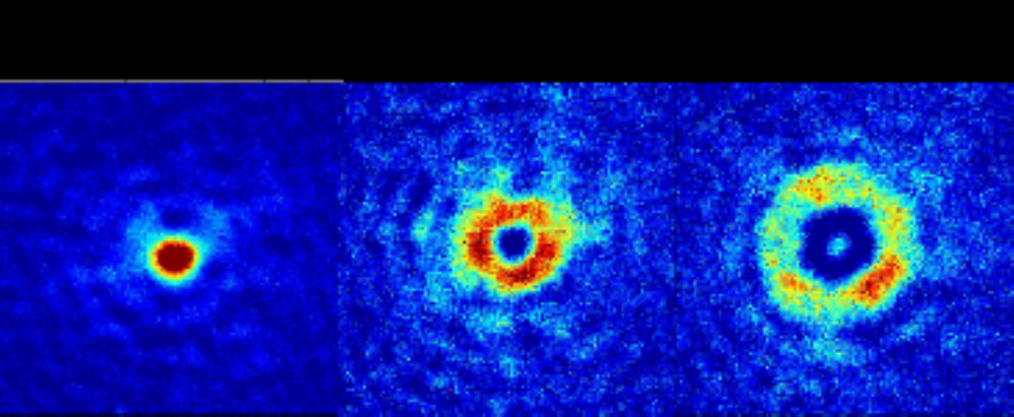}
\caption {\label{tofrings}Time-of-flight images of (a): a non-rotating ring, (b) and (c): rotating rings with different circulations. The rotation is imparted by a rotating $\SI{7}{\micro\meter}$-waist blue-detuned vertical Gaussian beam.}
\end{figure}

 The gas can be set into rotation by different procedures, {using either magnetic or optical means.} The first method, used in our experiment in Ref.~\cite{degoer2020preparation}, consists in slightly deforming the bubble trap with an ellipsoidal anisotropy, rotate this {magnetic} deformation at a given fixed frequency and finally restore the circular symmetry. In a second method (Fig.~\ref{tofrings}), the rotation is induced by a rotating optical defect~\cite{wright2013driving,degoer2020preparation} {driven by a dual-axis acousto-optic modulator system as described in Section~\ref{sec:RF-Dressing}. Well-controlled circulation} could also be imparted by direct optical phase imprinting onto the ring trap~\cite{kumar2018producing}.

Above some critical rotation frequency depending on the excitation strength, one observes, after a time-of-flight imaging procedure, a hole in the atomic distribution. The hole is absent when the ring is non rotating, and is thus evidence for a non-zero circulation of the superfluid in the ring trap (Fig.~\ref{tofrings}). The hole area grows for increasing rotation rates, and shrinks with time when one lets the cloud rotate freely in the trap.  In future experiments, optical barriers created by spatial light modulators could be imposed onto the ring and dynamically modulated in height and position. This would create the equivalent of Josephson junctions in superconductors and allow us to simulate models of non-equilibrium quantum systems and emulate new setups in mesoscopic superconductivity~\cite{ryu2015integrated,eckel2014hysteresis}. {This hybrid ring is very promising for the study of 1D superfluid dynamics, for example shock waves induced by rotation in the presence of a static barrier \cite{polo2019oscillations,Dubessy2020}. Increasing again the ring confinement towards the one-dimensional regime with fermionization of the atoms\cite{girardeau1960relationship} could lead to NOON states more robust against decoherence \cite{hallwood2010robust}}.

\subsection{\label{sec:outlook-ring}{Concluding remarks and outlook}}

The development of technology for controlling electronic systems and {generating complex optical fields} is giving ever greater control of ultra-cold atoms and condensates of atoms. 
The ring trap remains of particular interest because of the topology, the possibility for self-interference, circuital currents, Sagnac interferometry and so on. In a way, it is its own primitive atomtronic circuit. %
{For optical ring traps, painted optical potentials and digital micromirror devices have demonstrated high level of configurability and dynamic control over the condensate, allowing state-independent trapping, and the ability to introduce junctions, moveable barriers into the atomtronic ring. The ring systems based on RF dressed magnetic traps are also extremely flexible because of the level of electronic control. Atoms can be accelerated and rotated around ultra-smooth waveguides, simply by varying or introducing additional control frequencies with time.}
The future challenges for the technology, after this development, will be to create particular atomtronic applications and test the limits of technology for creating large scale structures and structures which possibly have some 3D features.  
 In the future we will undoubtedly see more control complexity and more hybrid approaches. Where surface interactions are less of a problem we can also envisage atomtronic circuits 
based on atom-chip technology, where rings, and complex guided circuits, may be enabled by the design of wire structures and the fields they produce from static and AC currents.

{\it Acknowledgments}.
The UQ group acknowledges funding by the ARC Centre of Excellence for Engineered Quantum Systems (project number CE1101013), and ARC Discovery Projects grant DP160102085.
WK would like to acknowledge the contribution of the AtomQT COST Action CA16221 and of HELLAS-CH (MIS 5002735) implemented under ``Action for Strengthening Research and Innovation Infrastructures," funded by the Operational Programme ``Competitiveness, Entrepreneurship and Innovation" (NSRF 2014-2020) and co-financed by Greece and the European Union (European Regional Development Fund).
HP and LL acknowledge financial support from the ANR project SuperRing (Grant No. ANR-15-CE30-0012) and from the R\'egion \^Ile-de-France in the framework of DIM SIRTEQ (Science et Ing\'enierie en R\'egion \^Ile-de-France pour les Technologies Quantiques), project DyABoG.
BMG would like to acknowledge support from the UK EPSRC grant EP/M013294/1.
MGB acknowledges support from the US DOE through the LANL LDRD Program.

\section{ATOMTRONIC CHIPS AND HYBRID SYSTEMS}
\label{hybrid}
\vspace*{-0.5cm}
\par\noindent\rule{\columnwidth}{0.4pt}
{\bf{\small{C. Hufnagel, M. Keil, A. G\"unther, R. Folman, J. Fortagh,  R. Dumke}}}
\par\noindent\rule{\columnwidth}{0.4pt}

During the last decade atom chip approaches to quantum technology have become a powerful platform for scalable atomic quantum-optical systems,\cite{folman2002microscopic,fortagh2007magnetic,keil2016fifteen} with applications ranging from sensor and imaging technologies to quantum processing and memory.
Atom chips coupled to solid state-based quantum devices, e.g. superconducting qubits or nitrogen vacancy centers, are thereby paving the way for promising quantum simulation and computation schemes.\cite{patton2013hybrid,yu2016charge,kurizki2015quantum}.
Along this research line, several groups around the world have developed versatile atom chip configurations, which allow trapping of ultracold atomic clouds and degenerate Bose-Einstein condensates (BECs) close to chip surfaces and well-defined manipulation of their internal and external degrees of freedom.
Atom chips provide a very relevant technology for the emerging field of atomtronics,\cite{keil2016fifteen,amico2017focus,japha2016suppression,charron2006theoretical,stickney2007transistorlike,caliga2016transport,sinuco2014inductively} for which dynamic tunnelling barriers are required.\cite{japha2007using,salem2010nanowire} Such barriers may be formed on atom chips with $\mu$m-scale widths, matching the length scale dictated by the atomic deBroglie wavelength. The atom chip offers the ability to realize guides and traps with virtually arbitrary architecture and a multitude of novel architectures,\cite{beguin2019advanced} with a high degree of control over atomic properties, like interactions and spin, enabling new quantum devices.\cite{keil2016fifteen,amico2017focus}

Here we review progress in our groups in Beer Sheva, T{\"u}bingen and Singapore on recent developments in atom chip technologies.

\subsection{Progress towards on-chip interferometry}
The Ben-Gurion University of the Negev (BGU) Atom Chip Group (\url{http://www.bgu.ac.il/atomchip}) is promoting the idea of atomtronics without light. This entails circuits for atoms based on electric and magnetic traps, guides and tunneling barriers. The vision is for a complete circuit, including particle sources and detection, that makes no use of gravity, e.g. no time-of-flight for the development of interference fringes. This requirement means that a future technological device could work at any angle relative to gravity.

As a basis for this effort we use the Atom Chip technology developed over the past 20 years.\cite{folman2002microscopic,keil2016fifteen} An example of a circuit design we plan to implement is a continuous-wave, high-finesse Sagnac interferometer, where the multiple turns enabled by the guiding potential allow miniaturization of the loop while maintaining sensitivity to rotation.\cite{japha2007using} In the following we briefly present some of the work that has been done to advance the atomtronics technology.

To begin with, a stable tunneling barrier (in terms of instabilities of tunneling rates) should be no wider than the de-Broglie wavelength, which is on the order of 1 $\mu$m. Since the resolution with which we can tailor fields is on the order of the distance from the field source, one must construct the atomic circuit at a distance of no more than a few micro-meters from the surface of the chip.\cite{salem2010nanowire} At these very small atom-surface distances, several problems must be avoided:
\begin{enumerate}
  \item \textbf{Johnson noise}. This is a hindering process as it may cause spin-flips (reducing the trap/guide lifetime), as well as decoherence. In several papers we have shown ways to combat both effects either by the geometry or by the choice of material.\cite{dikovsky2005reduction,david2008magnetic,dikovsky2009superconducting,petrov2009trapping,salem2010nanowire} We have also measured Johnson noise and calculated its interplay with phase diffusion caused by atom-atom interactions.\cite{japha2016suppression}
  \item \textbf{Finite size effects}. As the atom-surface distance becomes smaller, so should the current-carrying wire width, or else the magnetic gradients will be severely undermined. Narrow wires require high-resolution fabrication\cite{salem2010nanowire} or thin self-assembled conductors such as carbon nanotubes.\cite{petrov2009trapping} 
  \item \textbf{Casimir-Polder and van der Waals forces}. As the atom-surface distance becomes smaller, the magnetic barrier between the atoms and the surface should be strong enough to avoid tunneling of the atoms to the surface. This has been calculated for sub-micron distances.\cite{petrov2009trapping,salem2010nanowire}
  \item \textbf{Fragmentation}. Due to electron scattering in the current-carrying wires (e.g. due to rough wire edges), the minimum of the trap, or guide, is not smooth and the atomic ensemble may split and exhibit a non-uniform density along the wire axis. This was studied by us both experimentally and theoretically.\cite{japha2008model,aigner2008long,zhou2014phase}
  \item \textbf{High aspect ratios}. {As the atom-surface distance becomes smaller, the trap or guide exhibits much higher transverse frequencies compared to the longitudinal frequency.} This brings about low dimensionality and can cause different problems such as phase fluctuations in a 1D BEC. Alternative wire configurations allow more flexibility for adjusting the trap aspect ratio.
\end{enumerate} 

In a proof-of-principle experiment\cite{zhou2016robust} we were able to avoid all the above hindering effects, and showed that spatial coherence could be maintained for at least half a second at an atom-surface distance of just 5 $\mu$m.

Another important problem that needs to be overcome is that of atom detection at very small atom-surface distances. At these distances of a few micrometers, the stray light from the nearby surface makes it very hard to achieve a reasonable signal-to-noise ratio for {\it in situ} detection with typical optical elements. As a solution, and also to avoid on-resonance spin-flips and decoherence, we studied the possibility of off-resonant atom detection with high-Q micro-discs.\cite{rosenblit2004single,rosenblit2006simultaneous,rosenblit2007design}

\begin{figure}
\includegraphics[width=0.45\textwidth]{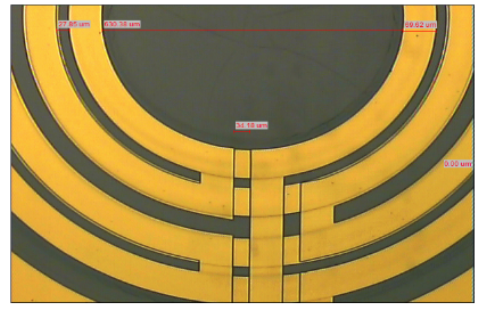}
\caption{\label{fig:BenGurion}{A multi-layer current-carrying chip produced at BGU for a Sagnac experiment, in co-operation with Thomas Fernholz and Peter Kr{\"u}ger at the University of Nottingham. Concentric gold conducting rings ($2\mu\textrm{m}$ thick, upper layer) are $70\mu\textrm{m}$ wide, with intervening gaps of $20-50\mu\textrm{m}$. The lower layer is also $2\mu\textrm{m}$ thick gold, with a $3\mu\textrm{m}$ thick SU8 insulating layer planarized to better than $0.4\mu\textrm{m}$. The entire device's outer diameter is 1.37mm. }}
\end{figure}

With the above tools we are now preparing to go forward with our vision for a Sagnac circuit,\cite{japha2007using} where as a first stage we have the goal of observing spatial coherence of atoms after one, and then several, turns in a guiding loop. The guiding potential is made in two alternative ways. The first method, using RF potentials, is being led by Thomas Fernholz of the University of Nottingham. It requires multi-layer chips (4 layers of currents), which are fabricated at BGU. Two such layers are shown in Fig.~\ref{fig:BenGurion}. The second effort also requires a unique chip. The guiding potential will be based on a repulsive permanent magnet potential in combination with an attractive electric field produced by a charged wire. The first experiments will be pulsed, whereby a BEC will be loaded onto the loop at the beginning of every cycle. Later on we will move towards realizing a continuous-wave version. We will first conduct the experiment in the pulsed mode by loading a thermal cloud, and later on use a 2D MOT as a continuous source.

Finally, let us note that quite a few groups around the world have realized free-space matter wave interferometry. It is now an important challenge to adapt these interferometers to the framework of atomtronics. Specifically, the BGU Atom Chip group has made significant steps in this direction by realizing, in the last 5 years, several types of novel interferometers which are not based on light. These interferometers are based on the magnetic splitting force (Stern-Gerlach) and they have already enabled the observation of spatial fringes,\cite{machluf2013coherent,margalit2019analysis} spin population fringes, \cite{margalit2018realization} unique T$^3$ phase accumulation, \cite{amit2019stern} clock interferometry,\cite{margalit2015self,zhou2018quantum} and the observation of geometric phase.\cite{zhou2019experimental} 

\subsection{Precision Sensing} 
\label{sec:Precision Sensing}


Precise sensors are one of the most important elements in applied and fundamental science. The use of quantum properties in sensing applications promises a new level of sensitivity and accuracy.\cite{degen2017quantum} Using cold atoms on atom chips as probes will enable many interesting applications.   

In the laboratories at the University of T{\"u}bingen we are working with atom chips that host one or two layers of lithographically implemented wire patterns. They allow the creation of spatially and temporally varying magnetic fields, as used for three-dimensional positioning and manipulation of cold atomic quantum matter.\cite{gunther2005combined} We typically use wire patterns made of gold in room temperature environments\cite{kraft2005atom} and superconducting patterns of Niobium in 4K and mK surroundings.\cite{bernon2013manipulation}

With such a 'carrier chip' for cold atoms on hand, we established a dual-chip process, where a second chip hosting nanostructured solid state systems is attached on top of the carrier chip.\cite{gunther2005combined} In this way, cold atoms can be efficiently coupled to other quantum systems and hybrid systems can be realized.

\begin{figure*}
\includegraphics[width=0.9\textwidth]{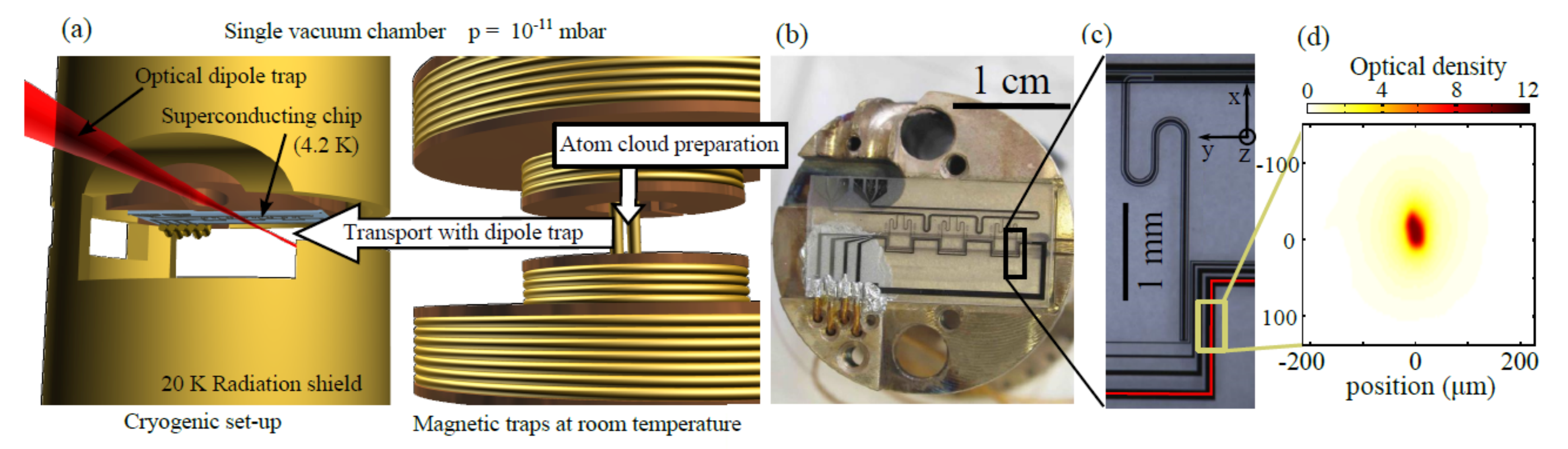}
\caption{\label{fig:Tubingen}{Superconducting atom chip at the University of T{\"u}bingen.} a) In vacuo trap setup with electromagnets for cold atom preparation (right) and cryostat with a superconducting chip at 4.2 K (left). b) Photograph of the chip holder and the superconducting atom chip. c) The chip features superconducting wire components for magnetic trapping and positioning of atomic clouds and a coplanar microwave cavity. d) Bose-Einstein condensate of 3x10$^5$ $^{87}$Rb atoms in a 15 ms time-of-flight image, released from the superconducting atom chip. Reprinted with permission from S. Bernon, H. Hattermann, D. Bothner, M. Knufinke, P. Weiss, F. Jessen, D. Cano, M. Kemmler, R. Kleiner, D. Koelle, and J. Fortágh, Nat. Commun.4, 2380 (2013). Copyright 2013 Nature Publishing Group.}
\end{figure*}

We have used this scheme to develop a novel cold-atom scanning probe microscope (CASPM), which uses ultracold atoms and BECs as sensitive probe tips for investigating and imaging nanoscale systems.\cite{gierling2011cold} Similar to an atomic force microscope (AFM), the probe tip is scanned across the surface of interest, while static and dynamical properties of the probe tip are monitored. Evaluating changes within the cold-atom tip density and motion then gives access to basic interactions and serves as a novel imaging and sensor technique. In contrast to conventional AFMs with their 'heavy and rigid' solid state tips, our CASPM uses a dilute gas of atoms, which not only allows for non-destructive measurements, but also for much higher sensitivity to external forces and fields. Inspired by conventional AFMs, we have been able to demonstrate several modes of operation.\cite{gunther2013cold} These include not only a contact mode, where we measure position-dependent losses of the probe tip, but also a dynamic mode, where we initiate a center-of-mass oscillation of the cold-atom tip and monitor the position-dependent changes of the probe tip oscillation frequency.\cite{gierling2011cold} Based on the latter, we have used cold-atom force spectroscopy to unveil anharmonic contributions in near-surface potentials. As in atomic force microscopy, this may be used to reconstruct the surface potentials. Moreover, we have developed a novel operation mode, not accessible to conventional AFMs, where we bring the dilute probe tip into direct overlap with the nano-object of interest. By measuring time-dependent probe tip losses, we have then been able to deduce the underlying van der Waals (Casimir-Polder) interactions.\cite{schneeweiss2012dispersion,jetter2013scattering} We have demonstrated and characterized all different operating modes of CASPM by measuring individual free-standing carbon nanotubes grown on a silicon chip surface. {Here we have shown that CASPM extends the force sensitivity of conventional AFMs by several orders of magnitude down to the yN regime, and the working distance up to several micrometers.}\cite{gunther2013cold} This makes CASPM a powerful tool for investigating fragile nano-objects with ultra-high force sensitivity.

While first measurements with CASPM suffered from long measurement times, we have just lately extended the microscope by a powerful single atom detection scheme. \cite{gunther2009observing,stibor2010single} It is based on continuous sub-sampling of the probe tip via a multi-photon ionization process in conjunction with temporally resolved ion detection and high quantum efficiency. This allows real-time monitoring of the probe tip dynamics and density while losing only few atoms from the probe tip.\cite{stibor2010single,menold2016dynamic} This not only speeds up probe tip oscillation frequency measurements by at least three orders of magnitude,\cite{stibor2010single} but also enables new applications for CASPM.

In one of these applications we proposed a quantum galvanometer to detect local currents and current noise in nanoscale mechanical quantum devices.\cite{kalman2012quantum,darazs2014parametric} Measuring the current noise would then give access to the quantum properties of the device. We successfully demonstrated the principal operating scheme of this galvanometer by coherently transferring artificially generated magnetic field fluctuations via a Bose-Einstein condensate onto an atom laser and investigating its single-atom statistics.\cite{federsel2015spectral,federsel2017noise} Employing second-order correlation analysis, we could not only extract the microwave power spectral density (current noise spectrum) but also the noise correlations within the bandwidth of the BEC, which will give access to the quantum noise properties of the current source. This will extend CASPM to a promising quantum sensor, not only for detecting local forces and force gradients, but also for currents as well as electric and magnetic fields (AC and DC), including their specific noise spectra.

\subsection{Cryogenic Atom Chips and Hybrid Quantum Systems}
\label{sec:CryoChip}

Atom chips made from superconducting circuits offer certain advantages over normal metal devices. The coherence properties of trapped atoms are improved by orders of magnitude due to reduction of magnetic noise coming from the surface of the chip. Moreover, superconductors can be operated in the mixed state, where vortices can be used to generate self-sufficient atom traps. In addition to that, working in cryogenic environments offers the possibility to interface atoms with solid state devices to form hybrid quantum systems.\cite{dikovsky2009superconducting}    

Besides atom chip experiments in room-temperature environments, the group in T{\"u}bingen also operates superconducting atom chips with trapped BECs of rubidium atoms. \cite{bernon2013manipulation,weiss2015sensitivity,cano2011experimental} As shown in Fig.~\ref{fig:Tubingen}, condensates are routinely transferred into coplanar cavity structures\cite{bothner2013inductively} and the measured coherence time between hyperfine ground state superpositions reaches several seconds. Microwave dressing is used to suppress the differential shift of state pairs with the “double-magic point” being the optimum working point for quantum memories \cite{sarkany2014controlling}. We have successfully demonstrated coherent coupling of a hyperfine state pair through a driven superconducting coplanar microwave cavity,\cite{hattermann2017coupling} which paves the way for future cavity-based quantum gate operations.

\begin{figure*}[t]
\includegraphics[width=0.8\textwidth]{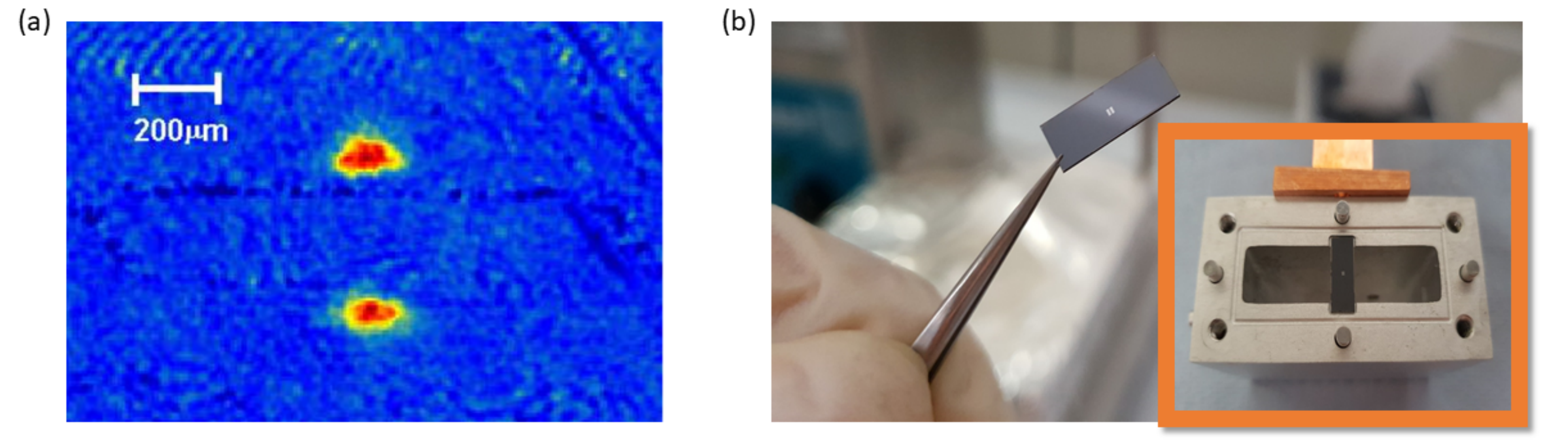}
\caption{\label{fig:Singapore}(a){Image of a thermal atomic cloud of $10^3$ $^{87}$Rb trapped in a self-sufficient quadrupole magnetic trap generated by vortices on a 1mm x 1mm square of superconducting YBCO (National University of Singapore).} The lower part of the image shows a mirror image of the cloud, caused by the reflection of the imaging beam from the chip surface.   Reprinted with permission from M. Siercke, K. S. Chan, B. Zhang, M. Beian, M. J. Lim, and R. Dumke, Phys. Rev. A 85, 041403 (2012). Copyright 2012 American Physical Society. (b) Setup for a hybrid quantum system of atoms and superconductors. The image shows a transmon qubit fabricated on a silicon substrate. Inset: the transmon installed in a 3D microwave cavity. The cavity allows to read-out and manipulate the transmon with RF radiation. In the future we plan to bring atoms into the cavity to form a hybrid quantum system.}
\end{figure*}

In addition to manipulating ground-state atoms we have successfully implemented two-photon Rydberg excitation in a cryogenic environment near the superconducting chip.\cite{mack2015all} We have developed techniques for optical detection of Rydberg populations and coherences \cite{karlewski2015state} and measured the increased lifetime of Rydberg states in cryogenic environments.\cite{mack2015all} In preparation for interfacing Rydberg atoms with superconducting circuits, we have obtained high-resolution spectra of rubidium Rydberg states in a field-free vapor cell as reference,\cite{mack2011measurement} and in precisely controlled electrostatic fields\cite{grimmel2015measurement} near surfaces at room and cryogenic temperatures. These studies add to our understanding of electrostatic fields of surface adsorbates that build up during experiments with cold atoms at chip surfaces. \cite{hattermann2012detrimental,hermann2014long,tauschinsky2010spatially} Based on the measured data, quasi-classical quantum defect theory,\cite{sanayei2015quasiclassical} Stark-map calculations,\cite{grimmel2015measurement} suitable dressing techniques,\cite{sarkany2014controlling,jones2013rydberg} and numerical methods developed for simulating quantum operations in the presence of thermal cavity photons,\cite{sarkany2015long} we are currently focusing on the coherent manipulation of Rydberg atoms and quantum computation schemes in the presence of inhomogeneous fields at the surface of superconducting coplanar cavities. 

The realization of hybrid quantum systems based on atoms and superconducting qubits requires truly cold temperatures in the 10 mK range, as dictated by the otherwise fast decoherence of the superconducting qubit. The great advantage is that at this temperature the number of microwave photons in the cavities that mediate the interaction between the solid state and the atomic system is near zero. The price to pay is a highly complex experimental system combining cold-atom technologies with a $^3$He/$^4$He dilution refrigerator.\cite{jessen2013trapping,landra2019design} Our dilution refrigerator consists of several temperature-shielded volumes (stages), of which we use the 6K-stage and the 1K-stage for cold atom experiments. The 1K-stage includes a cold plate with a nominal base temperature of 25 mK.\cite{weiss2015sensitivity} We routinely operate a magneto-optical trap at the 6K-stage, from which we transport magnetically trapped, ultra-cold rubidium clouds at 100 $\mu$K to the 1K-stage. The 1K-stage has a sufficiently large volume (several liters) to accommodate microwave cavities, such as coplanar waveguide cavities, and has convenient optical access for optical traps and laser beams for spectroscopic measurements. This experimental setup is currently being extended for studying the fully quantum regime of cold-atom superconductor hybrid systems.


In the Singapore group we are working in two directions. One is the exploration of superconducting atom chips using high-temperature superconductors and another is the development of coherent interfaces between superconducting circuits and ultracold atoms.

High temperature superconductors have various distinct properties when implemented as atom chips. First of all, the technical demands are lower due to the higher working temperatures, which can be reached with liquid nitrogen instead of liquid helium. Moreover, high temperature superconductors are type-II superconductors and allow the storage of magnetic fields in the remanent state. We have shown experimentally and in simulations, that these trapped fields can be used to generate novel traps for ultracold atoms.\cite{muller2010programmable,zhang2010design} Ramping a magnetic field perpendicular to a planar structure of YBCO we were able to generate various magnetic traps for cold atoms (see Fig. \ref{fig:Singapore}(a)).\cite{Muller2010trapping} These traps can be generated either by using external magnetic fields together with vortices or in a completely self-sufficient way, where the trap is solely created by vortices. In the latter case, low noise potentials can be generated, as there is no technical noise coming from external power sources and the noise coming from the movement of vortices is expected to be an order of magnitude less than Johnson noise in normal conductors.\cite{dikovsky2009superconducting}

Another property of vortices in superconducting thin films is that their distribution can be manipulated with light. Heating parts of the superconductor will result in a force on the vortices, which shifts the position of the vortices and consequently changes the vortex distribution. We have used this effect to generate various trap patterns with a thin square of superconducting YBCO, using light patterns generated by a spacial light modulator. \cite{tosto2019optically} The advantage of this technique is that multiple trap geometries can be generated with the same chip architecture in-situ, without the need of changing the chip and breaking the vacuum of the ultra-high vacuum chamber. 

Aside from using superconducting chips exclusively to manipulate cold atoms, we are also working on interfaces between cold atoms and superconducting qubits fabricated on the superconducting chip. What we envision here is the coherent transfer of quantum states between cold atoms and qubits made of superconducting integrated circuits. These hybrid systems will have many application, like the transduction of quantum states between the microwave and optical regime or the creation of universal quantum computing devices.

As mentioned before in this article, the practical implementation of a hybrid atom-superconducting qubit system is technically challenging. In Singapore we decided to bring cold atoms inside the dilution refrigerator by magnetically transporting them from a room temperature vacuum chamber directly to the mK stage of the refrigerator. With this technique we are able to bring clouds of 5x10$^8$ $^{87}$Rb atoms close to the mK stage, at a base temperature of 70 mK.\cite{landra2019design} Trapped inside the mK stage, the atomic cloud exhibits an exceptional lifetime of 13 minutes, which is a promising starting point for future experiments.   

In order to couple atoms and superconducting circuits a few scenarios are possible, which can be categorized in indirect and direct coupling. {Also, the state of the atoms i.e. ground state or highly excited (Rydberg), will have a significant influence on the experimental parameters.} When indirectly coupled, the qubit and atom are individually coupled to a resonator, which mediates the interaction. In this case the coupling of the resonator to the qubit is easily implemented and can reach the strong coupling regime. Coupling ground state atoms to a planar resonator is an ambitious task. It was shown that the coupling strength of a single atom is only 40 Hz at a resonator-atom distance of 1 $\mu$m.\cite{verdu2009strong} In order to reach strong coupling one consequently needs to collectively couple an ensemble of ~10$^6$ atoms to the waveguide, which is experimentally challenging. Using Rydberg states can considerably relax these requirements. We have shown that for Rydberg states strong coupling can be achieved with even a single atom. \cite{yu2016superconducting,yu2017superconducting} {The strong coupling can even be reached with atom-resonator distances of tens of micro-meters, when using the fringe field of the capacitive part of the resonator to couple the atom.}

When using Rydberg atoms, even directly coupling of atoms to charge qubits can be realized. 
A neutral atom placed inside the gate capacitor of a charge qubit acts as a dielectric medium and affects the gate capacitance, resulting in a modulation of the charge-qubit energy bands. Moreover, the local quasi-static electric field strongly depends on the charge-qubit state, leading to different DC Stark shifts of atomic-qubit states. We have shown that in such a setup quantum states can be transferred between the two qubits and CNOT and Hadamard gates can be realized.\cite{yu2016quantum} Schemes for Rydberg atoms interacting with flux qubits have been theoretically proposed to realize quantum memories\cite{patton2013ultrafast}.

We think that we now have the tools at hand to interface cold atoms with superconducting circuits. In the near future we would like to first couple atoms to 3D transmons, see Fig. \ref{fig:Singapore}(b). For this we designed and tested superconducting 3D cavities that have free space access for the transport and optical manipulation of cold atoms. First experiments to transport atoms inside the cavity are currently under way. At the same time we are developing our own fabrication for superconducting qubits. First chips have already been fabricated and tested. With both systems at hand we can then go forward to build hybrid systems of cold atoms and superconducting circuits.

\subsection{Concluding remarks and outlook}
\label{sec:Outlook-hybrid}

In this review we have described applications of atom chips in atomtronics, precision sensing and quantum information. We illustrated the state of the art in these topics and touched upon future prospects and utilizations. In this zoomed-in view, we omitted many other excellent activities in the field, due only to unavoidable space limitations. Here, we would like to bring up other achievements that will shape the future of the atom chip platform.

Most of the experimental studies described above used bosonic rubidium atoms. In fact, many other species, like fermions, molecules and ions are used in atom chips.\cite{keil2016fifteen} Fermions are another one of the fundamental building blocks of matter and therefore highly interesting objects to study, including low dimensional physics, the interaction of fermions with different species, or spin physics.\cite{extavour2011atom} 

Molecules, as the bridge between physics and chemistry, are an additional compelling candidate for many studies. Implementations range from fundamental science, like the measurement of the electric dipole moment and parity violation, to applied science in quantum processing. A 'Lab on a Chip' for molecules is thus a sought-after goal. Recently, the trapping of simple molecules on microchips was realized,\cite{meek2009trapping,santambrogio2015trapping} opening the way for many interesting investigations.

Trapped ions are one of the most promising candidates for practical quantum computing. In order to control and measure a large number of ions it will be necessary to fabricate surface-electrode traps on miniaturized microchips. The development and integration of these chips is currently ongoing and will be a major part in the future development of scalable quantum computer architectures with ions.\cite{bruzweicz2019trapped} 

Using the wave nature of atoms, atom chips will in future be used as precise sensors for material research and fundamental science. So-called 'quantum gas microscopes' have been developed for room-\cite{aigner2008long} and cryogenic\cite{yang2017scanning}-temperature environments and are ready to be used in the nontrivial studies of unique materials. At the same time, matter waves are being employed for precision measurements in atomic interferometers. By analogy to the optical interferometer, the splitting and recombination of matter waves on atom chips are, for instance, being used to test theories in quantum thermodynamics\cite{schmiedmayer2018thermodynamics}, quantum many-body physics\cite{schweigler2017experimental}, and find applications in gravitational sensing\cite{mdeangelis2011isense}.  

Intimately connected with precision sensing is the field of fundamental science. Many studies will be possible with atom chips, including tests of the Weak Equivalence Principle,\cite{herrmann2012testing} interactions of antihydrogen with matter and gravity,\cite{leefer2016investigation} non-Newtonian gravity, and the search for a fifth fundamental force.

All these examples show that atom chip technology has a bright future ahead. Combined with further integration and miniaturization, atom chips will play a role in many areas, both in fundamental research, as well as practical measurements. 

{\it Acknowledgments}
All groups are very thankful to their colleagues who have been working with them on the experiments and their interpretation.

The BGU work has been supported by the Israel Science Foundation, the Deutsche Forschungsgemeinschaft German-Israeli DIP program, the FP7 program of the European Commission, the Israeli Council for Higher Education, and the Ministry of Immigrant Absorption (Israel).

The T{\"u}bingen research team gratefully acknowledges financial support from the Deutsche Forschungsgemeinschaft through SPP 1929 (GiRyd) and through DFG Project No. 394243350 and 421077991.

The Singapore team is grateful for administrative assistance and financial support from the Centre for Quantum Technologies (CQT) in Singapore. 
\section{QUENCH DYNAMICS OF INTEGRABLE  MANY-BODY SYSTEMS}
\label{NonEqDyn}
\vspace*{-0.5cm}
\par\noindent\rule{\columnwidth}{0.4pt}
{\bf{\small{N. Andrei and C. Rylands}}}
\par\noindent\rule{\columnwidth}{0.4pt}


The study of non-equilibrium quantum physics is currently at the intellectual forefront of con-
densed matter physics. One-dimensional systems in particular provide an exciting arena where over
the last decade significant advances in experimental techniques have allowed very precise study of an
array of nonequilibrium phenomena and where a number of powerful theoretical tools were developed
to describe these phenomena. Here we give a brief account of a few systems that are described by
one dimensional integrable Hamiltonians, the Lieb-Liniger model and the Heisenberg chain and how
integrability gives access to the study of some of their local and global nonequilibrium properties.

While the principles of equilibrium statistical mechanics
are well understood and form the basis to describe a variety of phenomena,
there is no corresponding framework for the non-equilibrium dynamics,
although efforts to fully understand the underlying principles
extend back to Boltzmann and beyond. Solving particular models numerically or analytically and comparing to experiments may illuminate bits of the puzzle.

Here, is an extended version of talks given by the first  author at  Atomtronics 2019 at Benasque where  some aspects of the questions were discussed. It is based on a review article \cite{rylands2020nonequilibrium} written with Colin Rylands and builds on work carried out with several collaborators:  Deepak Iyer,  Garry Goldstein,  Wenshuo Liu, Adrian Culver, Huijie Guan  and Roshan Tourani to whom we are  very grateful for many enlightening and useful discussions.

\subsection{Quench Dynamics}
A convenient  protocol  to observe  a system out of equilibrium  is  to prepare it  in some initial state $\ket{\Phi_i}$, typically an eigenstate of an initial Hamiltonian $H_i$, and then allow it  to evolve in time using another Hamiltonian, $H$ for which $\ket{\Phi_i}$ is not an eigenstate
\cite{calabrese2002evolution, mitra2018quantum, andrei2016quench}.
One  then follows  the correlations of local observables,
\begin{eqnarray}\label{Psit}
   \langle  \Phi_i| \,    e^{iHt} \; \{  \mathcal{ O}_1(x_1) \mathcal{O}_2(x_2) \dots \} \,e^{-iHt} \, |\Phi_i\rangle
\end{eqnarray}
as they evolve. One may be interested  to  know  what new properties characterize the system, whether a dynamical phase transition occurs at some point in time \cite{heyl2013dynamical} or how its entanglements evolve. A particularly important question that arises in this context is whether the system thermalizes.
{Namely, can the system act as a bath to a small subsystem, here the small subsystem is the segment that contains  the local operators $ \mathcal{ O}_j(x_j)$.  In the long time limit (to thermalize it is necessary that $v t \gg L$ where $L$ is the size of the system, and $v$ a typical velocity)  one needs to show,}
\begin{eqnarray}\nonumber
\lim_{t \to \infty}     \langle  \Phi_i| \,    e^{iHt} \; \{  \mathcal{ O}_1(x_1) \mathcal{O}_2(x_2) \dots \} \,e^{-iHt} \, |\Phi_i\rangle  \\\label{therm}
= \Tr \;  \;e^{- \beta H}   \{   \mathcal{ O}_1(x_1) \mathcal{O}_2(x_2) \dots\}/Z
\end{eqnarray}
with the final inverse temperature $\beta$ determined by the initial energy, $E_0= \langle  \Phi_i| H |\Phi_i\rangle = \Tr \, e^{-\beta H}H /Z $.
 
Also global  properties are of interest. These are commonly studied via the  Loschmidt amplitude (LA),  the overlap between the initial state with its time evolved self, conveniently expressed using a complete set of energy eigenstates, $|n\rangle$: 
\begin{eqnarray}\label{LA}
\mathcal{G}(t)= \langle \Phi_i\,| e^{-iHt}\, |\,\Phi_i\rangle=\sum_n |\braket{n\, |\,\Phi_i}|^2e^{-i E_nt}
\end{eqnarray}
and its Fourier transform,
\begin{eqnarray}\label{P}
\mathcal{P}(W)&=&\int_{-\infty}^\infty\frac{\mathrm{d}t}{2\pi} e^{iWt}\mathcal{G}(t)= \nonumber\\
&=& \sum_n\delta(W- E_n)) |\langle n| \Phi_i\rangle|^2
\end{eqnarray}
which measures the work distribution done during the quench \cite{goold2018role}.
 
\subsection{Evolution under integrable Hamiltonians}

We shall consider   evolutions  effectuated by post-quench Hamiltonians that  are integrable, namely  Hamiltonians admitting a complete set of  eigenstates $| n \rangle$   and eigen-energies  $E_n $  given  by the  Bethe Ansatz.  The ability to obtain these follows from the existence of an infinite set of local  charges, $\{Q_n, n=1...\infty \}$,  that commute with the Hamiltonian and  constrain the time evolution leading to  a generalized Gibbs ensemble $e^{-\sum_n \beta_n  Q_n}$ with the final inverse temperatures $\beta_n$ determined by the initial values $ q_n^0=  \langle  \Phi_i| Q_n |\Phi_i\rangle$ \cite{rigol2008thermalization}. 

Thus  some features of integrable time evolution are  non generic. It  turns out however that   many  features observed in integrable models can also be observed when integrability is broken. An example is  the ''dynamical fermionization"  of  repulsively interacting bosons in the integrable Lieb-Liniger Hamiltonian,  discussed below. We showed this feature  can be also observed in the bose-Hubbard model, the  lattice version of the Lieb-Liniger model, which is not integrable \cite{iyer2013exact}. Further,  many  systems, in particular  ultra-cold atom systems, are actually described by integrable Hamiltonians and can therefore be studied as such. Here we discuss two of them. 

\subsubsection{The Lieb-Liniger model}
The model describes systems of  ultracold  gases of neutral bosonic atoms moving in one dimensional  traps and interacting  with each other via a local density  interaction of strength $c$ which can be repulsive $c>0$ or attractive $c<0$. Aside from being an excellent description of the experimental system, it is one of the simpler Hamiltonians for which there exists an exact solution via Bethe Ansatz. The Lieb-Liniger Hamiltonian reads
\begin{equation}
H=-\int\mathrm{d}x\,\Psi^\dag(x)\frac{\partial_x^2}{2m}\,\Psi(x)+c\int\mathrm{d}x\,\Psi^\dag(x)\Psi(x)\Psi^\dag(x)\Psi(x)
\end{equation}
(setting $\hbar=1$). Here $\Psi^\dag(x),~\Psi(x)$  create and annihilate  bosons  of mass $m$. The exact $N$-particle eigenstate is given by  \cite{lieb1963exact, lieb1963exact2},
\begin{eqnarray}\nonumber
\ket{\{k_j\}}=\int\mathrm{d}^Nx\,&&\prod^N_{\substack{i,j=1\\ i<j}}  [  \theta(x_i -x_j) +s(k_i,k_i)    \theta(x_j -x_j)] \times  \\
\label{Bethestates} &&\quad  \quad \times\prod_{l=1}^Ne^{ik_lx_l}\Psi^\dag(x_l)\ket{0}.
\end{eqnarray}
Here $s(k_i,k_i)=\frac{k_i-k_j+ic}{k_i-k_j+ic}= e^{i\varphi(k_i-k_j)}$ is the two particle  scattering matrix,  $\varphi(k)=2\arctan{(k/c)}$ is the  phase shift.  The  single particle momenta  $k_j$ are unrestricted in the infinite volume limit  while with periodic boundary condition on a line segment $L$  they must satisfy the Bethe Ansatz equations:  $k_iL= \sum_{j=1}^N \varphi(k_i-k_j)+ 2 \pi n_i$, with the integers  $n_i$ being the quantum numbers of the state.  The single particle momenta are related to the conserved charges by $q_n=\sum_{j=1}^Nk_j^n$, in particular the energy  is given by,  $q_2 = E=\sum_{j=1}^Nk_j^2$.  

This set of eigenstates allows the study of time evolution through the partition of the unity, $
 \mathbb{1}_N=\sum_{k_1... k_N}  \frac{\ket{\{    k_j\}}\bra{\{k_j\}}}{\mathcal{N}(\{k_j\})}$.
 Here  $  \mathcal{N}(\{k\})=\det\left[\delta_{jk}\left(L+\sum_{l=1}^N\varphi'(k_j-k_l)\right)-\varphi'(k_j-k_k)\right]$ is a normalization factor. 
 
 In terms of  the partition identity, the time evolved wavefunction is given by,
 \begin{eqnarray}
| \Phi(t)\rangle&=&e^{-iHt} \,|\Phi_i\rangle  \nonumber \\
&=&\sum_{k_1... k_N}    e^{-i t\, (\sum_{j=1}^N   k_j^2)}    \ket{\{    k_j\}}  \bra{\{k_j\}} \Phi_i \,\rangle
\end{eqnarray}
with the initial state $|\Phi_i\rangle$ encoded in the overlaps $ \bra{\{k_j\}} \Phi_i \,\rangle$.  These overlaps  have been studied by many groups, see e.g.\cite{caux2012constructing}
and are typically very difficult to calculate. Once these overlaps are known they can be put in exponential form and combined with matrix elements of a given operator to yield
a quench action which is typically evaluated in the saddle point
approximation \cite{caux2013time}.
\begin{figure}
\includegraphics[trim=2cm  2cm 2cm 0cm ,clip=true, width=.45\textwidth]{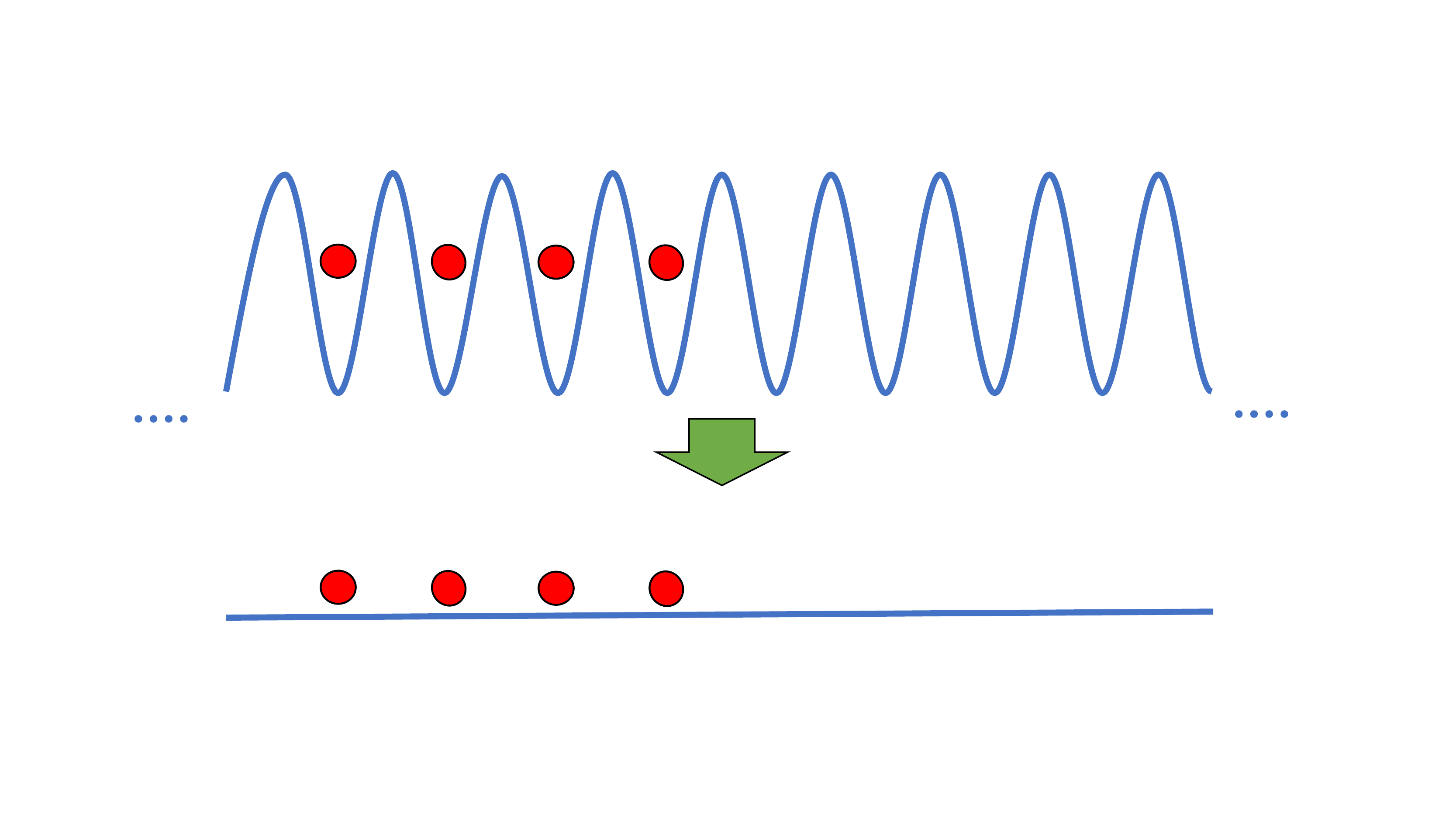}
\caption{The domain wall initial state :  A cold atom gas is held in the left part of a deep optical lattice, extending from $j=-\infty$ to $j=0$. This is then removed and the gas is allowed to expand.
The system is open - its size is large compared to time of evolution}\label{FigDomain}
\end{figure}  

\bigskip


\noindent{\bf Beyond overlaps:}
One may get around the difficulty of computing overlaps  by choosing an alternate form of the partition identity  obtained by
exchanging  the ordering in momentum space  for ordering in coordinate space leading to  the Yudson  representation of the partition of the unity  \cite{yudson1984rigorous, goldstein2013equilibration}, or equivalently choosing appropriate trajectories for integrating over the momenta, see \cite{iyer2013exact},
\begin{eqnarray}\label{IdentityY}
\mathbb{1}_N=\sum_{k_1,\dots,k_N}\frac{\ket{\{k_j\}}\bray{\{k_j\}}}{\mathcal{N}(\{k_j\})}.
\end{eqnarray}
Here we have introduced the notation $\kety{\{k_j\}}$ (notice the parenthesis replacing the ket) to denote  the Yudson state,
\begin{eqnarray}\label{Yud}
\kety{\{k_j\}}=\int\mathrm{d}^Nx\,\theta(\vec{x})\prod_l^N e^{ik_lx_l}\, \Psi^{\dagger}(x_l)\ket{0}
\end{eqnarray}
with $\theta(\vec{x})$ denoting a Heaviside function which is non zero only for the ordering $x_1>x_2>\dots >x_N$. The Yudson state is  simpler to work with than the full eigenstates of the model and its overlaps with the initial state can be  readily calculated, particularly  if the initial state is ordered in coordinate space.

\smallskip

\noindent{\bf The domain wall initial state:  }   As an example  we consider an initial state in the form of a  domain wall and quench it with $c>0$ Lieb-Liniger Hamiltonian.  Its time evolution can be studied analytically and several interesting phenomena will be shown to emerge:
 
 {\it 
  
-  Nonequilibrium Steady state (NESS)

-  RG flow in time

-  Evolution along space-time rays

- Hanbury Brown-Twiss  effect

- Dynamical fermionization}

  \bigskip
  
\begin{figure}
 \includegraphics[trim=4cm  9cm 4cm 0cm ,clip=true, width=.55\textwidth]{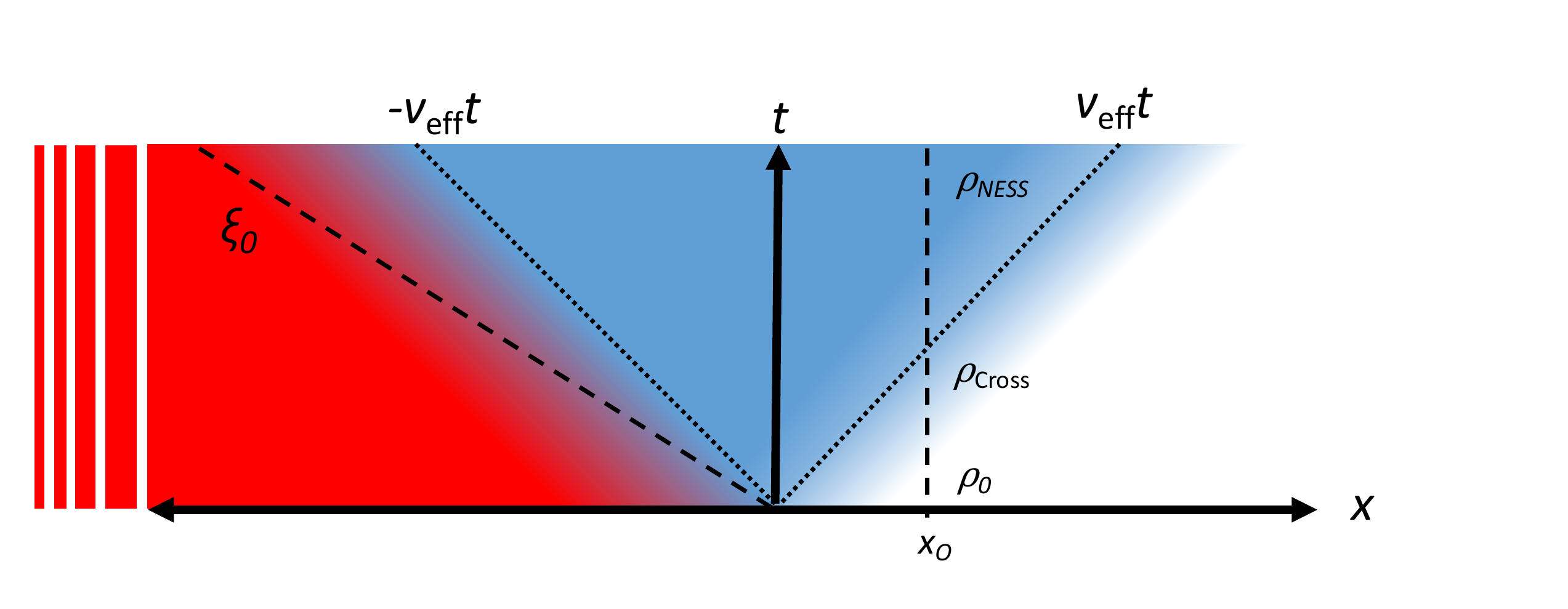}
\caption{ At long times a non-equilibrium steady state (NESS) is established as depicted on the right. Measuring the density at $x=x_0$  one will see the initial density $\rho_0$ change to the crossover regime $\rho_\text{Cross}$ at intermediate times  ending up as time and space independent value $\rho_\text{NESS}$  which encodes the interaction and the initial quench. Reprinted with permission from C. Rylands and N. Andrei, Annu. Rev. Condens. Matter Phys. 11, 147–168 (2020). Copyright 2020 Annual Reviews.} \label{Figlightcone}
\end{figure}

   The  initial state, as depicted in Fig. \ref{FigDomain},  consists $N$  cold atom bosons  held in a very deep optical lattice of length $L$ with $N,L \to \infty$ and  $  \delta  =L/N$ held fixed.  The lattice site  $\bar{x}_{j}=j \delta, \; 
   j= -\infty...-1, 0, 1... +\infty $, are filled with  
one boson per site  in the left half of the lattice:     $j=-\infty$ to $j=0$,  and none in the half to the right,
\be
\label{PSI0}
\ket{\Phi_i}= \int\mathrm{d}^Nx\prod_{j=-\infty}^0 \left[\frac{\omega}{2\pi}\right]^{\frac{1}{4}}e^{-\frac{\omega}{4}(x_j-\bar{x}_j)^2}\Psi^\dag(x_j)\ket{0}.
\ee
The quench consists of suddenly releasing the trap and allowing  the bosons to interact and  evolve under the Lieb-Liniger Hamiltonian. Time evolving the system and using the Yudson representation  we find,
\begin{equation}\label{PSIt}
\ket{\Phi_i(t)}=  \left[\frac{8\pi}{\omega}\right]^{\frac{N}{4}}\sum_{k_1,\dots,k_N}\frac{e^{-\sum_{j=1}^N\left[\frac{1 }{\omega}\left(1+i\omega t\right)k_j^2+ik_j\bar{x}_j\right]}}{\mathcal{N}(\{k_j\})}\ket{\{k_j\}}.
\end{equation}
When the lattice is removed the gas  expands and the particle density will become nonzero between the lattice sites and also to the right of the domain wall. 
In the vicinity of the domain wall particles will begin to vacate the left hand side of the system and populate the right hand side, see Fig. \ref{Figlightcone}.  The effects of this  quench can only be felt within a "light-cone"  centered at the edge  and determined by a finite effective velocity, $v^\text{eff}$ which depends upon $\omega$. On the right, $x\gg v^\text{eff}t$ the density will remain zero while to the left, $x\ll -v^\text{eff}t$, the average density will remain $1/\delta$ -  the effects  of the quench are still felt as the initially confined bosons will expand and begin to interact with each other. 

 We first examine the local portion of the quench around the domain wall. Since to the left there is an infinite particle reservoir and to the right an infinite particle drain the system will never equilibrate, however at long times a non-equilibrium steady state (NESS) consisting of a left to right particle current is established.
This can be investigated by  computing the expectation value  of the density $
\rho(x,t)=\matrixel{\Phi_i(t)}{\Psi^\dag(x)\Psi(x)}{\Phi_i(t)}$.
Utilizing the known formulae for the matrix elements of the density operator with Bethe eigenstates  \cite{korepin1997quantum} this can be calculated exactly. To the right of the domain wall, at long times and to leading order in $1/c\delta$ three regions emerge \cite{goldstein2013equilibration}
\begin{widetext} 
\begin{eqnarray}
\rho(x,t) =\begin{cases}
\rho_\text{NESS}=\frac{1}{2\delta}-\frac{4\pi}{c\delta^2}
&\frac{1}{\sqrt{\omega}}\ll x\ll v^\text{eff} t\\
\rho_{\text{Cross}}(x)=\frac{1}{\delta}f
+\frac{16}{\pi c\delta^2}\left[e^{-\frac{x^2}{\sigma}}\frac{x\sqrt{\pi}}{\sqrt{\sigma}}f-\frac{1}{2}e^{-2\frac{x^2}{\sigma}}+\frac{\pi}{2}f(1-f)\right]&x\sim v^\text{eff}t\\
\rho_0=0&x\gg v^\text{eff} t
\end{cases}
\end{eqnarray}
\end{widetext}
where $f=f(x,t)=\frac{1}{2}\text{erfc}\left(\frac{x}{\sqrt{\sigma(t)}}\right)$ and $\sigma(t)=\frac{t^2\omega}{2}+\frac{2}{\omega}$. Far to the right $x\gg v^\text{eff}t $ we see that the density vanishes while closer to the light-cone some complicated crossover behavior occurs. Since the model is Galilean rather than Lorentz invariant the light-cone is not sharp giving instead this crossover regime. {Most interesting is the region deep inside the light-cone in which the density becomes independent of $x,t$, signifying the appearance of the NESS, Nonequilibrium Steady State. We note that the particle density in this regime, $\rho_\text{NESS}=\frac{1}{2\delta}-\frac{4\pi}{c\delta^2}$, is reduced as compared to the equilibrium value, $\rho_\text{EQL}=\frac{1}{2\delta}$, the value a closed system would have reached  after a quench from a domain wall state. This is a nonequilibrium effect of an open system which allows the bosons to expand further to the right  in response to the repulsive interactions among the bosons. It follows from the order of limits with the size of the system $L$ satisfying  $L \gg v t$, to be contrasted with the behavior in a closed system, with the opposite order of limits.
Within this region all local properties of the system can be calculated by taking the expectation value with respect to this NESS, $\left<\mathcal{O}(x,t)\right>=\matrixel{\Psi_\text{NESS}}{\mathcal{O}}{\Psi_\text{NESS}}$ where $\ket{\Psi_\text{NESS}}$ can be determined by taking the appropriate limit of \eqref{PSIt}.}

\begin{figure}
  \includegraphics[width=.48\textwidth]{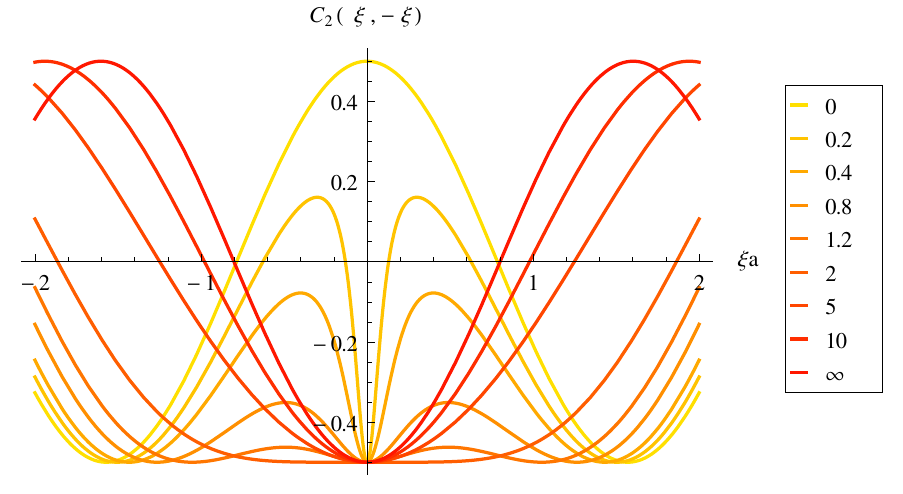}
\caption{ The noise correlation function $C_2(\xi,-\xi)$ , as a function of $\xi=x/\tau$ at long times for a quench from a lattice like initial state. For arbitrary values of $c>0$, with $\delta$ fixed, the system develops a distinct fermionic dip at the origin. Reprinted with permission from D. Iyer, H. Guan, and N. Andrei, Phys. Rev. A 87, 053628 (2013). Copyright 2013 American Physical Society.} \label{FigNoise}
\end{figure}

On the left portion of the lattice   $x\ll -v^\text{eff}t$ we are outside the light-cone, the system is unaffected by the domain wall portion of the quench and the lattice  translational invariance is restored. At long times  the density  within this region is,
\begin{eqnarray}
\rho(x,t)=\frac{1}{\delta}\left[1+\sum_{s=1}^\infty e^{-\sigma(t)\frac{\pi^2s^2}{\delta^2}}\cos{\left(\frac{2\pi s x}{\delta}\right)}\right]
\end{eqnarray}
which describes small oscillation about a uniform density of $1/\delta$. 

This result  coincides with what one would expect for a quench from a lattice initial state of the Tonks-Girardeau (TG) gas, the $c\to\infty$ limit of the  LL model. To understand this one should go beyond the density and compute the normalised noise correlation function $C_2(x,x')=\frac{\rho_2(x,x',t)}{\rho(x,t)\rho(x',t)}-1$ where
\be
\rho_2(x,x',t)=\matrixel{\Phi_i(t)}{\Psi^\dag(x)\Psi(x)\Psi^\dag(x')\Psi(x')}{\Phi_i(t)}.
\ee
This correlation function is related to the Hanbury Brown-Twiss  effect and will detect the nature of the interactions between particles, a peak indicating bosons while a dip indicates fermions \cite{hanburybrown1956test, jeltes2006hanbury}. Computing the  noise correlation function   $\rho_2(x,-x,t)$    by inserting two copies of the identity and evaluating the integrals at long time by saddle point method \cite{iyer2012quench}  one finds  it becomes a function only of the ray variables $\xi=x/t$ (measured with respect to $\xi_0=x_0/t$  see Fig. \ref{Figlightcone}). For sufficiently long times $\xi\sim 0$ a distinct fermionic dip is seen for arbitrary $c>0$  while $c=0$ shows a bosonic peak, the turn over to the dip occurring on the time scale, $t\sim c^{-2}$, see
Fig. \ref{FigNoise}.  This turn over results from an increase in time  of the  effective coupling constant $c$ -  starting from any initial repulsive value it will flow to strong coupling in the long time limit \cite{jukic2010reflection,  jukic2008free}. {This follows elegantly from the Yudson representation of the time evolving wave function \cite{iyer2012quench}: rewriting  the dynamic factor in Eq. (\ref{Bethestates}), $\theta(x_i -x_j) +s(k_i,k_i)    \theta(x_j -x_j)$, as   $\frac{k_i-k_j- ic\, sgn(x_i-x_j)}{k_i-k_j- ic}$,  we note it tends to   $sgn(x_i-x_j)$ upon  rescaling $k^2_j t \to k_j^2$.  Therefore, the product of  bosonic fields with the  dynamic factors, $\prod_{i<j} sign(x_i-x_j) \prod_j \Psi^\dag(x_j)$, behaves fermionically. The physical argument underlying the mathematical manipulations is also simple. In the long time limit only the slow bosons remain around $x,x'$ in the noise correlation function $\rho_2(x,x',t)$ and they interact via the effective S-matrix $S^{ij} \to -1$. The system in the long time limit will then behave as if it consisted of non interacting fermions. This dynamical fermionization, the development of fermionic-like  correlations,  was  subsequently observed in experiment both in the integrable Lieb-Liniger system (the Weiss group 2020) and previously in the  corresponding lattice version, the Bose-Hubbbard model (the Greiner group 2015) \cite{wilson2020observation}.}

\begin{figure}
  \includegraphics[width=.5\textwidth]{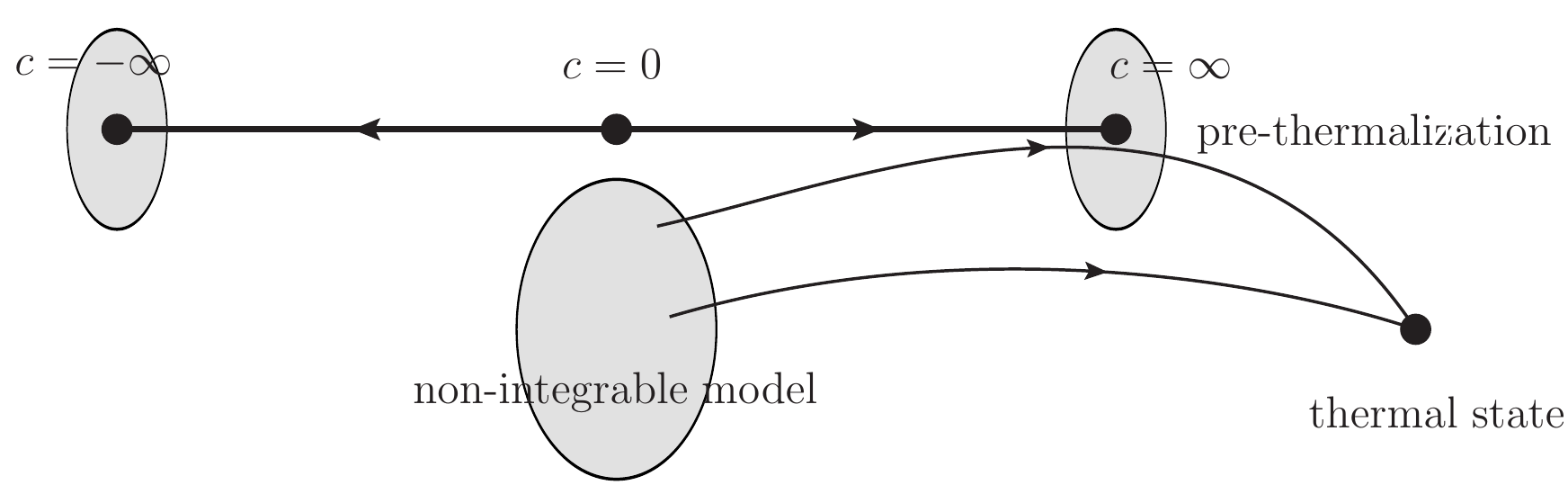}
\caption{ Prethermalization in the Bose Hubbard model. Reprinted with permission from D. Iyer, H. Guan, and N. Andrei, Phys. Rev. A 87, 053628 (2013). Copyright 2013 American Physical Society. } \label{FigDynamicalRG}
\end{figure}

	The flow of the coupling naturally leads to the concept of renormalisation group (RG) flow  in time $t$.  By analogy with conventional  RG ideas, increasing time plays the role of reducing the cut off  with $c=\infty$ being a strong coupling fixed point. For comparison we  recall that in the usual RG picture $c$ has scaling dimension 1 and so also flows to strong coupling. Subsequently, similar behavior was also seen in strongly coupled impurity models \cite{vasseur2013crossover, kennes2014universal}. 
Extending the dynamical RG analogy  one can envisage that other Hamiltonians close to the Lieb-Linger  will flow close the neighborhood of the same strong coupling fixed point, prethermalize in other words, only to end up thermalized on  longer time scales if the model is not integrable, see Fig. \ref{FigDynamicalRG}. An example is provided by  the lattice version of the Lieb-Liniger model,  the non integrable  Bose-Hubbard model which also exhibits dynamical fermionization  \cite{iyer2013exact}. 

\bigskip

\begin{figure}
  (a) 
  \includegraphics[width=.45\textwidth]{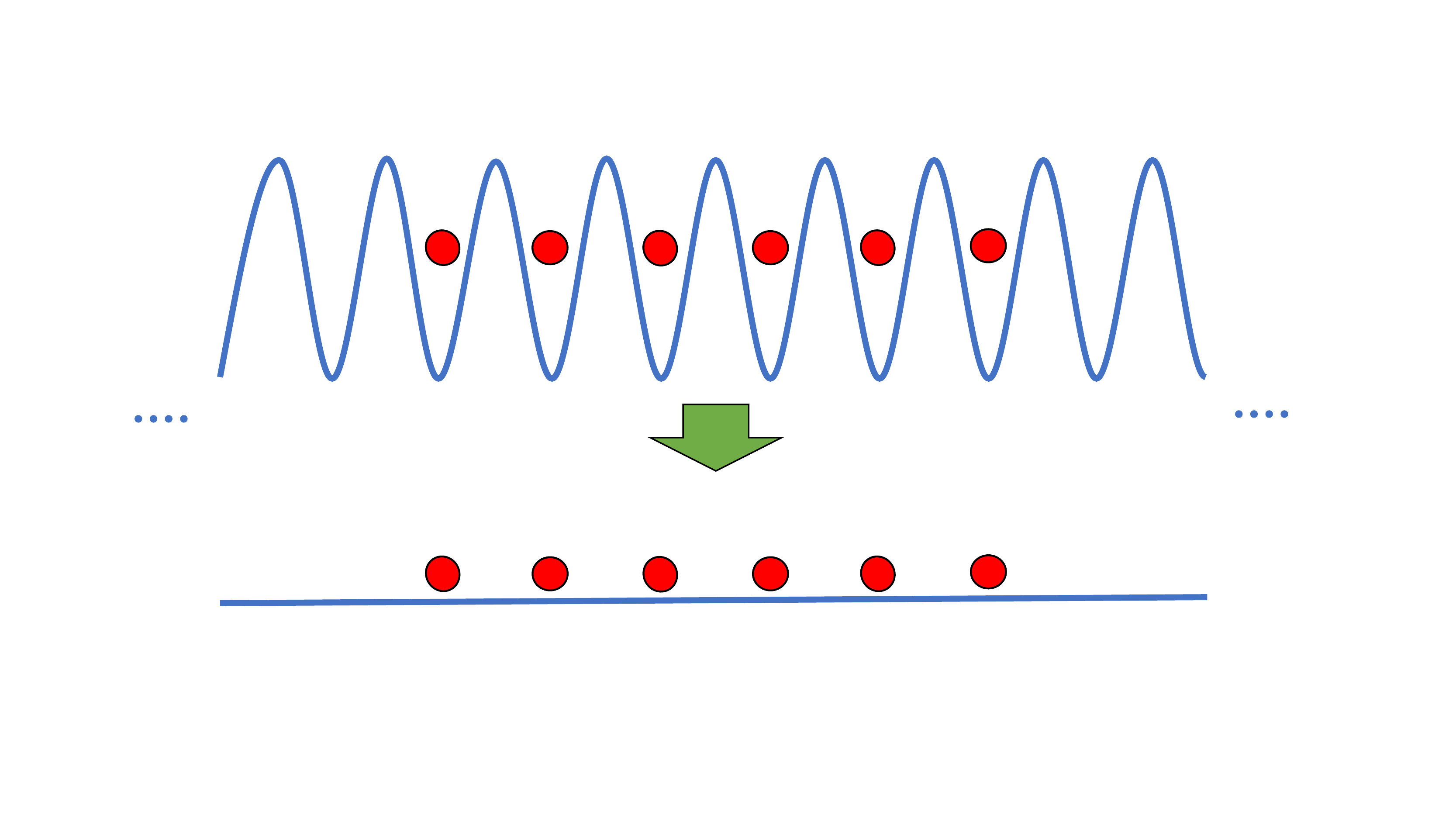}
  (b)
 \includegraphics[width=.45\textwidth]{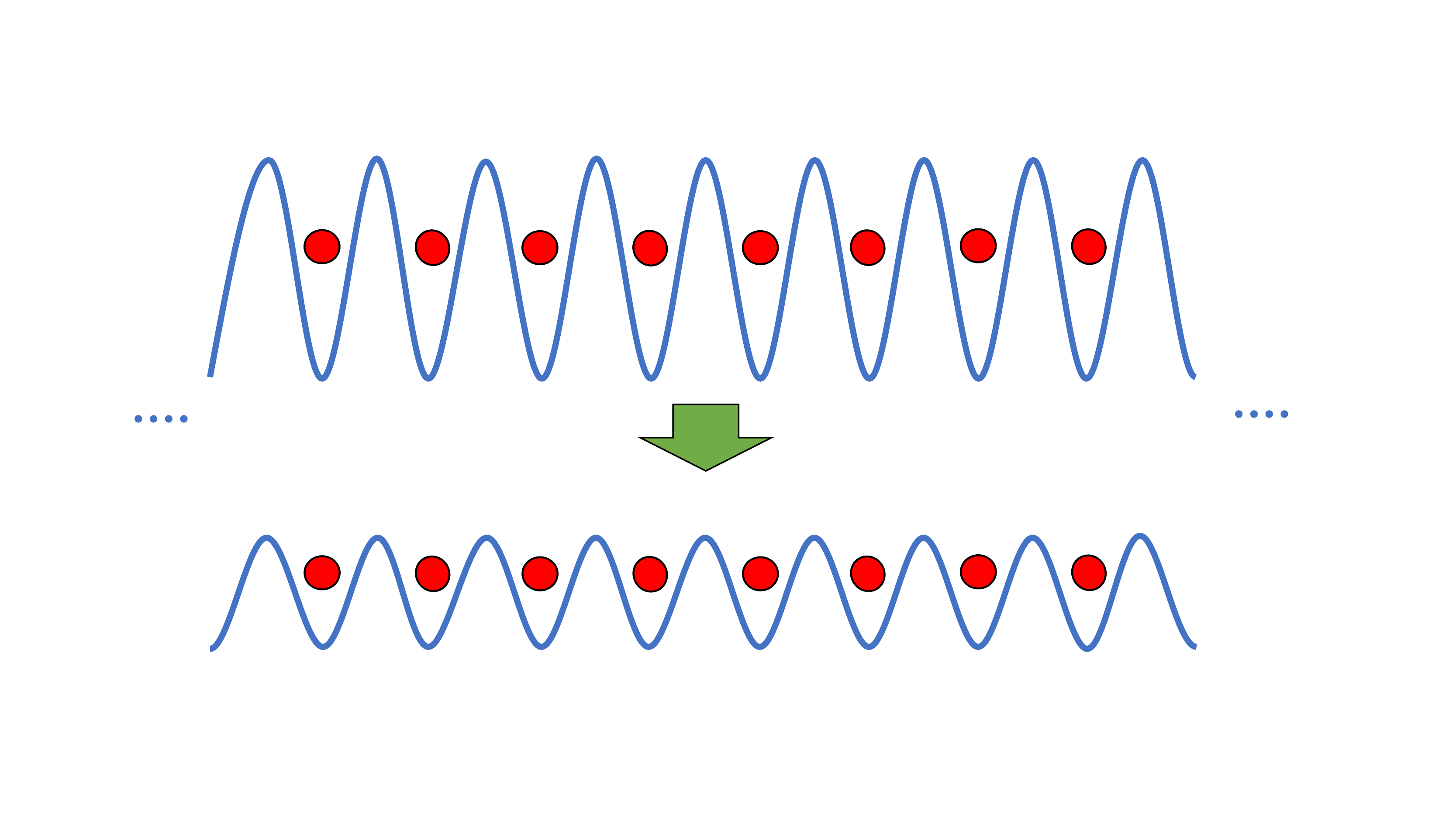}
\caption{ The ultracold atom gas is initially held in a deep optical lattice which is  (a) completely  removed - post quench dynamics described by  the Lieb-Liniger Hamiltonian or (b) merely  lowered  -  the post quench dynamics  given by Sine-Gordon Hamiltonian.  Quench (b) is discussed  in \cite{rylands2019loschmidt}.}\label{FigCartoon}
\end{figure}

We turn now to study  the global properties of the post quench system through the Loschmidt amplitude \eqref{LA} and the work distribution function \eqref{P} focusing on the experimentally relevant case of a cold atom gas initially held in a deep optical lattice which is then  removed entirely  in the quench, see
Fig. \ref{FigCartoon}.  The system is translationally invariant and described by the Lieb-Liniger model.

\begin{figure}
 \includegraphics[width=.45\textwidth]{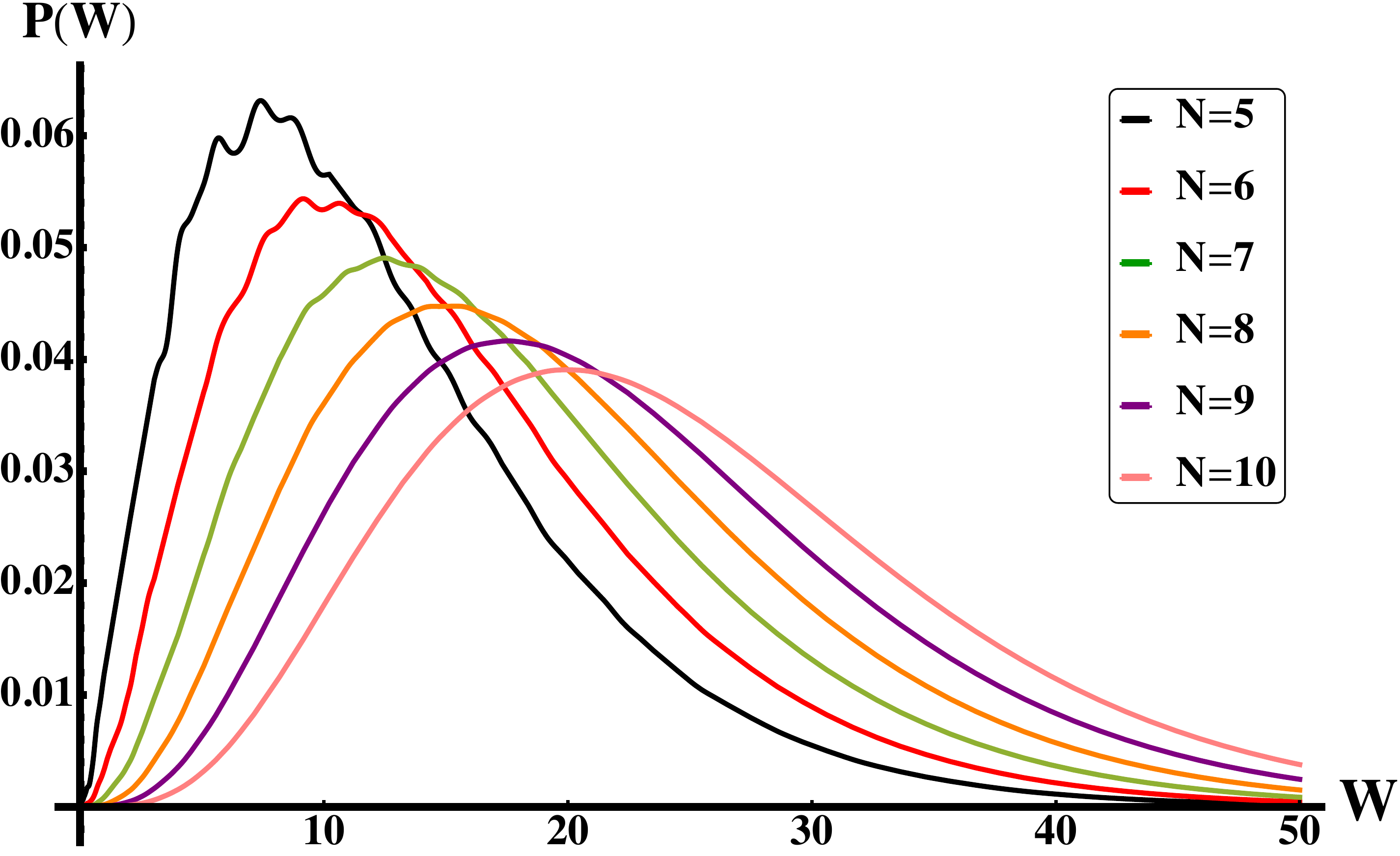}
  \includegraphics[width=.45\textwidth]{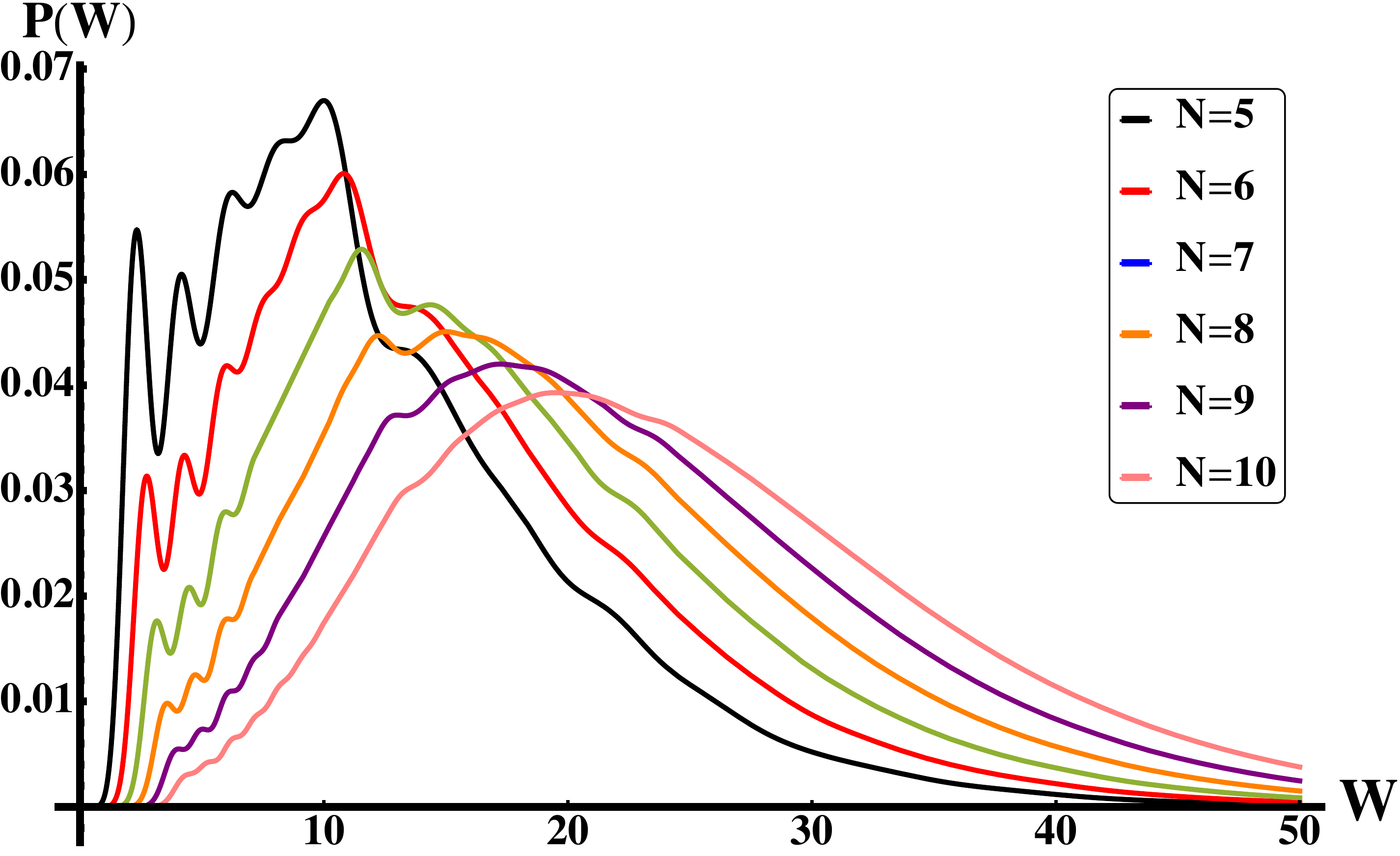}
\caption{The work distribution function, $\mathcal{P}(W)$, for different numbers of bosons released from an optical lattice  with $\delta/m=2$ and $\omega=10$. We measure the work from $\epsilon_i$ the initial state energy. On top we show the distribution for non interacting bosons while on the bottom we show the same quantity for interacting bosons, $c>0$. Reprinted with permission from C. Rylands and N. Andrei, Phys. Rev. B 100, 064308 (2019). Copyright 2019 American Physical Society.}  \label{FigWork}
\end{figure}

We consider  $N$ bosons on a circle of length $L$  initially described by the state \eqref{PSI0}  with  $N$ consecutive sites  filled, with $N\delta \ll L$ so that 
  the unfilled part of the lattice is taken to be much larger than the filled portion to  avoid complications arising from the boundary conditions. 
Employing the Yudson resolution of the identity, the Loschmidt amplitude can be determined to be \cite{rylands2019quantum},
\begin{eqnarray}\label{Gexact}
\mathcal{G}(t)&=&
\left[\frac{8\pi}{\omega}\right]^{\frac{N}{2}}\!\!\sum_{n_1,\dots,n_N}e^{-\frac{2}{\omega}\left[1+i\frac{\omega }{2}t\right]\sum_{j=1}^N\lambda_j^2}\frac{G(\{n\})}{\mathcal{N}(\{n\})}
\end{eqnarray}
where $G(\{n\})=\det{\left[e^{-i\lambda_j(\bar{x}_j-\bar{x}_k)-i\theta(j-k)\varphi(\lambda_j-\lambda_k)}\right]}$ and $\theta(j-k)$ is a Heaviside function. Using the same $1/c\delta$ expansion as before the Fourier transform of this can be explicitly found and analytic expressions for the work distribution, $\mathcal{P}(W)$ obtained. We plot this for both non interacting and strongly interacting bosons $c\delta\gg1$ in Fig. \ref{FigWork} for different particle number and see some commonalities as well as striking differences. Notice that the average work in both cases is the same, $\left<W\right>=N\omega/4$ as is the large $W>\left<W\right>$  behavior. The former statement can be understood from the fact that bosons are initially in non overlapping wavefunctions and $\left<W\right>=\matrixel{\Psi_0}{H}{\Psi_0}$. In comparison, the small $W\ll \left<W\right>$ behavior is strongly affected by the presence of interactions. Large resonant peaks are present in the interacting work distribution and  can be attributed to the scattering of strongly repulsive excitations in the post quench system. Those peaks which are closest to $\left<W\right>$ involve fewer scattering events while those $W=0$ involve more. As the particle number is increased these fluctuations are suppressed like $1/\sqrt{N}$ \cite{sotiriadis2013statistics, smacchia2013work}.
For large systems of bosons the most interesting behavior therefore occurs in the region of $W\sim 0$ where the effects of the interaction are most keenly felt. In this region it can be shown that the distribution decays as a power law with the exponent drastically differing between the free and interacting cases. For the former we have $\mathcal{P}_{c=0}(W)\sim W^{\frac{N}{2}-1}$ whereas in the latter it is $\mathcal{P}_{c>0}(W)\sim W^{\frac{N^2}{2}-1}$, the presence of interactions in the system causing a dramatically faster decay of the work distribution. Behavior such as this will be seen in the next section also when the excitations are gapped as well as interacting.  

We can use our knowledge of $\mathcal{P}(W)$ to investigate the global behavior of the post quench system. As a consequence of the large $W$ agreement between the distributions for the interacting and non-interacting  systems we can determine that at short times $|\mathcal{G}(t)|^2$ is independent of the interactions. This corresponds to the initial period  of expansion  from the lattice in which the particles do not encounter one another. On the other hand,   small $W$ behavior
provides insight to the long time dynamics, the power law decay of $\mathcal{P}(W)$ near the origin translating to the long time power law decay of the LE. Fourier transforming the distribution for free bosons we find that as $t\to \infty$,  $|\mathcal{G}(t)|^2\to 1/t^N$ while in the interacting case  we have instead   $|\mathcal{G}(t)|^2\to 1/t^{N^2}$, a much faster decay. 
We attribute this dramatic difference in the decay away from the initial state to the fact that the large repulsive interactions acting on each other  forcing them to spread out into the one dimensional trap, thereby decreasing their overlap with $\ket{\Psi_i}$.
We should note that this is true regardless of the strength of the interactions and highlights the strongly coupled nature of even weakly interacting systems in low dimensions. As we saw earlier, in the long time limit any repulsive coupling flows in time strong coupling, therefore  the exponent is independent of the initial strength of $c$,  in the TG limit ($c=\infty$) one finds the same power law behavior at long times as for the finite $c$ case. This is the dynamical fermionization discussed in the previous section.

\bigskip

The attractive regime is of significant interest.  The properties of the attractive model both in and out of equilibrium are much less studied than its repulsive counterpart.
This dearth of theoretical results stems from the increased complexity of the Bethe Ansatz solution in the attractive model.  When $c<0$ the model supports bound states and the ground state consists of a single bound state of all $N$ particles \cite{maguire1964study}. While the eigenstates given by \eqref{Bethestates}  remain valid, complex values of $k$ which correspond to bound states are allowed. The resolutions of the identity \eqref{IdentityY} also remains formally valid provided these complex valued solutions are accounted for. A stumbling block however remains as the normalization of the Bethe states in the attractive regime is not known in closed form. 

In the low density limit however it has been shown that for both repulsive and attractive interactions the spatially ordered identity \eqref{IdentityY} becomes
 \cite{yudson1984rigorous, iyer2012quench, iyer2013exact}
\begin{eqnarray}\label{IdentityIyer}
\mathbb{1}_N=\int_\Gamma\frac{\mathrm{d}^Nk}{(2\pi)^N}\ket{\{k\}}\bray{\{k\}}.
\end{eqnarray}
The contours of integration, $\Gamma$, lie on the real line for repulsive interactions and are spread out in the imaginary direction for the attractive case with Im$(k_{j+1}-k_{j})>|c|$. 

Making use of this here in conjunction  with the same $|c|\gg m\omega$ expansion we find that  the work done in the attractive regime separates into two contributions,
\begin{eqnarray}\label{Patt}
\mathcal{P}_{c<0}(W)&=&\mathcal{P}_{\text{free}}(W)+\mathcal{P}_\text{bound}(W).
\end{eqnarray}
 The first term $\mathcal{P}_{\text{free}}(W)$ is the contribution from particles which do not form bound states, 
it is identical to the expression in repulsive case  only now $c<0$. The major difference imposed by this is that the effective distance between the particles is smaller $\delta_\text{eff}<\delta$, the attractive interactions promoting the clustering of particles.  

The  simple analytic continuation to negative coupling of the first term is  reminiscent of the the super Tonks-Girardeau gas. This highly  correlated state of the LL (Lieb-Liniger) model is created by preparing a repulsive LL gas in the Tonks-Girardeau limit, $c\to\infty$ and then  abruptly changing the interaction strength from the being large and positive to large and negative. The result is a metastable nonequilibrium state which exhibits enhanced correlations. Many of the properties of this state can be shown to emerge from a simple analytic continuation of the coupling to large negative values.
In effect the negligible overlap of each particle of our initial state mimics the density profile of the TG gas and so super-TG like behaviour is not unexpected. We should stress  that the expression \eqref{IdentityIyer} is valid at arbitrary negative values $c$ and so not limited to super-TG regime.

The second term  $\mathcal{P}_\text{bound}(W)$ is entirely different. It is due to the bound states and is  calculated by deforming the contours in \eqref{IdentityIyer} to the real line and picking up  contributions due to the poles at $k_i-k_j=i c$ present in \eqref{Bethestates}.  
An $n$-particle bound state can be shown to contribute $\mathcal{P}_{n-\text{bound}}(W)\propto |c|^{n-1}e^{-n|c|\delta}$ with factors from multiple bound states being multiplicative. 

This exponential factor means that the probability that the initial state transitions to one containing bound states is highly suppressed and in the true super-TG limit vanish entirely.  
Despite this, for finite $|c|$ the bound states have a strong signature in work distribution function. Since forming a bound state will lower the energy of the system \cite{maguire1964study} the work distribution becomes non vanishing at negative values of $W$. There is a non zero probability that work can be extracted from the system. Importantly this does not violate the 2nd law of thermodynamics as the average work remains positive $\left<W\right>$ \cite{jarzynski1997nonequilibrium, jarzynski2011equalities}.  In fact, it has been  observed recently that the probability of extracting work from a single electron transistor can be as high as 65\% whilst still satisfying the 2nd law \cite{maillet2018optimal}.  

To see this we examine the leading term of $\mathcal{P}_\text{bound}(W)$ which arises due to the formation of a single two particle bound state 
 \begin{eqnarray}\label{Pbound}
   \mathcal{P}_\text{bound}(W)\approx N\sqrt{\frac{2\pi\omega}{m}}\frac{e^{-|c|\delta-\frac{2W}{\omega}}}{\Gamma\left(\frac{N}{2}-1\right)}\left[\frac{2(W+\frac{|c|^2}{4m})}{\omega}\right]^{\frac{N}{2}-2}.
 \end{eqnarray}
which is non vanishing for $-|c|^2/4m<W$. Determining the full bound state contribution is a straightforward yet involved calculation which we we will not deal with here.

\subsubsection{The XXZ Heisenberg spin chain}

\begin{figure}
  \includegraphics[trim=0 0 11cm 0cm,clip=true, width=.48\textwidth]{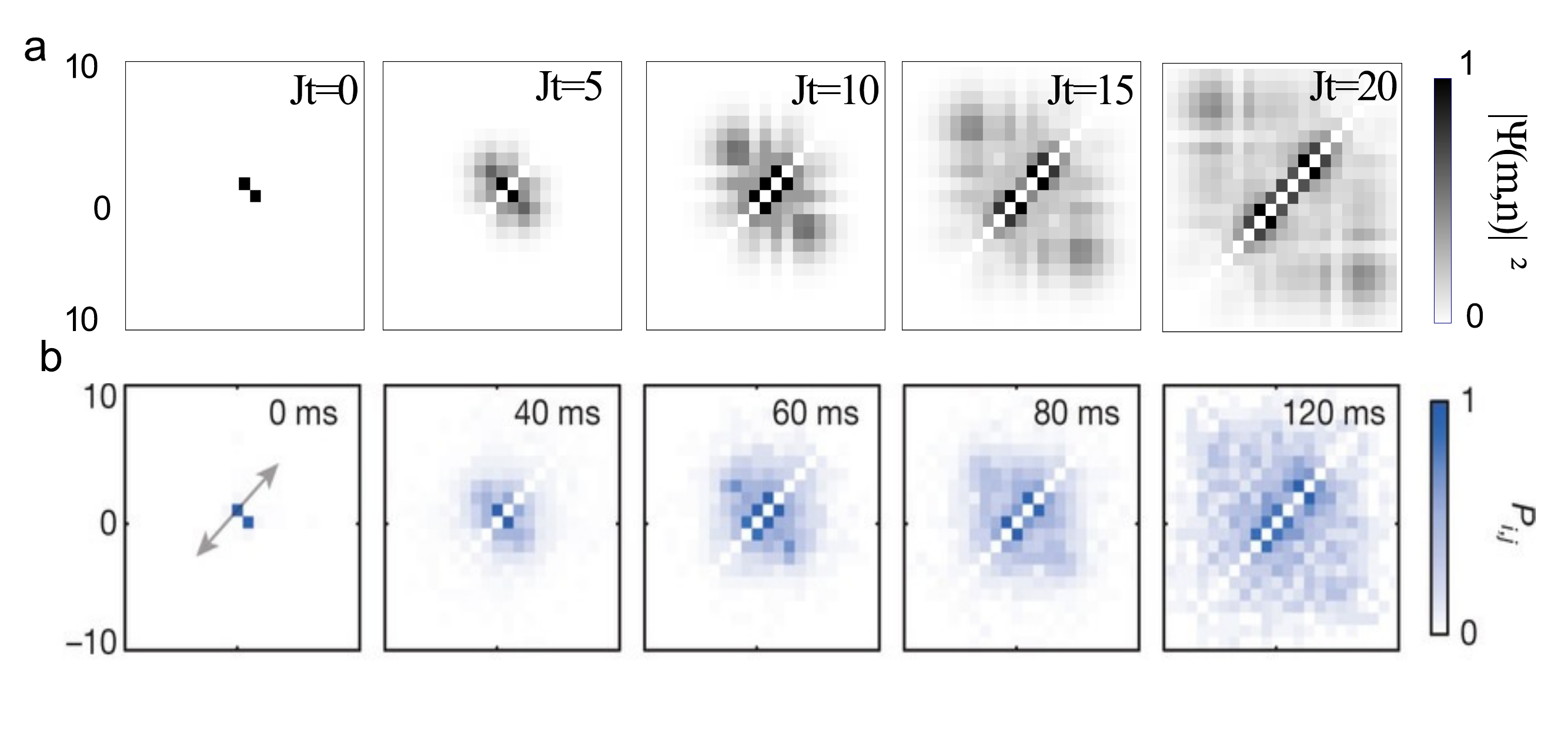}
  \caption{(a) The norm of the wavefunction  $|\Psi(m,n,t)|^2$ at different times for two flipped spins initially at $m=1,n=0$. (b) The joint probabilities at different times of two spins at sites $i$ and $j$ initially at $i=1,j=0$, measured experimentally in \cite{fukuhara2013microscopic}.
  Reprinted with permission from W. Liu and N. Andrei, Phys. Rev. Lett. 112, 257204 (2014). Copyright 2014 American Physical Society. }\label{FigXXZ1}
  \end{figure}

\begin{figure}
 \includegraphics[trim=0 0 0 9.05cm,clip=true, width=.48\textwidth]{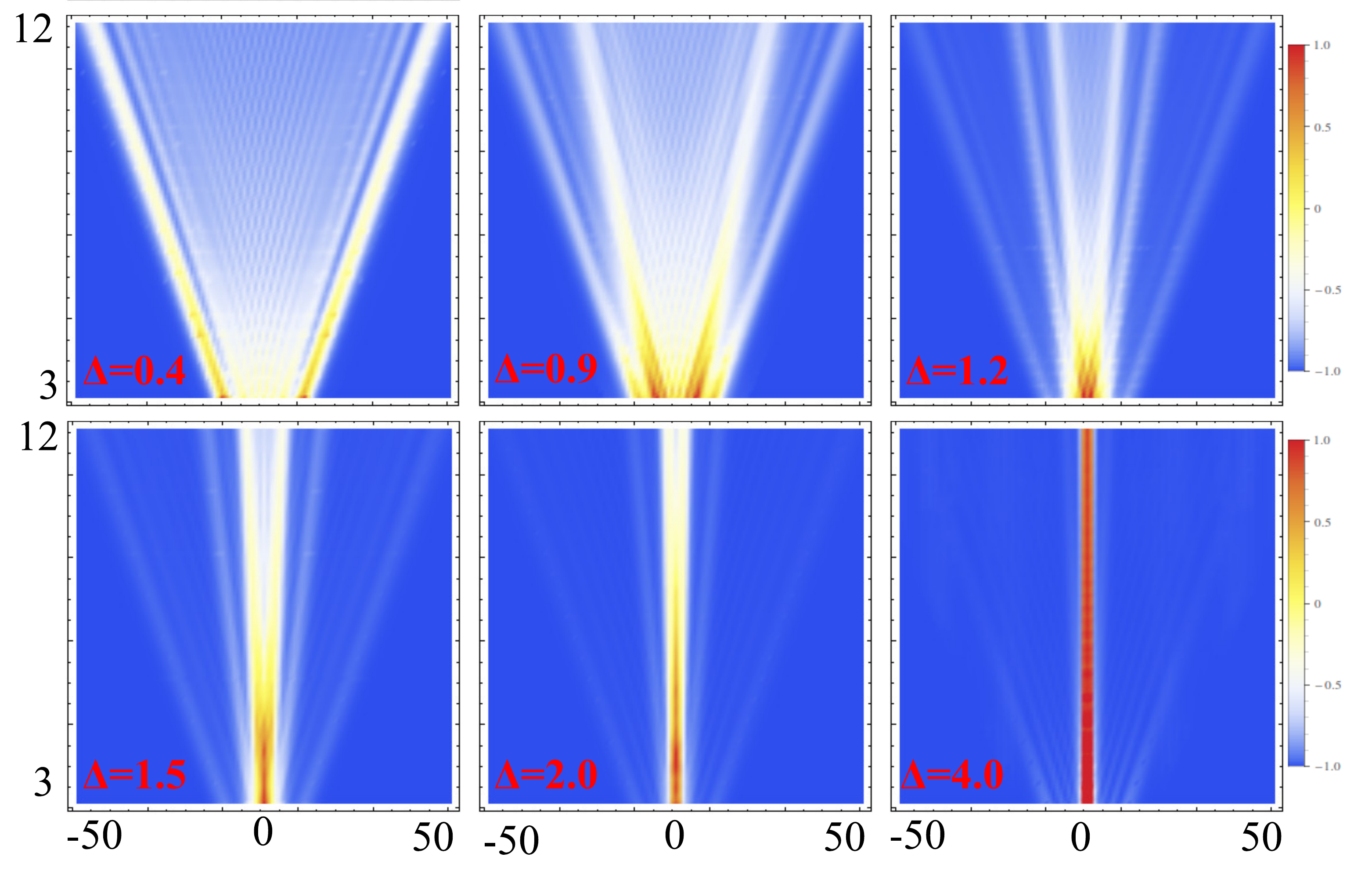}
  \caption{The local magnetization after  a quench from an initial state of 3 flipped spins at the origin for different values of the anisotropy $\Delta$. Time, the vertical direction, is measured in units of the exchange coupling $J$.
  Reprinted with permission from W. Liu and N. Andrei, Phys. Rev. Lett. 112, 257204 (2014). Copyright 2014 American Physical Society. }\label{FigXXZ}
  \end{figure}

The XXZ Heisenberg chain provides another example of an experimentally relevant integrable model. The Hamiltonian
\begin{eqnarray}\label{Eq:Heisenberg}
H=J \,\sum_{j=1}^N \{ \sigma^x_j \,\sigma^x_{j+1} +\sigma^y_j\, \sigma^y_{j+1} +\Delta \,\sigma^z_j \,\sigma^z_{j+1} \}
\end{eqnarray}
models a linear array of spin interacting via anisotropic spin exchange. 
The isotropic case $\Delta=1$ is $SU(2)$ invariant and enjoys the distinction of being the first model solved by Bethe  by means of the approach that bears his name \cite{bethe1931zur}. The generalization to the anisotropic case was  given by Orbach \cite{orbach1958linear}. The eigenstates are again characterized by a set of Bethe momenta $\{k_j\}$  describing the motion of $M$ down-spins in a background of $N-M$ up-spins, and are given by:

\begin{eqnarray}\nonumber
|\vec{k}\rangle = \sum_{\{m_j\}}\prod_{i<j}\,[\theta(m_i-m_j) +s(k_i,k_j)\theta(m_j-m_i)] \\
\times\prod_j e^{ik_j m_j}\,\sigma^-_{m_j}\,|\Uparrow\rangle~
\end{eqnarray}
where $m_j$ the position of the $j$th down spin is summed from 1 to $N$ (the length of the chain), and the S-matrix is given by,
\begin{eqnarray}\label{Heisenberg S matrix}
 s(k_i,k_j)= -\frac{1+e^{i k_i+ ik_j} -2\Delta e^{ik_i}}{1+e^{i k_i+ik_j} -2\Delta e^{ik_j}}.
\end{eqnarray}
The Heisenberg chain exhibits a complex spectrum which includes bound states in all parameter regimes. To carry out the quench dynamics for the model one needs to construct the appropriate  Yudson representation  and use it to time evolve any initial state \cite{liu2014quench}. Here we display in Fig. \ref{FigXXZ1} the time evolving wavefunction of two adjacent flipped spins in the background of an infinite number of unflipped spins and compare it to the experimental results (no adjustable parameters are involved.)  The time evolution of the magnetization from an initial state of three flipped spins for different values of the anisotropy $\Delta$ is given in Fig. \ref{FigXXZ}.  We see that excitations propagate outward after the quench forming a sharp light-cone in contrast to the Lieb-Liniger model. The boundary of the light-cone arises from the propagation of free magnons which travel with the maximum  velocity allowed by the lattice. Rays within the light-cone are the propagation of spinon bound states.  As the anisotropy $\Delta$ is increased the bound states slow down and more spectral weight is shifted to them. Due to the integrability of \eqref{Eq:Heisenberg} these excitations have infinite lifetime which prevents any dispersion of these features. The introduction of integrability breaking terms can therefore be expected to alter this picture, for example through spinon decay \cite{groha2017Spinon}.

\bigskip

\subsection{Concluding remarks and outlook}
 In this chapter we have explored some aspects of the far from equilibrium behavior of integrable models. After a broad overview of the current status of the field we investigated some particular phenomena through a number of  illustrative examples. We saw that the Bethe Ansatz solution of the Lieb-Liniger and  Heisenberg  models provided us with a powerful tool with which to study both the local and global, non-equilibrium behavior of these strongly coupled systems. The quench dynamics of more complex models such as the Gaudin-Yang model \cite{gaudin2014bethe, yang1967some}  describing multi-component gases  has also been accessed via the Yudson approach  \cite{guan2018quench}  allowing the study of  phenomena such a quantum Brownian motion or the dynamics of FFLO (Fulde-Ferrell-Larkin-Ovchinnikov) states \cite{fulde1964super, larkin1964nonuniform}. Similarly the quench dynamics of other models such as the Kondo and Anderson models are currently  studied via  the Yudson approach \cite{culverandrei, touraniandrei}. They  give access to such quantities as  the time evolution of the Kondo resonance or  of the charge or heat currents in voltage or temperature driven two lead quantum dot system.

 These methods we discussed could be thought as being microscopic, starting from the exact eigenstates of the system. Recently these problems have been studied from a macroscopic perspective by combining integrability and ideas from hydrodynamics \cite{castroalvaredo2016emergent}. Generalized hydrodynamics (GHD) provides a simple description of  the non-equilibrium integrable models on long length scales and times. It has been utilized in studies of domain wall initial states in the Lieb-Linger and the emergence of light-cones in quenches of the XXZ model \cite{bulchandani2017solvable, bulchandani2018bethe}.  This method allows  the incorporation  integrability breaking effects within the formalism, but is limited to ``Euler scale" dynamics. It would be of great interest compare the results and expectations of GHD with the methods and results presented here to further understand the limitations of both the microscopic and macroscopic approaches.

\section{NON-EQUILIBRIUM PROTOCOLS FOR ONE DIMENSIONAL BOSE GASES IN ATOMTRONIC CIRCUITS}
\label{DynProt}
\vspace*{-0.5cm}
\par\noindent\rule{\columnwidth}{0.4pt}
{\bf{\small{L. Piroli and A. Trombettoni}}}
\par\noindent\rule{\columnwidth}{0.4pt}

{A promising line of research in atomtronics
is the realization of configurations where several waveguides in which ultracold atoms move are merged to form circuits
\cite{amico2017focus}. Among the
challenges one has to
face, an important one is the tailoring of the circuits in a way to reduce
transverse instabilities during the dynamics of ultracold matter wavepackets
\cite{ryu2015integrated}. This would allow for the possibility of stable motion
of the matter wavepackets across the whole circuit,
including the passage through junctions and in the regions
where the waveguides composing the
circuit have to bend. Since transverse instabilites are suppressed in
one-dimensional geometries, the lines of research of atomtronics
and one-dimensional ultracold atoms have been developing
tight connections in the last decade.
On the one hand, the study of circuits made of one-dimensional waveguides open new
directions of investigation for the community working on one-dimensional
integrable systems, such as the study of junctions
of one-dimensional waveguides: an example is given in
\cite{buccheri2016holographic}, where a junction of three Tonks-Girardeau
gases is studied, and connected to the literature of
coupled/intersecting nanowires. On the other hand, the amount
of available results in the field of one-dimensional integrable models
provides an extremely useful basis for the characterization of ultracold
matter wavepackets on such geoemtries, which has been at center
of significant discussions in the Atomtronics@Benasque conference series
\cite{amico2017focus}.}

{One-dimensional interacting bosons are well described by the integrable
Lieb-Liniger model, which was
  extensively studied since its introduction in the sixties, also in connection
  with other one-dimensional integrable systems. Extensions and generalizations
  of the Lieb-Liniger model may apply to one-dimensional fermionic
  systems and mixtures, including Bose-Bose and Bose-Fermi mixtures. 
  Therefore, the field of atomtronics circuits made of (possibly connected)
  one-dimensional ultracold systems is a natural arena to apply
  such a body of knowledge,
  and at the same time calls for new ideas and investigations using
  integrability techniques.}

One-dimensional systems provide {\it per se} an exciting arena 
where, over the past decade, significant experimental technical
advances have allowed for very precise studies of a series of non-equilibrium
phenomena. At the same time, a number of powerful theoretical tools
were developed to describe them. The study of one-dimensional systems plays
a role as well
in the field of atomtronics and in particular in atomtronics circuits, 
where matter-wave packets can be controlled and moved. When the transverse
dimensions of the waveguides in which atoms move are small enough to create
one-dimensional tightly confined traps and the energies involved are negligible
with respect to the excitation energies of transverse degrees of freedom,
then one enters the one-dimensional regime. Ultracold bosons are then
effectively described by the Lieb-Liniger model \cite{yurovsky2008collisions,bouchoule2009atom,cazalilla2011one},
belonging to the family
of integrable theories. In such one-dimensional regimes quantum fluctuations
play a prominent role and a general issue is whether and for what applications
such one-dimensional features hamper or at variance make it easier to realize
atomtronics tasks.

{Here we give an account of some interesting properties of the Lieb-Liniger model
and how integrability gives access to the study of some of its
local and global non-equilibrium properties. The following contribution
focuses on the theoretical study of two of the more useful protocols to control the quantum dynamics  
of the Lieb-Liniger model: {\em i)}
integrable dynamics after a quench; {\em ii)} Floquet engineering. They are
relevant for atomtronics applications, both for the possibility
to have quenches and time-periodic potentials as a tool to control the dynamics
and induce desired dynamical regimes and for the remarkable progress in
experimental techiniques enabling the possibility to vary interaction
strenghts, geometry of the trap and the time-dependence of the potentials acting
on the atoms in one-dimensional ultracold systems  \cite{yurovsky2008collisions,bouchoule2009atom,cazalilla2011one}.}
In the present contribution, L.P. wrote Section \ref{II}, while A.T. wrote Section \ref{III}.

\subsection{Quench dynamics in the Lieb-Liniger model}
\label{II}


In the early noughties, a series of cold-atomic experiments contributed to the emergence of a growing theoretical interest in the non-equilibrium dynamics of isolated quantum integrable systems~\cite{bloch2008many,polkovnikov2011colloquium}. For instance, in the famous ``quantum Newton’s cradle'' experiment~\cite{kinoshita2006quantum}, out-of-equilibrium arrays of trapped one-dimensional ($1D$) Bose gases were shown not to reach thermal equilibrium within the accessible time scales. This peculiar behavior was attributed to the approximate integrability of the system: indeed, in the idealized situation where longitudinal confining potentials are neglected, a $1D$ gas of $N$ bosons with mass $m$ and point-wise interactions can be described by the integrable Lieb-Liniger Hamiltonian~\cite{lieb1963exact}. Denoting by $L$ the length of the
system, the Hamiltonian can be written as 
\begin{equation}
  H=\int_{0}^{L}\,\mathrm{d}x \left(\frac{\hbar^2}{2m}\partial_x\Psi^{\dagger} \partial_x\Psi+
  c\Psi^{\dagger}\Psi^{\dagger}\Psi\Psi\right),
\label{hamiltonian}
\end{equation}
where $\Psi$, $\Psi^\dagger$ are bosonic creation and annihilation operators satisfying canonical commutation relations.
Here, the interaction strength is related to the one dimensional scattering length $a_{\rm 1D}$ through
$c=-\hbar^2/ma_{\rm 1D}$
\cite{olshanii1998atomic}
and can be varied via Feshbach resonances \cite{inouye1998observation} to take either positive or negative values. 

Given its relative simplicity and experimental relevance, in the past decade a large number of studies have focused on  the non-equilibrium dynamics in the Lieb-Liniger gas, especially within simplified protocols such as the one of a quantum quench~\cite{calabrese2006time,calabrese2007quantum}: in this setting one considers the ground state of some Hamiltonian $H(c_0)$ ($c_0$ denotes an internal parameter), which is suddenly changed at time $t=0$ by an abrupt variation  $c_0\to c$. These studies have played an important role for the development of a general theory of integrable systems out of equilibrium~\cite{calabrese2016introduction}. In this section, we provide a review of a selected number of them, focusing exclusively on the simplest case of homogeneous settings (see Section \ref{sec:outlook-protocol} for recent further developments in the presence of confinement potentials and inhomogeneities). 

\subsubsection{The quench problem}

Physical intuition suggests that after a quench an extended system should act as an infinite bath with respect to its own finite subsystems, and that local properties should relax to stationary values described by a thermal Gibbs ensemble. While for generic models this picture turns out to be correct
\cite{rigol2008thermalization,cazalilla2010focus,d2016quantum}, a quite different scenario emerges in the presence of integrability, due to the existence of an extensive number of local conservation laws which strongly constrain the dynamics. In this case, it was proposed in Ref.~\cite{rigol2007relaxation} that the correct post-quench stationary properties are captured by a generalized Gibbs ensemble (GGE), which is written in terms of all higher local conservation laws beyond the Hamiltonian~\cite{rigol2007relaxation,vidmar2016generalized,essler2016quench}. It was later discovered that quasi-local conservation laws must also be taken into account~\cite{fagotti2014conservation,ilievski2015complete,ilievski2016string,ilievski2016quasilocal,ilievski2017interacting,pozsgay2017generalized} and the validity of the GGE is now widely accepted.

Despite the established conceptual picture, computations based on the GGE are hard, and more generally the characterization of the post-quench dynamics remains extremely challenging in practice. In order to explain the difficulties involved, it is useful to consider the time evolution of a physically relevant observable for  the $1D$ Bose gas, namely so-called pair correlation function~\cite{kinoshita2005local}
\be
g_2=\frac{\braket{\Phi|\Psi^{\dagger 2}(x)\Psi^2(x)|\Phi}}{D^2},
\label{eq:g_2}
\ee
where $D=N/L$ is the particle density, with $L$ the system size, while $\ket{\Phi}$ is the state of the system. Physically, $g_2$ quantifies the probability that two particles occupy the same position.
For a quantum quench, we have the formal expression (setting $\hbar=1$)
$$
\langle\Phi(t)|\Psi^{\dagger 2}(x)\Psi^2(x)| \Phi(t)\rangle=\sum_{m,n} \braket{\Phi(0)|n} \braket{m|\Phi(0)}$$
\bea
\times \langle n|\Psi^{\dagger 2}(x)\Psi^2(x)| m\rangle e^{-i\left(E_{n}-E_{m}\right) t}.
\label{eq:spectral_sum}
\eea
Here we denoted by $\ket{n}$, $E_n$ the energy eigenstates and eigenvalues respectively, while $\ket{\Phi(t)}$ is the state of the system evolved at time $t$ after the quench. For the Lieb-Liniger model the Bethe Ansatz~\cite{korepin1997quantum} is a very efficient tool to obtain most of  the ingredients appearing in Eq.~\eqref{eq:spectral_sum}, including the matrix elements of the local operator $\Psi^{\dagger 2}(x)\Psi^2(x)$~\cite{pozsgay2011local,piroli2015exact}. However, due to the complicated form of the energy eigenfunctions, there appears to be no simple way to compute the overlaps $\braket{\Phi(0)|n}$ for general initial states. Furthermore, Eq.~\eqref{eq:spectral_sum} involves the evaluation of  a double sum over all the eigenstates of the Hamiltonian, which is currently out of reach in most of the physically interesting situations.

Due to the above difficulties, initial studies in the Lieb-Liniger model were restricted to the limit of either vanishing~\cite{mossel2012exact,sotiriadis2014validity} or infinitely repulsive post-quench interactions~\cite{gritsev2010exact,muth2010dynamics,muth2010fermionization,kormos2013interaction,collura2013quench,collura2014stationary,kormos2014analytic}, where the Hamiltonian can be mapped onto free fermions through a Jordan-Wigner transformation. While these works already made it possible to explore in some detail interesting phenomena such as local relaxation~\cite{kormos2013interaction} and ``light-cone'' spreading of correlation functions~\cite{mossel2012exact,kormos2013interaction}, it remained as an open problem to provide predictions in the case of finite values of the interactions. 

\subsubsection{The Quench Action}

A conceptual and technical breakthrough came with the introduction, by Caux and Essler, of the so-called Quench Action method~\cite{caux2013time,caux2016quench}, which proved to be a powerful and versatile approach to the quench dynamics in integrable systems (other methods, that will not be discussed here, have also been developed, including a Yudson-representation approach, which is also suitable to study inhomogeneous initial states, see Refs.~\cite{iyer2012quench,rylands2019quantum} and the contribution
of N. Andrei and C. Rylands).

It is well known that, in the thermodynamic limit, each eigenstate of an integrable system is associated with a distribution function $\rho(\lambda)$, where $\lambda$ are the quasi-momenta of the (stable) quasi-particle excitations~\cite{korepin1997quantum}. Based on physical arguments, it was proposed in Ref.~\cite{caux2013time} that this description could be exploited to replace the double sum in Eq.~\eqref{eq:spectral_sum} with a functional integral over all distribution functions $\rho(\lambda)$. This approach is particularly powerful to investigate the late-time limit, for which one can write (in the thermodynamic limit)~\cite{caux2013time,caux2016quench}
$$
\lim_{t\to \infty}\langle\Phi(t)|\Psi^{\dagger 2}(x)\Psi^2(x)| \Phi(t)\rangle=
$$
\be
=\int \mathcal{D}\rho\,e^{S[\rho]}  \langle \rho|\Psi^{\dagger 2}(x)\Psi^2(x)| \rho\rangle,
\label{eq:functional_int}
\ee
where $\ket{\rho}$ denotes an eigenstate corresponding to the distribution function $\rho(\lambda)$. Here we introduced the ``Quench Action'' $S[\rho]$, which can be determined based on the knowledge of the overlaps $\braket{\Phi(0)|n}$.  While, as we have already mentioned, it is not known how to obtain these in general, it turned out that they can be computed in several interesting cases~\cite{kozlowski2012surface,pozsgay2014overlaps,calabrese2014interaction,piroli2014recursive,piroli2017what,brockmann2014overlaps,brockmann2014gaudin,brockmann2014neel,horvath2016initial,horvath2017overlaps,horvath2018overlap,brockmann2017universal,de2015one,buhl2016one,foda2016overlaps,de2016ads,deLeeuw2017_s05,deleeuw2018scalar,pozsgay2019integrable,jiang2020exact,de2020spin,linardopoulos2020solving}.

Given $S[\rho]$, the functional integral can be treated exactly by saddle-point evaluation, so that the r.h.s. of Eq.~\eqref{eq:functional_int} can be replaced by $\langle \rho_s|\Psi^{\dagger 2}(x)\Psi^2(x)| \rho_s\rangle$, where $\delta S[\rho_s]/\delta \rho=0$. Crucially, the saddle-point distribution function $\rho_s(\lambda)$ determines all the post-quench local expectation values (which can be explicitly computed via exact Bethe Ansatz formulas~\cite{pozsgay2011local,kormos2009expectation,kormos2010one,bastianello2018exact,bastianello2018sinh}), and thus represents an effective characterization of the late-times steady state. 
 
 The Quench Action approach was first applied in the Lieb-Liniger model for quenches from zero to positive values of the interactions, $c_0=0\to c>0$~\cite{denardis2014solution}. It was found that the steady state displays quantitative different features from a thermal state, unequivocally proving the absence of thermalization. The same approach also allowed for the computation of the full time evolution of $g_2$~\cite{denardis2015relaxation} (see also~\cite{denardis2014analytical,vandenberg2016separation}), unveiling a quite general power-law decay to stationary values for local observables, and for a detailed study of the statistics of the work  performed by the quench~\cite{perfetto2019quench,silva2008statistics,gambassi2012large}.

\begin{figure}
	\includegraphics[scale=0.45]{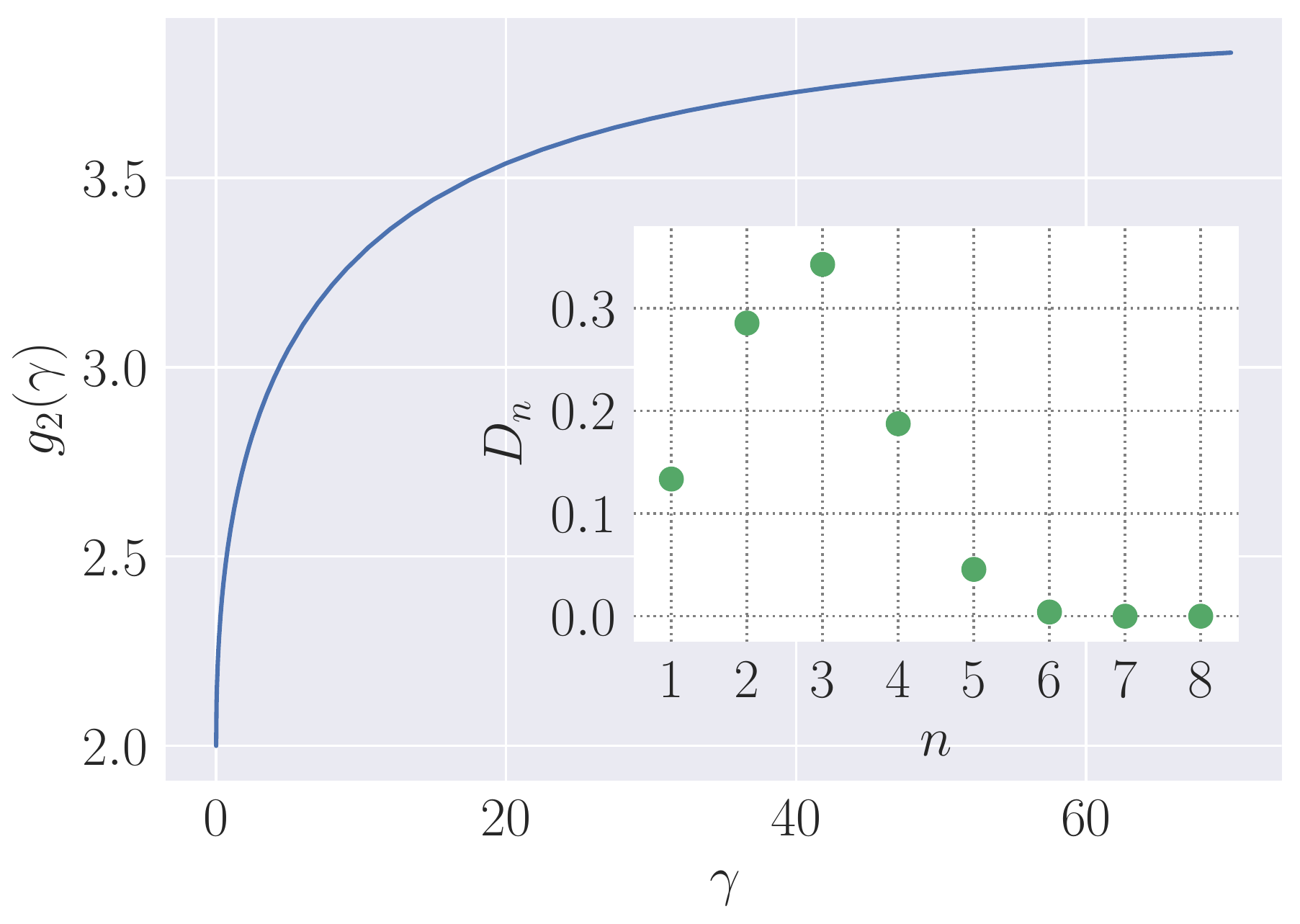}
	\caption{Main panel: Pair correlation function, as defined in Eq. \eqref{eq:g_2} in the steady state reached after a quench $c_0=0\to c<0$. The plot shows $g_2$ as a function of the rescaled interaction $\gamma=|c|/D$, and is computed using the results of Ref.~\cite{piroli2016multiparticle}.  Inset: Densities $D_n$ of the $n$-particle bound states for the same quench, and $\gamma=2$.}
	\label{eq:pair_correlation}
\end{figure}

\subsubsection{Quenches to the attractive regime} 

In the case of quenches to repulsive interactions, the late-time steady state appears to display features that are only quantitatively different from those observed at thermal equilibrium~\cite{denardis2014solution}. In this respect, an even more interesting picture emerges for quenches to the attractive regime. These were investigated in Refs.~\cite{piroli2016multiparticle, piroli2016quantum}, where the formalism of Refs.~\cite{caux2013time,denardis2014solution} was employed to study interaction quenches of the form $c_0=0\to c<0$. 

The main results of these works are arguably the prediction of the dynamical formation of $n$-boson bound states with finite densities $D_n$, and the characterization of the corresponding distribution of quasi-momenta $\rho_n(\lambda)$. Interestingly, it was shown that the value of $n$ for which the density $D_n$ is maximum decreases with the rescaled interaction $\gamma=|c|/D$. Although this result might appear counter-intuitive, there is in fact a simple physical interpretation: in the attractive regime, the bosons have a tendency to form multi-particle bound states. However, in the quench setup the total energy of the system is fixed by the initial state, while the energy of $n$-particle bound states increases, in absolute value, very rapidly with $\gamma$ and $n$~\cite{piroli2016multiparticle}. Therefore, $n$-particle bound states cannot be formed for large values of $\gamma$, while they become accessible as $\gamma$ decreases. 

We note that the structure of the stationary state predicted in Refs.~\cite{piroli2016multiparticle, piroli2016quantum} is qualitatively very different from the super Tonks-Girardeau gas, which is obtained by quenching the one-dimensional Bose gas from infinitely repulsive to infinitely
attractive interactions
\cite{astrakharchik2005beyond,batchelor2005evidence, haller2009realization, chen2010transition, muth2010dynamics, kormos2011local, tschischik2015repulsive}. Indeed, the latter features no bound
state, even though it is more strongly correlated than the traditional Tonks-Girardeau gas, as has been observed experimentally~\cite{haller2009realization}. The findings of Refs.~\cite{piroli2016multiparticle, piroli2016quantum} are thus also interesting because they show that the physics emerging at late times after a quench depends qualitatively on the initial state of the system.

Importantly, the formation of bound states after the quench have consequences on the local correlation functions. For instance, the value of $g_2$ at large times is always greater than $2$, and increases with $\gamma=|c|/D$~\cite{piroli2016quantum}. This is displayed in Fig. \ref{eq:pair_correlation}, and is once again qualitatively different from the case of the super Tonks-Girardeau gas. We note that these results are consistent with subsequent numerical calculations reported in Ref.~\cite{zill2018quantum} and based on the method developed in Ref. \cite{zill2015relaxation, zill2016coordinate}.

\subsection{Floquet Hamiltonian for the Periodically Tilted Lieb-Liniger Model}
\label{III}

{Another promising protocol for inducing and controlling interesting
instances of quantum dynamics is provided by the Floquet engineering.
In this scheme the original Hamiltonian -- in this Chapter the Lieb-Liniger
model -- is subject to a time-periodic driving $V$. The Floquet Hamiltonian
control then the time dynamics of the system when observed at stroboscopic
times, {\it i.e.} at times multiples of the period of $V$. The general goal
is to design $V$ in a way that the Floquet Hamiltonian is the one inducing
the desired quantum dynamics.}


{In general, when a periodic driving acts on an integrable model,
then the resulting
Floquet Hamiltonian is non-integrable.
In this Section we consider
the case of the Lieb-Liniger model subject to a potential periodic
in time and linear in space, which we refer to as a
periodic tilting \cite{colcelli2019integrable}.}  
The Floquet Hamiltonian of the integrable Lieb-Liniger model for such linear potential 
with a periodic time--dependent strength is integrable and its quasi-energies can be determined using
well known results for the undriven Lieb-Liniger model.

{We pause here to comment about the relevance of the investigation of Floquet engineering, and periodic tilting in particular, starting
from the Lieb-Liniger Hamiltonian for atomtronics
applications and perspectives. Controlling matter-wave dynamics in
waveguides and other atomtronics circuitry and components is in general
an interesting perspective to be discussed and studied. A time-independent potential
linear in space induces a motion in the atomtronics devices, and a time-dependent periodic
tilt can be used to control the motion across, to and fro,
a circuit. As discussed in the Introduction,
to reduce trasnverse excitations it may be convenient to use and merge one-dimensional waveguides,
and a natural question is what is the effect of a time-dependent periodic tilting in such
one-dimensional systems.}

{We then consider the periodic tilting 
$$V(x,t)=f(t) \, x$$
with $f(t)$ a periodic function with period $T$.}
The Lagrangian density of the system is 
\begin{eqnarray}
\mathcal{L} &=& \frac{i}{2} \left( \Psi^{\dagger} \p_t \Psi - 
h.c.\right)- \frac{1}{2m}
\p_x \Psi^{\dagger} \, \p_x \Psi 
\label{Lag_LL}- \frac{c}{2} \Psi^{\dagger} \Psi^{\dagger} \Psi \Psi   \nonumber \\
&&\,\,\,\,\,\,
-V(x, t)\Psi^{\dagger} \Psi,
\end{eqnarray}
where $h.c.$ denotes the hermitian
conjugate of the first term and $\Psi=\Psi(x,t)$. 

{When the potential $V$ is time-independent with $f(t)$ constant, then it is well known that one can gauge away the potential 
linear in space by moving to the center-of-mass accelerating frame. Notice that this property is valid in any dimension and also for interacting systems, as long as the two-body interaction depends only on the relative distance (for a pedagogical presentation see, {\it e.g.}, Ref. \cite{colcelli2020free}).}

{Let now come back to the case of $f(t)$ periodic in time.} Proceeding as one does for the single-particle and the two-particles cases \cite{colcelli2019integrable,colcelli2020dynamics}, one
can solve the Schr\"odinger equation of the many-body interacting model. To this aim, one introduces
the following gauge transformation 
\be
\label{field_gaugetrasf}
\Psi(x,t) \equiv e^{i \theta(x,t)} \varphi(y(t), t),
\ee
where $$y(t) = x - \xi(t),$$ with the functions $\xi(t)$ and $\theta(x,t)$
to be suitably determined in order to gauge away the potential term $V$ from the Lagrangian
density when rewritten in terms of the field $\varphi$.

The functions $\xi$ and $\theta$ are determined as it follows. We start by imposing
\be
\label{conditions_integrab_app_0}
\p_t \xi =  \frac{1}{m} \p_x \theta
\ee
and
\be
\label{conditions_integrab_app}
- \p_t \theta = \frac{1}{2m} \left(\p_x \theta \right)^2 + x f(t).
\ee
We now make the Ansatz 
\be
\label{ansatz_theta_app}
\theta(x,t) = m x \p_t \xi+ \Gamma(t),
\ee
finding the conditions
\be
\label{vANDdelta_onebody_app_0}
m \p_{t}^2 \xi = -f(t)
\ee
and
\be
\label{vANDdelta_onebody_app}
\p_t \Gamma =-\frac{m}{2} \left( \p_t \xi \right)^2, 
\ee
determining $\xi(t)$ and $\Gamma(t)$ in terms of $f(t)$. 
From the differential equations \eqref{vANDdelta_onebody_app_0}-\eqref{vANDdelta_onebody_app}
one gets \cite{colcelli2019integrable}
\be
\label{theta_f(t)_app}
\theta(x,t)=-x \, \int_0^t f(\tau)\,d\tau -\frac{1}{2m} \int_0^t \left[\int_0^\tau f(\tau')\,d\tau' \right]^2\,d\tau.
\ee
Notice that, with our choices of the initial conditions
[$\xi(0)\,= d\xi(0)/dt= 0$ and $\Gamma(0)=0$], one has $\theta(x,0)=0$ and 
$y(0)=x$. Using \eqref{theta_f(t)_app}, from 
\eqref{conditions_integrab_app_0}, 
$\xi$ can be readily determined. 

For the sake of simplicity we will discuss the case
\be
\label{condition_f(t)_app}
\int_0^T f(\tau) d\tau=0
\ee 
(referring to \cite{colcelli2019integrable} for a discussion about the case $\int_0^T f(\tau)\,d\tau \neq 0$).
The major simplification is that the gauge phase \eqref{theta_f(t)_app} 
does not depend anymore, at stroboscopic times, on the spatial variable, 
{\it i.e.} $\theta(x,T)\equiv\theta(T)$.

Provided the condition \eqref{condition_f(t)_app} holds, with the functions $\theta$ and $\xi$ previously
determined we can rewrite the Lagrangian density \eqref{Lag_LL}
in terms of $\varphi(y,t)$ which involves no longer the external potential: 
\be
\mathcal{L} =\frac{i}{2}\left( \varphi^{\dagger} \p_t \varphi - 
h.c.\right)- \frac{1}{2m} \p_y \varphi^{\dagger} \, \p_y \varphi - \frac{c}{2} \varphi^{\dagger} \varphi^{\dagger} \varphi \varphi.
\label{Lag_LL_free}
\ee
Notice that the outlined procedure also works for a more general potential of the 
form $V(x,t) = xf(t)+ g(t)$. 

To determine the Floquet Hamiltonian
we need to determine the function $\theta$ at the stroboscopic times: $t\equiv nT$,
with $n$ integer. One has
\be
\label{theta_final_app}
\theta(nT) =-\frac{1}{2m}\int_0^{nT} dt \left[\int_0^t dt'f(t') \right]^2
\ee
and 
\be
\label{v_final_app}
\xi(nT)=-\frac{1}{m}\int_0^{nT} dt \int_0^t dt'f(t').
\ee

Now we want show that the ratios $\frac{\xi(nT)}{nT}$ and $\frac{\theta(nT)}{nT}$ 
do not depend on time, {\it i.e.} on $n$. Let be $\mathcal{F}(t)$ a function such that 
$\frac{d\mathcal{F}}{dt}=f(t)$. The constant of integration is chosen to be 
such that $\mathcal{F}(0)=0$. From \eqref{condition_f(t)_app} one has
\be
\int_0^T f(t)dt=\mathcal{F}(T)=0,
\ee
so that one can see that $\mathcal{F}(t)$ is a periodic function of period $T$. 
Using the definition of the function $\mathcal{F}$, from
\eqref{v_final_app} one gets
\be
\xi(nT)=-\frac{1}{m}\,\int_0^{nT}\mathcal{F}(t)dt=-\frac{I}{m}n,
\ee
where $I\equiv\int_0^T\,\mathcal{F}(t)dt$. 
It follows that the ratio $\frac{\xi(nT)}{nT}$ is $n$--independent. 
The same reasoning applies for the gauge phase $\theta$, since it is 
\be
\theta(nT)=-\frac{I'}{2m} n,
\ee
where $I' \equiv\int_0^T\mathcal{F}^2(t)dt$.

We are now able to write the Floquet Hamiltonian, which is found to be
\be
\label{Floq_HamG_manybody}
H_F=\sum_{j=1}^N \left( \frac{\hat{p}_j^2}{2\,m}+\frac{\xi(T)}{T}\hat{p}_j -
\frac{ \theta(T)}{T}\right)+c \sum_{j<i} \delta(x_j-x_i).
\ee

{We observe that the previous derivation of the Floquet Hamiltonian is valid 
not only for the one-dimensional Lieb-Liniger Hamiltonian, but also for a generic interacting system in any dimension 
subject to a periodic tilting. In the one-dimensional Lieb-Liniger this result 
implies the main point of this Section, relevant 
for atomtronics application}:  
the Floquet Hamiltonian \eqref{Floq_HamG_manybody} is {\it integrable}, as it can be immediately seen.
{A further important comment, on which we are going to comment more in the following, is that the derivation of \eqref{Floq_HamG_manybody} is valid for translational invariant systems.}

One
can then apply the standard Bethe Ansatz techniques
(see \cite{lieb1963exact,korepin1997quantum})
to compute the quasi-energies and the eigenfunctions. More precisely,
one has to compute the pseudo-momenta
$k_j$ obeying the Bethe equations.
If the system is subjected to periodic boundary conditions,
the pseudo-momenta $k_j$ are determined in terms of the following
Bethe equations
\be
\label{Bethe_eqs}
k_j\,L+2\sum_{i=1}^N {\rm \arctan}\left(\frac{k_j-k_i}{m c}\right)=2\pi\left(j-\frac{N+1}{2}\right),
\ee
for $j=1,\dots,N$, where $L$ is the circumference of the ring in which the system is confined.

Using the previous results one can write the many-body states at the stroboscopic times. A
multiparticle $\ket{\psi}$
state for the Lieb-Liniger model read (apart from the normalization factor)
\barray
\label{oold_50}
\ket{\psi}& \,=\,& \int \mathrm{d}^Nx \, \chi(x_1,\dots,x_N,t)
\nonumber \\
& &\times \Psi^{\dagger}(x_1,t)\dots\Psi^{\dagger}(x_N,t)\ket{0}
\earray
where $\chi(x_1,\dots,x_N,t)$ is the $N$--body wavefunction.

The wavefunction $\chi$ in \eqref{oold_50} is a solution of the Schr\"odinger equation 
\be
\label{se_1}
i \p_t \chi(x_1,\dots,x_N,t)=H \chi(x_1,\dots,x_N,t),
\ee
with $H$ being the Lieb-Liniger Hamiltonian in first quantization
\be
H=-\frac{1}{2m} \sum_{j=1}^{N} \p_{x}^2 + c \sum_{j<i} \delta(x_j -x_i) + \sum_{j=1}^N V(x_j,t).
\ee

Using \eqref{field_gaugetrasf} one can write for the periodically driven model
\barray
\ket{\psi}& =& \int \mathrm{d}^Ny \,\eta(y_1,\dots,y_N,t)
\nonumber \\
& &\times \varphi^{\dagger}(y_1,t)\dots\varphi^{\dagger}(y_N,t)\ket{0}.
\earray
The relation between the functions $\chi$ and $\eta$ is given by 
\be
\label{multipart_states}
\chi(x_1,\dots,x_N,t) \equiv \prod_{i=1}^{N} e^{i \theta(x_i ,t)} \eta(y_1,\dots,y_N,t),
\ee
where $\eta$ is the solution of the same 
Schr\"odinger equation \eqref{se_1} but with {\it no} external potential ($V\,=\,0$).

{An important comment is related to the fact that, as mentioned, the treatment of the Lieb-Liniger model in a periodic tilting presented 
in this Section is valid only for translational invariant systems. However, for setups relevant 
for atomtronics one has to consider the effect 
of their particular boundary conditions. As an example, one can consider circuits obtained merging one-dimensional waveguides. The derivation presented here does not longer applies and the results of this Section provides only 
a first step towards the determination of the correct Floquet Hamiltonian, a study which in the opinion of the author is a deserving subject in view of the possible obvious applications in atomtonics components and circuits. Separate considerations apply to ring geometries. One can think to periodically rotate the ring, with now the angle $\varphi$ playing the role of the coordinate $x$ since one can show \cite{colcelli2019integrable} that {\it in the comoving reference frame} (and under the assumption that $f(t)=0$ at the stroboscopic times) the Floquet Hamiltonian has the form \eqref{Floq_HamG_manybody}. This is analogous of what occurs for a two-dimensional harmonic potential in rotation, where the Hamiltonian in the rotating frame has the form $H-\Omega L_z$ 
\cite{cooper2008rapidly} and one can slightly deform the isotropic potential to break translational invariance. The equivalent of this 
in a periodically rotating ring geometry could be 
the addition of an out-of-plane periodic rotation 
component.}

\subsection{Concluding remarks and outlook}
\label{sec:outlook-protocol}

The past few years have witnessed very rapid developments within the theory of integrable systems out of equilibrium. Arguably, the most important piece of progress pertains the introduction of the so-called generalized hydrodynamics (GHD)~\cite{bertini2016transport,castroalvaredo2016emergent}. This is a very powerful framework, which builds upon the techniques developed in the idealized case of homogeneous systems, and allows one to provide exact predictions also for inhomogeneous settings, although only at hydrodynamic scales.

While a review of these results is beyond the scope of the present article, we note that recent works have shown that GHD is more than adequate to tackle exactly experimentally relevant set-ups of  repulsive $1D$ Bose gases, including systems with confining potential~\cite{doyon2017large,doyon2017note,caux2019hydrodynamics,ruggiero2020quantum}, spatial inhomogeneities~\cite{bastianello2019generalized} and dephasing noise~\cite{bastianello_generalised_2020}. In fact, quite remarkably, GHD predictions have now also been experimentally verified by monitoring clouds of bosonic cold atoms trapped on an atom chip~\cite{,schemmer2019generalized}. 

It would be extremely interesting to extend some of these recent results to inhomogeneous $1D$ Bose gases with attractive interactions, where the study of homogeneous quantum quenches have already revealed unexpected new features. More generally, a promising route is to analyze the out-of-equilibrium dynamics of even more complicated inhomogeneous integrable quantum gases, such as multicomponent mixtures of fermions and bosons~\cite{guand2013fermi,pagano2014one}, for which the emergence of interesting phenomena at the hydrodynamic scale has been already pointed out in simple settings~\cite{mestyan2019spin,peng2019gruneisen}.
 
Going beyond quench protocols, the effect of a time-periodic tilting in the Lieb--Liniger model with repulsive interactions {has been discussed}.
It was shown that the corresponding Floquet Hamiltonian is integrable, by studying the spectrum of the quasi-energies and the dynamics of the system at stroboscopic times. Importantly, the analysis presented for the Lieb-Liniger model can be extended to other $1D$ integrable systems in time-periodic linear potentials
such as, for instance, the Yang-Gaudin model for fermions. In the future, 
it would be very interesting to study the effect of periodic tilting in more general configurations. 
{A main issue to be studied starting 
from the results presented here is that of the  boundary conditions of the specific atomtronics 
system of interest when subject to a periodic tilting. 
Among the different cases that would provide a worthwhile investigation is that of atomtronics circuits periodically tilted and their application to atomtronics tasks.}

\emph{Acknowledgments:} We are grateful to P. Calabrese,
A. Colcelli, F. Essler, G. Mussardo and G. Sierra
for several enlightening discussions on the subjects presented here 
and on related topics.
Further acknowledgments go to the participants to the first week
of the 2019 Atomtronics conference in Benasque, and in particular to V. Ahufinger, V. Bastidas,
D. Cassettari, J. Mompart, N. Proukakis, F. Scazza and W. von Klitzing.

\section{PERSISTENT CURRENTS AND VORTICES IN ATOMTRONIC CIRCUITS} 
\label{Persistent_toroid}
\vspace*{-0.5cm}
\par\noindent\rule{\columnwidth}{0.4pt}
{\bf{\small{T. Bland, M. Edwards, N. P. Proukakis, A. Yakimenko}}}
\par\noindent\rule{\columnwidth}{0.4pt}


Atomtronics relies on the flow of coherent matter waves in the form of atomic Bose-Einstein Condensates (BECs) in closed circuits, such as in the form of closed toroidal traps, or more extended, race-track-like, potentials.
Persistent currents in such geometries enable fundamental studies of superfluidity and may lead to applications in high-precision metrology and atomtronics \cite{amico2017focus,turpin2015blue}. The question of the generation and stability of the atomic persistent currents -- which in the absence of external driving should be topologically protected -- is of fundamental importance; thus it has been the subject of numerous experimental and theoretical investigations
\cite{edwards2013probing,ramanathan2011superflow,wright2013driving,yakimenko2015vortices,yakimenko2015vortex,yakimenko2015generation,hadzibabic2013persistent,yakimenko2013stability,paraoanu2003persistent,rooney2013persistent,mathey2014decay,wright2000toroidal,wright2008optical,moulder2012quantized,ryu2007observation,ryu2013experimental,eckel2014hysteresis,jendrzejewski2014resistive,eckel2016contact,corman2014quench,aidelsburger2017relaxation,das2012winding,bland2020persistent}

The quantized circulation in a ring effectively corresponds
to an $m$-charged vortex line pinned at the center of the
ring-shaped condensate, where the vortex energy has a local
minimum. Noting that there is no condensate density at that location, we can think of this as a `ghost' vortex 
-- in the sense that, at some radial distance from the centre of the closed loop where there is non-negligible superfluid density, the arising phase profile is identical to that corresponding to a vortex located at the centre, where the superfluid density is practically zero. 
{Since the vortex line energy increases with condensate density such a vortex turns out to be bounded by the potential
barrier, that is why even multicharged ($m > 1$) metastable vortex states can be very robust.} The generation and decay of a persistent current is governed by the dynamics of these quantum vortices, which can be deterministic, or random, depending on the particular setting considered.

Specifically, 
{as modelled theoretically and observed experimentally,}
persistent currents can form in toroidal BECs by stirring the condensate with an optical paddle potential, imparting angular momentum 
in the ring through the generation of vortices and through the decay dynamics after an external perturbation\,{\cite{ryu2007observation, ramanathan2011superflow, wright2013driving, ryu2013experimental, eckel2014hysteresis, jendrzejewski2014resistive, eckel2016contact,brand2001stirring,rooney2013persistent,mathey2014decay,yakimenko2015vortices,yakimenko2015vortex, paraoanu2003persistent}.}
 They can also be induced by transmitting angular momentum from a Laguerre-Gaussian beam\cite{wright2000toroidal,wright2008optical,moulder2012quantized}.
Moreover, persistent currents can also spontaneously form in toroidal BECs as phase defects appearing after a quench into the BEC state\cite{corman2014quench,aidelsburger2017relaxation,das2012winding,bland2020persistent}. 
{Persistent currents also arise in multi-component condensates in a ring geometry \cite{beattie2013persistent,yakimenko2013stability,gallemi2015coherent,white2017odd,chen2019immiscible,yakimenko2013stability}.
}

Coupled persistent currents of ultracold atomic gases provide a possibility to investigate the interaction of the superflows in a tunable and controllable environment, providing the possibility for precision measurements and even potentially controllable quantum gate operations. Previous theoretical studies \cite{brand2009tunneling, brand2018nonlinear, brand2018nonlinear, aghamalyan2013effective} have drawn considerable interest to
systems of coupled circular BECs. Using accessible experimental techniques, it is possible to consider a
variety of physical phenomena in this setting: from Josephson effects in~the~regime of weak interactions (where the superflow decays by inducing phase slips reviewed in Section VIII) to quantum Kelvin--Helmholtz instability for
merging rings.

In this contribution we review recent developments in the understanding of the formation and dynamics of persistent currents in such closed geometries 
{based on mean-field, dissipative, and stochastic simulations.} We start by considering the mechanism of formation of persistent currents in a racetrack BEC, induced by a stirring potential (Section~\ref{sec:Edwards}) (which also encompasses ring-trap geometries as a special case).
We then discuss more complicated atomtronic architectures, focusing on BECs trapped in two coupled toroidal potentials which are either embedded within a single plane, or are linked transversally.
Specifically, we firstly review (Section~\ref{sec:Proukakis})
recent work \cite{bland2020persistent} discussing the spontaneous formation of persistent currents in co-planar double ring geometries. We then present a brief overview (Section~\ref{sec:Yakimenko}) of recent investigations \cite{oliinyk2019tunelling,oliinyk2019symmetry,oliinyk2020nonlinear} of the dynamics of quantum vortices in a pair of vertically stacked atomtronic circuits. 
We end with some concluding statements.


\begin{figure}[b]
\centering 
\includegraphics[scale=0.30]{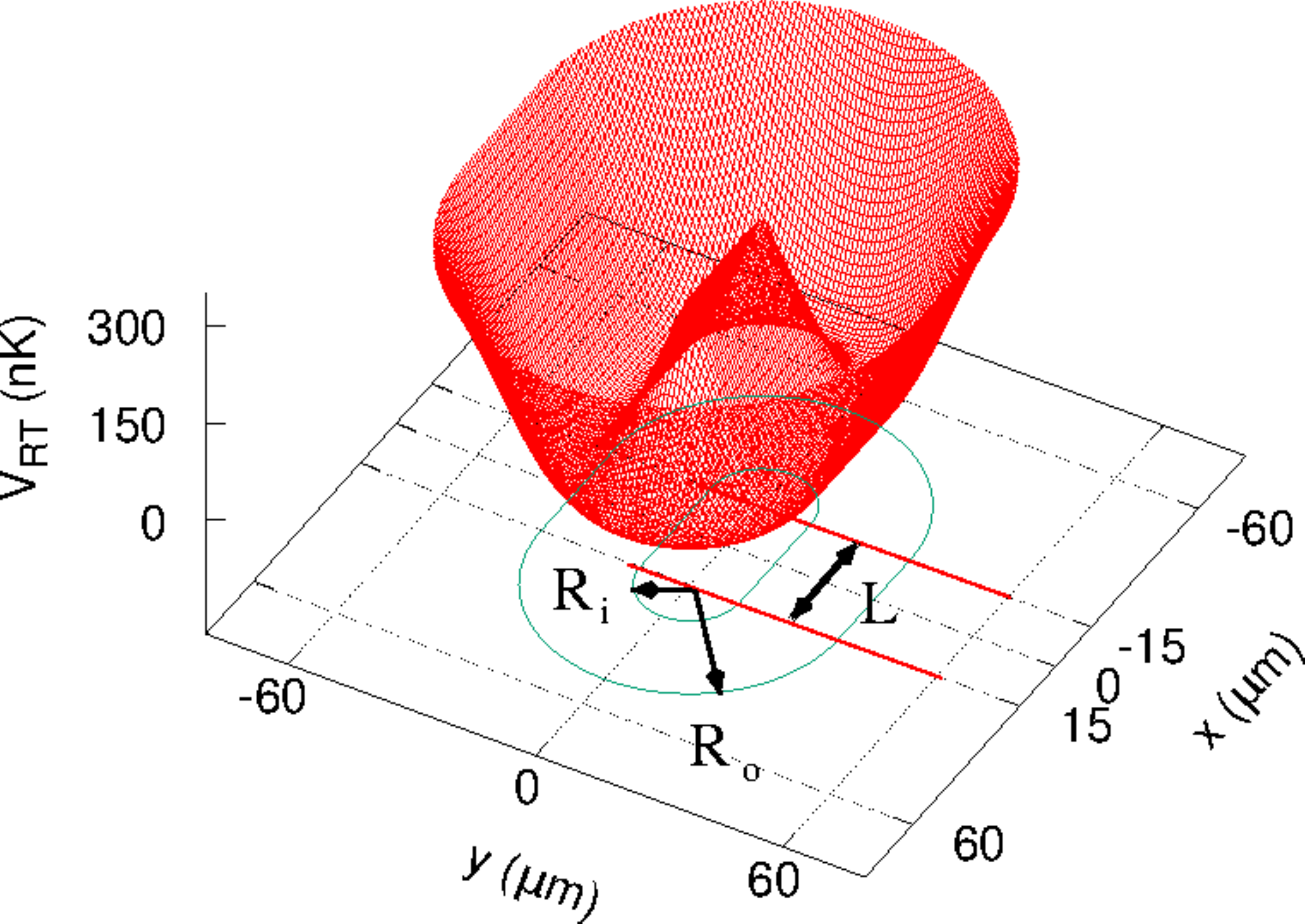}
\caption{The atom--circuit racetrack potential consists of two semi--circular endcaps (inner radius $R_{i}=12\,\mu$m, outer radius $R_{o}=36\,\mu$m) separated by straightaways of length $L$.  {This figure was created using data reproduced from Ref.\,\cite{PhysRevA.102.063324}.}}
\label{racetrack}
\end{figure}
%
%

\subsection{\label{sec:Edwards} Mechanism for producing flow in a racetrack  atom--circuit BEC by stirring}


%
%
\begin{figure*}[!t]
\centering 
\includegraphics[scale=0.35]{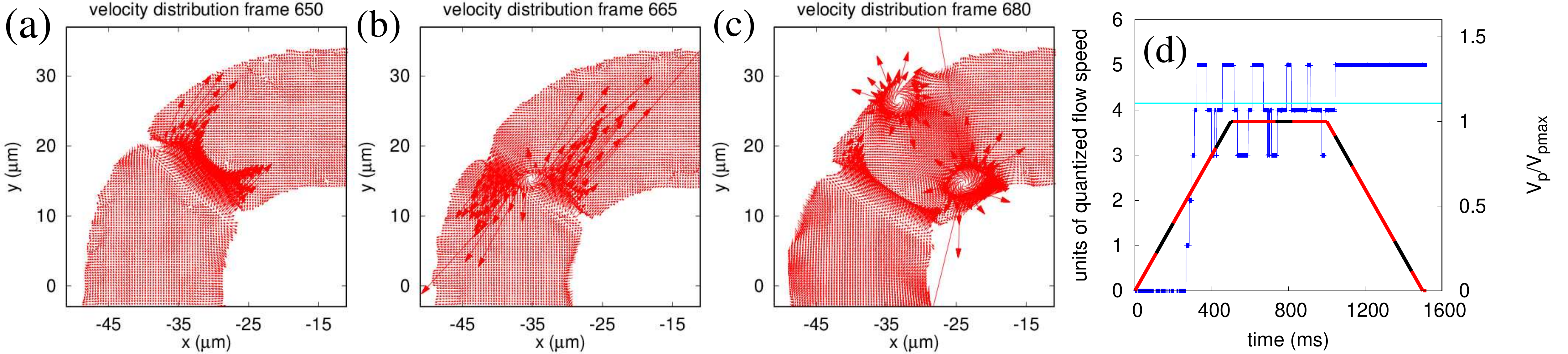}
\caption{Velocity distributions of the racetrack condensate during the ramp--up of the stirring. (a) A backflow plus vortex/antivortex pair develops in the barrier region, (b) the vortex/antivortex pair swap, (c) the the vortex/antivortex pair move away form the barrier in the anti--stir direction while a compression wave moves off in the stir direction. (d) {Circulation (blue line) around the midline track versus time for this case. The red and black curve gives the barrier height versus time.  The curve is colored red when the barrier is on the curved parts of the racetrack and colored black when it is on the straight parts. The quantity $V_{\rm p,mas}$ is the maximum energy height of the barrier during the stir. The straight cyan curve shows the stirring speed of the barrier in units of the quantized flow speed, $v_{flow}=2\pi\hbar/(Ms)$, where $s$ is the arc length of the racetrack midline. Parameters: $L=30\,\mu$m, $v_{\rm stir}=339\,\mu$m/s. Results shown are based on zero--temperature Gross-Pitaevskii mean-field theory. Panel (d) of this figure was created using data reproduced from Ref.\,\cite{PhysRevA.102.063324}.}}
\label{vortex_swap}
\end{figure*}

\subsubsection{How stirring a racetrack atom circuit produces flow }
\label{mechanism}

We start by presenting a picture of how macroscopic flow is produced in a BEC confined in a racetrack atom--circuit by stirring with a wide rectangular barrier within the Gross--Pitaevskii model.  The atom--circuit BEC is strongly confined to a horizontal plane and the 2D racetrack channel potential (see Fig.\,\ref{racetrack}) consists of two half--circles separated by straightaways of length $L$; we note that in the limit $L=0$, this reduces to a ring potential, so the present discussion fully encompasses that setting~\cite{PhysRevA.102.063324}.

It is well--established that flow is accompanied by production and motion of vortices\,\cite{yakimenko2015vortices, yakimenko2015vortex, ramanathan2011superflow, wright2013driving} or dark solitons\,\cite{cominotti2014optimal} in the condensate.  Here we describe how and where vortices form, how they move thereafter, how circulation localized to a vortex is converted into macroscopic flow around the entire racetrack, and what conditions lead to the final amount of flow.

\subsubsection{Creation of a single unit of flow: vortex swap}
\label{one_flow_unit}

Flow can be induced in a racetrack BEC atom circuit by stirring with a weak--link potential barrier. As the stirring barrier moves and strengthens it produces a region of lowered density. This region of depressed density causes a backflow (flow opposite the stir direction) to develop in this region.  This backflow spawns a vortex (circulation same as the stir) located on the outside of the channel and an antivortex (anti--stir circulation) on the inside.  At a critical value of the barrier height the two vortices swap positions.  This event generates two disturbances that move away from the barrier in opposite directions at the average speed of sound.  The first is the vortex/antivortex pair that moves off in the anti--stir direction and the second is a compression wave moving in the stir direction.

This backflow is illustrated in Fig.\,\ref{vortex_swap}(a). In the full figure we have plotted a series of snapshots of the velocity distribution from shortly before until shortly after the creation of the first unit of flow.  It is easy to see that the backflow speed is greatest at the inner and outer edges of the channel where the racetrack plus barrier potential is largest.  As the barrier moves the condensate in front of the barrier must migrate to the back of the barrier.  The atoms at the inner and outer channel edges must move faster to avoid the regions of high potential.  In this way vortices are formed by stirring with a barrier that is much wider than the stirred condensate.

When the height of the barrier reaches a critical value, the vortex migrates from the outside to the inside of the channel as can be seen in Fig.\,\ref{vortex_swap}(b). Shortly after this vortex swap two disturbances are generated.  The first is the vortex/antivortex pair, located on the inside and outside of the channel respectively, move away from the barrier in the anti--stir direction.  This vortex pair causes atoms on the anti--stir side of the barrier to flow in the stir direction between the vortices.  The second disturbance is a compression wave that propagates away from barrier region in the stir direction.  This compression wave also moves atoms in the stir direction.  Both disturbances move at a speed that is approximately the local speed of sound ($c({\bf r})=\sqrt{gn_{c}({\bf r})/m}$) averaged over the cross section of the condensate. These disturbances are the mechanism by which the localized circulation in the form of a vortex is converted into macroscopic flow around the entire racetrack.

\subsubsection{Final flow production: flow overtakes the barrier}
\label{ring_final_flow}

The final circulation produced can be roughly predicted as the number of units of quantized flow that lies closest to the speed of the stirring barrier.  The exact number depends on the details of the stirring and the geometry of the racetrack as we describe below.  When vortices inside the racetrack potential are generated the circulation they provide is localized near their cores.  As stirring proceeds this circulation is converted into a nearly constant tangential velocity component around the midline track by the pair of disturbances generated each time a vortex swap occurs. 

The circulation as a function of time is shown in Fig.\,\ref{vortex_swap}(d) for the case where $L=30\,\mu$m, $v_{\rm stir}=339\,\mu$m/s and, $T=0$ nK).  This graph shows that the circulation (shown as the blue curve) is zero until a succession of vortex--swap events produces enough flow so that the flow generated is greater than the speed of the stirring barrier (shown as the cyan horizontal line in the figure in quantized flow speed units).  

In this case, five units of flow exceeds the barrier speed by almost a full flow speed unit.  When the disturbance pair generated by the first vortex swap travels around the racetrack and comes back to sweep through the barrier region again they cause a {\em forward flow} to develop in the barrier region.  At this moment an inverse vortex swap event can occur causing the total circulation to decrease by one unit.  In this way the circulation can oscillate around the number of units that makes the flow closest to the stir speed of the barrier.

Another circulation--changing mechanism that is only present in the non--ring racetrack case occurs when the barrier transitions from straight parts of the racetrack to curved parts or vice--versa. The times when the barrier is on straight or curved parts are indicated in Fig.\,\ref{vortex_swap}(d) by the red-- and black--colored curve that depicts the barrier height.  This graph is colored red for times when the barrier is on the curved parts of the racetrack and black--colored when it is on the straightaways. Careful examination of the circulation graph shows that, when the barrier transitions from curved to straight (red to black) racetrack parts, the circulation increases by one unit.  When the barrier transitions from straight to curved parts the circulation decreases by one unit.  We also note that this only happens when the barrier strength is above a critical value.

The general mechanism for flow production in the racetrack by stirring with a rectangular barrier in the context of the Gross-Pitaevskii equation can thus be summarized as follows.  The stirring barrier both moves and increases in strength.  This generates a backflow in the region of depressed density.  This backflow is fastest at the inner and outer channel edges in this region.  This flow spawns a vortex/anti--vortex pair at the outer and inner edges, respectively.  Eventually these two vortices swap locations generating a moving vortex (now on the inner channel edge)/anti--vortex (now at the outer edge) pair that moves away from the barrier in the anti--stir and also generating a compression wave that moves away from the barrier in the stir direction.  These disturbances both move at the average speed of sound.  The total amount of flow produced is roughly the number of flow--speed units closest to the speed of sound.

\subsection{\label{sec:Proukakis}Persistent Currents in Co-Planar Double-Ring Geometries}



Having identified the key mechanism for flow production in the context of a pure $T=0$ BEC, it is natural to also consider the role of phase fluctuations and dissipation on the (spontaneous) emergence of supercurrents, and what happens when multiple ring-trap geometries are coupled.



\subsubsection{Spontaneous Persistent Current Formation in a Ring Trap}
\label{proukakis-single-ring}

The formation of persistent currents in a ring trap can also proceed spontaneously; it is well-known that the generation of a superfluid in such a geometry can carry with it a randomly-generated winding number, which is expected to be statistically distributed about the most probable `zero' (0) value [which corresponds to the absence of a persistent current] \cite{das2012winding,corman2014quench}. This is because phase coherence forms locally in a ring, and the size and width of the toroidal geometry, along with the rate of the actual quench leading to the formation of the ring-trap condensate, control the maximum winding number that can spontaneously emerge \cite{paraoanu2003persistent,bland2020persistent}. 

This is already well-known in the context of the Kibble-Zurek mechanism \cite{kibble1976topology,zurek1985cosmological}, which relates the generated winding number to the quenching rate of the driven phase transition, an effect already discussed and observed across different physical systems.
Among those, this effect has also been confirmed in ultracold atoms in ring-trap geometries through a controlled gradual cooling rate quench, producing an experimentally observed distribution of winding numbers \cite{corman2014quench}, in agreement with numerical and theoretical expectations \cite{bland2020persistent}. In fact, the local nature of such coherence evolution implies that  this effect of spontaneous persistent current generation already manifests itself even in the limit of very rapid (or instantaneous) quenches towards a coherent superfluid regime. The existence of phase fluctuations and non-zero winding numbers can, for example, affect the dynamics of otherwise deterministically-generated dark solitons in ring-trap geometries \cite{gallucci2016engineering}.

%
%
\begin{figure*}[t]
\centering 
\includegraphics[scale=1.30]{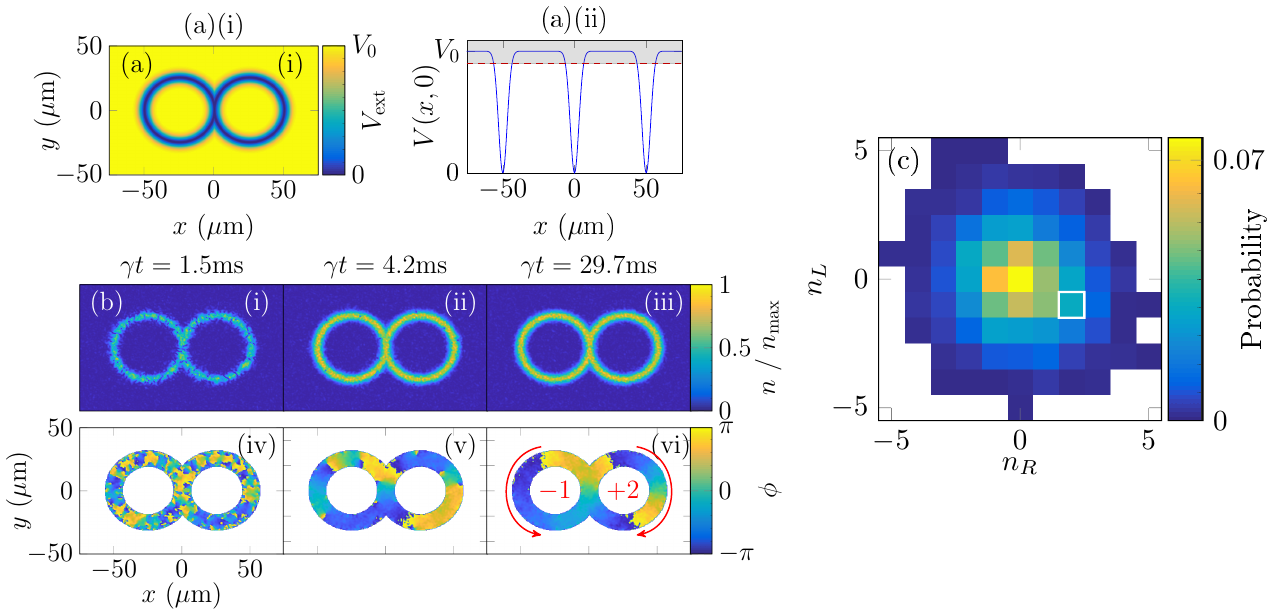}
\caption{
(a) Simplest co-planar connected double-ring geometry: Shown are (i) the 2D potential, and (ii) the cut through $y=0$. Note that the interfacial potential considered here is identical to that of the outer unconnected regions of the rings.
(b) Formation dynamics of characteristic numerical realization with different winding numbers across the two rings: shown are growing density (from a noisy initial configuration; top row (i)-(iii)) and corresponding phase evolution (bottom row, (iv)-(vi)), for a characteristic example with distinct winding numbers $n_{\rm left} = -1$ and $n_{\rm right} = +2$ across the left and right rings respectively.
(c) Histogram of all possible steady-state winding number contributions, when performing numerical quenches from an initial noisy configuration. The indicated white box highlights the case $(n_{\rm left}, \, n_{\rm right}) = (-1,\, +2))$ considered in (b). We have explicitly verified that once such an unequal winding number contribution forms after sufficient system relaxation/equilibration, it does remain stable for all subsequent evolution (with this feature also found to be insensitive to the precise choice of the `growth' parameter $\gamma$ in the stochastic simulations).
Parameters: $N \sim 2 \times 10^5$ $^{87}$Rb atoms, $T=10$nK, effective 2D interaction strength $\tilde{g} = 0.077$; Tight harmonic transverse confinement with $\omega_{z} = 2 \pi \times 1000$Hz, with the 2D potential defined by 
$V(x,y)=V_0\,\text{min}\left(1-\exp\left[-2\left(\rho(x-R,y)-R\right)^2/w^2\right],
\,\, 1-\exp\left[-2\left(\rho(x+R,y)-R\right)^2/w^2\right]\right)\,,$
where $\rho(x,y)=\sqrt{x^2+y^2}$, $V_0 = 1.1 \mu$, the ring radius is $R = 25 \mu$m and its width is $w = 6 \mu$m. 
{Adapted with permission from T. Bland, Q. Marolleau, P. Comaron, B. Malomed, and N. P. Proukakis, J. Phys. B. 53, 115301 (2020). Licensed under a Creative Commons Attribution 4.0 license \cite{bland2020persistent}. }
\label{proukakis-fig-1}
}
\end{figure*}
%
%

\subsubsection{Spontaneous Persistent Current Formation in Co-Planar Connected Ring Traps}
\label{proukakis-connected-rings}

Next, we turn our attention to the dynamics of winding numbers in connected geometries, focusing here on the simplest possible such co-planar example, based on the 2D geometry shown in Fig.~\ref{proukakis-fig-1}(a). 
Utilizing state-of-the-art numerical simulations based on the stochastic (projected) Gross-Pitaevskii equation \cite{stoof2001dynamics,gardiner2003stochastic,bradley2008bose,proukakis2008finite,blakie2008dynamics,das2012winding,rooney2012stochastic,proukakis2013quantum,berloff2014modeling,bland2020persistent}, we find that -- as in the case of a single  ring trap -- coherence within the double-ring trap forms locally during condensate growth, as shown in Fig.~\ref{proukakis-fig-1}(b). 

After a quench, the phase develops locally along and across the ring circumference \cite{das2012winding}, with an early such example of typical evolution shown by the phase profile of Fig.~\ref{proukakis-fig-1}(b)(iv). Although at the common interface around $x \sim 0$ (where the trap depth has the same depth and width as the outer double-ring edges) the phase of the superfluid is constrained to be the same in both connected rings, this does not nonetheless dictate the behaviour of the emerging phase in the remaining `unconnected' regions forming the bulk of the ring's spatial extent. Specifically, the `unconnected' regions in the double-ring geometry are free to randomly establish their own phase dynamics (constrained by the size and width of the unconnected regions), thus often leading to non-zero winding numbers with varying (nonlinear) phase gradients across the ring circumference. 

An example of such long-term behaviour with winding numbers $-1$ and $+2$ across the left and right rings respectively, is shown in Fig.~\ref{proukakis-fig-1}(b)(vi), with a positive winding number referring arbitrarily, by convention\cite{bland2020persistent,rooney2013persistent}, to clockwise rotation.
In the double-ring case, we also find a distribution of winding numbers about the most probable value of zero net winding number, as shown in Fig.~\ref{proukakis-fig-1}(c). In fact, when integrating over the winding numbers of the other ring, we find that (in our chosen, experimentally-relevant, geometry), the distribution of winding numbers in each ring actually exactly matches that of the single ring trap with the same radius, width and depth \cite{bland2020persistent}. We expect this to be true for the majority of experimentally-relevant potentials, for which the ring radius typically largely exceeds any transversal width (and motion is frozen out in the third, transverse, direction).

Remarkably, our previous work \cite{bland2020persistent} has shown such features to be largely independent of the exact details of the connected geometry, provided it does consist of two (2D) planar-connected closed geometries with a unique single (extended) connected region. For example, we have verified all above conclusions to be also valid in a figure-of-eight (`lemniscate') potential, where there is a real crossing in the path of propagating ultracold atoms \cite{bland2020persistent}. By extension, we would therefore expect similar features to hold in extended or `flattened' geometries, such as connected race-track geometries, as the underlying physics is that of how much winding can be supported by the combination of loop radius $R$ and width $w$, which are found to obey the winding number relation $\langle | n_w | \rangle \sim \sqrt{2 \pi R/w}$ \cite{zurek1985cosmological,das2012winding,bland2020persistent}.

Given the potential independence of winding numbers supported across two identical connected ring traps, it is interesting to enquire about the stability of such features. Essentially, one can think of a winding number  of, say, $\pm n$ (where $n=0,1,2,\cdot$) around a closed loop (whether exactly ring-shaped, or not), as being due to the existence of a `ghost' vortex trapped in the middle of the closed loop. 
%
Using such an intuitive interpretation, the winding number of a ring trap will change by an integer unit if such a `ghost' vortex is allowed to leave its enclosure, mapped out by the underlying trap potential. 
{Due to the topological protection of the winding number, such an effect can be achieved by deforming the system topology through a change in the trap potential: for example, in the single ring-trap case, this could be achieved by opening a small hole in the potential, such that the enclosed ‘ghost’ vortex can escape to the region outside of the ring. The related topic of phase slips in the presence of fluctuations is discussed in the next Chapter \ref{PhaseSlips}. 
}

In the double-ring geometry, we have explicitly verified that the transfer of the winding number from one side of the double-ring geometry, to the other can be facilitated by allowing for a zero-potential region to connect the two sides. Such a transfer can be reasonably controlled by the specific details of the potential deformation, even potentially leading to the annihilation of oppositely-oriented superflows (corresponding to `ghost' vortices of opposite circulation, which can hence annihilate), a topic of active ongoing research investigations to harness potential atomtronic applications.
Our present work for interacting superfluids adds to that of tunnelling angular momentum states considered at the single-particle level in single-component condensates \cite{pelegri2017single,pelegri2019topological,pelegri2019topological2} 
  and also for two-component condensates \cite{gallemi2015coherent}.

Although work discussed here has been restricted to a coupled geometry with a single extended interface, once such transfer process becomes reasonably controlled for multi-particle systems, one may envisage possible extensions to multiple connected closed-loop geometries (whether ring-shaped, race-track, or related), with the aim of deterministic transfer of winding numbers across a multiple-loop atomtronic architecture. Research into this promising direction is currently very active by the present authors.


Next, we discuss coupled persistent current dynamics in an alternative geometry of two transversally stacked ring-trap potentials connected by tunneling.

\subsection{\label{sec:Yakimenko} 
Persistent Currents in Transversally
coupled atomtronic circuits}



\begin{figure}[t]
\includegraphics[width=1.0\columnwidth]{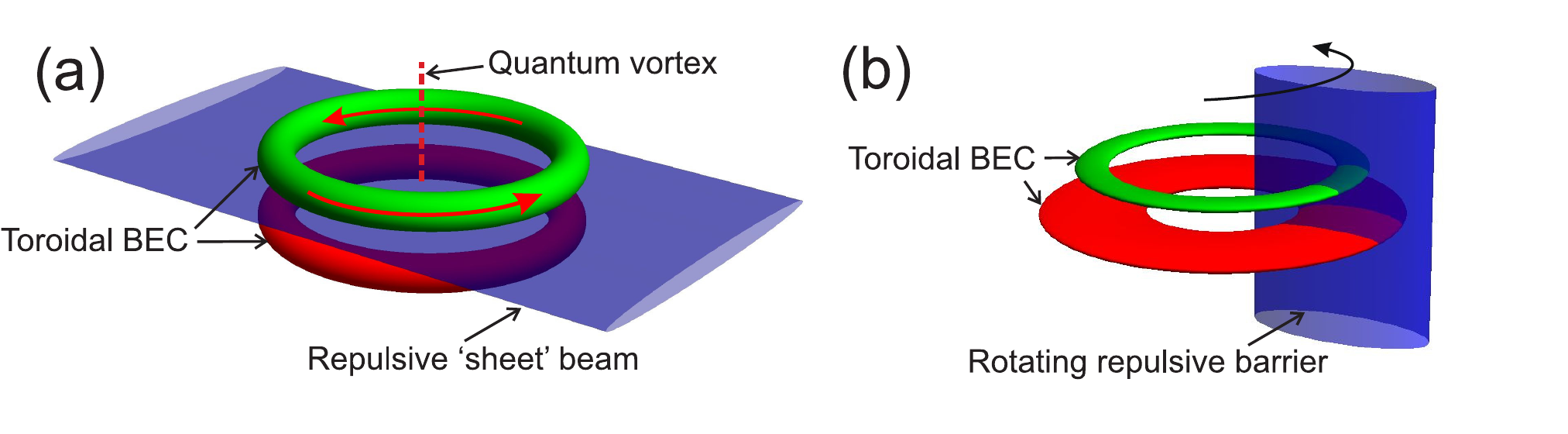}
\caption{(a) Schematics of the coupled ring-shaped condensates. Coaxial rings separated by repulsive potential, allowing investigation regime of  tunneling coupling (long Josephson junction) and regime of merging rings (when the barrier is eliminated). (b) Schematics of  preparation of the state with different angular momenta  in double-ring system.  Coaxial rings with different number of atoms are stirred by a rotating potential barrier. A persistent current is generated in a less populated ring (shown by green) while more populated toroidal condensate (shown by red) remains in non-rotating state.}
\label{fig:Schematics_Double_Ring}
\end{figure}

Here we briefly overview our recent findings \cite{oliinyk2019tunelling,oliinyk2020nonlinear,oliinyk2019symmetry} on dynamics of quantum vortices in two coupled vertically stacked toroidal condensates with persistent currents (see Fig. \ref{fig:Schematics_Double_Ring} (a)).

{In practice,} the double-ring system with different angular momenta in its top and bottom parts may appear spontaneously as a result of cooling,
with different momenta, $m_{1}$ and $m_{2}$, being frozen into the two rings after the transition into the BEC state, similar to spontaneous persistent current formation in co-planar coupled rings, described in Sec. \ref{sec:Proukakis}. {We note that creation of ring currents in a double ring system by cooling \cite{BrandPRL13} and instability of nonrotating tunnel coupled annular Bose-Einstein condensates \cite{Klitzing_2007, brand2010prl} have been discussed in literature.}

The asymmetry of the density distribution in the top and bottom rings makes it possible to excite the vorticity also by applying a stirring laser beam, similar to {the} mechanism described in Sec. \ref{sec:Edwards}. Generating {a} vortex in {the} lower-populated ring only, keeping the higher-populated one in the zero-vorticity state is illustrated in Fig. \ref{fig:Schematics_Double_Ring} (b). {A detailed analysis of the methods for persistent current generation in the system of coupled rings is under progress and will be published elsewhere.}

\begin{figure}[t]
\includegraphics[width=1.0\linewidth]{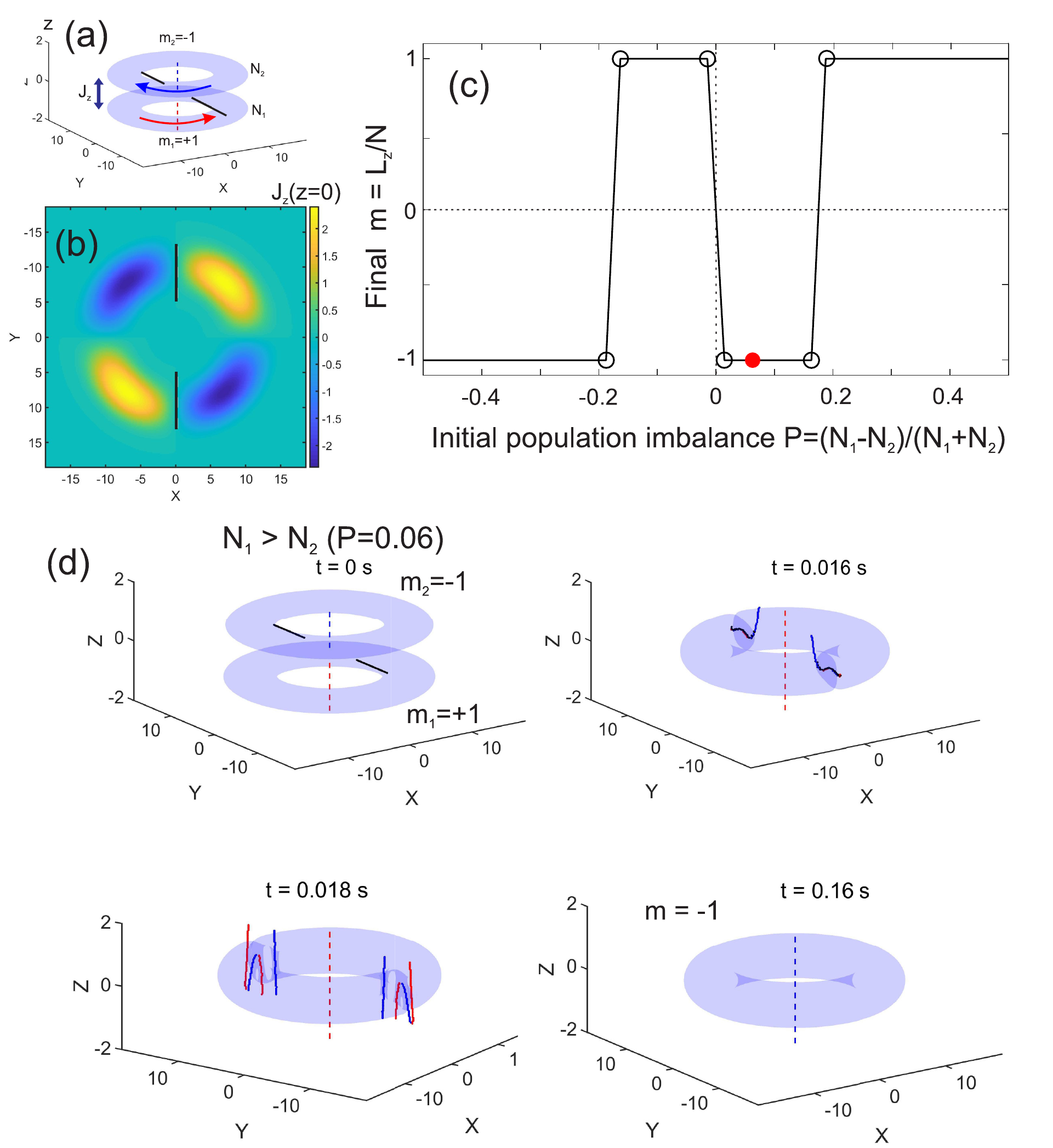}
\caption{Coupled coaxial superfluid atomic circuits with counter propagating persistent currents.  (a) Hybrid vortex stationary states with hidden vorticity. Vertical red and blue dashed lines designate cores of the counter-propagating persistent currents in the two rings. The cores of the Josephson vortices are indicated by solid black lines.  {(b)  $z$-component of the corresponding tunnel-flow density distribution through the barrier, $J_z(x, y, z = 0)$.  (c) The final value of the total angular momentum per particle $L_z/N$ for the merging rings with initial vorticities $(+1,-1)$ as a function of initial population imbalance $P=(N_1-N_2)/(N_1+N_2)$.}  	(d)	An example of evolution of the merging rings in oblate trapping potential. The barrier separating two rings is switched off at $t > t_d = 0.015$ s, dissipative parameter $\gamma =0.03$. Red (blue) lines indicate positions of the vortex (antivortex) core. The population of the bottom ring, with vortex $m_1 = +1$, is slightly larger than in the top one, with antivortex $m_2 = -1$ [initial imbalance parameter, $P = 0.06$, is indicated by filled {red circle in (c)}]. The final state has $m = -1$. The symmetric drift of two diametrically opposite antivortices towards the central hole leads to subsequent annihilation of the central vortex and relaxation of the toroidal condensate into a final \emph{antivortex} $m=-1$ state,  {as described in Ref. \cite{oliinyk2019symmetry}}.}
\label{fig:Coaxial}
\end{figure}

In our recent work \cite{oliinyk2019tunelling} it was demonstrated that the azimuthal
structure of the tunneling flows in double-ring system with topological charges $m_1$, $m_2$ implies formation of $|m_1-m_2|$ Josephson vortices, also known as rotational fluxons. The azimuthal structure of the tunneling flow (see the inset in Fig. \ref{fig:Coaxial} (a)) implies zero net (integral) current through the junction for states, built of persistent currents with different topological
charges in coupled rings ($m_1\ne m_2$). {In particular, these include the case of opposite
topological charges ($m_1=-m_2)$ -- considered in Ref. \cite{driben2014three} and called 'hybrid
vortex soliton'. These structures host two different types of the vortices: vertical vortex lines and horizontal Josephson vortices.}  It turns out that the fluxons' cores rotate and bend,
following the action of the quench, i.e. formation of tunnel junction with chemical potential difference.
It was found in Ref. \cite{oliinyk2019tunelling,oliinyk2019symmetry,oliinyk2020nonlinear}, as the barrier decreases, and the effective coupling between the
rings respectively increases, the Josephson vortices accumulate more and more energy. When the persistent currents merge the relaxation process to new equilibrium state is driven by 3D dynamics of interacting Josephson vortices and vortex lines of the persistent currents (see for example Fig. \ref{fig:Coaxial} (b)). 

In our simulation of the merging rings we have used the dissipative Gross-Pitaevskii equation in the form: 
\begin{equation}
(i-\gamma )\hbar \frac{\partial \psi }{\partial t}=-\frac{\hbar ^{2}}{2M}%
\nabla ^{2}\psi +V_{\text{ext}}(\mathbf{r},t)\psi +g|\Psi |^{2}\psi -\mu
\psi ,  \label{GPE3D}
\end{equation}%
where $g=4\pi a_{s}\hbar ^{2}/M$ is the coupling strength, $M$ is the atomic
mass ($M=3.819\times 10^{-26}$ kg for $^{23}$Na atoms), $a_{s}$ is the $s$%
-wave scattering length (positive $a_{s}=2.75$ nm, corresponding to the
self-repulsion in the same atomic species, is used below), $\mu $ is the
chemical potential of the equilibrium state, and $\gamma \ll 1$ is a
phenomenological dissipative parameter. This form of the dissipative GPE has
been used extensively in previous studies of vortex dynamics (see, e.g.,
\cite{proukakis2008finite,yakimenko2015vortices,yakimenko2015generation,yakimenko2013stability,yan2014exploring}). 
%
%
Note that  main results of our work, concerning the role of the symmetry breaking in the interacting superfluids rings are not sensitive to the weak dissipative effects. We demonstrated in our works \cite{oliinyk2020nonlinear,oliinyk2019symmetry} that the symmetry of the system is the key feature explaining remarkable properties of the interacting quantized superflows. Certainly, the symmetry is in no way related to details of the dissipative terms.  We have found that the subsequent relaxation process is determined by the initial stage of the evolution of the merging ring, in the course of several first microseconds after the barrier was switched off. Obviously, an effect of the weak dissipation on these fast processes is practically negligible.  The dissipation plays a significant role in the course of subsequent temporal evolution of the condensate. In fact, in most experiments in-situ observation of the vortices is not possible, and only the final state can be analyzed after the completion of the relaxation.  We include the dissipative effects in our model to investigate the final states of the merging superflows, which can be directly compared with expected experimental observations.

\begin{figure}[htb]
\includegraphics[width=1.0\linewidth]{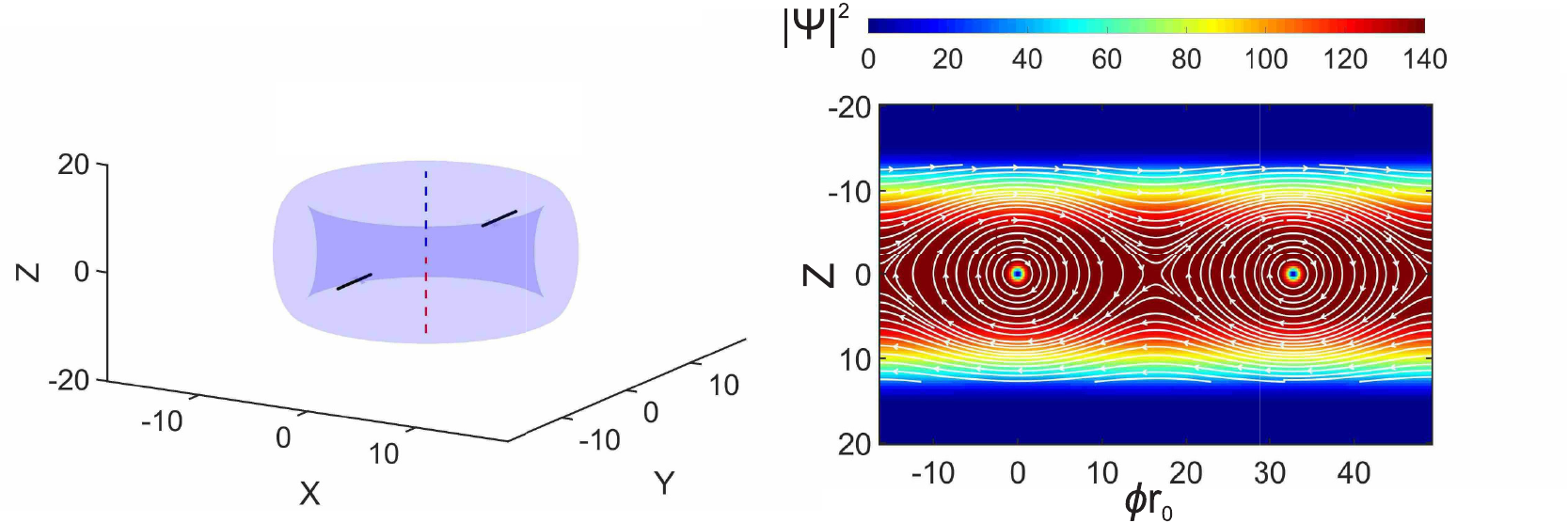}
\caption{A long-lived hybrid complex is produced by the evolution of merging strongly elongated toroidal condensates, { as described in Ref.  \cite{oliinyk2020nonlinear}}. Shown are  3D isosurface with constant condensate density (left) and maps of the distribution of the density (right) on the cylindrical surface at the radius corresponding peak density of the condensate, $\phi$ being the angular coordinate.}
\label{fig:hybrid}
\end{figure}


The relaxation of the merging rings is driven by substantially 3D nonlinear dynamics of the vortex lines corresponding to persistent currents and Josephson vortices, as illustrated in Fig. \ref{fig:Coaxial} (b). It turns out that the final state of the condensate crucially depends on an initial population imbalance in the double-ring set, as well as on the shape of the 3D trapping potential, oblate or prolate \cite{oliinyk2020nonlinear}.  In the oblate (axially squeezed) configuration, a ring with non-zero angular momentum can impose its quantum state onto the originally non-rotating ring only above a well-defined critical value of the population imbalance.

{It is apparent that two merging classical counter-propagating flows with zero total angular momentum evolve to the ground (non-rotating) state.
Surprisingly,} merging counter-rotating quantized flows in the axially-symmetric trap \emph{never evolve towards the non-rotating ground state}, with $L_z = 0$, even for small imbalances, $P\ll 1$ (see Fig. \ref{fig:Coaxial}). {It is particularly remarkable} that the vorticity of the final state is imposed by the \emph{less populated component} if $P < P_{cr}\approx 0.1755$, and by the stronger component only if $P > P_{cr}$. {These counter intuitive properties of merging superflows are illustrated in Fig. \ref{fig:Coaxial} (c),(d). In this example number of atoms in the ring with  topological charge $m_1=+1$ is moderately  greater than number of atoms in antivortex state ($m_2=-1$). The symmetric drift of two diametrically opposite antivortices towards the central hole leads to subsequent annihilation of the central vortex and relaxation of the toroidal condensate into a final antivortex state, i.e. the final topological charge of the merger is imposed by less populated ring as is seen in Fig. \ref{fig:Coaxial} (d). A remarkable role of the the symmetry of this system for vortex dynamics is investigated in Ref. \cite{oliinyk2019symmetry}.}

Instead of the development of the classical Kelvin-Helmholtz instability at the interface of the merging persistent currents in a prolate potential trap, sufficiently elongated in the axial direction, we observe the formation of nonlinear robust hybrid vortex structures (as illustrated in Fig. \ref{fig:hybrid} and explained in Ref. \cite{oliinyk2020nonlinear}).

Thus,  the ring-merging process and topological charge of the final state can be controlled
by the perturbation of the trapping potential, specially adapted for the initiation of symmetry-breaking
of the system, and by tuning of the initial population imbalance \cite{oliinyk2019symmetry,oliinyk2020nonlinear}.

\subsection{\label{sec:conclusions}Concluding remarks and outlook}

We have reviewed our current understanding of the spontaneous and controlled formation and stability of persistent currents in basic atomtronic circuits consisting of single or coupled ring-trap potentials and extended racetrack potentials. We have found that on--demand persistent flow can be created in a racetrack potential by stirring. The flow speed can be set to any value by adjusting the stir speed and/or the racetrack geometry. We discussed how persistent currents can also be generated spontaneously after crossing the BEC phase transition. In co-planar geometries we showed that the spontaneous generation of persistent flow is unaffected by the density overlap of the two rings, taking the first step in understanding ring-ring interactions and opening the possibility of many-ring arrays in the future. 
We have discussed our recent findings on dynamics of quantum vortices in two coupled vertically stacked toroidal condensates with persistent currents. It turns out that evolution of weakly coupled superfluid rings and merging quantized superflows with different topological charges is determined by complex dynamics of rotational Josephson vortices located between persistent currents.

The control over quantum topological excitations in such geometries offers an outstanding route to emerging quantum technologies 
{
with wide-ranging applications, such as topologically protected fault-tolerant quantum computation and quantum sensors for acceleration and rotation. These critically rely on minimising decoherence and dissipation and optimising the engineering of such components. For example, flexible sensor operation would require a rapid generation of the desired initial state, with further reduction in shot-to-shot atom number fluctuations crucial for sensor accuracy. Other areas where further theoretical and experimental work is needed (and currently well underway) include atomtronic transport, gate-like manipulation of quantum topological excitations  and readout mechanisms in atomtronic circuits.
}

{\it Acknowledgments}.
ME acknowledges substantial contributions from Ben Eller and Olatunde Oladehin. This work was supported in part by the US National Science Foundation grant no.\,PHY--1707776. TB and NP acknowledge contributions by Paolo Comaron, Quentin Marolleau, and Boris Malomed, and discussions with Veronica Ahufinger, Jordi Mompart, Jerome Beugnon, Jean Dalibard, Muntsa Guilleumas and Fabrizio Larcher.  TB would like to thank support from the EPSRC Doctoral Prize Fellowship Grant No.~EP/R51309X/1, while TB and NP  acknowledge financial support from the Quantera ERA-NET cofund project NAQUAS through the EPSRC, Grant No.~EP/R043434/1. AY acknowledges support from National Research Foundation of Ukraine through grant No. 2020.02/0032 and substantial contributions from Boris Malomed, Artem Oliinyk, and Igor Yatsuta.



%

\section{PHASE SLIP DYNAMICS ACROSS JOSEPHSON JUNCTIONS }
\label{PhaseSlips}
\vspace*{-0.5cm}
\par\noindent\rule{\columnwidth}{0.4pt}
{\bf{\small{A. Minguzzi, A. P\'erez-Obiol, J. Polo, N. P. Proukakis, K.~Xhani}}}
\par\noindent\rule{\columnwidth}{0.4pt}



The  phenomenon of superfluidity and its consequences can reveal itself in different ways in a quantum gas. One of the most remarkable manifestations of superfluidity is the frictionless motion of particles within the fluid, which is in direct relation with the macroscopic quantum phase coherence of the fluid. However, in certain circumstances this frictionless motion can be broken, with dissipation taking over. Phase slips represent one of the fundamental mechanisms leading to dissipation in superfluid systems \cite{varoquaux2015anderson,anderson1966considerations,levy2007ac,mckay2008phase,wright2013driving,jendrzejewski2014resistive,valtolina2015josephson}. 

%
{ Phase slips correspond to jumps in the phase structure of the wavefunction of a quantum fluid.
%
They can arise in dynamical superflow through a barrier in distinct manners, as summarized below:
In the case of a coherent superfluid, dynamical flow through a barrier can trigger excitations that lead to phase slips. The form of such excitations depends on the dimensionality and the geometry of the system, and can take the form of solitonic or vortex excitations, with associated acoustic emission.
The presence of fluctuations in the system -- whether of thermal, or quantum nature -- creates an additional mechanism of `incoherent' phase slips, thus giving rise to richer dynamics.
Atomtronic circuits typically consist of one (or more) Josephson junctions, embedded within a closed, typically ring-shaped and low-dimensional, geometry.
Therefore, understanding phase slip processes can prove crucial for the development of quantum technologies and, in particular, Atomtronic devices.
}


Currents in ring geometries are ideal candidates for the study of superfluidity in interacting quantum gases. In particular, these currents are metastable states that can maintain the flow of particles even when no external field or force is applied. However, these metastable states can decay in different scenarios. 
{
The decay of current states in one dimensional rings
corresponds to a sequence of phase slips
associated to the loss of angular momentum by the system. Such type of event may occur as triggered eg by thermal fluctuations or other type of fluctuations, and in such case they correspond to
the aforementioned
incoherent phase slips, as well as in quantum coherent manner, ie as an oscillation among different angular momentum states \cite{mooij2006superconducting}.
}
%
%
Examples of dissipative motion
have been observed in hysteresis dynamics \cite{wright2013threshold,wright2013driving,eckel2014hysteresis,yakimenko2015vortex,munoz2015persistent} where thermal activation plays an important role \cite{kumar2017temperature}.

In addition, phase slips can also be triggered by an external mechanism, for instance a weak link can catalyze the production of vortices at zero temperature \cite{piazza2009vortex} {and solitons \cite{dubessy2020universal}}. It has been shown that thermally activated phase slips can become dominant in the damping dynamics of some observables, at relatively low temperatures \cite{mathey2014decay,kunimi2017thermally}.

Recent studies have demonstrated the connection between the dissipative motion observed in Josephson systems and phase slips \cite{valtolina2015josephson,burchianti2018connecting,xhani2020critical,polo2019oscillations}. In this case, vortex nucleation can be triggered through the weak link producing the Josephson-like junction\cite{gauthier2019atomtronic} or through thermal activation, depending on the range of parameters. In a similar case, the connection between low-energy excitations and dissipative motion was proven to be the main mechanism \cite{polo2018damping} leading to the damping of Josephson oscillations \cite{pigneau2018relaxation}.

In this chapter, we present recent developments in the topic of phase slips and their role in the dissipative dynamics of observables such as population imbalance of Josephson systems and current dynamics in ring potentials. Special attention is given to the nonlinear excitations of the different systems, such as vortex rings and dark solitons. In the following sections we summarize different studies 
{ performed by the authors} that illustrate how phase slips can emerge in Atomtronics devices and isolated quantum systems.

Section~\ref{sec:vortex_dynamics} is devoted to the study of nucleation of vortex rings in a weakly linked three-dimensional elongated superfluid. In Sec.~\ref{sec:Damping_Josephson_oscillations} we consider the damping of Josephson oscillations in a one-dimensional (1D) strongly interacting Bose gas. Section~\ref{sec:stirred_BEC} is dedicated to the excitation spectrum of a 1D stirred Bose gas. Section~\ref{sec:Bose_gas_ring} focuses on the dynamical phase slips occurring in a phase imprinted Bose gas trapped in a ring potential. Finally, in Sec.~\ref{sec:conclusions-phase} we present the conclusions and outlook, summarizing and discussing how phase slips play a crucial role on Atomtronic-based devices.

\subsection{Critical transport and vortex dynamics in a thin atomic Josephson junction}
\label{sec:vortex_dynamics}

{
In this section, we give a detailed and intuitive 
picture of the emergence of phase-slips and dissipation across a single Josephson junction in a full three-dimensional (3D) atomtronic geometry.
Although there exist phenomenological models which can account for dissipative effects in such a setting -- such as the extension of the two-mode model of Josephson junctions~\cite{smerzi1997quantum} to include a damping term~\cite{marino1999,bidasyuk2018,pigneur2018} or models based on the analogy to a resistively-shunted junction (RSJ) circuitry~\cite{tinkham2004introduction} -- such models can only offer limited insight into the microscopic characterization of the observed dissipation.
For a more complete discussion, this section therefore focuses on the case of a junction embedded within an elongated harmonically-trapped superfluid, as a paradigmatic example of the arising dynamics.
Such choice is based on the existence of a carefully characterised experiment \cite{valtolina2015josephson,burchianti2018connecting}, detailed {\em ab initio} numerical analysis of which~\cite{xhani2020critical,xhani2020njp} has enabled not only qualitative connections to be made, but also facilitated direct links between microscopic and macroscopic observables and manifestations, directly relating these to the experimental observables. The discussion below is thus based on our recent works~\cite{xhani2020critical,xhani2020njp}, conducted at both zero and finite temperatures, which have fully analysed all aspects of the arising microscopic dynamics.
}



{The relevant experiment 
focussed upon here}
was conducted in Florence \cite{valtolina2015josephson,burchianti2018connecting}, in the context of an elongated $^{6}$Li {\em fermionic} superfluid,
separated by a thin Gaussian barrier induced by a focussed laser beam located at $x=0$ and of $1/e^2$ width $w \sim 4 \xi \sim 2 \mu$m $\ll R_x \sim 110 \mu$m, where $\xi$ ($R_x$) denotes the superfluid healing length (axial system size).
The experiment probed all regimes of values of $(k_F a)^{-1}$, where $k_F$ denotes the Fermi wavevector and $a$ the atomic s-wave scattering length. Although relevant and subtle differences were observed when transitioning from the BEC to the BCS superfluid regimes -- mainly associated with different critical velocities and spatial extents due to the changing interaction dependence and the increasing importance of the fermionic degrees of freedom --, the key underlying physical process leading to dissipation of superflow was found to be the same in all regimes, as outlined below.
{
(See also Chapter~\ref{TransportFermi}. for a discussion of transport and dissipation in ultracold Fermi gases.)
}

The presence of an initial population imbalance across the two wells separated by the barrier (initiated by moving the superfluid relative to the barrier at $t=0$) induced a neutral current flow in the negative $x$-direction, leading to the transfer of particles from the right to the left well. As expected,  small values of initial population imbalance were found to lead to symmetric Josephson `plasma' oscillations about a zero population imbalance, and associated oscillations in the relative phase \cite{smerzi1997quantum,raghavan1999coherent}. 
%
%
%
Nonetheless, when the initial fractional population imbalance exceeded a critical value, the system population dynamics transitioned to a different regime. Based on earlier experiments with ultracold Josephson 
junctions\cite{albiez2005direct,levy2007ac,spagnolli2017crossing}, one may have expected a transition to a so-called macroscopic quantum self-trapping regime, in which the population transferring oscillations proceed around a non-zero value of the population imbalance (i.e.~one side of the junction always has a higher population than the other), and with a running phase\cite{smerzi1997quantum,raghavan1999coherent}; the existence of such a regime has been argued to be related to the presence of a vortex ring in the barrier region, 
{
which
annihilates within the weak-link region (but outside the region of observable condensate density) \cite{piazza2011instability,abad2015phase}.
Interestingly, a different regime was observed in the recent Florence experiment, characterised by the observation of vortices in the superfluid bulk. 
Such a dissipative regime can arise when the emergent vortex ring acquires sufficient energy to overcome the barrier and penetrate the bulk superfluid. 
This is the regime analysed in our recent work~\cite{xhani2020critical}.
}
Specifically in our case -- and in particular during the first part of the dynamical evolution when particle flow is still in a single direction -- and for sufficiently large population imbalances, the very narrow nature of the barrier~\cite{xhani2020critical} was found to induce a local superfluid flow which accelerates in time and exceeds the local critical velocity for vortex excitation which, in this geometry, takes the form of a vortex ring excitation. As a result, the presence of a Josephson current flow led to the generation of a vortex ring, associated with a phase jump of $2 \pi$ across the axial direction, and a flow which is no longer dissipationless. 
Such a phase slippage process is well documented in related contexts of Josephson junctions in superfluids and superconductors \cite{anderson1966considerations,varoquaux2015anderson}, where it can be described in terms of {\em phenomenological} models.

\begin{figure*}
\centering 
\includegraphics[scale=0.3]{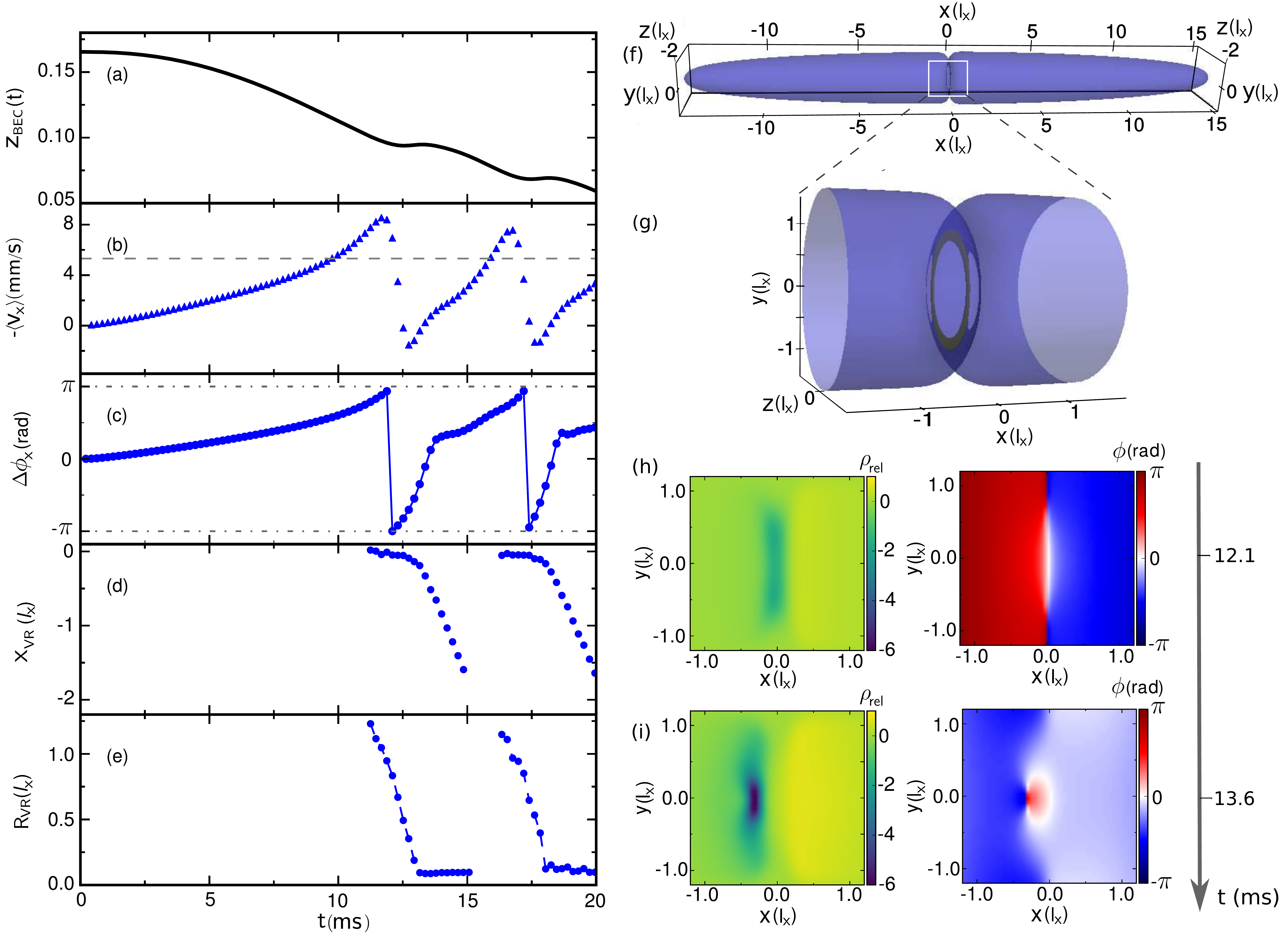}
\caption{
Characterization of phase slip process and subsequent vortex ring dynamics in a weakly-interacting elongated 3D condensate:
(a) Fractional population imbalance $z_{\rm BEC} = (N_{\rm R} - N_{\rm L}) / (N_{\rm R} + N_{\rm L})$ across a thin Josephson junction against time.
(b) Induced superfluid velocity along $x$, weighted over the
transverse density in the $x = 0$ plane.
(c) Induced superfluid relative phase along $x$, evaluated at $z=0$, $y=0.81 l_x$, corresponding to the vortex ring semi-axis along the $y$ direction at $t = 12.1$ms (see subplots (g)-(h)).
(d)-(e) Corresponding position $x_{\rm VR}$ and mean radius $R_{\rm VR}$ of the generated vortex rings as a function of time.
(f)-(g) 3D density profile (density isosurface taken at 0.002 of maximum density) at $t = 12.1$ms,
revealing the superfluid geometry and narrow barrier, along with a zoom-in to the highlighted central region (enclosed within the white rectangle in (f)) where the generated vortex ring (green near-circular structure) becomes clearly visible.
(h)-(i) Corresponding planar ($z = 0$) 2D snapshots of the condensate density (left) after substracting the background density and scaled to its maximum value,  and phase profiles (right) revealing the emergence and early dynamics of the first generated vortex ring at the indicated times. Spatial axes are plotted in terms of the harmonic oscillator length along the $x$-axis, l$_x \simeq 7.5 \mu$m. 
Parameters for this figure \cite{xhani2020critical}: $N_{\rm BEC} = 60,000$ bosonic $^{6}$Li molecules, $1/(k_Fa) \simeq 4.6$, $z_{\rm BEC}(t=0) = 0.17$, $\omega_x\simeq2 \pi \times 15$ Hz,  $\omega_y \simeq2 \pi \times 187$Hz, $\omega_z\simeq 2 \pi \times 148$Hz (cigar-shaped trap), based on a double-well potential defined by 
$V_\mathrm{trap}(x,y,z)=(M/2) ({\omega_x}^2 x^2+ {\omega_y}^2 y^2+ {\omega_z}^2 z^2)+V_0 \, e^{-2x^2/w^2}$, where $M=2m$ is the molecular mass, $V_0 = 0.8 \mu$ is the height of the Gaussian barrier and $w \approx  2.0 \mu$m is the barrier $1/e^2$ width.
{
Figure similar in spirit to individual plots shown (for other population imbalances) in Ref~\cite{xhani2020critical}. 
}
\label{proukakis-phaseslip-fig}
}
\end{figure*}

{Numerical analysis has shed more light on this process}
in the highly controlled environment of an ultracold atomic gas, modelled by the full 3D Gross-Pitaevskii equation describing the low-temperature regime of a weakly-interacting condensate.
Such analysis has been conducted based on the experimental parameters of Refs.~\cite{valtolina2015josephson,burchianti2018connecting}  in the molecular BEC regime.
Key findings reported in \cite{xhani2020critical}
are clearly summarized  in Fig.~\ref{proukakis-phaseslip-fig}.
{
Interestingly, the observed dissipation arises as a combination of the transfer of incompressible kinetic energy from the particle flow to the vortex and the acoustic (phonon) emission (with the latter potentially amounting to a significant fraction of the total flow energy).}
More details about this phase-slip process and the associated acoustic emission can be found in \cite{xhani2020critical}.
%
{
For completeness -- before proceeding further with such characterisation -- we note that 
while such a dissipative regime 
was found  in the limit of relatively low and narrow barriers (height $\lesssim O(\mu)$, and width four times the system healing length, for the parameter space probed in the Florence experiments~\cite{valtolina2015josephson,Burchianti2018}), a transition to a self-trapped regime also exists in the limit of broader and/or higher barriers; such analysis, providing a unified overview of the distinct dynamical regimes across a single Josephson junction was presented in \cite{xhani2020njp}. The stability of this self-trapping regime is directly affected by thermal and quantum fluctuations  \cite{Leggett_1998,raghavan1999,franzosi2000,gati2006,bidasyuk2018finite,imamoglu1997,milburn1997} and by
higher order tunnelling processes \cite{Zwerger_2001}, which are known to gradually destroy such a state, eventually leading to oscillations about a zero population imbalance.
}


%
%
%
%

{
We proceed here by reviewing the phase-slip-related transition to the dissipative regime probed in the Florence experiments~\cite{valtolina2015josephson,Burchianti2018}. Consider the case when
}
the fractional population imbalance $z_{\rm BEC}$ across the junction starts at a positive value (i.e. right well has higher population than left well): the induced Josephson dynamics leads to an (initial) superflow towards the left side of the junction (Fig.~\ref{proukakis-phaseslip-fig}(a)), thus causing an initial decrease in the fractional population imbalance. 
However, as such imbalance decreases at an increasing rate, implying an increase in the superfluid velocity (and corresponding superfluid current), there comes a point when the magnitude of the superfluid velocity exceeds some threshold value (loosely set by the 
mean speed of sound shown by the horizontal dashed line in Fig.~\ref{proukakis-phaseslip-fig}(b)), acquires a temporally local maximum value, as a result of which it becomes energetically favourable for a vortex ring excitation to be generated at the barrier at $x = 0$. Such a process is associated with an abrupt jump of $\sim 2 \pi$ in the condensate relative phase, as shown in Fig.~\ref{proukakis-phaseslip-fig}(c). The vortex ring generation instantaneously opposes the population transfer (leading to the flattening of $z_{\rm BEC}(t)$ visible in Fig.~\ref{proukakis-phaseslip-fig}(a)), and can even lead to a reversal of the background superflow (i.e.~$-\langle v_x \rangle$ changing sign in Fig.~\ref{proukakis-phaseslip-fig}(b)) due to the additional `swirling' 
velocity of the induced vortex ring.
The vortex ring, initially generated (as a `ghost' vortex) in the low density region outside the local transversal spatial extent of the BEC, remains initially within the axial barrier region $x_{\rm VR} \sim 0$ (Fig.~\ref{proukakis-phaseslip-fig}(d)), shrinking transversally (Fig.~\ref{proukakis-phaseslip-fig}(e)) and entering the Thomas-Fermi radius. After a short time, the accelerating vortex ring reaches the axial edge of the barrier (the superfluid density maximum is located at $|x_{\rm VR}| \sim 2w$) and starts travelling at a constant speed (linear part of decreasing $x_{\rm VR}(t)$), while maintaining its radius. A detailed instructive visualization of the overall superfluid geometry and the narrow nature of the barrier region  can be found in Fig.~\ref{proukakis-phaseslip-fig}(f)-(i), which also displays the vortex ring generation and initial dynamics.

The long-term dynamics after the generation of a vortex ring from the decay of the superflow depends critically on the system parameters. If the initial population imbalance is relatively weak (but still above the required threshold for defect-inducing decay of superflow), 
{a single vortex-ring may be generated, whose lifetime and subsequent dynamics outside the barrier depends on the value of the barrier height, as shown in~\cite{xhani2020njp}}.
However, in cases of larger initial population imbalance, after the first vortex ring has been generated and left the central region,
the background superflow due to the remaining population imbalance, i.e. chemical potential difference, picks up its pace (around $t  \sim 13$ms in Fig.~\ref{proukakis-phaseslip-fig}(b)), 
until at some time later, when the previously generated vortex ring has already travelled a  (potentially significant) axial distance from the barrier region, it once again exceeds the local critical speed and a second vortex ring is generated (around $t \sim 16.3$ ms). This process can repeat itself, leading to even more vortex ring generation, 
until (due to the decreasing population imbalance) the background flow weakens to the point that it can no longer exceed the critical velocity. The resulting sawtooth-like profile of
$-\langle v_x \rangle$ (Fig.~\ref{proukakis-phaseslip-fig}(b)) is typical of phase slippage phenomena seen in superfluid helium \cite{anderson1966considerations,Avenel1985,Sato2011,Hoskinson2006}.
A generated vortex ring eventually decays either by shrinking into a rarefaction pulse during its axial propagation (as relevant for the case considered here), or by interacting with the transversal condensate boundaries as the transversal spatial extent decreases during its propagation towards the axial condensate edge \cite{xhani2020critical}. 
In cases of  
high initial population imbalance, the time window between successive vortex ring generation events (depending on $h/\Delta \mu$ with $\Delta \mu$ being the chemical potential difference between the two wells) can be shorter than the vortex ring lifetimes,  
thus allowing the co-existence of multiple sequentially-generated vortex rings; 
{such rings may further interact both indirectly (through their respective emitted acoustic waves), and directly (vortex-vortex interactions), potentially leading (for very high initial population imbalances) to reconnection processes, `leap-frogging' (sequential passage of one vortex ring through the other), or even a `turbulent-like' regime (already discussed, for example, in 2D geometries \cite{griffin2020});
this, in turn, leads to a highly complicated long-term dynamics of the population imbalance.
}
 
The experimental observations \cite{valtolina2015josephson,burchianti2018connecting} are consistent with the picture described here. More concretely, the experiments led to the observation of one, or more, individual vortices, seen after removing the barrier (an added experimental complication required for imaging purposes), allowing the system to evolve and expand. This is consistent with the underlying picture described above, upon detailed consideration of the transversally asymmetric nature of the potential (which leads to excited, non-circular, vortex rings exhibiting Kelvin wave excitations), inherent fluctuations (which lead to asymmetric generation, propagation and decay of the vortex rings, and can thus explain the presence of single/odd-number-of defects in experimental expansion images) and dynamical barrier removal (which is found to significantly extend the lifetime of generated vortex rings) \cite{xhani2020critical}. 

{
Ref~\cite{xhani2020critical}
also considered }
the role of temperature and thermal fluctuations by means of a self-consistent (`ZNG') kinetic theory \cite{griffin2009bose,proukakis2008finite,proukakis2013quantum,berloff2014modeling}, in which the condensate is described by a dissipative Gross-Pitaevskii equation which explicitly includes friction and collisional population transfer with the thermal cloud, the latter being treated by a quantum Boltzmann equation.
This demonstrated
that the presence of small thermal fluctuations does not significantly influence the above vortex generation process, although we have observed that a high enough temperature can in fact induce {\em additional} thermally-activated vortex rings; examples of the latter behaviour in the one-dimensional context are discussed by means of a different finite-temperature model in Sec.~\ref{sec:Bose_gas_ring} below.
Beyond the initial generation process, thermal effects were found\cite{xhani2020critical} to have a significant role on the long-term vortex dynamics, where they act both to destabilize the otherwise symmetric motion of the vortex ring through the introduction of fluctuations, and to damp the motion through a mutual friction damping mechanism \cite{jackson2009finite,allen2013observable}.

The process we have discussed here is generic, and applies to any geometry and dimensionality, even though specific details will vary. For example, in the case of one-dimensional systems (see Sec.~\ref{sec:Damping_Josephson_oscillations} and Sec.~\ref{sec:Bose_gas_ring}), the underlying defects generated are dark solitons, with the corresponding dynamics in ring traps of direct relevance to atomtronics discussed in Sec.~\ref{sec:stirred_BEC}. 

{
Although the above discussion focussed primarily on reviewing the emergence 
of dissipation across a single 3D junction embedded within a harmonic trap, it is pertinent to highlight here the important related work of quantum transport across an atomtronic `dumbbell' circuit, consisting of two reservoirs connected by a configurable linear channel of variable length and width \cite{gauthier2019atomtronic}: this was both studied experimentally, and analysed theoretically in the context of a superfluid acoustic model, a phase-slip model for the conductance, and via mean-field simulations. 
Such work also highlighted the existence of Josephson plasma  and dissipative dynamical regimes, with the transition between such regimes observed (for given channel) at variable initial population imbalance. 
Moreover, Gauthier {\em et al}~\cite{gauthier2019atomtronic} noted that the relative importance of sound  and vortex energy as the origin of dissipative dynamics depends sensitively on the details of the geometry of the channel: for example, small channel widths which cannot support vortex dipoles, lead to the generation of unstable topological excitations which decay rapidly to compressible excitations.
Thus for small channels they found the origin of dissipation to be sound-dominated.
This is qualitatively consistent with the findings of \cite{xhani2020njp} which characterised 
the dominant dissipation mechanism for sufficiently high barriers as the propagation of emitted sound waves.
}


\subsection{Bose-Josephson junction among two one-dimensional atomic gases:  a quantum impurity problem}
\label{sec:Damping_Josephson_oscillations}

The one-dimensional (1D) geometry in ultracold Bose gases provides an ideal physical platform for the study of the quantum dynamical behavior of Bose-Josephson junctions, as the low dimensionality of these systems leads to the enhancement of quantum fluctuations and correlations. Recent experiments have realized and studied the 1D strongly interacting regime by using quasi-one-dimensional cigar-shaped potentials in which the transverse motion of the particles is effectively frozen \cite{paredes2004tonksgirardeau,kinoshita2004observation,hofferberth2007nonequilibrium,hofferberth2008probing,haller2009realization,yang2017quantum,pigneau2018relaxation}. One-dimensional systems present features that clearly separate them from the higher dimensional ones, especially in the intermediate and strongly interacting limit where the motion of the particles is defined  by its collective behavior. This collective motion is tightly connected to the low-energy excitation spectrum of the gas.

%
{
One dimensional systems are characterized by specific thermalization properties (eg to Generalized Gibbs Ensemble for integrable systems), which has been a topic of continuous interest \cite{rigol2008thermalization,gring2012relaxation,kinoshita2006quantum,ueda2020quantum}. 
Phase slips play a crucial role in the dissipative dynamics of quantities such as population imbalance in Josephson systems and current dynamics in ring potentials. For instance, phase slips are the only mean to change angular momentum in one-dimensional rings, as such rings cannot host vortices in the transverse direction, and they occur at the position of a localized barrier.  In one-dimensional wires, phase slips  occur when a soliton is formed or destroyed  upon hitting the barrier giving rise to the junction \cite{dubessy2020universal}.
Hence, Josephson junctions in one-dimensional systems
 are an appealing physical platform to investigate 
 such damping phenomena and one of the simplest yet complete many-body systems displaying thermal and quantum phase slips. 

Recent studies have investigated phase slips in different contexts: for instance in \cite{ruggiero2020thermalization} they investigated two tunnel-coupled one-dimensional tubes placed side-by-side and characterized their low-energy physics described by unequal Luttinger liquids. Other approaches are also being investigated; e.g. in \cite{mennemann2020relaxation} they attribute the short-time evolution to multi-mode dephasing, while for longer times, they relate the relaxation to the nonlinear dynamics of the system. 


%
%
The following subsection presents a study of the microscopic origin of phase slips in 1D bosonic Josephson junctions. Specifically, the analysis is performed in the strongly interacting regime by considering two weakly coupled one-dimensional wires in a head to tail configuration. The results and discussion presented here are adapted from \cite{polo2018damping}.
}

\paragraph*{Model:}

The intermediate and large interaction regimes of a 1D Bose gas are difficult to treat, both numerically and analytically, due to the many-body character of the system. Using the Luttinger liquid (LL) theory ~\cite{cazalilla2004bosonizing} one can calculate the low-energy dynamical response of two strongly interacting one-dimensional bosonic fluids confined within an effective 1D waveguide of length $L$, tunnel-coupled through a weak link created by a barrier. In particular, {in  \cite{polo2018damping} we have studied} the system's response to a quench in the particle number difference between the two subsystems. By using a mode expansion of the density fluctuation and phase field operators from the LL theory, and by defining the relative coordinates for the field operators, we identified the zero modes $\hat{N}$ and $\hat{\phi}_0$ as the relative population and phase differences between the two coupled wires, and $\hat{Q}_\mu$ and $\hat{P}_\mu$ as the relative coordinates for the excited modes. 
The resulting Hamiltonian reads:
\begin{eqnarray}
\hat{H}^{rel}_{T}&=&
\frac{\hbar^2}{2ML^2}(\hat{N}-N_{\text{ex}})^2
-E_J\cos\left(\hat{\phi}_{0}\right)
\label{eq:Hamiltonian_trans}\\
&+&
\sum_{\mu\ge1}\left[
\frac{1}{2M}\left( \hat{P}_{\mu} +\frac{\sqrt{2}\hbar}{L}(\hat{N}-N_{\text{ex}})\right)^2
+\frac{1}{2}M\Omega_{\mu}^2\hat{Q}_{\mu}^2
\right]
\nonumber
\end{eqnarray}%
with effective mass $M=\hbar K/2\pi vL=K^2m/2\pi^2N_0$, $N_0$ being the average particle number in each tube and $N_{\text{ex}}\ll N_0$ the excitation imbalance, which may be tuned by a suitable choice of the initial conditions. It is worth mentioning that the center-of-mass coordinates are completely decoupled from the relative ones and they simply take the form of a harmonic oscillator, which can be readily diagonalized, not playing any role in the observable of interest. 


We identify in Eq.~(\ref{eq:Hamiltonian_trans}) three terms: (i) a {\it quantum impurity particle} term corresponding to the two collective variables $\hat{N}$ and $\hat{\phi}_0$, (ii) a bath of harmonic oscillators formed by the excited modes, and (iii) a coupling term $\propto \hat{P}_{\mu}\hat{N}$, obtained by expanding the second line of Eq.(\ref{eq:Hamiltonian_trans}). Hamiltonian \eref{eq:Hamiltonian_trans} has the same structure as that of the Caldeira-Leggett model \cite{caldeira1981influence,caldeira1983quantum,schon1990quantum}. However, it is important to remark that in our model the bath of harmonic oscillators is intrinsic to the microscopic model, while in the Caldeira-Leggett model it is phenomenologically introduced.
The energy scales $E_Q$ and $E_J$ depend on interactions, the latter being renormalized by quantum fluctuations \cite{cominotti2014optimal}. Concomitantly, the sound velocity and the Luttinger parameter vary with the interaction strength as described in \cite{cazalilla2004bosonizing}. The first two terms of \eref{eq:Hamiltonian_trans} correspond to the familiar Josephson Hamiltonian, where two regimes can be identified depending on the ratio of the Josephson energy, $E_J$, and kinetic energy, $E_Q=\hbar^2/M L^2= 2\Delta E/K$, with $\Delta E=\hbar\pi v/L$ being the level spacing among the phonon modes of the bath.

Case 1: Let us first consider the case $E_J\gg E_Q$, i.e. the Josephson potential term $- E_J \cos(\hat{\phi}_0)$ dominates upon the kinetic energy term in \eref{eq:Hamiltonian_trans}. In particular, starting from an initial particle imbalance among the two wires, its dynamical evolution was obtained from the Heisenberg equations of motion \cite{polo2018damping}, leading to the quantum Langevin equation of motion with three dominant parameters: the Josephson frequency $\omega_J=\sqrt{\omega_0^2-\gamma^2}$ where ${\omega_0=\sqrt{E_JE_Q}/\hbar}$, the memory-friction kernel $\xi_{N}(t)$ whose large temperature properties are given by $\langle\xi_{N}(t)\rangle=0 $ and $\langle\xi_{N}(t)\xi_N(t')\rangle= 2E_J^2k_BT\!/\hbar^2\! MLv \delta(t-t')$ and a damping rate given by $\gamma=\pi E_J/\hbar K$ (assuming a large frequency cut-off for the LL theory):
\begin{equation}
\ddot{\hat{N}}+\omega_0^2\cos(\hat{\phi}_0)\hat{N}+\int_{0}^{t}dt'\,\gamma_{N}(t,t')\dot{\hat{N}}(t') =\xi_{N}(t)\label{eq:eq_motion_n}
\end{equation}%
Within this Josephson regime, two different behaviors depending on interactions exist. In the weakly interacting limit, where $K\sim 1/\sqrt{g_{1D}}$ and $v_s\sim \sqrt{g_{1D}}$ with $g_{1D}$ being the 1D interaction strength, the predictions of the two-mode model in its small-oscillation limit are recovered, i.e. $E_Q\propto g_{1D}$ and $\gamma/E_Q$ vanish for $g_{1D}\rightarrow 0$, yielding undamped Josephson oscillations. However, for strong interactions $E_Q$ increases, as it is related to the compressibility of the system, and $E_J$ decreases, since it is renormalized by increasingly larger phase fluctuations. Thus, by inspecting the dimensionless damping rate $\gamma_Q\equiv\gamma/\omega_0 = \pi\sqrt{E_J}/\sqrt{E_Q}K$, one can predict that the Josephson oscillations will be more and more damped at increasing interactions.
%
%
\begin{figure}[h!]
\includegraphics[width=\linewidth]{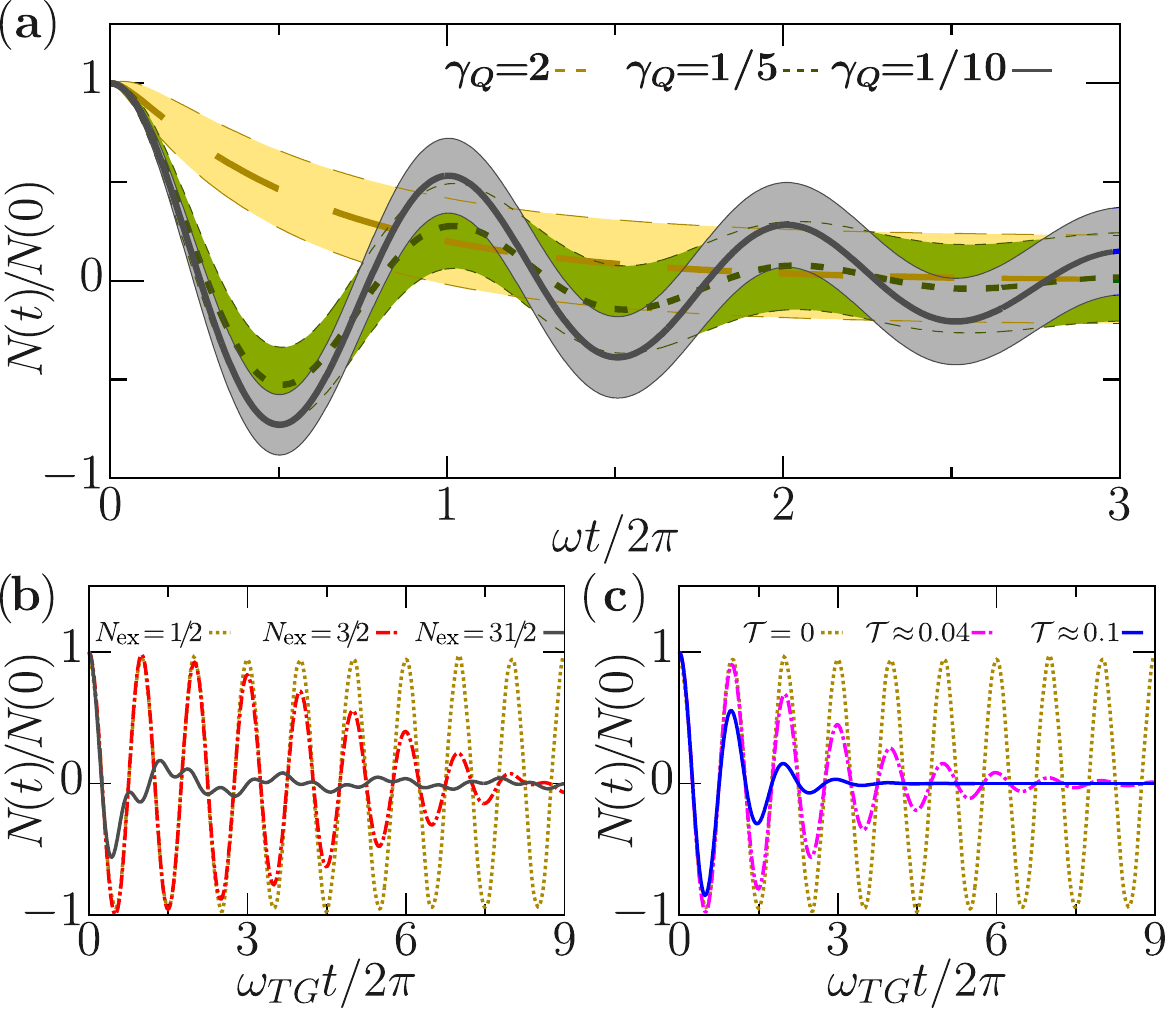}%
\caption{\label{fig:POLO_LL_TG_JJ} 
(a) Relative number dynamics $N(t)/N(0)$ (dimensionless) in two tunnel-coupled wires (LL approach) for various values of $\gamma_Q\!=\!\gamma/ \omega_0$. Stochastic noise uncertainties are indicated in shaded areas.
(b-c) Relative-number oscillations (TG regime) following a quench of the step potential $\delta V_0$ creating the initial imbalance: (b) at zero temperature for $\delta V_0/E_F = 0.07$ (yellow-dotted line), $0.14$ (magenta dashed line) and $0.72$ (blue solid line), with $E_F$ the Fermi energy; (c) at finite temperature for $\delta V_0/E_F = 0.07$.
{
Reprinted with permission from J. Polo, V. Ahufinger, F. W. Hekking, and A. Minguzzi, Phys. Rev. Lett. 121, 090404 (2018). Copyright 2018, American Physical Society \cite{polo2018damping}.
}
}

\end{figure}

Case 2: In the $E_J\ll E_Q$ limit, the phase is only weakly pinned and therefore it will display large fluctuations. In this regime, it is more convenient to use the Fock basis for the relative number. In this case, the energy levels of the quantum particle in \eref{eq:Hamiltonian_trans} can be described as a function of the number of excitations $N_{\text{ex}}$, which now plays the role of quasi-momentum in the effective crystal, taking the form of parabolas $\varepsilon_n(N_{\text{ex}})=E_Q(n-N_{\text{ex}})^2/2$, with $\hat N|n\rangle=n |n\rangle$. These parabolas present gaps of amplitude $E_J$ opening at semi-integer values of $N_{\text{ex}}$. If we focus on the anticrossing points $N_{\text{ex}}=\pm 1/2, \pm 3/2,...$ the system effectively behaves as a two-level model and the Josephson dynamics correspond to the Rabi oscillations of the quantum particle, with frequency $E_J/\hbar$. 
Note that the large value of $E_Q$ fixes the scale of bath-modes level-spacing. This, creates a large gap between the level-spacing of the bath and that of the quantum particle, $\hbar\omega$.
In the strongly interacting limit there exists an exact solution for infinitely repulsive interactions, corresponding to $K=1$ in the LL theory, known as the Tonks-Girardeau (TG) regime \cite{girardeau1960relationship,girardeau2000dark,yukalov2005fermibose}. In particular, in Fig.~\ref{fig:POLO_LL_TG_JJ}~(b) we {show} that for a small initial imbalance, i.e. $N_{\text{ex}}=1/2$, undamped oscillations occur with a frequency $\omega_{TG}=\epsilon_{N+1}-\epsilon_N$. This is where the correspondence between the LL and TG regime can be made as $E_J=\hbar \omega_{TG}$ and $E_Q=\hbar^2 \pi^2 N/mL^2$. Hence, the oscillations observed in the exact solutions at small $\delta V_0$ are the undamped Rabi oscillations of the quantum particle predicted by the LL model. For a larger imbalance, and thus beyond the low-energy description given by the LL theory, an effective damping appears due to the high energy excitations produced by the quench. 
The finite temperature regime {was also}  addressed within the exact TG regime, as shown in \fref{fig:POLO_LL_TG_JJ}~(c) for small-imbalances. Unlike in the LL predictions, in the TG exact solution { damped oscillations were found}. In order to pinpoint the origin of this damping, we {computed}  the spectral function of the system at finite temperature \cite{polo2018damping}.  We {found} that the exact spectral function contains multiple particle-hole excitations while the LL model assumes a linear excitation spectrum. In fact, {we found that} the exact spectral function also contains several low-energy excitations with frequencies of the order of $E_J$, which are associated with the presence of a finite barrier and are responsible for the observed damping. 

In summary, {our work showed that} the LL model for two tunnel-coupled atomic gases can be mapped on a quantum impurity problem in the presence of a bath. The exact TG solution validates the frequency of the Josephson oscillations predicted in the LL model, and that the oscillations may in fact be damped by an intrinsic bath made out of low-energy excitations, but points out the existence of other modes that are beyond the LL model and that also provide damping of the excitations. 

\subsection{Bose-Einstein condensate confined in a 1D ring stirred with a rotating delta link}
\label{sec:stirred_BEC}

Analyzing the spectrum of BECs trapped in ring settings
is an important step towards understanding the generation and decay of supercurrents.
The spectrum of a BEC in a 1D ring stirred by a rotating link can be first
illustrated in the mean field limit, at zero temperature,
and with a Dirac delta potential rotating at constant speed.
{Here we follow these assumptions and base this section on 
\cite{perezobiol2019bose,perezobiol2020current}.}
This approach has the advantage that the stationary solutions in the delta comoving
frame are the ones of the free 1D GPE,
and the effect of the moving potential is relegated to fixing specific boundary conditions.
This is in contrast to models with finite width potentials \cite{fialko2012nucleation,li2012nonlinear,munoz2019nonlinear}, and a generalization
of a static point like impurity \cite{seaman2005effect,perezobiol2019stationary}. 
{It allows for analytical expressions of the solitonic trains dragged by a rotating
weak link and for
the critical velocities at which the condensate becomes unstable and decays.
The metastability of each excited state can be readily studied through a Bogoliubov
analysis, and the hysteresis cycles observed in stirring experiments~\cite{eckel2014hysteresis} 
can be qualitatively understood in terms of this model.}

\paragraph*{Model:}
The stationary solutions are given by the condensate
wave function, $\phi(\theta)$, $\theta\in[0,2\pi)$,  and the corresponding chemical potential,
$\mu$, in the delta comoving frame.
Using natural units, $\hbar=M=R=1$, with $M$ the mass of the
atoms and $R$ the radius of the ring,
the GPE and boundary conditions read,
\begin{eqnarray}
\label{eq:gpe}
-\frac12 \phi''(\theta)+g|\phi(\theta)|^2\phi(\theta)=&\mu\,\phi(\theta),
\\
\label{eq:bc1}
\phi(0)-e^{i 2\pi\Omega}\phi(2\pi)=&0,
\\
\label{eq:bc2}
\phi'(0)-e^{i2\pi\Omega}\phi'(2\pi)=&\alpha\,\phi(0),
\end{eqnarray}
where $g>0$ is the reduced 1D coupling, assumed to be repulsive,
$\frac{\alpha}{2}>0$ and $\Omega$ the
strength and velocity of the Dirac delta, and $\phi$ is normalized to
$\int_0^{2\pi}d\theta|\phi(\theta)|^2=1$.
The spectrum is thus determined by three parameters, $g$, $\alpha$, and $\Omega$.
A general solution, $\phi(\theta)=r(\theta)e^{i\beta(\theta)}$, can be written 
in closed form in terms
of one of the twelve Jacobi functions \cite{seaman2005effect}. These functions contain two free parameters,
the elliptic modulus, $m$, which generalizes the trigonometric functions into the Jacobi
ones, and a frequency $k$. Any set of values for $k$ and $m$ entails
a solution that satisfies Eqs.~(\ref{eq:gpe})-(\ref{eq:bc2})
for a specific strength $\alpha$, velocity $\Omega$, and chemical potential $\mu$.

\begin{figure}[t]
\centering
\includegraphics[width=.48\textwidth]{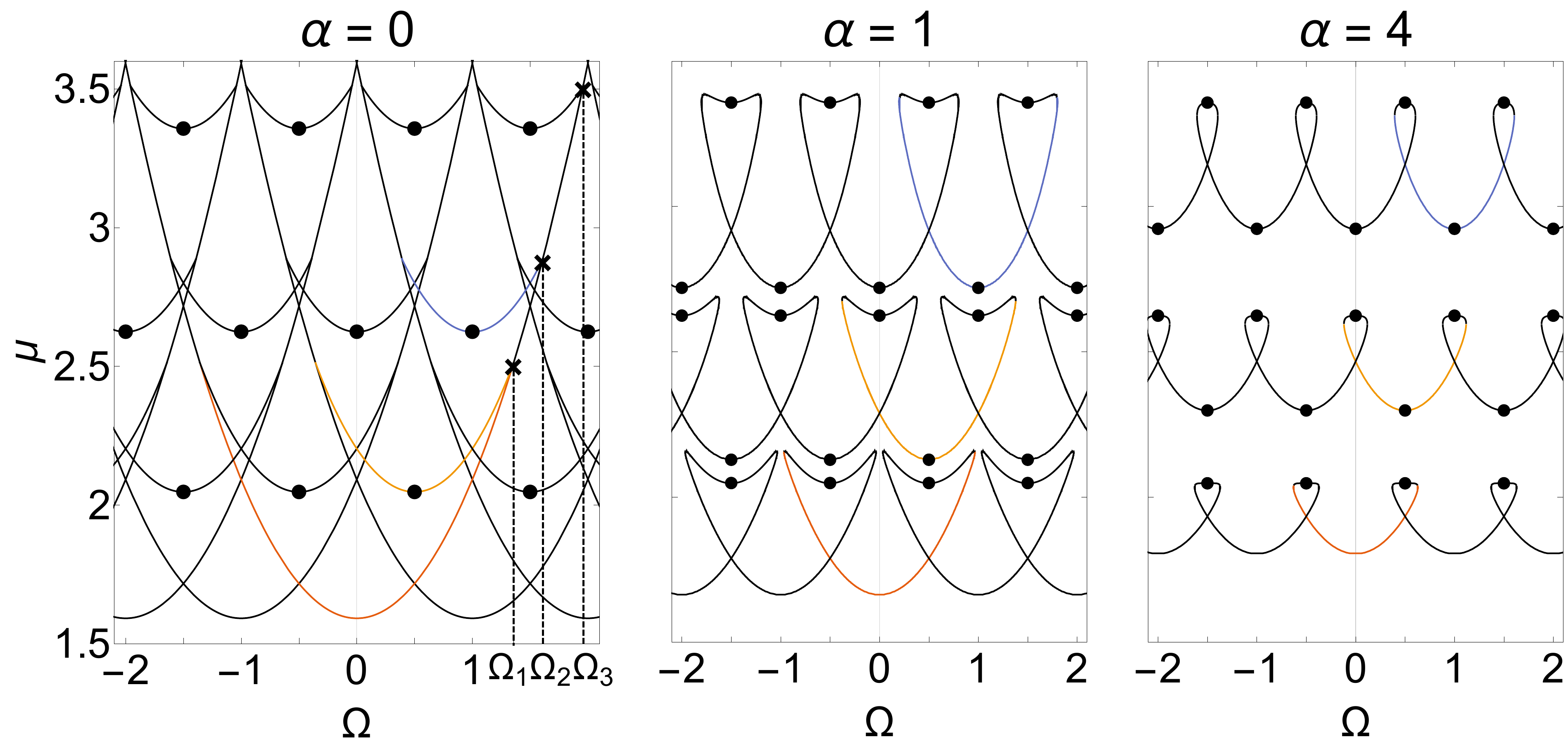}
\caption{
Lower part of the spectrum $\mu(\Omega)$, in the reference frame of a Dirac delta 
of strengths $\alpha=0,1$, and $4$,
moving at constant velocities $\Omega\in(-2,2)$, and where $g=10$.
Dots mark the velocities and energies of dark solitonic trains.
Crosses in the left panel mark the velocities $\Omega_n$
at which gray solitonic trains merge into the ground state for $\Omega>0$. {Reprinted with permission from A. Pérez Obiol and T. Cheon, Phys. Rev. E 101, 022212 (2020). Copyright 2020, American Physical Society~\cite{perezobiol2019bose}.}
}
\label{fig:st}
\end{figure}

\paragraph*{Spectrum:}
The free and stationary solutions, $\Omega=\alpha=0$, consist in plane waves, real
symmetry breaking solutions, and complex symmetry breaking solutions \cite{carr2000stationary}.
They correspond to vortex states, dark solitonic trains with an even number
of zeros, and gray solitonic trains. The latter is a generalization
of the former two, plane waves representing the limit in which gray solitons
become infinitely shallow, and dark solitons the limit in which the minima
of gray solitons become zero.
All these solutions 
are found by imposing a phase jump
$\beta(2\pi)-\beta(0)=2\pi n$, with $n$ an integer.
If instead one constrains an arbitrary phase difference of $2\pi\Omega$,
the obtained solitonic trains move at velocity $\Omega$
---and are stationary in the frame of reference rotating at $\Omega$.
The spectrum of stationary solutions from the point of view of an observer
moving at $\Omega$ is plotted in the left panel of Fig.~\ref{fig:st}.
These solutions also include plane waves under a boost of $\Omega$.
Dark solitonic trains with an even number of zeros comove with the condensate
at $\Omega=l$, while trains with an odd number of zeros travel
at $\Omega=l+\frac12$, where $l$ is an integer. Waves moving at velocities departing from $\Omega=\frac{l}{2}$
consist in gray solitonic trains, with shallower solitons the larger $|\Omega-\frac{l}{2}|$.
At $\Omega=\frac{l}{2}\pm|\Omega_n-\frac{n}{2}|$, with $\Omega_n=\sqrt{\frac{g}{2\pi}+\frac{n^2}{4}}$
and $n$ indicating the number of dark solitons in the original train, the amplitudes become constant,
and the gray soliton solutions merge into plane waves (parabolas in Fig.~\ref{fig:st}).

Once a barrier is created, the rotational symmetry is broken,
and gray and dark solitonic train solutions are split into two.
The energy spectrum, as observed in the Dirac delta comoving frame, is split into
a set of swallowtail (ST) diagrams, see middle and right panels of Fig.~\ref{fig:st}.
This looped structure implies that each wave train with a fixed number of dips in the density can be dragged
only at a certain range of velocities.
This range is limited by a pair of critical velocities, one $\Omega<\frac{l}{2}$
and another $\Omega>\frac{l}{2}$, beyond which stationary solutions do not exist
for the particular ST centered at $\Omega=\frac{l}{2}$.
These pair of velocities are marked by the tips in each swallowtail,
 and depend on the magnitude of the weak link.

\begin{figure}[t]
\centering
\includegraphics[width=.48\textwidth]{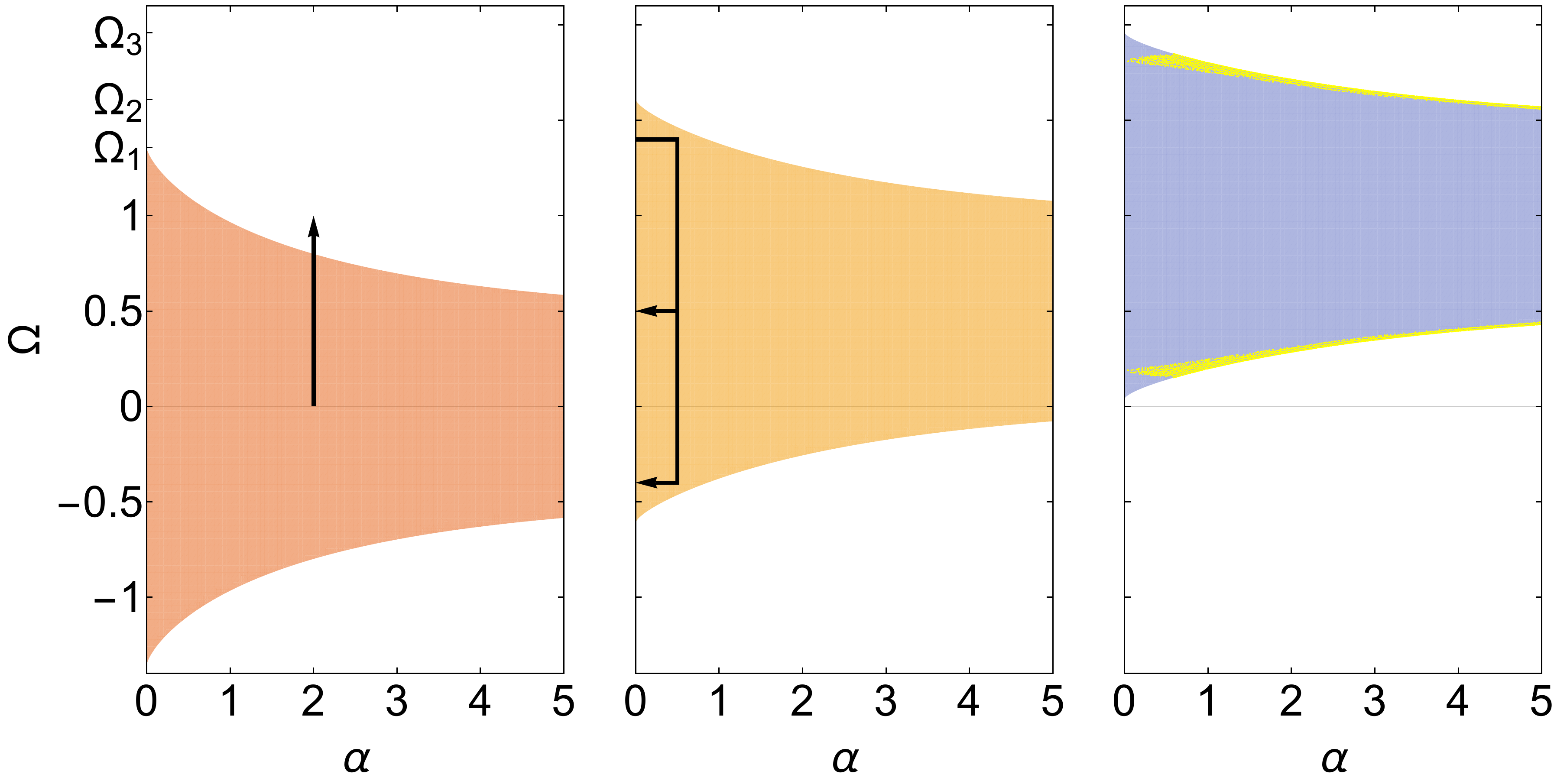}
\caption{
Regions
where stationary solutions exist corresponding
to the bottom of the first three ST centered at $\Omega=0,\frac12,1$
and for $g=10$. They are colored as the corresponding sections
in Fig.~\ref{fig:st}.
{All these solutions are stable against perturbations
except for the regions in yellow in the right plot.}
A path which explicitly leaves the stationary region is drawn on 
the left panel. The paths in the middle panel represent cycles to obtain
a dark soliton and a vortex with one quantum of angular momentum. {Reprinted with permission from A. Pérez Obiol and T. Cheon, Phys. Rev. E 101, 022212 (2020). Copyright 2020, American Physical Society~\cite{perezobiol2019bose}}
}
\label{fig:regions}
\end{figure}

\paragraph*{Metastability:}
{The critical velocities define the regions in parameter space where stationary
solutions exist, and the possible stirring protocols, or paths $(\alpha(t),\Omega(t))$
through which solitonic trains can be dragged, see Fig.~\ref{fig:regions} for
a sample of these regions.
Bogoliubov analysis show that the solutions corresponding to these regions are mostly
stable under perturbations, while the solutions corresponding to top parts of swallowtails
are completely unstable. This metastability analysis does not qualitatively change for
a wide range of nonlinearities ($g=1$ to $g=50$).
In general weaker atomic interactions and larger $\alpha$ imply narrower ranges of velocities
at which solitons can be dragged without dissipation. Moreover, for small $g$,
the swallowtail loops become smaller, and at half-integer velocities,
pairs of dark solitonic states are energetically very close to the ground states.}

\begin{figure}[t]
\centering
\includegraphics[width=.48\textwidth]{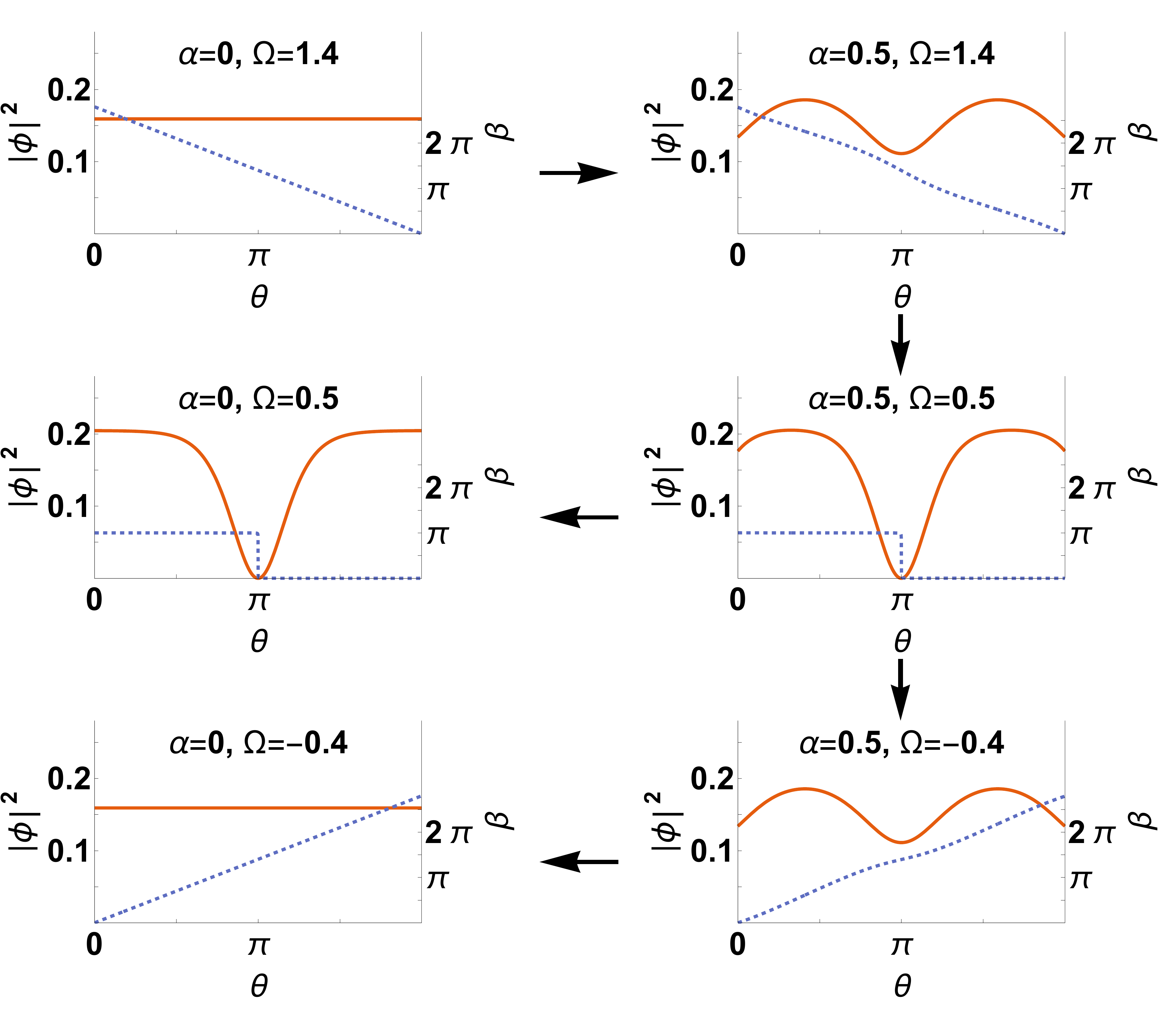}
\caption{
Densities (solid lines) and phases (dashed lines) in the Dirac delta comoving frame
corresponding to the key points of the dark solitons and vortex cycles
in Fig.~\ref{fig:regions} (where $g=10$). 
The first steps in both of them consist in setting a delta
on the ground state (top left plot) while rotating at $\Omega=1.4$,
and then lowering its velocity to $\Omega=\frac12$ (top and middle right plots).
At this point, one can unset the delta and obtain a dark soliton,
or further decrease the velocity down to $\Omega=-0.4$, and then unset
the delta, in which case the state of one vortex is reached.
{Reprinted with permission from A. Pérez Obiol and T. Cheon, Phys. Rev. E 101, 022212 (2020). Copyright 2020, American Physical Society~\cite{perezobiol2019bose}}
}
\label{fig:eigenfunctions}
\end{figure}

With the general features of the spectrum and its metastability laid out,
one can devise adiabatic paths that avoid both, the critical lines delimiting the
tips of the ST, and the metastable regions.
Setting a weak link at zero velocity and then accelerating it, only
produces currents of angular momentum $J\lesssim 1$ before the condensate
becomes unstable.
If instead the weak link is set while rotating at a finite velocity
$\Omega_i\in(\Omega_n,\Omega_{n+1})$, $n\ge1$, and then
 slowed down, any number of dark solitons or vortex states with
any number of quanta of angular momentum can be obtained.
As an example, cycles to obtain one dark soliton and a vortex are drawn in the middle
panel of Fig.~\ref{fig:regions}. The densities and phases
corresponding to the vertices of this rectangular path are plotted
in Fig.~\ref{fig:eigenfunctions}.

One can also devise paths which explicitly cross the critical line separating
the stable and unstable regions. One such path is schematically plotted in the left panel of Fig.~\ref{fig:regions}, where a weak link is set in the condensate at rest,
and then accelerated passed the critical velocity. In this case, one can expect the condensate to enter an
unstable state and decay to the immediate lower state,
corresponding to the lower branch of the swallowtail diagram.
{If  then the opposite path is taken, in which the weak link is slowed down, 
the previous states are not recovered, and the condensate
is left with an increased angular momentum, producing hysteresis.
To recover the initial state, the weak link velocity has to be further decreased
to reach the other critical point, so that the condensate decays again and
the hysteresis cycle can be closed. Therefore, the swallowtails and critical velocities 
also provide a basis model to understand the hysteresis cycles observed
in experiments. In particular, in~\cite{eckel2014hysteresis}, the cycle widths and critical velocities
decrease as stronger weak links are rotated. This qualitatively agrees with
the spectrum studied here, where the hysteresis widths $\Delta \Omega$
are defined by the widths of the swallowtails, and decrease as weaker interactions $g$
or larger link strengths $\alpha$ are considered.
Moreover, the model also predicts hysteresis cycles coupling states with different
angular momentum, corresponding to paths along upper swallowtail diagrams.
On the other hand, these downward swallowtails are characteristic
of repulsive interactions, and it can be shown that such hysteresis
cycles cannot happen for attractive interactions~\cite{perezobiol2020current}}.

The above stirring mechanisms, deduced analytically from the spectrum,
are corroborated by numerical simulations of the time-dependent GPE
where a Gaussian potential is rotated, instead of a Dirac delta.
These simulations also allow to test the stability of stirring protocols
involving more excited states not studied in the analytical case.
Indeed, for Gaussian widths of about 2\% of the ring perimeter,
dark solitonic trains with various zeros and vortices with up to a few quanta
of angular momenta are produced following the protocols provided by
the model with a rotating Dirac delta.
Similarly, the condensate is able to sustain
stable solutions when stirred by a Gaussian link, up to a certain velocity.
This critical velocity decreases with the Gaussian height,
as expected from the regions of stationary solutions in Fig.~\ref{fig:regions}.

{This model offers a new approach to study metastability
and vortex and soliton nucleation in ring condensates
with a rotating weak link.
The processes described in this section can also be understood in terms of 
a rotating trap and a fixed weak link or defect.
Analytical expressions for the dragged solitonic trains and critical velocities
allow to study ground states as well as excited states,
and to understand how they are coupled among them.
At half-integer velocities and for strong enough weak links
dark solitonic states with zero current could be easily excited from the ground
states. They could also be produced if a defect appears in a rotating condensate,
which is then slowed down.
Bogoliubov analysis indicate that the spectrum is roughly divided
in stable and unstable regions (bottom and top parts of swallowtails).
At the same time,
the main features of hysteresis cycles
 can be qualitatively understood in terms of the swallowtail structure of the spectrum.
This analysis complements phase slip mechanisms studied
in 2D and 3D traps,
where richer dynamics involving vortex excitation is possible,
see Sec.~\ref{sec:vortex_dynamics},
and numerical works in 1D where either stronger interaction
regimes are studied, see Sec.~\ref{sec:Damping_Josephson_oscillations},
or explicit thermal activation is considered,
as discussed next in Sec.~\ref{sec:Bose_gas_ring}.}

\subsection{Thermal and quantum phase slips in a one-dimensional Bose gas on a ring}
\label{sec:Bose_gas_ring}


{
This section's results and discussions are adapted from \cite{polo2019oscillations} and give an example of a one-dimensional quantum system where both coherent and incoherent phase slips appear at different regimes of bosonic interactions and temperatures. The section focuses on phase slips between different angular momentum states occurring in a one-dimensional Bose gas trapped in a ring potential. The current is induced through phase imprinting \cite{dobrek1999optical,zheng2002classical,kumar2018producing} and due to the dimensionality of the system vortex nucleation is forbidden within the ring. Therefore, one-dimensional phase slips require the existence of a different microscopic mechanism.

Previous studies have investigated the origin of phase slips and how these can lead to a decay of the superfluid current in different scenarios. For instance, in \cite{kunimi2017thermally} they introduced a shallow lattice to trigger such phase slips and investigated the system using a combination of techniques based on the mean field Gross-Pitaevskii equation, including a Bogoliubov analysis as well as phenomenological noise and dissipation term in the mean field description. In a more recent work it has been shown that phase slips can also be driven by acoustic waves in higher dimensional system \cite{kuriatnikov2020phase}, which indicates that low energy excitations can be one of the triggering mechanisms of phase slips. 

The work presented below covers all interaction range using different approaches and models. It also considers an experimentally realistic scenario where the currents are introduced through phase imprinting and takes into account how phase slips are originated in each regime. Therefore, it provides a good example of the current techniques and observations found currently in the field.  

}

\paragraph*{Model and methods:}

{ As presentend in \cite{polo2019oscillations}, we }%
consider $N$ bosons of mass $m$ with repulsive contact interactions on a ring of circumference $L$ with periodic boundary conditions. The coupling between different angular momentum states is triggered by the presence of a barrier, a procedure analogous to the experimental implementations of phase imprinting \cite{dobrek1999optical,zheng2002classical,kumar2018producing}. 

In order to investigate phase-slips in the system, let us start from the equilibrium state $\Psi_0$ in which a static barrier is present, and then induce a quench in the many-body wavefunction such that $\Psi_0(x_1,...x_N)\to\Psi_1(x_1,...x_N)=\Psi_0\times e^{i2\pi\ell \sum_j x_j/L}$.  The current dynamics is then obtained from:
\begin{equation}
J(t)=-i \frac{\hbar}{2m}\frac{1}{N}\int_0^L \frac{dx}{L}\,\mean{\hat{\Psi}^\dagger\partial_x\hat{\Psi}-\left(\partial_x\hat{\Psi}^\dagger\right)\hat{\Psi}}.
\end{equation}
The dynamical response is found using different methods depending on the interaction strength, $\gamma=mg/\hbar^2 n$, and temperature regimes: (i) the Gross-Pitaevskii equation (GPE) and analytical two-mode model adapted from~\cite{smerzi1997quantum} at $T=0$ for a weakly interacting gas ($\gamma\ll1$);  (ii) the projected Gross-Pitaevskii equation (PGPE) at $T>0$ and $\gamma\ll1$ ~\cite{davis2001simulations,blakie2008dynamics,berloff2014modeling} and (iii) the time-dependent Bose-Fermi mapping at $\gamma\gg1$, describing the infinitely strong interaction Tonks-Girardeau (TG) limit for the entire temperature range~\cite{girardeau1960relationship,girardeau2000dark,yukalov2005fermibose}.
Throughout ths section, a quench inducing a circulation $\ell=1$ is considered, although the results can be generalized to larger circulations. Depending on the model, two types of barriers are considered: a delta potential $V(x)=\alpha\delta(x)$, for which analytical results can be obtained, and a Gaussian potential $V(x)=V_0\exp{\left(-\frac{x^2}{2 w^2}\right)}$, which is more realistic from the experimental point of view. Both cases are compared using a dimensionless parameter for the barrier strength: $\lambda_{\rm GP}=V_0/\mu_0$ for weak interactions, with $\mu_0=gn$ being the chemical potential of the homogeneous annular gas, and $\lambda_{\rm TG}=V_b/E_F$ for strong interactions, with $V_b=\alpha n$ being the energy associated to the barrier and $E_F=\hbar^2n^2\pi^2/2m$ the Fermi energy.

\paragraph*{Discussion:}

\begin{figure}[h!]
\includegraphics[width=\linewidth]{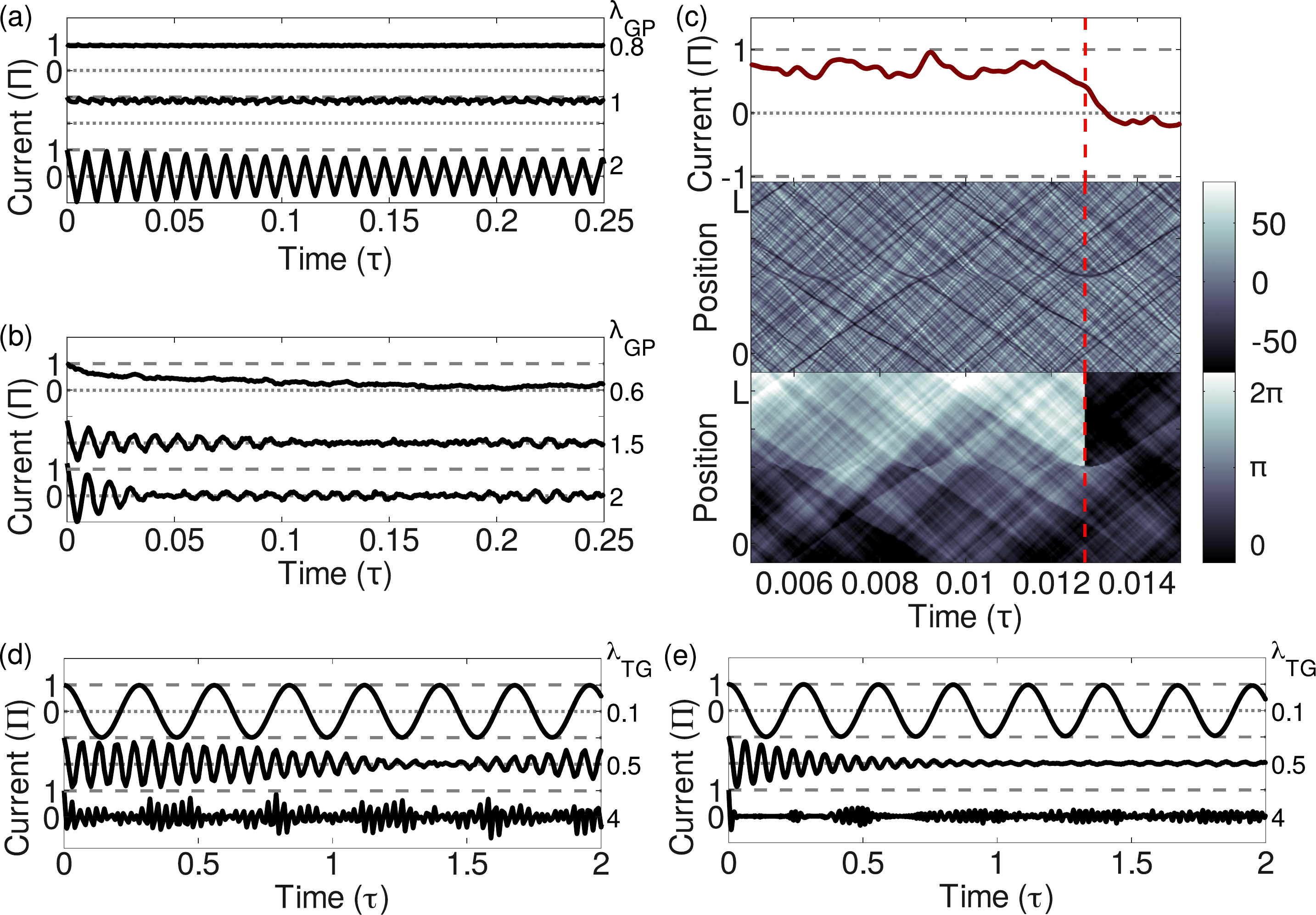}%
\caption{\label{fig:Fig_currents} 
(a-c) Classical field simulations of the quench dynamics in the mean-field regime for $\gamma=0.02$. 
(a) Average current per particle (solid lines, unit: $\Pi=\hbar/(Nm)$) as a function of time (unit: $\tau=mL^2/\hbar$) at $T=0$. Top to bottom:  $\lambda_{\rm GP}=\{0.8,1,1.05,2\}$.
(b) Current at $T=\mu_0/k_B$ for barrier strengths $\lambda_{\rm GP}=\{0.6,0.9,1.5,2\}$.
(c) Zoom on a single classical field trajectory at $T=\mu_0/k_B$ and $\lambda_{\rm GP}=0.6$, illustrating a phase slip. This consists in a jump in the current (top panel), corresponding to the reflection of a slow soliton on the barrier, also visible in the density deviation map (middle panel) and appearing as a singularity in the phase profile (bottom panel).
(d-e) Exact solutions in the Tonks-Girardeau regime.
(d) Average current per particle vs.~time after the quench for $N=23$, at $T=0$, for barrier strength $\lambda_{\rm TG}=\{0.1,0.5,1,4\}$.
(e) Current at $T=E_F/k_B$ (solid lines) for  $\lambda_{\rm TG}=\{0.1,0.5,1,4\}$ from top to bottom.
(a-e) Horizontal dotted (dashed) lines indicate the values for $J=0$  ($\pm1$).  
{
Reprinted with permission from J. Polo, R. Dubessy, P. Pedri, H. Perrin, and A. Minguzzi, Phys. Rev. Lett. 123, 195301 (2019). Copyright 2019, American Physical Society~\cite{polo2019oscillations}.
}
}
\end{figure}

Within the considered system, the current dynamics depends on interaction and temperature regimes \cite{polo2019oscillations}. Figure~\ref{fig:Fig_currents}(a-c) shows the results in the weakly interacting regime. At zero temperature, the current remains very close to the initial quenched circulating state for weak to moderate barriers, up to $\lambda_{\rm GP}\sim1$. Above this critical value, a fast decay of the current appears, followed by oscillations around the 0 value. This behavior is found to correspond to the transition of the currents from  self-trapping to Josephson oscillations, in analogy to the well known Josephson effect for particle imbalance predicted in \cite{smerzi1997quantum}.

For temperature $T=\mu_0/k_B$~\cite{kheruntsyan2003pair}, the dynamics of the current are quite different. At low barriers, i.e. $\lambda_{\rm GP}\leq0.5$, we observe an exponential decay of the current, while for larger barriers one observes damped oscillations.
In this regime, thermal phase slips  occur deterministically at the position of the barrier where the density vanishes. In order to elucidate the mechanisms behind the current decay, \fref{fig:Fig_currents}(c) shows  a \emph{single} classical field trajectory, showing  many spontaneous thermal gray solitons~\cite{karpiuk2012spontaneous}. While most of the solitons present a small density dip, thus being fast and transmitted through the barrier~\cite{bilas2005dark}, notice that the current undergoes discrete jumps each time a soliton reflects on the barrier. In this case, the density profile vanishes when the soliton reaches zero velocity, allowing for a phase slip to occur.
This corresponds to the adiabatic process indicated by the dashed red line in \fref{fig:Fig_currents}(c).
The observed exponential decay of the average current can be understood as an intrinsically stochastic process occurring when the barrier couples the soliton dynamics to the long wavelength sound excitations~\cite{bilas2005dark}. 
%
%
%

The strongly interacting regime $\gamma\gg 1$, where the classical picture does not apply, is described using the exact Tonks-Girardeau solution. We show that the dynamics of the current microscopically corresponds to quantum coherent oscillations between different angular momentum states.
At zero temperature, in contrast to the weakly interacting regime, it can be seen that for weak barriers ($\lambda_{\rm TG}\ll1$) there is no self-trapping (see \fref{fig:Fig_currents}(d)). Rather, the current undergoes Rabi-like oscillations. These oscillations correspond to coherent quantum phase slips due to backscattering  induced by the barrier, which breaks the rotation symmetry thus coupling different angular momentum states \cite{moij2006superconducting,cominotti2014optimal}. Microscopically, it corresponds to dynamical processes involving the whole Fermi sphere, i.e. multiple-particle  hole excitations where each particle coherently undergoes oscillations of angular momentum from $L_z=\hbar$ to $L_z=-\hbar$. At increasing barrier strength, an envelope appears on top of the current oscillations, degrading the Rabi oscillations. This envelope  originates from  the population of  higher-energy modes, each transition being characterized by a different frequency (see \cite{polo2019oscillations}), leading to a mode-mode coupling and dephasing, and more complex current oscillations.
At finite temperatures the quench dynamics of the current involves  high-energy excitations with an amplitude  weighted by the Fermi distribution \cite{polo2019oscillations}.  The resulting dynamics corresponds to an effective damping of the current oscillations with an exponential decay (see \fref{fig:Fig_currents}(e)), due to the effect of incoherent phase slips.  The revivals observed for large  barriers  at zero temperature are highly suppressed due to the  thermal excitations. We identify the oscillation frequency as Josephson oscillations, in which at increasing barrier strength the frequency   crosses over from a Rabi-like regime with $\omega=\pi^2N \lambda_{\rm TG}$ to a Josephson-like regime with $\omega\propto \sqrt{\lambda_{\rm TG}}$. This is in agreement with the  predictions of  the low-energy Luttinger liquid theory (see \cite{polo2018damping} and Sec.~\ref{sec:Damping_Josephson_oscillations} above).

In summary, in this section, we have presented a study of the dynamics of a one-dimensional ring pierced by a localized barrier, following a phase imprinting. From a static point of view, a localized barrier can lead to solitonic excitations as seen in Sec.~\ref{sec:stirred_BEC}. However, these can also be thermally activated or created dynamically by quenching a current in the system. 
Within the mean field regime, the self-trapping behavior of the current prevents coherent phase slips, but at finite temperatures incoherent phase slips are observed. Their microscopic origin is related to the coupling between the soliton dynamics and the long wavelength sound excitations which are intrinsically stochastic, leading to an exponential decay of the average current.
A similar microscopic behavior is found in higher dimensions, however the excitations take other specific forms such as vortex rings as seen in Sec.~\ref{sec:vortex_dynamics}.
When considering the strongly interacting regime, coherent phase slips dominate the dynamics. Finally, at finite temperatures, incoherent dynamics appears due to thermally occupied high-energy excitations that lead to an average decay of the current. 


\subsection{Concluding remarks and outlook}
\label{sec:conclusions-phase}

We have discussed different regimes of ultracold atomic gases in which phase slips play a crucial role on the dissipative motion of certain macroscopic observables, and discussed their connection to low-lying and macroscopic excitations in such systems. Understanding these processes in detail is crucial to harnessing future atomtronic applications.

To understand the microscopic origin of such mechanisms we have considered the diverse settings of 3D harmonic traps, and 1D systems in harmonic and ring traps with weak and strong interactions. We have also considered the effect of thermal fluctuations on such dynamics.

In the context of a weakly-interacting harmonically-confined 3D ultracold quantum gas, with a weak link creating a Josephson-like junction, the phase slips are related to the generation of vortex rings and associated sound emission, with increasing population imbalances leading to sequential ring generation, even potentially opening up an avenue for a turbulent-like regime. Our analysis was performed for a bosonic system, but the relevant experiment is actually performed across different  superfluid regimes of an ultracold fermionic gas. As such, our results only strictly apply to the BEC side, and numerous interesting open questions remain on how the explicit nature of the fermionic statistics affects this picture as one moves towards the unitary and BCS superfluid regimes\cite{valtolina2015josephson,burchianti2018connecting,park2018critical}

In a 1D strongly correlated Bose gas, it is low-energy excitations within the bulk that provide the underlying mechanism leading to the dissipative motion across the junction. Although in this regime we also observe that damping of large particle imbalances proceeds through higher-energy modes. 
On the other hand, in 1D ring potentials the relevant excitations leading to the decay of the current at weak interactions are dark solitons, although low-energy excitations are also involved in the decay mechanism. In our studied regime, dark solitons were thermally activated, however they can also have other origins, e.g.~being triggered by the presence of an impurity. In that case, the specific soliton or solitonic train generated depends on the size of the impurity and its velocity relative to the current.
The microscopic origin of the phase slippage is then related to the coupling between the soliton dynamics and the long wavelength sound excitations (which are intrinsically stochastic).
We note that,
while the 1D systems considered in Secs.~\ref{sec:Damping_Josephson_oscillations}-\ref{sec:Bose_gas_ring} allow to identify the microscopic origin of excitations, the barriers and topologies considered in these microscopic theories are significantly simplified ones, and extending this work to more realistic scenarios could bring new insight regarding the energy scales at which the damping of oscillations occurs. Indeed, present-day experiments use finite width barriers and external confinements to trap strongly correlated atoms, which could influence the damping.  Moreover, the results shown in Sec~\ref{sec:Damping_Josephson_oscillations} rely on the low-energy theory given by the Luttinger liquid model. Beyond the low-energy model, one should approach the problem numerically. However, numerical simulations of strongly correlated systems are highly complex. Therefore, developments in this field could prove of great importance in corroborating and extending the dynamics of strongly correlated Josephson coupled systems for strong quenches. 
Finally, we note that the spectrum and role of impurities presented in Sec.~\ref{sec:stirred_BEC} and the microscopic mechanism leading to damping can notably depend on the dimensionality.


One of the main challenges for quantum technologies is to control the system's quantum state while maintaining its quantum coherence for longer times \cite{acin2018quantum}. Thus, reducing dissipative motion becomes crucial for the development of atomtronic devices. Moreover, controlling and understanding the mechanisms involving coupling to low-energy excitations can also lead to a reduction of this dissipation. In addition, the initial quantum state can also be of consequence to the system's final stability, as the projection to high energy excitations can lead to complex damped dynamics.
From these results we can draw some insight regarding future directions for improving and reducing dissipative behavior. Integrable or quasi-integrable systems, in which many conserved quantities exist compared to the system's degrees of freedom, have been shown to present a long-lived coherence and dissipation-free dynamics \cite{kinoshita2006quantum,rigol2007relaxation,rigol2008thermalization}. Also, several recent studies have focused on topologically protected states \cite{hasan2010colloquium,cooper2019topological} as the main building blocks for future atomtronic devices, as these states prove very robust against perturbations.

{\it Acknowledgements} Juan Polo and Anna Minguzzi would like to especially thank: Ver\`onica Ahufinger - for her contribution to the work on the 1D Josephson junction; the late Frank Hekking - who greatly helped us in understanding Josephson junctions; Romain Dubessy and Paolo Pedri - for their contributions to the theoretical development and numerical simulations as well as for fruitful discussions regarding dynamical phase slips in rings. In addition, we thank Maxim Olshanii and Jook Walraven for stimulating discussions. We also acknowledge financial support from the ANR project SuperRing (Grant No.  ANR-15-CE30-0012). LPL is a member DIM SIRTEQ (Science et Ing\'enierie en R\'egion \^Ile-de-France pour les Technologies Quantiques).
Axel P\'erez-Obiol thanks Taksu Cheon for his contribution on the topic of stirred BECs
on 1D rings. Klejdja Xhani and Nick Proukakis acknowledge contributions to the theoretical modelling and understanding by Carlo Barenghi, Luca Galantucci, Kean Loon Lee, Andrea Trombettoni and to the experimentalists Alessia Burchianti, Elettra Neri, Giacomo Roati, Francesco Scazza and Matteo Zaccanti. Nick Proukakis acknowledges financial support from the Quantera ERA-NET cofund project NAQUAS through the Engineering and Physical Science Research Council, Grant No. EP/R043434/1.




\section{ATOMTRONICS ENABLED QUANTUM DEVICES AND SENSORS} \label{DeviceSensors}
\vspace*{-0.5cm}
\par\noindent\rule{\columnwidth}{0.4pt}
{\bf{\small{D. Anderson, V. Ahufinger, 
A. S. Arnold, G. Birkl, M. Boshier,  S. A.  Gardiner, B.M. Garraway, J. Mompart}}}
\par\noindent\rule{\columnwidth}{0.4pt}


%
%
%
%
%
%
%



\label{sec:level1}
In this section we will discuss some example cases where the atomtronics approach is leading to novel components (which may be part of a larger device) and applications such as rotation sensing and magnetic sensing.

\subsection{\label{sec:level2}Diodes, transistors, and other discrete components}


The terminology \emph{Atomtronics} suggests, correctly, but not exclusively,
an analogy between circuits for atomic matter and those based on standard
\emph{electronics}. The flow of electrons in an electric circuit can
be considered in a way analogous to the flow of cold neutral
atoms in an atomtronic circuit (At present, there is some flexibility
in this interpretation.). Important questions concern how to confine
the atoms to a circuit, how to control them, and what devices and
applications can arise.

It would be a gross over-simplification to suggest that an atomtronic
circuit should merely mimic an electronic circuit. Whilst this \emph{might}
be the case, it is by no means essential, and indeed, it is very much
intended that future atomtronic systems go beyond analogs of standard
electronic circuits. To be specific: Although we may start with
these basic analogs, the future hope is for devices that use the
properties of matter-wave coherence and other quantum properties of
matter to go beyond these direct analogs, and to even create devices with
no electronic counterpart because of the unique properties of the
quantum physics of matter.
Meanwhile, however, in this first part of Section \ref{sec:level1} we
will explore the progress made in formulating and implementing discrete
atomtronic components that are similar to electronic ones.

The basic electronic elements are often regarded to be batteries,
resistors, capacitors, diodes, transistors, and the like.  If we start with the battery, in
the atomtronic world it can be regarded as a reservoir of atoms. Clearly,
that is too simplistic and not enough on its own. So first steps are
to involve atoms contained in a reservoir and allowed to flow out of that reservoir into a
more complex circuit, or at least into another reservoir as in cases of
two-terminal flow \cite{krinner2017two}.
The current of neutral atoms is driven by the difference in the chemical
potential between two reservoirs, typically implemented by the two sides 
of a barrier, in a way that is
analogous to Ohm's law in conventional circuit theory.
The battery is later intended to supply power to an
atomtronic circuit via the transport of cold atoms.

\begin{figure}[t]
  \centering
  \includegraphics[width=0.85\columnwidth]{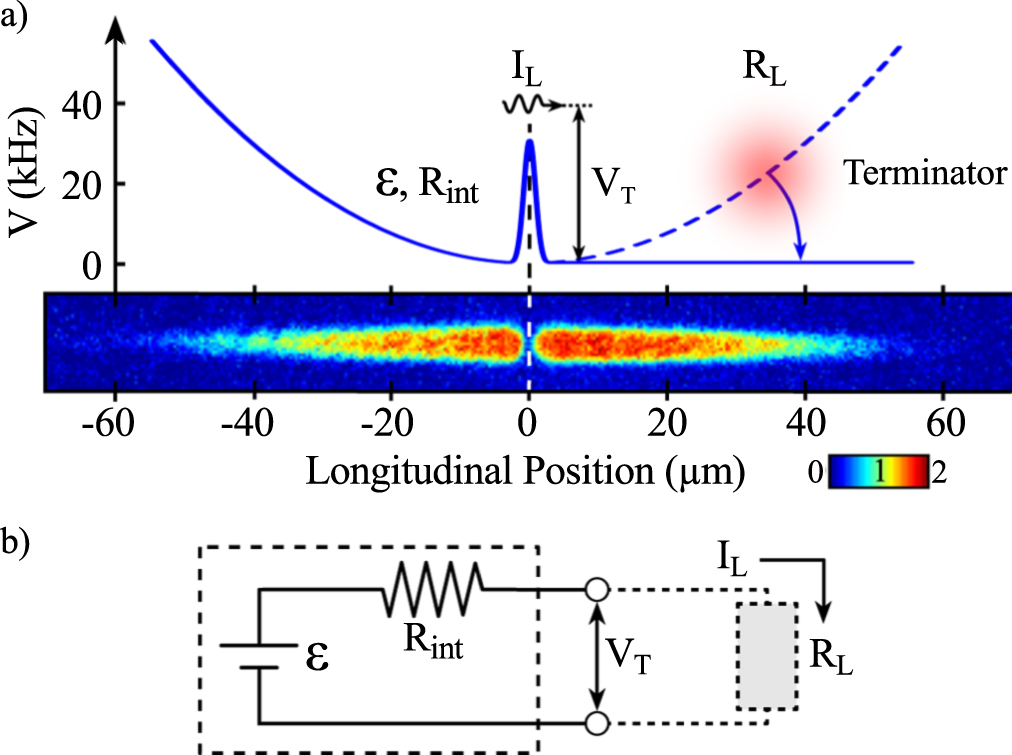}
  \caption{\label{fig:battery}An atomtronic battery. (a) The confining fields for the atoms
  are formed using magnetic and optical potentials. The image of atoms demonstrates the situation when they fully occupy the system with the terminator off. (b) An equivalent circuit for the atomtronic battery.
  Reprinted with permission from S. C. Caliga, C. J. E. Straatsma, and D. Z. Anderson, New Journal of Physics 19, 013036 (2017), under a Creative Commons Attribution 3.0 license.}
\end{figure}

Such a battery was demonstrated in Ref.\ \onlinecite{caliga2017experimental}
which was complemented by the respective theoretical description 
in Ref.\ \onlinecite{zozulya2013principles}. 
In this `battery' experiment a relatively large confining potential
for ultra-cold atoms is divided into two parts by introducing a
spatially narrow beam of blue-detuned light acting as a barrier
(see Fig.\ \ref{fig:battery}a).
Atoms can be
confined on one side of the combined potential, the other side, or both. During the experiment discussed here,
atoms are initially loaded on one side 
(the left side in Fig.\ \ref{fig:battery}a)
and then allowed to flow to the right side as controlled by the
sharp barrier which essentially controls the `resistance'.
{
  This flow can be classical, or quantum, depending on the temperature of the loaded atoms.
  (Tunneling does not play a role in Fig.\ \ref{fig:battery}a if the main contribution to flow is from over-the-barrier atoms and if the left well re-thermalizes sufficiently to maintain the distribution.)
}
To prevent the reflection of atoms from the second well, a `terminator' is
added to the system by means of an optical beam that pumps atoms in
the second well into untrapped states, such that they are lost from
the system. This terminator also represents a load-matched
impedance (see Fig.\ \ref{fig:battery}b) in the analogous electric circuit. 

{ A cold atom system with ballistic atoms has been used to demonstrate the dynamics of a linear} $RLC$ circuit
\cite{lee2013analogs}.  In this case, a light sheet trap is modified
with additional dipole beams to create 2D confinement of the atoms:
two reservoirs are generated that are joined by a narrow link. This `capacitor'-like systems is
charged by loading atoms into one of the reservoirs. Subsequently the
flow of atoms through the narrow channel discharges the capacitor. 
The channel possesses a finite resistance and appears to have inductance as well \cite{lee2013analogs}.


In analogy to solid state materials modifying the electronic wave function, 
optical lattices offer band-gap structures for cold atoms (In this context, see e.g.\ Ref.\
\onlinecite{damon2015band}.). 
These allow the creation of diodes and transistors by changing the
base-line potential of the lattice across a discontinuity or
junction in just the same way as for semi-conductors across a NP-\ or
PN-type junction \cite{seaman2007atomtronics,pepino2010open}.
However, it is interesting to note that
the atomtronic diode can display its functionality with just a few lattice sites (i.e.\ with just a few potential minima) and
the atomtronic transistor is proposed to be functional with just three potential wells \cite{caliga2016transport}.
    It can be constructed in the same way as the battery experiment discussed above \cite{caliga2017experimental}, but with an additional blue-detuned dipole beam adding one additional barrier to the passage of the atoms. The transmission of atoms through the double-barrier system is now dependent on the chemical potential between the two barriers \cite{stickney2007transistorlike,caliga2016principles}, giving a transistor-like behaviour with 'source', 'gate', and 'drain' assigned to the three regions around the barriers.
Furthermore, by cascading transistor junctions, i.e.\ by
adjusting the sets of lattice potentials, a logic gate (AND gate) has been proposed
consisting of just five lattice sites \cite{pepino2010open}. 

One direction for future extensions is the development of more exotic circuit elements. For example,
asymmetric double well potentials made by optical dipole beams have
been used to create a Josephson junction for an atomtronic system
\cite{gati2006realization,gati2007bosonic}. 
%
%
In the direction of increasing the complexity of circuits, for example, the proposal for an AND gate starts to open the way for a very unusual
type of quantum logic which is based on the flow of neutral atoms. 
Here, we can imagine going from AND gates to NAND gates, which are universal
gates, and then by further increasing the complexity, a universal
matter-wave quantum computer is accessed---at least in principle.

\subsection{\label{sec:level3}Atomtronic SQUIDs}


Quantum interference has a high importance in atomtronics which is particularly true in the 
case of \emph{atomtronic SQUIDS}.
Atomtronic SQUIDS are \emph{not} superconducting devices, but are named for their analogy with SQUIDS.
They are sometimes denoted AQUIDs for Atomtronic QUantum Interference Devices \cite{amico2017focus}.

A conventional SQUID can be built from a superconducting ring with one or two `weak links'
which form Josephson junctions. At each junction the current and the junction phase are closely
related in such a way that magnetic field strengths can be determined from the oscillatory behaviour
of the voltage drop across the junction. The resulting device makes an excellent magnetic field sensor \cite{fagaly2006superconducting}.

%

\begin{figure}[t]
  \centering
  \includegraphics[width=0.85\columnwidth]{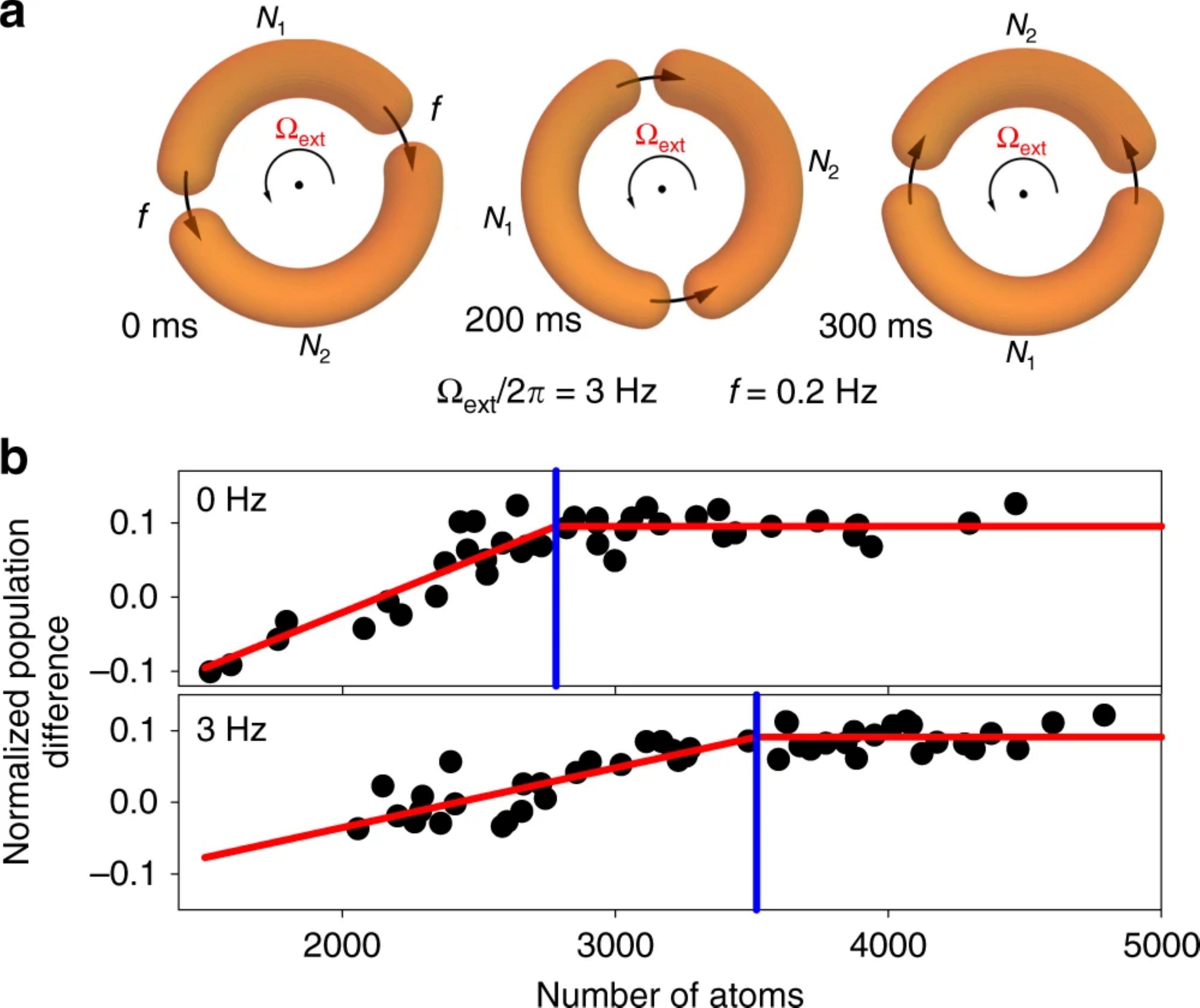}
  \caption{\label{fig:squid}(a) We show how the `Josephson junction' barriers move in the atomtronic SQUID in order to observe a synthetic external rotation. The two barriers move at different rates $\Omega_\text{ext} \pm 2\pi f $.
The number of atoms on either side of the barriers is $N_1$ and $N_2$.
(b) The normalised population difference $( N_2 - N_1 ) / ( N_2 + N_1 ) $ is plotted as a function of the number of atoms in each experimental run. The blue line indicates the point at which a critical atom number is reached, where the system switches from AC to DC Josephson regimes. The critical atom number varies with the `rotation' rate, being larger in the lower panel of (b).
Reprinted with permission from C. Ryu, E. C. Samson, and M. G. Boshier, Nat. Commun. 11, 3338 (2020), under a Creative Commons Attribution 4.0 International License.}
\end{figure}

A basic atomtronic SQUID consists of a ring waveguide for ultra-cold neutral atoms with two barriers
inserted \cite{ryu2013experimental,ryu2020quantum} (see Fig.\ \ref{fig:squid}a). In the analogy, the ring waveguide replaces the
superconducting ring, and the barriers replace the Josephson junctions.   The current-phase relation for atomtronic tunnel junctions and weak links was explored in refs \cite{ryu2013experimental,eckel2014hysteresis}.
Each barrier clearly affects the phase of the wavefunction and, just like the super-conducting analogs, there is 
a critical current: in the atomtronic case, when the flow is too fast it breaks up into vortex - antivortex pairs \cite{ramanathan2011superflow,wright2013driving}.
In a conventional SQUID, current flow in the superconducting loop is established by changing the magnetic flux through the loop.  In the atomtronic SQUID, current flow in the waveguide loop is created by rotation of the system. It follows that the device's behavior is sensitive to rotation of the SQUID (or equivalently rotation of the barriers). This principle, including quantum interference effects, has been recently demonstrated in Ref.\  \onlinecite{ryu2020quantum}. 
Figure \ref{fig:squid}b shows the transition from the AC Josephson regime (AC current through the junctions with non-zero chemical potential difference across them) to the DC Josephson regime (DC tunnelling current flowing through the junctions with no chemical potential) marked with the vertical blue line as a function of the number of atoms for two different rotation rates. This transition point becomes oscillatory as a function of the rotation rate (for a fixed number of atoms) or as a function of the number of atoms (at different rotation rates as shown here).  These oscillations can be used to determine the rotation rate ($\Omega_\text{ext}$ in Fig.\ \ref{fig:squid}).  Another recent work has shown that the atomtronic SQUID also exhibits hysteresis behaviour analogous to the conventional SQUID \cite{eckel2014hysteresis}. 

The recent experiments with ring traps and barriers \cite{eckel2014hysteresis,jendrzejewski2014resistive,wright2013driving,ramanathan2011superflow,ryu2013experimental,ryu2020quantum}
are based on optical dipole potentials. The ring trap and weak link can be created with a Laguerre-Gauss beam (with a hole) and a focussed Gaussian beam\cite{ramanathan2011superflow}. For the ring part, potentials created by conical refraction are possible as well \cite{turpin2015blue}. Alternatively, a very flexible approach is to use
`painted' potentials\cite{henderson2009experimental,ryu2015integrated}, where an optical dipole beam is rapidly scanned around the region of interest. As the beam is scanned, the intensity of the light is modulated so that a two-dimensional image is formed which produces a rather flexible 2D potential. Confinement in the third dimension is provided by a light sheet and the dipole potential from that. The painted potential can include a ring, and the `weak link' barriers which can be moved around the ring at will.

The atomtronic SQUID has the clear potential of being a central building block for atomtronic devices: e.g.,\ for rotation or magnetic field sensors. However, it may
also play other important roles in atomtronic circuits. It could be a component in a more complex circuit
where matter wave interference is essential, or allow for the storage of quantum information in the quantum states of the atomtronic SQUID.

\subsection{\label{sec:level4}Sagnac interferometry and rotation sensing}


Rotation detection devices come in many forms, including micro-electromechanical systems (MEMS), hemispherical resonator gyros (HRGs), ring laser gyros (RLGs), and fiber-optic gyroscopes (FOGs). They are ubiquitous in the sense that mobile phones contain low-sensitivity gyroscopes that users can access via apps such as \href{https://phyphox.org/}{Phyphox}. These more `traditional' sensors have been reviewed with viewpoints that are mainly academic \cite{schreiber2013large} and industrial  \cite{passaro2017gyroscope}.

Atoms can also be used as gyroscopes, for example via nuclear spins in co-magnetometers \cite{kornack2005nuclear}. Alternatively, demonstration of atom-optical manifestations of the Sagnac effect date back to 1991 \cite{riehle1991optical}, and already in 1997 short-term sensitivities of $2\times 10^{-8}$ (rad/s)/$\sqrt{\mathrm{Hz}}$ were demonstrated in an atom-based Sagnac interferometer \cite{gustavson1997precision,lenef1997rotation}; since then there have been many technical advances in such ``free space'' interferometers, \cite{barrett2014sagnac} commonly based on a Mach--Zehnder-type configuration. 

From an atomtronic standpoint we wish to consider ``closed-path''  guided configurations, which have been used for cold molecules \cite{crompvoets2001prototype}, cold atoms \cite{sauer2001storage,dumke2002interferometer}, and Bose-Einstein condensates \cite{arnold2006large,gupta2005bose} (BECs) since the early 2000s, and have been recently considered for chip-scale development \cite{sinuco2014inductively,garridoalzar2019compact}. Theoretical and experimental guiding geometries include optical dipole, magnetic and Stark confinement using constant, time-averaged, inductive or dressed potentials \cite{arnold2004adaptable,wu2004bidirectional,andersen2006quantized,morizot2006ring,lesanovsky2006adiabatic,courtade2006dark,jain2007quantum,bhattacharya2007lattice,ryu2007observation,wu2007demonstration,olson2007cold,fernholz2007dynamically,lesanovsky2007time,frankearnold2007optical,heathcote2008ring,schnelle2008versatile,griffin2008smooth,houston2008reproducible,baker2009adjustable,henderson2009experimental,sherlock2011time,bruce2011smooth,pritchard2012demonstration,arnold2012extending,moulder2012quantized,beattie2013persistent,cominotti2014optimal,ryu2015integrated,marti2015collective,gauthier2016direct,bell2016bose,henderson2016comparative,henderson2020optical}. Highly supersonic superfluid flow is also possible \cite{arnold2006large,guo2020supersonic,pandey2019hypersonic}, however whilst the concept of closed-path cold atom configurations has long held traction, the sensitivity has yet to reach the high levels of free-space cold atom gyros \cite{dutta2016continuous}. There has been some important recent experimental progress on this front \cite{moan2019quantum,boshier2019talk}, i.e.,\ all the necessary experimental tools appear in principle to be present. 

This means we consider Bose-condensed atoms held within an appropriate (e.g., toroidal, as used in atomtromic SQUIDs) trapping potential $V$. This is necessarily assumed to be in a rotating frame defined by the angular velocity vector $\boldsymbol{\Omega}$, manifesting as an additional term $i\hbar\boldsymbol{\Omega}\cdot(\mathbf{x}\times\nabla)$ in either the single-particle Schr\"{o}dinger equation, or the Gross--Pitaevskii equation (GPE) when describing mean-field dynamics of Bose-condensed atoms. A coordinate system where $\boldsymbol{\Omega}$ points along the $z$ axis (dynamics viewed from within a frame rotating anti-clockwise around the $z$ axis with angular frequency $\Omega$) simplifies this term to $i\hbar\Omega(x\partial_{y} -y\partial_{x})$. A good general starting point for describing the dynamics within a variety of such systems is the following system of GPEs:
\begin{equation}
\begin{split}
i\hbar\frac{\partial}{\partial t}\Psi_j
=& 
-\frac{\hbar^2}{2m}\nabla^2\Psi_j
+i\hbar\Omega\left(x\frac{\partial}{\partial y}\Psi_j -y\frac{\partial}{\partial x}\Psi_j\right)
\\&
+V_{j}\Psi_j
+\sum_{j'}g_{jj'}|\Psi_{j'}|^{2} 
\Psi_j.
\end{split}
\label{Eq:GeneralGPE}
\end{equation}
The $j$ index labels different internal atomic states, the $V_{j}$ incorporate energy differences between the internal states and any internal state dependences of the trapping potential, and the $g_{jj'}$ quantify the strengths of   $s$-wave scattering terms (ignoring the possibility of internal-state-changing collisions).

We first consider a tight toroidal trapping potential, reducing our treatment to a radius $R$ one-dimensional ring geometry, leaving only the polar angle $\phi$ free, reduce to a single internal state, and neglect all interactions. The GPE then becomes
\begin{equation}
i\hbar\frac{\partial}{\partial t}\psi(\phi)
= \left(-\frac{\hbar^2}{2mR^{2}}\frac{\partial^2}{\partial\phi^{2}}
+i\hbar\Omega\frac{\partial}{\partial\phi}\right)\psi(\phi).
\label{Eq:PolarGPE}
\end{equation}
Note the interactions are genuinely insignificant if the gas is very dilute, or if interactions are tuned away using an appropriate Feshbach resonance. The evolution of an initial, localized matter wave split into an equal superposition with opposite velocity splitting products can, at the simplest level, be considered without explicit mention of the initial wave packet. {At this level, an initial state $\psi(0)$  evolves as $\psi(t):$
\begin{eqnarray}\psi(0)& = &(e^{i\ell\phi} + e^{-i\ell\phi})/\sqrt{4\pi}=\cos(\ell\phi)/\sqrt{\pi} \nonumber\\  \psi(t)&=&e^{-i\hbar\ell^{2}t/2mR^{2}}\cos(\ell[\phi+\Omega t])/\sqrt{\pi}, \nonumber
\end{eqnarray} yielding intensity fringes $\propto(2/\pi)\{\cos(2\ell[\phi+\Omega t]) +1\}$ multiplied by the number of atoms whenever the matter wave splitting products overlap in space. For a matter wave initially centred at $\phi = 0$ enclosed in a ring with area $A = \pi R^{2},$ this occurs at: $\phi=\pi\;\textrm{or}\;0$, when $t= (1\;\textrm{or}\;1) Am/\hbar \ell$, yielding a phase shift of $\Delta\phi=\Omega t = (2\;\textrm{or}\;4) A\Omega m/\hbar$, respectively.}

Using the speed $v=\hbar \ell/Rm$ and wavelength $\lambda = 2\pi R/\ell$ of the propagating atoms we express the latter phase shift as $\Delta\phi = 8\pi A\Omega/\lambda v$, the same form as the phase shift accumulated by an optical Sagnac interferometer, with $\lambda$ and $v$ replaced by the wavelength and speed of light, respectively. This highlights the promise of atom interferometry, in that $\lambda v$ can be made much smaller than its optical equivalent. Note also that if the initial state is literally $\psi(0)=\cos(\ell\phi)/\sqrt{\pi}$, i.e., there is no localizing ``envelope'' to the wave packet, the accrued fringe shift can be observed at any time, effectively increasing the enclosed area of the interferometer. This highlights an important feature, in that as the speed of light is a constant, it is necessary to increase $A$ in order to increase the interrogation time; with cold atoms this is not the case.

With typical repulsive interactions (positive $g$), such standing wave fringes will rapidly disperse, however in a two-component system with very similar scattering lengths ($\Rightarrow g_{11} \approx g_{12} \approx g_{22}$, as can be achieved in $^{87}$Rb),\cite{halkyard2010rotational} producing an equal superposition initial state such that $\psi_{1}(0)=\cos(\ell\phi)/\sqrt{2\pi}$, $\psi_{2}(0)=\sin(\ell\phi)/\sqrt{2\pi}$, the two components stabilise each other by making the total mean field potential $g(|\psi_{1}|^{2}+|\psi_{2}|^{2})$ essentially flat (hence, no gradients and no dispersive forces). Alternatively, initializing the system such that $\psi_{1}(0)= e^{i\ell\phi}/\sqrt{4\pi}$, $\psi_{2}(0)= e^{-i\ell\phi}/\sqrt{4\pi}$, followed by an evolution time $T/2$, a $\pi$ pulse swapping the internal states, and a second evolution time $T/2$, produces {
\begin{eqnarray}
\psi_{1}(T)=e^{-i\varphi T}e^{-i\ell(\phi+\Omega T)}/\sqrt{4\pi}, \nonumber \\ 
\psi_{2}(T)=e^{-i\varphi T}e^{i\ell(\phi+\Omega T)}/\sqrt{4\pi},\nonumber
\end{eqnarray} 
($\varphi$ is a global phase depending on $\ell^{2}$ and the values of $g_{11}$, $g_{12}$, and $g_{22}$). Repeating the initializing process produces 
\begin{eqnarray}
\psi_{1}(T)=-ie^{-i\varphi T}\sin(\ell[\phi+\Omega T])/\sqrt{2\pi}, \nonumber \\ 
\psi_{2}(T)=e^{-i\varphi T}\cos(\ell[\phi+\Omega T])/\sqrt{2\pi}.\nonumber
\end{eqnarray} }
The value of $\Omega$ can then be inferred from population measurements: $N_{1}=N[1+\cos(2\ell\Omega T)]/2$, $N_{2}=N[1-\cos(2\ell\Omega T)]/2$, where $N$ is the total particle number. The $\pi$ swap pulse at $T/2$ is carried out to counteract accumulation of relative phase due to differences in internal state energy and values of $g_{11}$ and $g_{22}$; if $g_{11}\approx g_{22}$ and the energy gap between the internal states is well known, this can be neglected, and a detailed experimental proposal based around $^{87}$Rb and magnetic vortex pumping has been determined by Helm \textit{et al.}.\cite{helm2018spin} This alternative has the advantage of there being no mean field gradients even when the scattering lengths are quite different, and turns a measurement of interference fringes into a measurement of relative population. Note, however, that optimum sensitivity of such a measurement is when the slope of the response curve is maximal, e.g., when $2\ell \Omega T \approx \pi/2$ \cite{helm2018spin,haine2016mean}; it may therefore be advisable to add a controlled relative phase in an experimental realisation.

\begin{figure}[!t]
  \centering
  \includegraphics[width=\columnwidth]{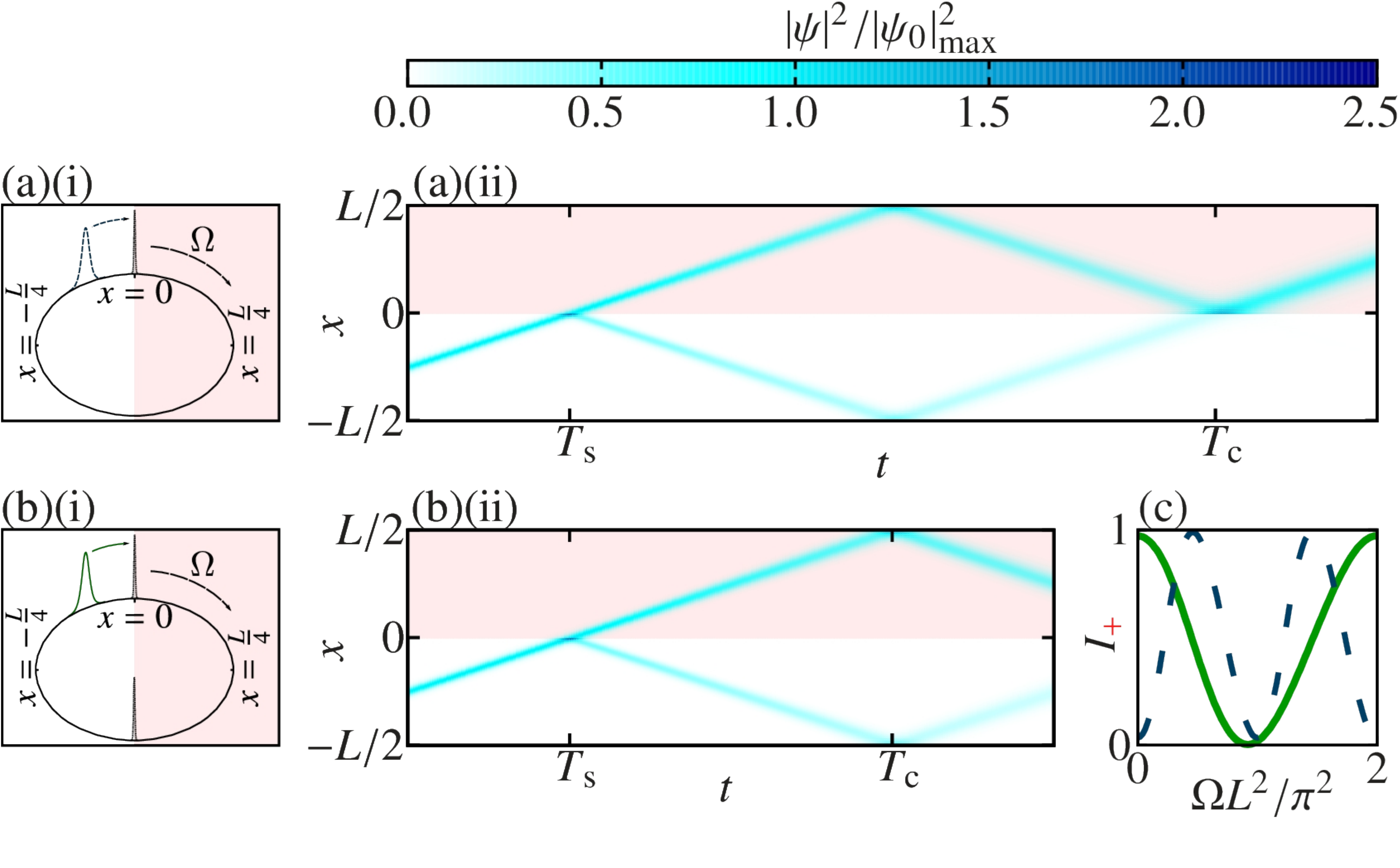}
  \caption{ An incoming soliton splits at time $T_{\mathrm{s}}$ on a barrier into two solitons of equal amplitude and opposite velocity.  At a time $T_{\mathrm{c}}$ the solitons recombine either at the same barrier (a), or a second barrier (b) antipodal to the first (the example value of $\Omega$ is the same in both cases).  The resulting phase difference is read out via the population difference in the final output products within the positive (shaded) and negative domains.  (c) Final population in the positive domain $I_{+}$ as a function of $\Omega$.  The sensitivity of the single barrier case (dashed line) is twice that of the double barrier case (solid line) because the interrogation time $T_{\mathrm{c}}-T_{\mathrm{s}}$ is doubled. Reprinted with permission from J. L. Helm, S. L. Cornish, and S. A. Gardiner, Phys. Rev. Lett. 114, 134101 (2015). Copyright 2015, American Physical Society.}
\label{Fig:Soliton_Sagnac_Overview} 
\end{figure}

A quite different approach, again in near 1D, is to have a single component condensate with attractive interactions, using, e.g., $^{85}$Rb or $^{7}$Li, again ideally in a ring geometry, as illustrated in  Fig.~\ref{Fig:Soliton_Sagnac_Overview}.\cite{helm2015sagnac} In this case, the GPE exhibits soliton solutions: stable, non-dispersive, localized wave packets which are robust to collisions, behaving something like classical particles. Given a sufficiently sharp barrier ---  ideally a $\delta$-function, more realistically a Gaussian barrier with width significantly smaller than the soliton's characteristic length\cite{helm2012bright,polo2013} formed e.g.\ by a focused off-resonant sheet of light \cite{wales2019splitting} --- an initial soliton can be split into two halves propagating with equal and opposite velocity if its incoming velocity is correctly calibrated to the barrier size. In an essentially similar way, the splitting products can accumulate a relative Sagnac phase, which could in principle be visualised through spatial interference fringes.\cite{mcdonald2014bright} The solitons' small size can make this a less suitable approach, however, than recombining them again on a barrier, where the relative phase manifests through the relative sizes of the wave packets emerging on either side of the barrier as the result of this second collision, i.e., again as relative population measurement, with an essentially similar (ideal) dependency on $\Omega$ as that outlined above. This second barrier interaction can take place either at a second barrier exactly opposite to the first,\cite{helm2014splitting} or, due to the fact that solitons are robust to collisions and therefore in some sense should ``pass through'' one another, back at the same barrier at which the initial splitting took place, following both splitting products having completely circumnavigated the ring.\cite{helm2015sagnac}. 

The roles of quantum noise and interaction for rotation sensing with bright solitons in the quantum regime were studied in \cite{haine2018quantum}. It was found  that  interaction and noise should be carefully considered in order the performances of the system are not spoiled.
{ In \cite{weiss2009creation} the scattering properties of a quantum matter wave soliton splitting in a barrier were studied.}
In addition,  the GPE analysis is of limited accuracy for the quantitative analysis of  the sensitivity of atom interferometry in the presence of interaction.  For other features of bright soliton interferometers, please see Chap. \ref{SolitonAtomtronics}.

Finally, we note that everything we have described is in a sense ``classical,'' in the sense of a classical field description of the BEC being completely adequate, and that more explicitly quantum elaborations have been proposed, exploiting spin squeezing \cite{nolan2016quantum} or ideas from quantum information \cite{Che2018phase}.

\subsection{\label{sec:level5}Magnetometry}

The development of compact highly sensitive magnetometers with high spatial resolution is one of the current challenges of Atomtronics. The capability of measuring with high precision and accuracy very weak magnetic fields is at the basis of numerous applications including bio-magnetism, geology, data storage and archaeology \cite{robbes2006highly}. Different approaches have been followed in the previous years to reach this goal,  mainly using superconducting quantum interference devices (SQUIDs),
nitrogen-vacancy {(NV) diamond} 
magnetometers, and atomic magnetometers.
{Currently, SQUIDs and spin-exchange relaxation-free magnetometers reach sensitivities at the fT/$\sqrt{\rm{Hz}}$ level and below \cite{kitching2018chip} whereas NV-magnetometers allow for pT/$\sqrt{\rm{Hz}}$ sensitivities \cite{barry2020sensitivity}. A valuable pictorial log-log plot of experimental device scale vs.\ magnetic sensitivity for a wide variety of magnetometer sensor technologies can be found in Fig.~2 of a recent review\cite{mitchell2020colloquium}.}


Atomic magnetometers can be classified depending on whether the magnetic field drives the internal or the external degrees of freedom of the atoms. The former are typically based on the measurement of the Larmor spin precession of optically pumped atoms either using thermal clouds or BECs. 
In the case of thermal clouds
double-resonance  optically  pumped  magnetometers  are  an  attractive  instrument  for  unshielded magnetic-field  measurements  due  to  their  wide  dynamic  range  and  high  sensitivity\cite{ingleby2018vector}.
In the BEC case, the use of stimulated Raman transitions has been reported \cite{terraciano2008single} as well as the separate probe of the different internal states of a spinor BEC after free fall \cite{hardman2016simultaneous}, or the measure of the Larmor precession in a spinor BEC \cite{isayama1999observation,vengalattore2007high,fatemi2010spatially,eto2013spin,eto2013control,muessel2014scalable}. Note also that a two-component BEC has been also investigated for magnetometry \cite{tojo2010controlling}.

An alternative approach to atomic magnetometry is based on 
encoding the magnetic field information in the spatial density profile of 
matter waves. Some examples of this approach are those based on detecting density fluctuations in a BEC due to the magnetic induced deformation of the trapping potential \cite{wildermuth2005microscopic,wildermuth2006sensing,yang2017scanning}. 
Recently, a different scenario has been explored and a quantum device for measuring two-body interactions, scalar magnetic fields and rotations based on a BEC in a ring trap has been proposed \cite{pelegri2018quantum}. To this aim, the BEC is prepared in an imbalanced superposition of the two counter-propagating Orbital Angular Momentum (OAM) $l = 1$ modes and due to quantum interference, a line of minimal atomic density appears. In the presence of non-linear interactions, this nodal line shows a soliton-like rotating motion. An analytical expression relating the angular frequency of the rotation of the minimal density line, $\Omega_{m}$, to the strength of the non-linear atom-atom interactions and the difference between the populations of the counter-propagating modes is derived: 
\begin{equation}
\Omega_{m}=\frac{U n_{1\pm}}{2(1+\frac{U}{\Delta})}.
\label{omeganode}
\end{equation}
where
$n_{1\pm}$ is the population imbalance between the $l=\pm 1$ modes, $U$ is a non-linear parameter proportional to the scattering length, and $\Delta$ is the chemical potential difference between the $l=3$ and $l=1$ modes, see\cite{pelegri2018quantum}. This expression constitutes the basis to use the physical system under consideration as a quantum sensing device by measuring the  rotation  frequency  of  the  minimum  density  line  by  direct imaging,  in  real  time,  the spatial density distribution of the BEC. In fact, a full experimental protocol based on direct fluorescence imaging of the BEC that allows to measure all the quantities involved in the analytical model is proposed.

\begin{figure}
\includegraphics[width=1.0\linewidth]{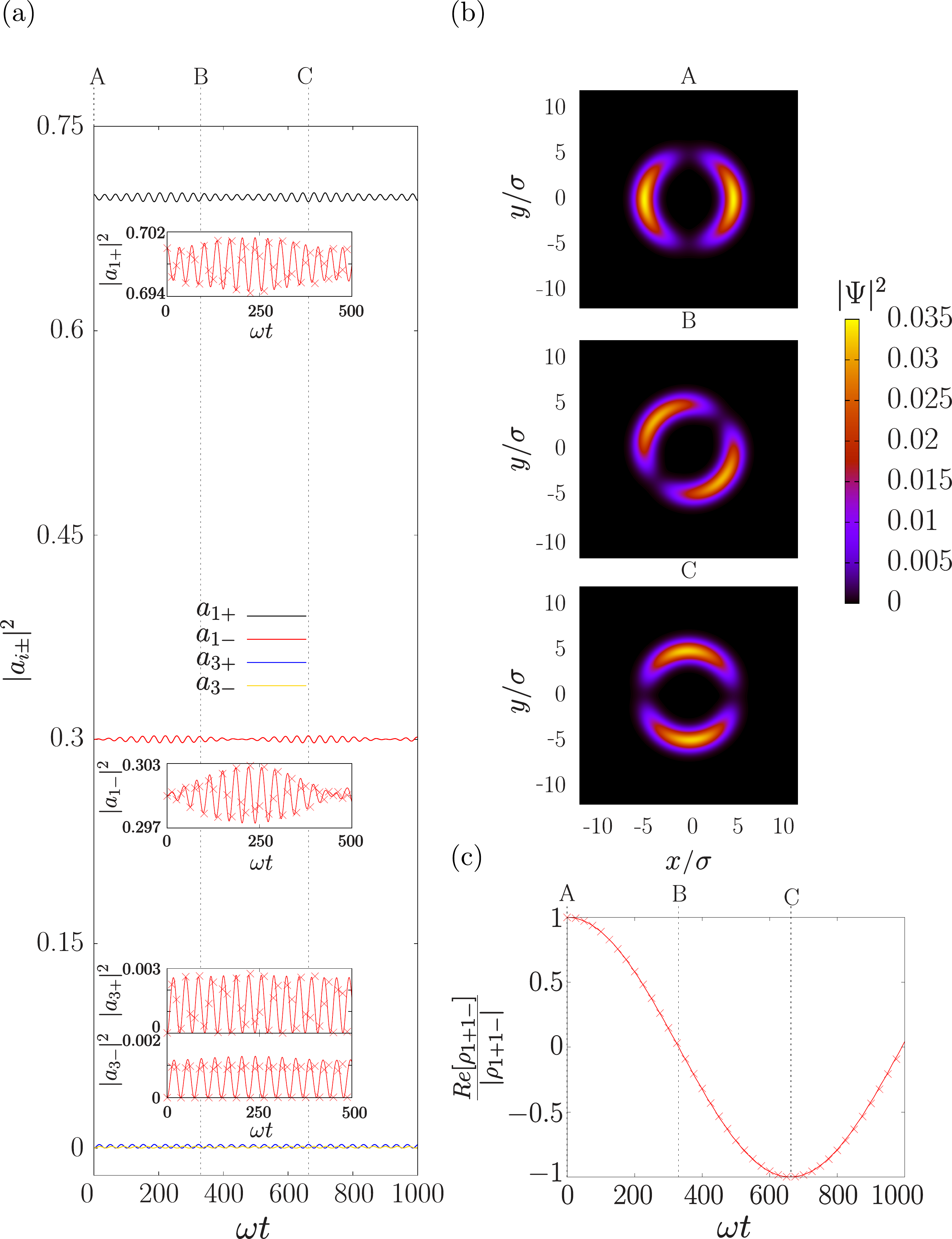}
\caption{(a) Time evolution of the population of the states involved in the dynamics. (b) Snapshots of the density profile for different instants of the dynamical evolution. (c) Time evolution of the real part of the coherence between the $\ket{1,+}$ and $\ket{1,-}$ states. The points correspond to the numerical simulation of the GPE, while the continuous lines are obtained by solving the FSM equations. The considered parameter values are $R=5$, $g_{2d}=1$, for which $U=0.0128$, ${\mu_1=0.529}$ and $\mu_3=0.699$, $a_{1+}(0)=\sqrt{p_{1+}(0)}=\sqrt{0.7}$ and $a_{1-}(0)=\sqrt{p_{1-}(0)}=\sqrt{0.3}$. Reprinted with permission from G. Pelegrí, J. Mompart, and V. Ahufinger, New Journal of Physics 20, 103001 (2018), under a Creative Commons Attribution 3.0 License.}
\label{dynamics_g1_p+07_v3}
\end{figure}

Let us assume that the lifetime of the BEC is $\tau$. Then, the condition $\Omega_{m} \omega \gtrsim 1/\tau$ being $\omega$ the ring trapping frequency must be fulfilled to be able to observe the rotation of the minimum density line for the time the experiment lasts. The upper limit of observable values of $\Omega$ is imposed by the regime of validity of the model, i.e., $\Omega_{m} < 0.025 $ to avoid the excitation of  states with OAM higher than $1$. A magnetic field produces, in general, a variation of the scattering length, which does not depend on the magnetic field orientation. Therefore, the presence of an external magnetic field will induce a variation of $\Omega_{m}$ and  the system could be used as a scalar magnetometer by relating changes on the frequency of rotation of the minimal line to variations of the modulus of the magnetic field. The sensitivity in the measurement of magnetic field variations increases with the number of condensed particles but keeping the scattering length small and having a strong dependence of the scattering length on the magnetic field modulus. Thus, close to a Feshbach resonance, these requirements could be meet. However, close to a Feshbach resonance the three-body losses may limit the lifetime of the BEC making difficult the measurement. Nevertheless, some atomic species such as $^{85}$Rb, $^{133}$Cs, $^{39}$K or $^7$Li have been reported to form BECs that are stable across Feshbach resonances with lifetimes on the order of a few seconds, so they could be potential candidates for using the system as a magnetometer. Taking into account that the trapping frequency $\omega$, is typically of the order of a few hundreds of Hz for ring-shaped traps, and considering typical values of $\Omega_{m} \omega \sim 1$Hz compatibles with typical times for the experiment of around $\tau\sim 1$s, the minimum density line would perform some complete round trips. Assuming that one could resolve angular differences on the order of $\sim 0.1$ rad, variations in the rotation frequency on the order of $10^{-2}$Hz could be measured. Thus, for the parameter values reported in \cite{wang2015resonant}, in principle, this magnetometer would allow to measure changes in the magnetic field on the order of a few pT at a bandwidth of 1 Hz.

{
\subsection{Concluding remarks and outlook}
\label{sec:outlook-enabled}
In this section we have presented some examples for engineering of atomtronic devices. We have discussed recent advances in the development of basic components for atomtronics such as batteries, diodes, and transistors. However, atomtronics applications are expected to go beyond analogs of electronic circuits with 
atoms by making use of the specific quantum properties of ultra-cold atomic matter. In this context, we have shown that by taking advantage of quantum coherence and interactions between ultra-cold atoms, it is possible to design atromtronic SQUIDS, matter-wave interferometers, as well as rotation and magnetic sensors with extremely high accuracy and resolution. All these applications open the door to the future development of completely new types of quantum devices, which might be integrated into complex and large-scale atomtronic circuits. 
}

{\it Acknowledgements}
V.A. and J.M. would like to thank Gerard Pelegrí for fruitful discussions and 
acknowledge financial support from the Ministerio de Economía y Competitividad, MINECO, (FIS2017-86530-P), from the Generalitat de Catalunya (SGR2017-1646), and from the European Union Regional Development Fund within
the ERDF Operational Program of Catalunya (project
QUASICAT/QuantumCat).
V.A., G.B., and J.M. thank the DAAD for financial support through contracts DAAD 50024895 and DAAD 57059126. 
B.M.G. and A.S.A. would like to acknowledge support from the UK EPSRC grant EP/M013294/1.



%

\section{TWO LEVEL QUANTUM DYNAMICS IN RING-SHAPED CONDENSATES AND MACROSCOPIC QUANTUM COHERENCE}
\label{MQD}
\vspace*{-0.5cm}
\par\noindent\rule{\columnwidth}{0.4pt}
{\bf{\small{D. Aghamalyan, M. Boshier, R. Dumke, T. Haug, A. Minguzzi, L.-C. Kwek, L. Amico}}}
\par\noindent\rule{\columnwidth}{0.4pt}


A qubit is a two state quantum system that can be coherently manipulated, coupled to its neighbours, and measured.
Several qubit physical  implementations have been proposed in the last decade, all of them presenting specific virtues and bottlenecks at different levels~\cite{clarke2008superconducting,bloch2008quantum,saffman2010quantum,blatt2008entangled,vandersypen2001experimental,petta2005coherent}.
In neutral cold atoms proposals the qubit is encoded into well isolated internal atomic states.
This allows long coherence times, precise state readout and, in principle, scalable quantum registers.
However, individual qubit (atom) addressing is a delicate point~\cite{bakr2010probing, sherson2010single}.
Qubits based on Josephson junctions allow fast gate operations and make use of the precision reached by lithography techniques~\cite{lucero2012computing}.
The decoherence, however, is fast in these systems and it is experimentally challenging to reduce it.
For charge qubits the main problem arises from dephasing due to background charges in the substrate;
flux qubits are insensitive to the latter decoherence source, but are influenced by magnetic flux fluctuations due to impaired spins proximal to the device~\cite{clarke2008superconducting}.

Here we aim at combining the advantages  of cold atom and Josephson junction based implementations.
The basic idea is  to use  the  persistent currents flowing through ring shaped optical lattices~\cite{amico2005quantum,amico2014superfluid,aghamalyan2013effective,aghamalyan2015coherent,aghamalyan2016atomtronic,cominotti2015scaling,hallwood2006macroscopic,nunnekamp2008generation} to  realize a cold atom analogue of the superconducting flux qubit (see~\cite{madison2000vortex,shaeer2001observation,lin2011synthetic,amico2005quantum,hallwood2010robust,cooper2012robust} for the different schemes that can be applied to induce  persistent currents). A barrier potential painted along the ring gives rise to a weak link, acting as a source
of back-scattering for the propagating condensate,
thus creating an interference state with the forward scattered current.
This gives rise to an atomic condensate counterpart of the celebrated
rf-SQUID---a superconducting ring interrupted by a Josephson junction~\cite{clarke2008superconducting,tinkham2004introduction},
namely an Atomtronics Quantum Interference Device (AQUID).
Due to the promising combination of advantages characterizing Josephson junctions and cold atoms,
the AQUID is now object of intense investigation~\cite{eckel2014hysteresis,jendrzejewski2014resistive}.
The first experimental realizations have been done by means of a Bose-Einstein condensate
free to move along a toroidal potential, except through a small spatial region,
where a very focused blue-detuned laser creates weak links,
namely an effective potential constriction~\cite{ramanathan2011superflow,ryu2013experimental,wright2013driving}.
By adapting the logic applied in the context of solid state Josephson junctions ~\cite{mooij1999josephson,clarke2008superconducting} to a specific cold atoms setup, a cold atom version of the SQUID can be created.
On the theoretical side, it has been demonstrated that the two currents flowing
in the AQUID can, indeed, define an effective two-level system,
that is, the cold-atom analog of flux qubits~\cite{solenov2010macroscopic,amico2005quantum,amico2014superfluid,aghamalyan2015coherent,aghamalyan2016atomtronic}.
The system is assumed to be driven by an effective flux piercing the  ring lattice.
The potential constriction breaks the Galilean invariance and splits the qubit levels,
that otherwise would be perfectly degenerate at half-flux quantum. By a combination of analytic and numerical techniques, one can demonstrate that the system can sustain a two-level effective dynamics ~\cite{amico2005quantum,amico2014superfluid,aghamalyan2015coherent,aghamalyan2016atomtronic}. We also review a physical system consisting of a Bose-Einstein condensate confined to a ring shaped lattice potential interrupted by three weak links~\cite{aghamalyan2015coherent}. 
By employing path integral techniques, we explore the effective quantum dynamics of the system in a pure quantum phase dynamics regime. By a combination of analytic and numerical techniques, it was demonstrated that the system can sustain a two-level effective dynamics giving other realization of atomtronic qubit.

{After outlining theoretical framework which leads to obtaining a single qubit, we go further by showing how single qubit and two-qubit gates can be implemented by using an effective action approach~\cite{amico2014superfluid,aghamalyan_effective_2013}. In order to achieve two-qubit gates we allow a non-vanishing hopping term between the different rings.}

{We also review the experimental realisation of ring lattices with one and three weak links 
performed at Nanyang Technological University in the experimental group
of Rainer Dumke in Singapore.  Indeed using spatial light modulator, they have experimentally realized~\cite{amico2014superfluid} both single ring lattice with a weak link as well  scaled ring-lattice potentials that could host, in principle, n ~ 10 ring-qubits, arranged in a stack configuration, along the laser beam propagation axis. Trapping potential of a ring-shaped optical lattice with three week links a $\sim 20 \mu m$ diameter using a spatial light modulator has been reported in the Ref.~\cite{aghamalyan2016atomtronic}. }

The current and type of state inside these atomic qubits can be read out via time-of-flight measurements~\cite{haug2018readout}. When the ring is interfered with a reference condensate, a spiral pattern appears in the time-of-flight, which indicates the magnitude and direction of the current. For low resolution images, these spirals can be read out from the density-density correlation images.
Furthermore, the type of superposition state can be measured from the noise in the time-of-flight images~\cite{haug2018readout}.

First progress towards an experimental realization has been made recently. {In a recent experiment  interference of persistent currents of AQUIDs have
been demonstrated at the Los Alamos National Laboratory in
the group of Malcom Boshier~\cite{ryu2020quantum}.} By inducing a bias current in a rotating atomic ring interrupted by two weak links, the interference between the Josephson current with the current from the rotation creates a oscillation in the critical current with applied flux. This oscillation is measured experimentally in the transition from the DC to the AC Josephson effect. This experiment has been performed within a dilute Bose-Einstein condensate that is well described within a mean-field description and thus entanglement of currents, which is a key ingredient for the atomic qubit, has not been demonstrated. Nonetheless,  it is a major step towards the implementation of the atomic qubit.





\subsection{The Atomtronic quantum interference device:AQUID}
We start by considering analytical models fore the confined one-dimensional many-body systems and use them to demonstrate an emerging effective two level dynamics of the system. Let us start by considering  $N$ interacting bosons at zero temperature,
loaded into a 1D ring-shaped optical lattice of $M$ sites. The discrete rotational symmetry of the lattice ring is broken by the presence
of a localized potential in one lattice site(later we also consider case of three weak links), which gives rise to a weak link.
The relevant physics of the system is captured by the Bose-Hubbard Model. The Hamiltonian reads
\begin{equation}
\mathcal{H}_{\text{BH}}  =  \sum_{i=1}^{M} \left[
\frac{U}{2} \bm{n}_{i}(\bm{n}_{i}-1)+\Lambda_i\bm{n}_{i}
- {J_i} \left(\expU{-i2\pi\Omega/M} \bm{a}_{i{+}1}^{\dag} \bm{a}_{i} + \text{h.c.} \right)
\right]
\label{BH}
\end{equation}
where $\bm{a}_i \, (\bm{a}^{\dagger}_i)$ are bosonic annihilation (creation) operators on the $i$th site of a ring with length $M$
and $n_{i}= \bm{a}_{i}^{\dag}\bm{a}_{i}$ is the corresponding number operator.
Periodic boundaries are imposed, meaning that $\bm{a}_{M} \equiv \bm{a}_0$.
%
\begin{figure}[htbp]
	\centering
	\includegraphics[width=0.45\textwidth]{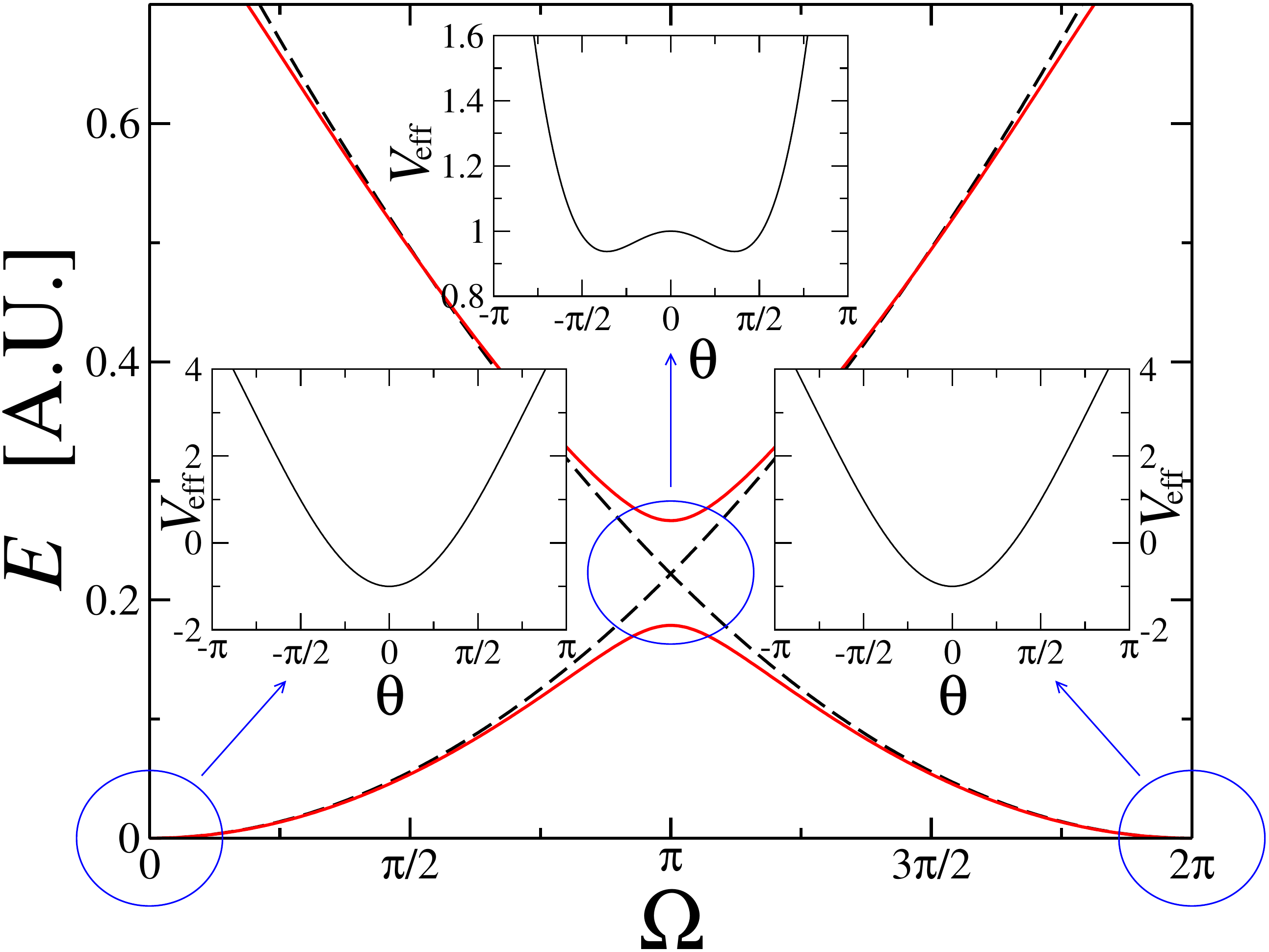}
	\caption{Main panel: sketch of the qubit energy splitting,
		due to the barrier $\Lambda$, for the two lowest-lying energy states
		in the many-body spectrum of model~\eqref{BH}.
		Black dashed lines denote the ground-state energy in absence of the barrier, as a function
		of the flux $\Omega$. Switching on the barrier opens a gap at the frustration
		point $\Omega = \pi$ (continuous red lines).
		The three insets show the qualitative form of the effective potential
		at $\Omega = 0, \, \pi, \, 2 \pi$.
		Note the characteristic double-well shape forming at $ \Omega=\pi$.
		The qubit, or effective two-level system, corresponds to the two lowest energy levels
		of this potential. In this figure the energies are plotted in arbitrary units. Reprinted with permission from D. Aghamalyan, M. Cominotti, M. Rizzi, D. Rossini, F. Hekking, A. Minguzzi, L.-C. Kwek, and L. Amico, New J. Phys. 17, 045023 (2015), under a Creative Commons Attribution 3.0 License. }
	\label{fig:Model_sketch}
\end{figure}
The parameter $U$ takes into account the finite scattering length for the atomic 
two-body collisions on the same site: $U=4 \pi \hbar^2 a_0\int dx |w(x)|^4/m$,  $w(x)$ being the Wannier functions of the lattice, $m$ the mass of atoms and $a_0$ the scattering length. To break the translational symmetry, there are two possible ways: Either, the  hopping parameters are all equal ${J_i=J}$ 
except in  one weak-link hopping $i_0$ where  ${J_{i_0}=J'}$. The other alternative, which we choose in this review, is to place a potential barrier at a single site ${\Lambda_i=\Lambda }$ and at all other sites the potential is set to zero, with $J_i=J,  \forall i$. The two options show qualitatively the same physics~\cite{aghamalyan2015coherent}.
The ring is pierced by an artificial (dimensionless) magnetic flux $\Omega$,
which can be experimentally induced for neutral atoms as a Coriolis flux by rotating the lattice
at constant velocity~\cite{ramanathan2011superflow,ryu2013experimental,wright2013driving}, 
or as a synthetic gauge flux by imparting a geometric phase
directly to the atoms via suitably designed laser fields~\cite{lin2011synthetic,dalibard2011colloquium}.
The presence of the flux $\Omega$ in Eq.(\ref{BH}) has been taken into account through 
the Peierls substitution: $J_i \rightarrow e^{-i2\pi\Omega/M} J_i$. 
The Hamiltonian~(Eq.\ref{BH-qubit}) is manifestly periodic in $\Omega$ with period $1$.
In the absence of the weak-link, the system is also rotationally invariant and therefore the particle-particle interaction energy does not depend on $\Omega$. 
The many-body ground-state energy, as a function of $\Omega$, is therefore given by a set of parabolas 
intersecting at the frustration points ${\Omega_{n} = (n + \frac{1}{2})}$\cite{caldeira1981influence,loss1992parity}. 
The presence of the weak-link breaks the axial rotational symmetry and couples different angular momenta states, thus lifting the degeneracy at $\Omega_{n}$. This feature sets the qubit operating point\cite{amico2014superfluid,aghamalyan2015coherent}.

It is worth noting that  the interaction $U$ and the weak-link strength induce competing physical effects: the weak-link sets an  healing  length in the density as a
further spatial scale; the interaction  tends to smooth out the healing length effect. As a result, strong interactions tends to renormalize the weak link
energy scale\cite{aghamalyan2015coherent,cominotti2014optimal,cominotti2015scaling}.

In the limit of a large number of bosons in each well $\bar{n}=N/M$, $a_i\sim \sqrt{ \bar{n}}e^{i\phi_i} $,  and the Bose-Hubbard hamiltonian (BHH) (\ref{BH}) can be mapped to the Quantum-Phase model
employed to describe Josephson junction arrays\cite{amico2000time,Fazio2001quantum}:
%
%
\begin{equation}
\mathcal{H} \ \ = \ \  \sum_{i=1}^{M}
\left[
\frac{U}{2} \bm{n}_{i}^2
- J_i  \cos\left(\bm{\phi}_{i{+}1}{-}\bm{\phi}_{i} - {\Omega}\right)
\right]
\label{H_qp}
\end{equation}
where  $\left [\bm{n}_i , \bm{\phi}_l\right ]=i \hbar\delta_{il}$ are canonically conjugate number-phase variables and $J_i\sim \bar{n} t_i$ are the Josephson tunneling amplitudes.

\paragraph{The rf-AQUID qubit.} In this case, a single weak link occurs along the ring lattice $t''=t$.
The presence of the weak link induces a slow/fast  separation of the effective (imaginary time) dynamics: the dynamical variables relative to the weak link are slow compared
to the 'bulk' ones, playing the role of an effective bath (nonetheless, we assume that the ring system is perfectly isolated from the environment). Applying the harmonic approximation to the fast dynamics and integrating it out,  the  effective dynamics
of the AQUID  is governed by(See for detailed derivation appendix material of~\cite{amico2014superfluid})

\begin{equation}
\mathcal{H}_{\text{eff}} \ \ =
\mathcal{H}_{\text{syst}}+
\mathcal{H}_{\text{bath}} +  \mathcal{H}_{\text{syst-bath}}
\label{H_eff}
\end{equation}
The slow dynamics is controlled by
\begin{equation}
\mathcal{H}_{\text{syst}}=U  \bm{n}^2
+ E_L \bm{\varphi}^2
- E_J \ \cos(\bm{\theta}-\Omega)
\label{jj}
\end{equation}
where $ \theta$ is the phase slip across the weak link,  with  $E_L=J /M$, and $E_J=J^{\prime}$.
For $\delta \doteq {E_J}/{E_L} \ge 1 $, $\mathcal{H}_{\text{syst}}$ describes a particle in a double well potential with the two-minima-well(See Fig.~(\ref{fig:Model_sketch})) separated from the other
features of the potential.  The two parameters, U and $t'/t$, allow control of the two level system. The  two local minima of the double well are degenerate for  ${\Omega=\pi}$.
The minima correspond to the clock-wise and anti-clockwise currents in the AQUID.The presence of a finite barrier, $\Lambda>0$, breaks the axial rotational symmetry
and couples different angular momenta,
thus lifting the degeneracy at the frustration points by an amount $\Delta E$, see Fig.~\ref{fig:Model_sketch}.
Provided other excitations are energetically far enough from the two competing ground-states,
this will identify the two-level system defining the desired qubit and its working point.
Because of the quantum tunneling between the two minima of the double well, the two states of the system (qubit)
are formed by symmetric and antisymmetric combinations of the two circulating current states.

The WKB level splitting is(see for detailed derivation Appendix C.3 \cite{aghamalyan2015atomtronics})
\begin{equation}
\Delta \simeq \frac{2 \sqrt{UE_J}}{\pi}\sqrt{ (1-\frac{1}{\delta})} e^{-12\sqrt{ E_J/U} (1-1/\delta)^{3/2} }\;.
\label{gap}
\end{equation}
From this formula we can see that the limit of a weak barrier and intermediate to strong interactions form the most favourable regime to obtain a finite gap between the two energy levels of the double level potential as depicted on Fig.~(\ref{fig:Model_sketch}).
Incidentally, we comment that the bath  Hamiltonian in  Eqs.(\ref{H_eff}), (\ref{jj}), is similar to the one
describing the dissipative dynamics of a  single Josephson junction in the framework of the Caldeira-Leggett model\cite{caldeira1981influence}. As long as the ring has finite size, however,
there are a finite number of discrete modes and no real dissipation occurs~\cite{rastelli2013quantum}.
In  the limit $N \rightarrow \infty$, a proper Caldeira-Leggett model is recovered. In agreement to the arguments reported above, the qubit dynamics encoded in the AQUID is less and less addressable by increasing the size of the ring~\cite{amico2014superfluid,aghamalyan2015coherent}.
\paragraph{Atomtronic flux-qubit:~Ring lattice interrupted with three weak links.} Here we consider  $N$ Bosons in an $M$~site ring described by the Bose-Hubbard Model. The Hamiltonian reads
\begin{equation}
\mathcal{H}_{\text{BHH}} \ \ = \ \ \sum_{i=1}^{M} \left[
\frac{U}{2} \bm{n}_{i}(\bm{n}_{i}-1)
- {t_i} \left(e^{i\Omega} \bm{a}_{i{+}1}^{\dag} \bm{a}_{i} + \text{h.c.} \right)
\right]~.
\label{BH-qubit}
\end{equation}
where $\bm{a}_i \, (\bm{a}^{\dagger}_i)$ are bosonic annihilation (creation) operators on the $i$th site
and $n_{i}= \bm{a}_{i}^{\dag}\bm{a}_{i}$ is the corresponding number operator.
Periodic boundaries are imposed, meaning that $\bm{a}_{M+1} \equiv \bm{a}_1$.
\begin{figure}[htbp]
\centering
\includegraphics[width=0.45\textwidth]{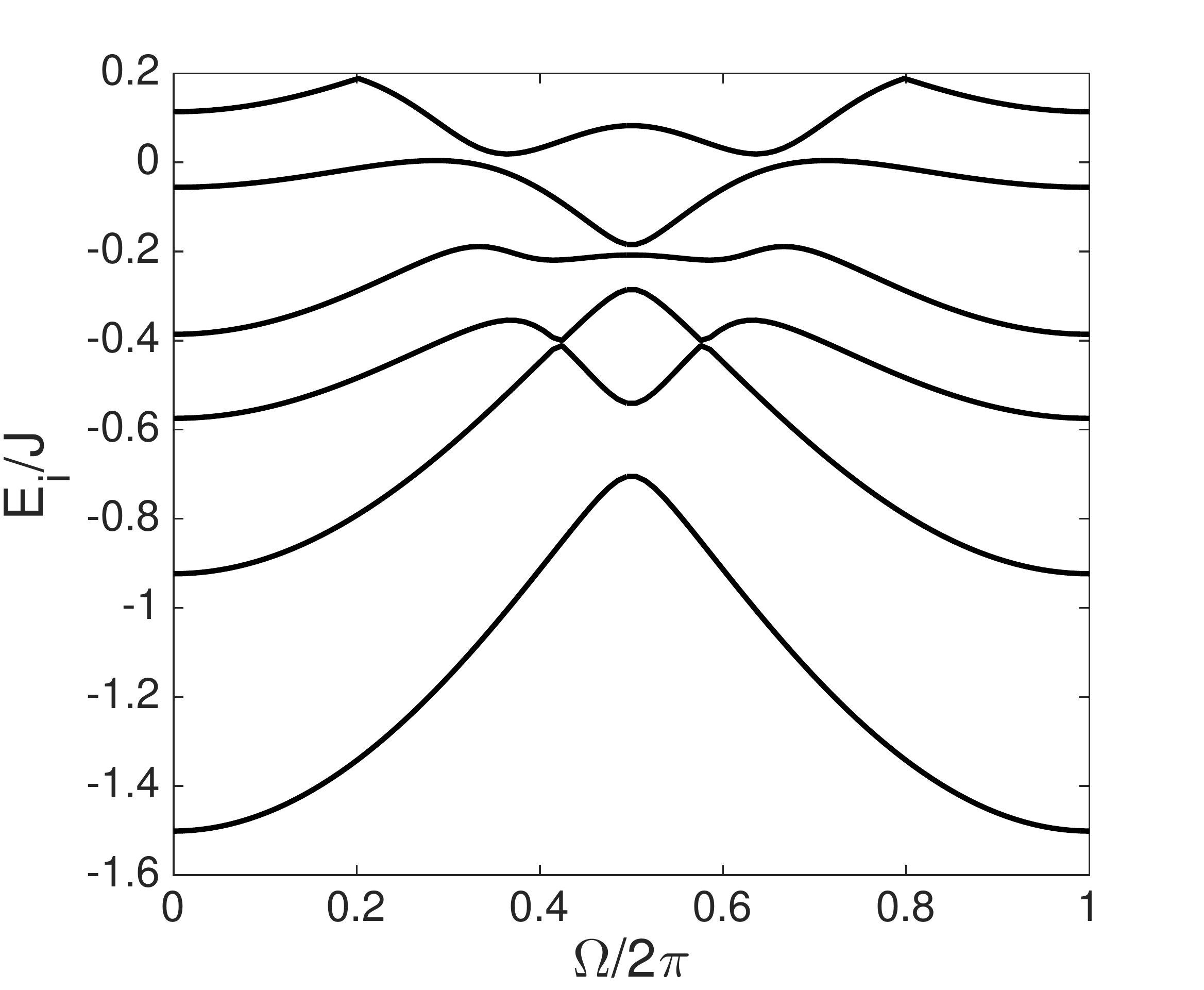}

\caption{Six first energy levels of the reduced system given by the effective potential Eq. (\ref{potential})   as a function of the dimensionless external flux $\Omega$. Here ${{J'}}=0.7J$, ${{J''}}=0.8J$, $U=0.5J$, and $\theta_1=-\theta_2$. Reprinted with permission from D.  Aghamalyan, N. Nguyen, F. Auksztol, K. Gan, M. M. Valado, P. Condylis, L.-C. Kwek, R. Dumke, and L. Amico, New J. Phys. 18, 075013 (2016), under a Creative Commons Attribution 3.0 License.}
\label{SpectrumQF}
\end{figure}
The parameter $U$ takes into account the finite scattering length for the atomic
two-body collisions on the same site. The  hopping parameters are constant $t_j=t$
except in the three weak-links lattice sites $i_0,i_1,i_2$ where they are  $t_{i_0}=t^{\prime}, t_{i_1}=t_{i_2}=t''$.
The ring is pierced by an artificial (dimensionless) magnetic flux $\Omega$,
which can be experimentally induced for neutral atoms as a Coriolis flux by rotating the lattice
at constant velocity~\cite{fetter2009rotating,wright2013driving},
or as a synthetic gauge flux by imparting a geometric phase
directly to the atoms via suitably designed laser fields~\cite{berry2002imprinting,lin2009synthetic,dalibard2011colloquium}.
The presence of the flux $\Omega$ in (Eq.\ref{BH-qubit}) has been taken into account through
the Peierls substitution: $t_i \rightarrow e^{-i\Omega} t_i$.
{The Hamiltonian~(Eq.\ref{BH-qubit}) is manifestly periodic in $\Omega$ with period $2\pi$; in addition  it enjoys  the symmetry $\Omega \leftrightarrow -\Omega$. The presence of the weak-link breaks the axial rotational symmetry and couples different angular momenta states, thus lifting the degeneracy at $\Omega_{n}$. This feature sets the qubit operating point\cite{amico2014superfluid,aghamalyan2015coherent}.}

%
%
%
Here again in the limit of a large number of bosons in each well $\bar{n}=N/M$, $a_i\sim \sqrt{ \bar{n}}e^{i\phi_i} $,  and the Bose-Hubbard hamiltonian (BHH) (Eq.\ref{BH-qubit}) can be mapped to the Quantum-Phase model and equivalent Hamiltonian is given by Eq.~(\ref{H_qp}).

%
The effective action for the quantum phase model reads (See for details~\cite{aghamalyan2016atomtronic})
\begin{eqnarray}
\hspace*{-2cm}S_{eff}&=&\sum_{\alpha=0,1,2} \int_0^\beta d \tau \left [ {\frac {1}{ U}}  \dot{\theta_\alpha}^2 +  V(\theta_\alpha) \right ]       \\
&-& \frac{J}{ U} \int d \tau d \tau' \theta_\alpha(\tau)G_\alpha(\tau-\tau') \theta_\alpha(\tau')
\label{effective_single}
\end{eqnarray}
where
\begin{eqnarray}
\hspace*{-1cm}V(\theta_\alpha)& \doteq&  J    c_\alpha \theta_\alpha^2 -{\frac{J'}{3}} \cos (\theta_1-\theta_2-\Omega)    \\  \nonumber
&-&{\frac{J''}{3}}(\cos\theta_1+\cos\theta_2)
\label{potential}
\end{eqnarray}
with  $\displaystyle{c_\alpha=  {\frac{1}{2}} \left ( {\frac{1}{2}}-U J \sum_{k=1}^{\frac{M-4}{2}} {\frac{{\zeta_\alpha}_k^2}{\omega_k^2}} \right )}$. Where $\theta_\alpha\doteq  \phi_{i_\alpha+1}-\phi_{i_\alpha} $.  We assume that the weak links are sufficiently spaced   to 
make the nearest neighbour phase differences in between them (fast variables) small. This implies that substantial phase  slips occur at the weak links with the constraint $\theta_0+\theta_1-\theta_2=0 \; mod(2\pi)$. The interaction between the fast and the slow modes is described by the kernel
\begin{equation}
\hspace*{-1cm} G_\alpha(\tau)=\sum_{l=0}^\infty  \sum_{k=1}^{\frac{M-4}{2}} {\frac{\omega_l^2 {\zeta_\alpha}_k^2}{\omega_k^2+\omega_l^2}} e^{i\omega_l \tau}\; .
\end{equation}

We observe that $V(\theta_\alpha) $ defines the  effective dynamics of the superconducting Josephson junctions flux qubits\cite{mooij1999josephson, chiodi2010microwave}, but perturbed by the $\theta^2$ terms; by numerical inspection, we see that  the corresponding coefficients are small in units of J, and decreases by increasing $M$. Moreover, on Fig.\ref{SpectrumQF} we introduce the numerical result for the spectrum of the quantum particle which moves in the potential given by Eq. (\ref{potential}) under the additional assumption that $\theta^2$ terms do not contribute. From this figure we clearly see that near the frustration point $\Omega=\pi$ two lowest energy levels are well separated from each other and from higher excitations, which means that effective dynamics of the system defines a qubit.
It is important to point out that quantum phase model is applicable in the limit of the high filling, however the results for the effective-two level description were demonstrated to hold in the limit of low filling by applying an exact diagonalization method for the Bose-Hubburd model as it has been demonstrated in Refs.~\cite{aghamalyan2015coherent,aghamalyan2016atomtronic}.

\subsection{ Demonstration of the one qubit and two qubit unitary gates}
The aim of this section is to show how the effective phase dynamics of optical ring-lattices with impurities serves to the  construction of  one- and two-qubit gates - a necessity for a universal quantum computation. Here, we adapt results which were obtained by Solenov and Mozyrsky~\cite{Mozyrsky2011cold} for the case of homogeneous rings with impurities. It results,  that a single ring optical lattice with an impurity is described by the following effective Lagrangian (see the Eq. (\ref{jj})):
\begin{equation}
L={\frac {1}{2 U}}  \dot{\theta}^2+\frac{J}{N-1} (\theta-\Phi)^2 -J' \cos \theta
\end{equation}
Then we introduce the canonical momentum P in a usual way:
\begin{equation}
P=\frac{\partial L}{\partial {\dot{\theta}}}=\frac {1}{ U}\dot{\theta}
\end{equation}
After performing a Legendre transformation we get the following Hamiltonian:
\begin{equation}
H=J' \left[ \frac {P^2}{2 \mu}-\frac{J}{J'(N-1)} (\theta-\Phi)^2 + \cos \theta  \right]\;,
\label{H_double}
\end{equation}
where $\mu=J'/U$ is an effective mass of the collective particle.
The quantization is performed by the usual transformation $P\rightarrow -d/d\theta$.
For $\delta=\frac{J'(N-1)}{2J}>1$ the effective potential in Eq. (\ref{H_double}) can be reduced to a double well; for $\Phi=\pi$, the two lowest  levels of such double well are symmetric and antisymmetric superpositions of the states in the left and right wells respectively. The effective Hamiltonian  can be written as:
\begin{equation}
H\simeq\varepsilon\sigma_z
\end{equation}
and the lowest two states are $|\psi_g\rangle=(0,1)^{T}$ and $|\psi_e\rangle=(1,0)^{T}.$ WKB estimate for the energy splitting $\epsilon$ of the qubit is given by the Eq.(\ref{gap}).

\subsubsection{ Single qubit gates}
For the realization of single-qubit rotations, we consider the system close to the symmetric double well configuration $\Phi\simeq\pi$. In the basis of the two-level system discussed before, the Hamiltonian takes the form:
\begin{equation}
H\simeq\varepsilon\sigma_z+\frac{\Phi-\pi}{\delta}\langle \theta \rangle_{01}\sigma_x,
\label{sigma}
\end{equation}
where $\langle \theta \rangle_{01}$ is the off-diagonal element of the phase-slip in the two-level system basis. It is easy to show that spin flip, Hadamard and phase gates can be realized by this Hamiltonian. For example, a phase gate can be realized by evolving the  state through the  unitary transformation $U_z(\beta)$ (tuning the second term of Eq.({\ref{sigma}) to zero by adjusting the imprinted flux)
\begin{equation}
U_z(\beta)=exp(i\varepsilon \tau\sigma_z)=\begin{pmatrix}e^{i \varepsilon \tau} & 0\\
0 & e^{-i \varepsilon \tau}
\end{pmatrix} \; .
\end{equation}
After tuning the gap energy close to zero (adjusting the barrier height of the impurity), we can realize the following rotation
\begin{equation}
U_x(\beta)=exp(i\alpha \tau \sigma_x)=\begin{pmatrix}\cos{\alpha} & i\sin{\alpha}\\
i\sin{\alpha} & \cos{\alpha}
\end{pmatrix}
\end{equation}
where $\alpha=\frac{\Phi-\pi}{\delta}\langle \theta \rangle_{01}\tau$. When  $\alpha=\pi/2$  and   $\alpha=\pi/4$ the NOT and  Hadamard gates are  respectively realized. \newline

\subsubsection{Two-qubit coupling and gates}
The effective dynamics for two coupled qubits, each realized as single ring with localized impurity (as in Fig.\ref{exp-onering}), is governed by the Lagrangian
\begin{eqnarray}
L&=&  \sum_{\alpha=a,b} {\frac {1}{2 U}}\dot{\theta_\alpha}^2  +\left [ \frac{J}{2(N-1)} (\theta_\alpha-\Phi_\alpha)^2 -J' \cos(\theta_\alpha) \right ] \nonumber  \\
&-& \tilde{J''}  \cos[\theta_a-\theta_b-{\frac{N-2}{N}}(\Phi_a-\Phi_b)]
\end{eqnarray}
Where $J''$ is the Josephson tunnelling energy between the two rings. When $\Phi_a=\Phi_b=\Phi$ and $J''\ll J'$ the last term reduces to $-J''\frac{(\theta_a-\theta_b)^2}{2}$  and the Lagrangian takes the form
\begin{eqnarray}
L&= & J'  [\sum_{\alpha=a,b} {\frac {1}{2 J' U}}\dot{\theta_\alpha}^2 +  [ \frac{J}{2J'(N-1)} (\theta_\alpha-\Phi_\alpha)^2 - \cos(\theta_\alpha)  ] \nonumber  \\
&+&\frac{J''}{J'}\frac{(\theta_a-\theta_b)^2}{2}] \; .
\end{eqnarray}
By applying the same procedure as in the previous section, we obtain the following Hamiltonian in the eigen-basis of the two-level systems of rings $a$ and $b$
\begin{eqnarray}
H&=&H_a+H_b+\frac{J''}{J'}\sigma_x^{1}\sigma_x^{2}\langle \theta \rangle_{01}^2  \;  ,\\
H_\alpha&=&\epsilon\sigma_z^{\alpha}+(\frac{\Phi-\pi}{\delta}+\frac{J''\pi}{J'})\langle \theta \rangle_{01}\sigma_x^{\alpha} \; .
\end{eqnarray}
From this equations it follows that qubit-qubit interactions can be realized using our set-up. If we choose the tuning $\varepsilon\rightarrow0$ and $\Phi \rightarrow \pi-\frac{\delta J''\pi}{J'}$ the natural representation of a $(SWAP)^{\alpha}$ gate\cite{fan2005optimal} can be obtained:
\begin{equation}
U(\tau)=exp[-i\frac{J''}{J'}\sigma_x^{1}\sigma_x^{2}\tau],
\end{equation}
where $\alpha=\frac{\tau J''}{ J'}$. A CNOT gate can be realized by using two $\sqrt{SWAP}$ gates~\cite{fan2005optimal}. It is well known that one qubit rotations and a CNOT gate are sufficient to implement a set of universal quantum gates~\cite{loss1998quantum}.
}

\subsection{Readout of atomtronic qubits}
For a controlled quantum system, it is essential to be able to read out the state of a prepared quantum state. For the atomtronic qubit, one can determine the properties of the qubit by reading out the current of the atoms.
The existence of the atomic current flowing  in AQUID can be detected by standard time-of-flight measurement of the ring condensate~\cite{aghamalyan2015coherent}. 
A more in-depth analysis can be performed by interfering the ring  condensate with a second condensate confined  in the center of a ring. This condensate sets a phase reference for the phase winding of the ring condensate. By {\it in-situ} measurement of the two interfering condensates the self-heterodyne detection of the phase of the wave function is realized. For weakly interacting continuous ring systems, where no entanglement is present, both  the orientation and the intensity of the current states have been detected~\cite{eckel2014interferometric,corman2014quench,mathew2015self, roscilde2016quantum, aidelsburger2017relaxation}.

For atomtronic qubits, this detection scheme has to be applied to the case of ring lattices with stronger interactions. This has been studied in \cite{haug2018readout} and the key results are reviewed below. 

\subsubsection{Interferometric detection of the current states}
To read-out the direction and the intensity of the current in the ring lattice, an approach originally carried out by the Maryland and Paris groups to map-out the circulating states in continuous ring-shaped condensates can be applied~\cite{moulder2012quantized,wright2013driving,eckel2014hysteresis,eckel2014interferometric,corman2014quench}.  Accordingly, the ring condensate is made to  interfere with a another condensate at rest, located at the center, fixing the  reference for the phase of the wavefunction. 
The combined wavefunction evolves in time, interferes with itself and finally is measured.  The number of spirals gives the total number of rotation quanta. 

In the actual experiment,  the condensate is imaged through in-situ measurements. In this way, the current direction and magnitude is well visible as a spiral pattern.
The position of the spirals depends on the relative phase between ring and the central condensate. 
In a single experimental run, the spirals will be visible for a condensate with high number of particles. 
However, if the number of particles is low or the atom imaging is inefficient, one has to average over multiple shots and take expectation values, which experimentally corresponds to take averaged results over  many experimental runs.
However, every realization of the experiment has a random phase in the phase of the spirals, which is averaged out over many repetitions.
As the relative phase between ring and central condensate is determined  randomly upon measurement, the expectation value of the density operator will average over different realizations of the spiral interference pattern, washing out the information on the current configuration structure. 
However, as we show below, the information about the spirals can be recovered using density-density correlations.

The expansion dynamics is modeled with the Bose-Hubbard model.
The ring wavefunction is calculated by solving the ground state of the Bose-Hubbard Hamiltonian, while the central condensate is simply a single decoupled site with $N_\text{c}$ particles. 
The dynamics of the density $\op{n}(\vc{r},t)=\op{\psi}^\dagger (\vc{r},t)\op{\psi}(\vc{r},t)$ is initialized assuming that the bosonic field operator of the system is ${\op{\psi}(\vc{r})=\sum_n w_n(\vc{r})\an{a}{n}}$, where  $w_n(\vc{r})$ are a set of Wannier functions forming a complete basis\cite{altman2004probing,gerbier2008expansion}. In our calculation, we approximate the full basis for wave functions living in the ambient space on which the condensate expands with the set of  Wannier functions  composed of Gaussians peaked at the ring lattice sites and at its  centre  (the Gaussian approximation for the Wannier functions is a well verified approximation for single site wavefunctions -- see \cite{slater1952soluble,chiofalo2000collective}). For the free evolution (we are indeed in a dilute limit) we  assume that each particle at site $n$ expands in two dimensions as  
\begin{equation}
w_n(\vc{r},t) = \frac{1}{\sqrt{\pi}}\frac{\sigma_n}{\sigma_n^2+\frac{i\hbar t}{m}}\expU{-\frac{(\vc{r}-\vc{r}_n)^2}{2\left(\sigma_n^2+\frac{i\hbar t}{m}\right)}} \;,\label{Gaussian}
\end{equation}
where $\sigma_n$ is the width of the condensate located at the $n$-th site. The dynamics of the condensates is then approximated as 
$\op{\psi}(\vc{r},t)=\sum_n w_n(\vc{r},t)\an{a}{n}$.
We observe that such approximation works well in the situations in which the optical lattice is assumed to be sufficiently dense in the space in which the condensate is released (as in the release from large three dimensional optical lattices). 

To observe the intereference, and thus the qubit properties, using averaging over multiple shots,  the interference pattern is measured with higher order density-density correlations. 
We calculate the density-density covariance\cite{castin1997relative,mullin2006origin,folling2005spatial,kang2014revealing}
\begin{equation}
\text{cov}(\vc{r},\vc{r'},t)=\avg{\op{n}(\vc{r},t)\op{n}(\vc{r'},t)}-\avg{\op{n}(\vc{r},t)}\avg{\op{n}(\vc{r'},t)}\;.
\end{equation}
We also define the root of the density covariance which has the same unit as the density to improve the contrast of the measured interference pattern
\begin{equation}
\sigma(\vc{r},\vc{r'},t)=\text{sgn}(\text{cov}(\vc{r},\vc{r'},t))\sqrt{\abs{\text{cov}(\vc{r},\vc{r'},t)}}\;.
\end{equation}

\begin{figure}[htbp]
\centering
\includegraphics[width=0.45\textwidth]{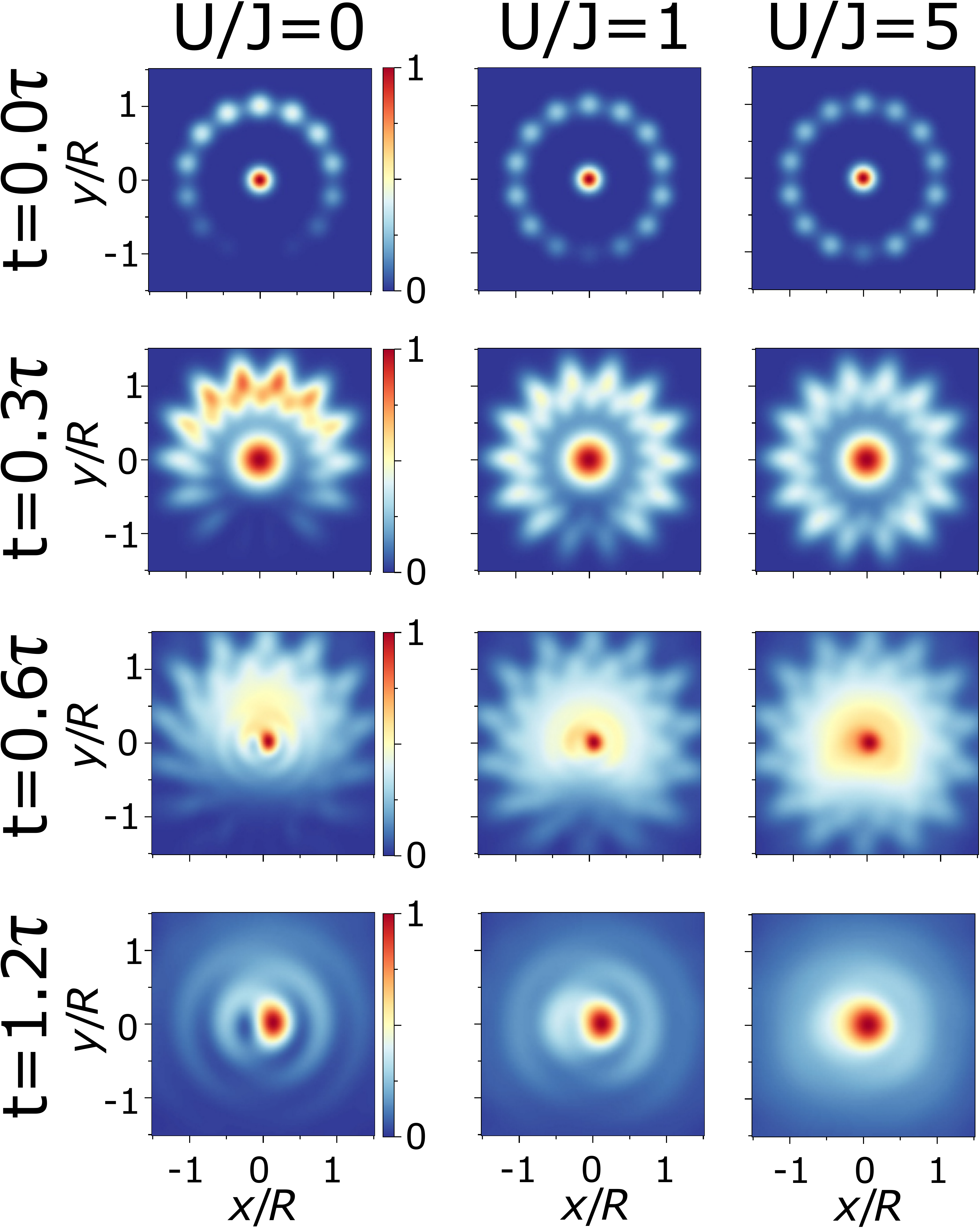}
\caption{Density of expanding atoms at times ${t=0,\,0.3\tau,\,0.6\tau,\,1.2\tau}$, with ${\tau=mR\sigma_r/\hbar}$. From left to right: ${U=0}$, ${U/J=1}$, ${U/J=5}$. Flux ${\Omega=\frac{1}{2}}$ at the degeneracy point.  At intermediate time, we observe some spiral-like structure at the edges. This is not the interference with the central condensate, but a residue of the ring lattice interfering with itself. Calculated using Bose-Hubbard model, no interaction during expansion. Data in color and normalized to one. Ring has 7 particles, ${M=14}$ ring sites, ring radius $R$. Width of central and ring cloud is ${\sigma_r=2 R/L}$ and potential barrier ${\Lambda=J}$, 25\% of atoms in central condensate. Barrier at ${x=0}$, ${y=-R}$. Reprinted with permission from T. Haug, J. Tan, M. Theng, R. Dumke, L.-C. Kwek, and L. Amico, Phys. Rev. A 97, 013633 (2018). Copyright 2018 American Physical Society. }
\label{densUtime}
\end{figure}

First, we plot the expectation value of the density of expanded atoms for different values of interaction at the degeneracy point $\Omega=1/2$ in Fig.\ref{densUtime}. 
The density of expanded atoms at longer times has some characteristic features depending on the interaction. For interaction energy smaller than the potential barrier, the center shows a characteristic bright and dark spot. For stronger interaction, it becomes a single, blurred spot. At the degeneracy point we observe a superposition of counter-flowing current states. Interaction modifies the many-body entanglement, which changes the characteristic time-of-flight pattern. After a long enough free expansion, the atom density assumes the initial momentum distribution. 
However, it is difficult to read out the exact state of the  current as the characteristic spirals are not visible in the expectation values of the density.

\begin{figure}[htbp]
\centering
\includegraphics[width=0.45\textwidth]{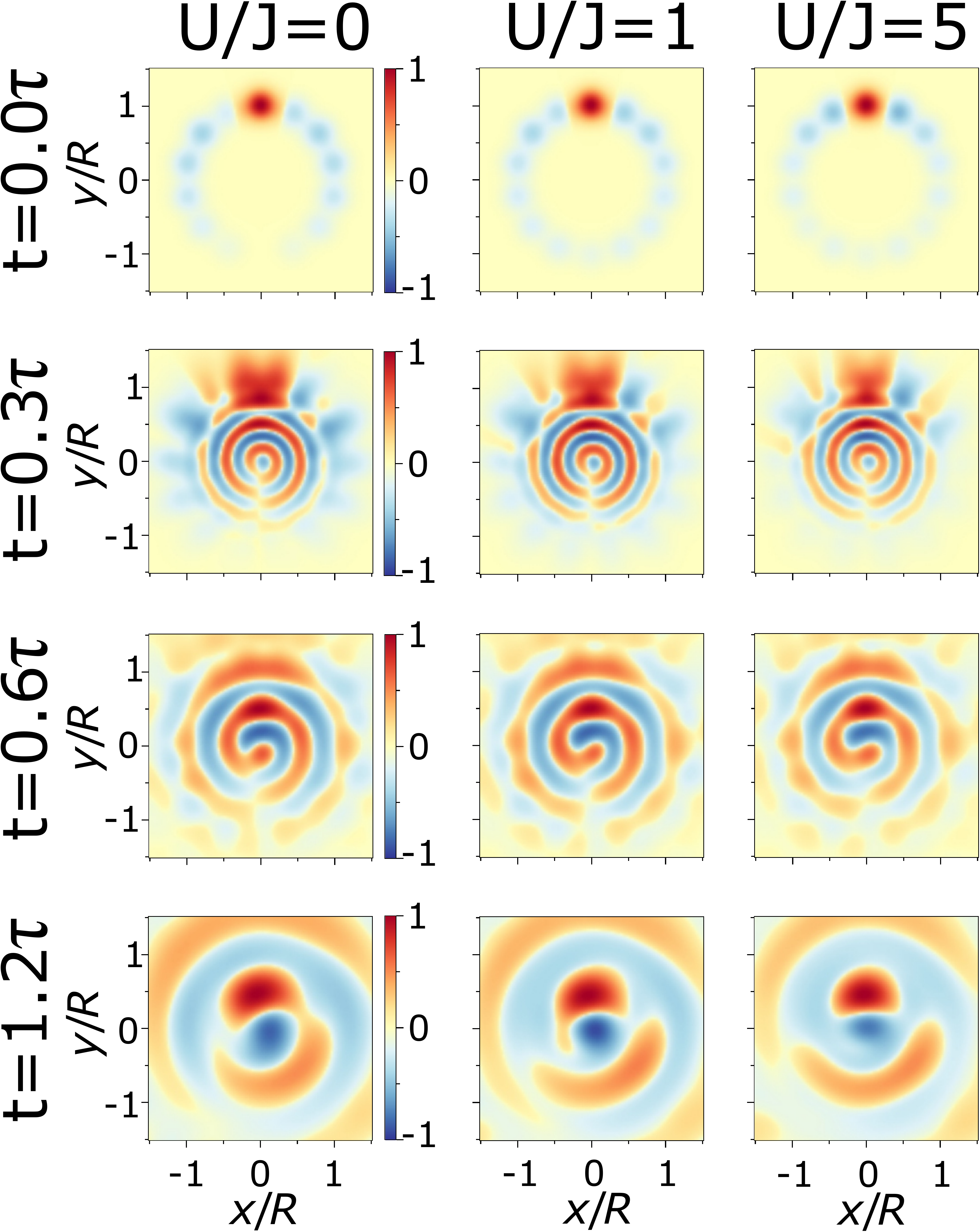}
\caption{Root of density-density covariance ${\sigma(\vc{r},{\vc{r'}=\{0,R/2\}})}$ of expanding atoms with flux ${\Omega=\frac{1}{2}}$ at the degeneracy point. The discontinuity in the bottom of the spirals at intermediate times $t=0.3\tau$ and $t=0.6\tau$ shows that the ring condensate is in a superposition of zero and one rotation quantum. Same parameters as Fig.~\ref{densUtime}. Reprinted with permission from T. Haug, J. Tan, M. Theng, R. Dumke, L.-C. Kwek, and L. Amico, Phys. Rev. A 97, 013633 (2018). Copyright 2018 American Physical Society.}
\label{corrUtime}
\end{figure}

Next, we show the  density-density covariance $\sigma(\vc{r},\vc{r'})$ in Fig.\ref{corrUtime}. A clear spiral pattern emerges here. In this case,  a step in the spirals at the weak link  site (here at the center bottom) is clearly visible for intermediate times. This indicates the appearance of a superposition of two winding numbers. 
Although the interferometric pictures can look similar, different interactions lead to current states that may be very different in nature. For ${U=0}$, the current is in a non-entangled superposition state, whereas for interaction ${U=J}$ in a highly entangled NOON state. 

\begin{figure}[htbp]
\centering
\subfigimg[width=0.23\textwidth]{a}{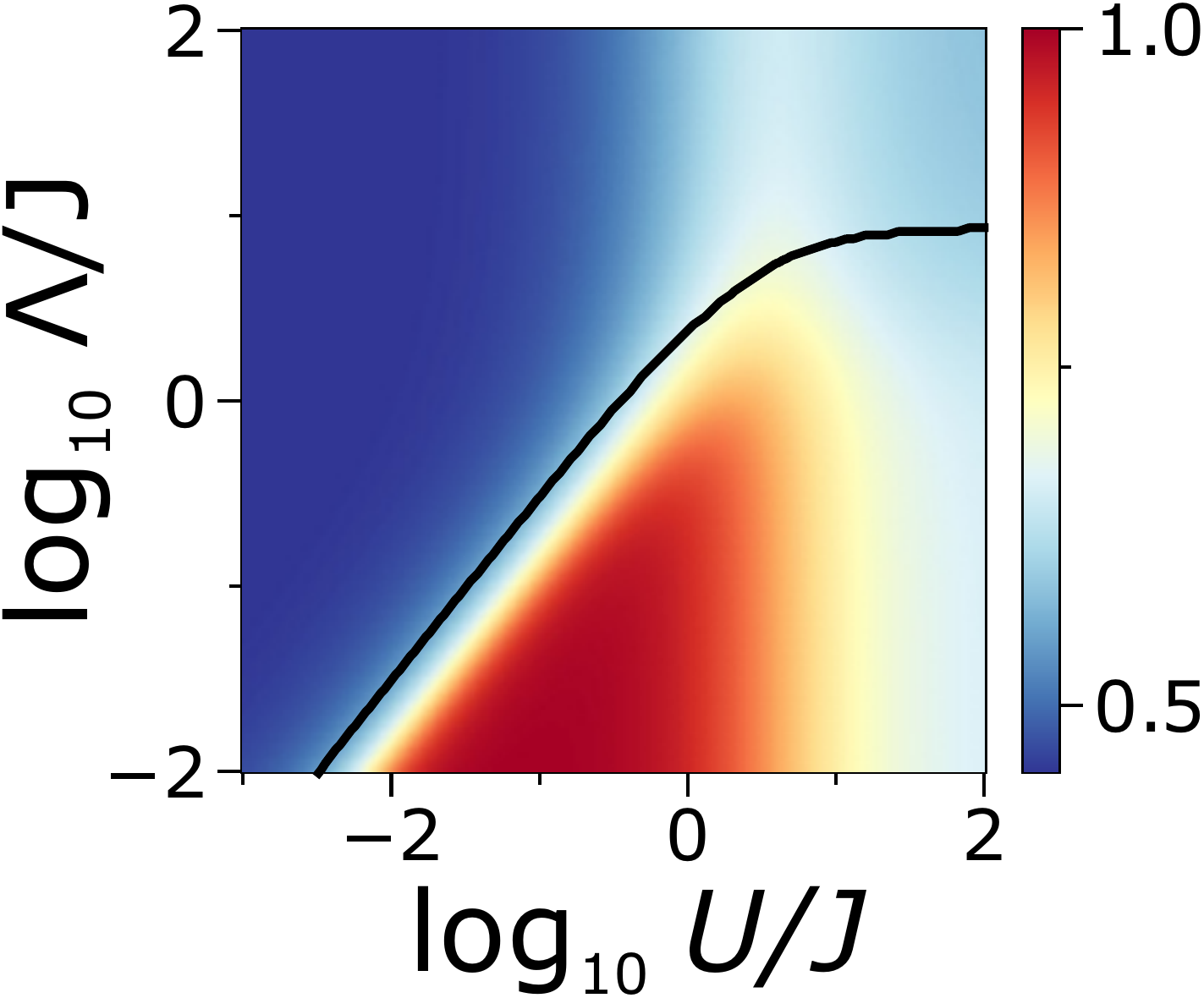}
\subfigimg[width=0.23\textwidth]{b}{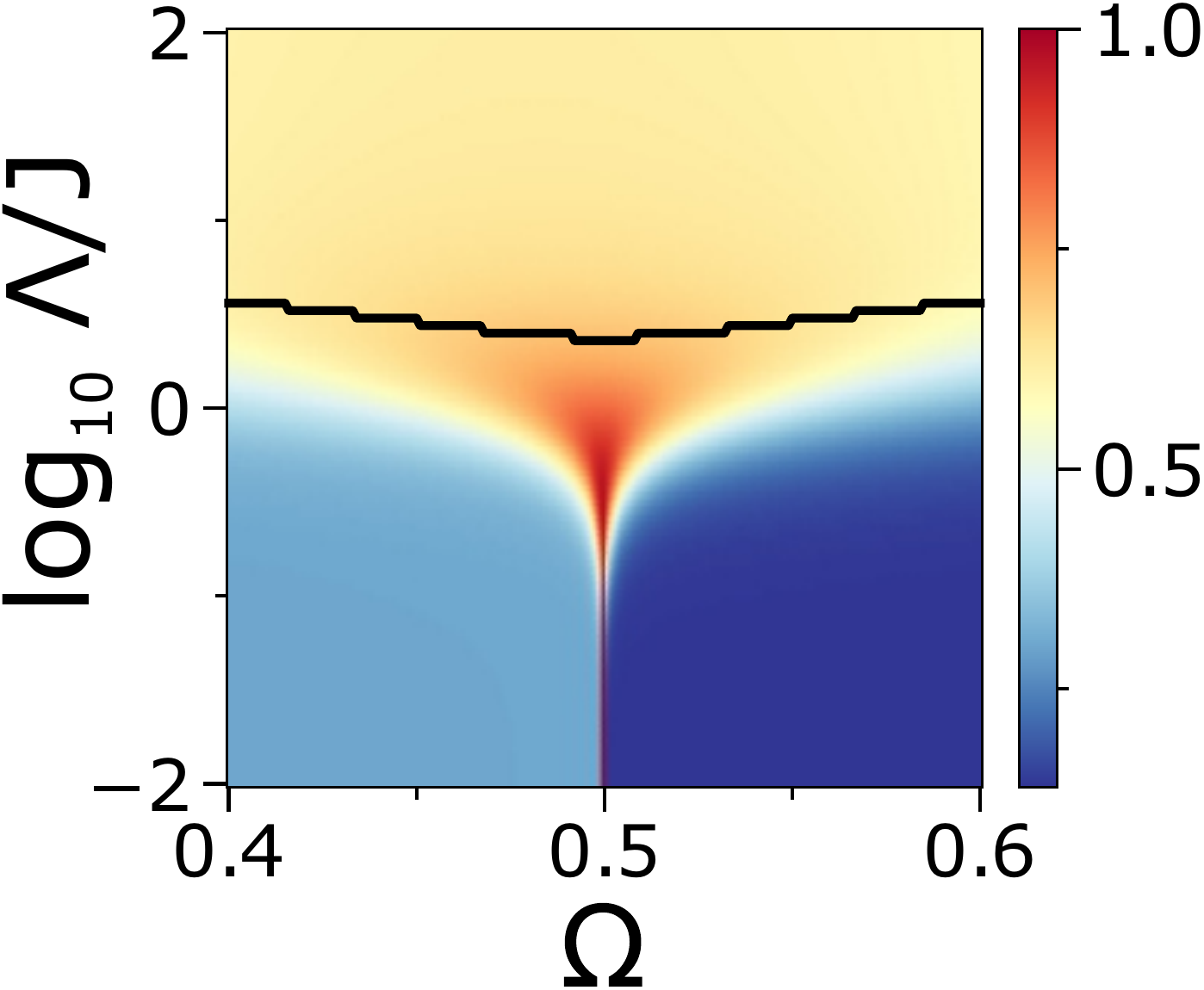}
\caption{Momentum noise $\sigma_k({\vc{k}=0})$ (in color, normalized to one) plotted for potential barrier $\Lambda$ against \idg{a} on-site interaction $U$ (${\Omega=\frac{1}{2}}$) and \idg{b} flux $\Omega$ (${U/J=1}$). Momentum noise is extracted from time-of-flight image after long expansion. Only ring is expanded, without central condensate. Black line shows the critical point where depletion at the potential barrier is 1\% of the average particle number per site. Above the line the potential barrier site is depleted. Other parameters are ${M=11}$ ring sites and 5 particles. Reprinted with permission from T. Haug, J. Tan, M. Theng, R. Dumke, L.-C. Kwek, and L. Amico, Phys. Rev. A 97, 013633 (2018). Copyright 2018 American Physical Society.}
\label{CorrUvsp}
\end{figure}

Below, we shall see how additional  information on the states can be grasped analysing the noise  in the momentum distribution of the ring condensate. Indeed, the  noise for zero momentum depends strongly on the specific entanglement between the clockwise and anti-clockwise flows. 
In the case of an entangled cat state all atoms have together either zero or one momentum quanta. A projective measurement will collapse the wavefunction to either all atoms in the zero or one momentum state. Averaging over many repeated measurements will result in erratic statics of the measurements.
In contrast, in non-entangled single-particle superpositions, each particle has independently either zero or  one momenta quanta. A single projective measurement will result in on average half the atoms having zero and half the atoms having one rotation quantum. Therefore,  fluctuations averaged over many measurements will be low.
We define the noise of the momentum distribution
\begin{equation}
\sigma_k(\vc{k})=\sqrt{\avg{\op{n}(\vc{k})\op{n}(\vc{k})}-\avg{\op{n}(\vc{k})}\avg{\op{n}(\vc{k})}}\;.
\end{equation}
Having in mind a time-of-flight experiment, 
the optimal point to measure the noise is at ${\vc{k}=0}$, as at this point the density is maximal for zero rotation quanta, and zero for one or more rotation quanta.
We plot the noise of the time-of-flight image at ${\vc{k}=0}$ without a central condensate in Fig.\ref{CorrUvsp}. 
First, the interaction $U$ and weak link $\Lambda$ is plotted in Fig.\ref{CorrUvsp}a.
We see that the momentum noise is minimal in the parameter regime ${U/J\ll1}$ and ${\Lambda/J>cU/J}$, where $c$ is some constant, which corresponds to the mean-field limit. As soon the interaction becomes larger than the energy gap induced by the potential barrier, the noise increases. 
Here, entangled phase winding states of zero and one winding quantum appear. For large interaction, the noise decreases again, however remains higher than in the mean-field regime.
With increasing interaction, we can define three regimes of entanglement\cite{nunnenkamp2011superposition}: At the degeneracy point ${\Omega=\frac{1}{2}}$,  for interaction smaller than the energy gap created by the weak link, we observe one-particle superposition states  $\ket{\Psi}\propto(\ket{{l=0}}+\ket{{l=1}})^N$, where $N$ is the number of particles and $l$ is the angular momentum of the atom. This regime is well described by the Gross-Pitaevskii equation. Here, the noise at ${\vc{k}=0}$ is minimal and is given by ${\sigma_k^\text{GP}(\vc{k}=0) \propto \sqrt{N}/4}$.
When the interaction and the weak-link energy gap is on the same order, the near-degenerate many body states mix and entangled NOON states are formed $\ket{\Psi}\propto\ket{{l=0}}^N+\ket{{l=1}}^N$. The noise is maximal and given by ${\sigma_k^\text{NOON}(\vc{k}=0) \propto N/4}$. The ratio of the minimal and maximal noise is $\sqrt{N}$. Thus, with increasing particle number the superposition and entangled states are clearer to distinguish. 
Increasing interaction further will fermionize the system. With interaction, angular momentum of each atom individually is not conserved, however the center of mass angular momentum $K$ of the whole condensate is. Then, the ground state is a superposition of  $\ket{\Psi}\propto\ket{{K=0}}+\ket{{K=N}}$. 

Next, the momentum noise is plotted against applied flux $\Omega$ in Fig.\ref{CorrUvsp}b. Due to the two level system effective physics, the noise in the time-of-flight of the ring condensate is particularly pronounced at the degeneracy points. This phenomenon allows to detect the degeneracy point  in the ring condensate, without resorting the heterodyne detection protocol. The noise is maximal at the degeneracy point and when barrier and interaction are on the same order. Changing the flux away from the degeneracy point decreases the noise.

Further information can be identified by looking at the density at the site of the weak link.
For zero on-site interaction, the site at the potential barrier is always depleted at the degeneracy point for any value of potential barrier strength. However, when  the interaction exceeds  a critical value,  particles start occupying the site~\cite{aghamalyan2015coherent}. This is plotted as black line in Fig.\ref{CorrUvsp}. For small interaction the critical value has a linear relationship between $U$ and $\Lambda$~\cite{aghamalyan2015coherent}. The filling of the potential barrier site indicates the onset of entanglement between different flux quanta. The depletion factor can be measured by a lattice-site resolved absorption measurements. 

\subsection{Experimental realization  of the ring-lattice potential with  weak links}
In this section, we provide the experimental details for the realization of ring-lattice potentials with weak-links.
Among the different architectures, the focus is on the structure that can be relevant for the construction of two level quantum systems.

\begin{figure}[tbp]
\centerline{
	\includegraphics[width=4. cm]{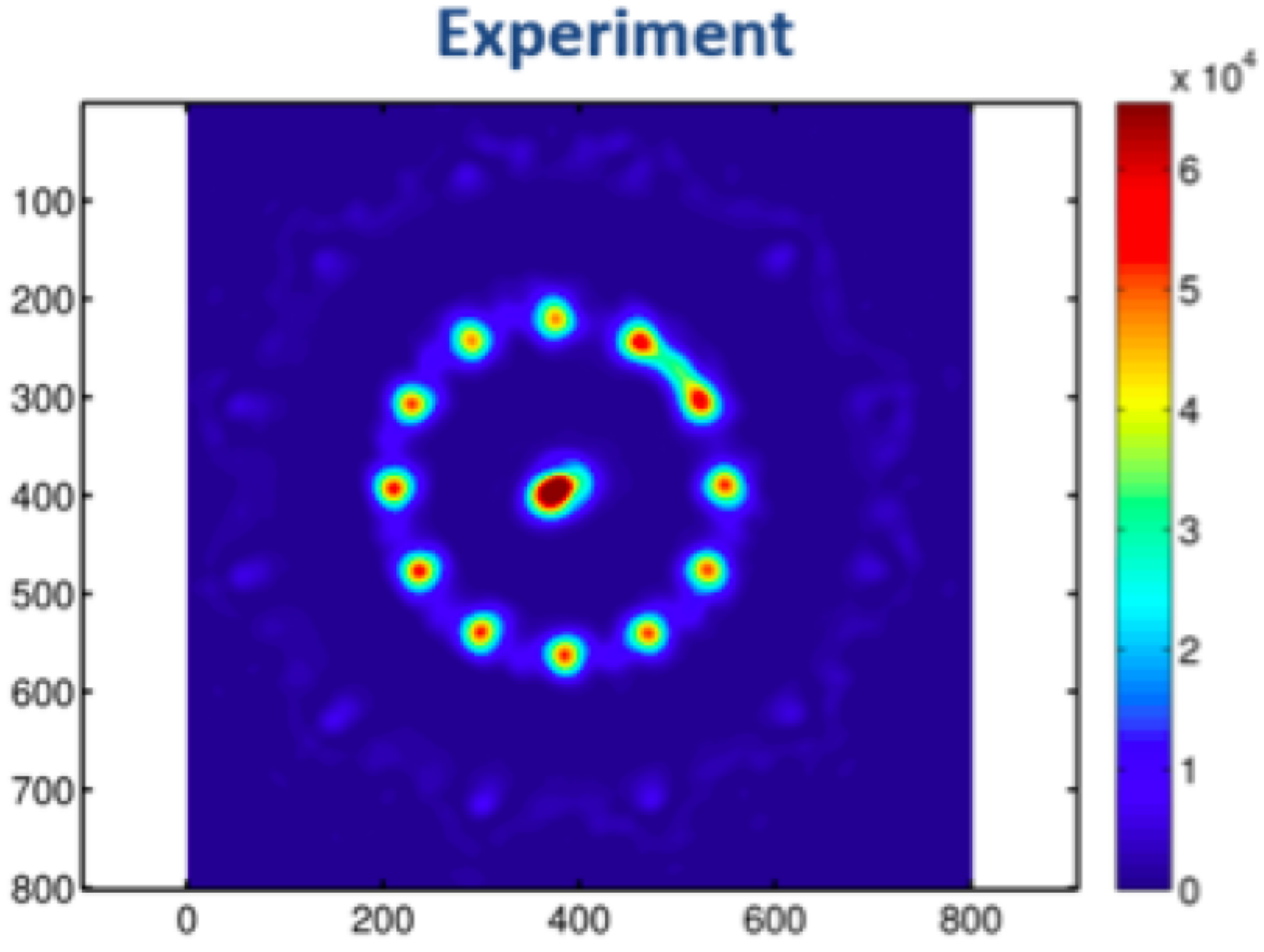}
	\includegraphics[width=4. cm]{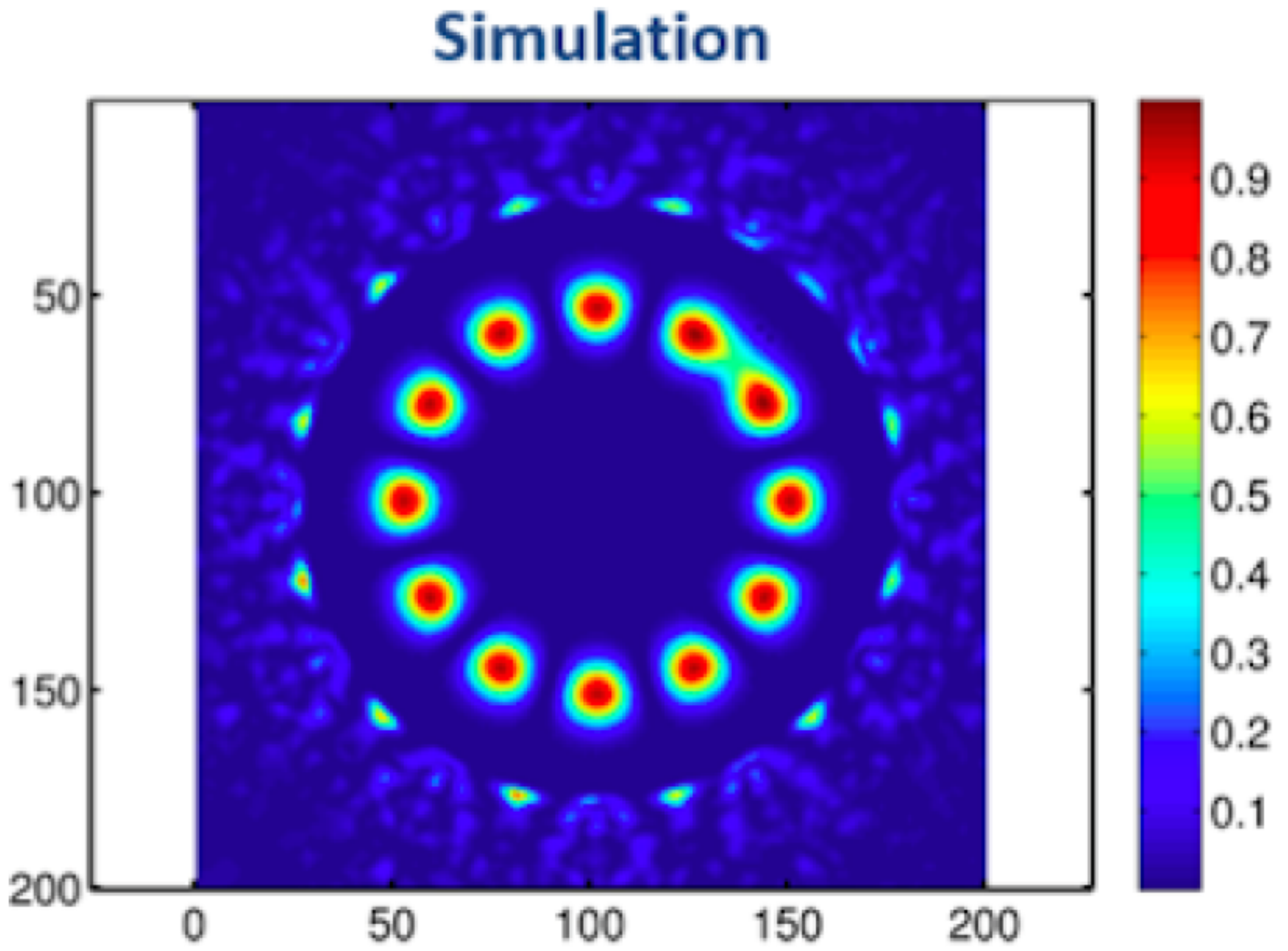}
}
\caption{Simulation(left) and experimentally realized(right) intensity distribution of a ring- lattice with a weak link between  two lattice sites.
Reprinted with permission from D. Aghamalyan, ``Atomtronics: Quantum technology with cold atoms in ring shaped optical lattices'' Ph.D. dissertation (Singapore, 2015).}
\label{Rings}
\end{figure}

\paragraph{A ring lattice with single weak link}
\label{One Ring}
The optical potential was created with a liquid crystal on silicon spatial light modulator (PLUTO phase only SLM, Holoeye Photonics AG) which imprints a controlled phase onto a collimated laser beam from a 532 nm wavelength diode pumped solid state (DPSS) laser.
The SLM acts as a programmable phase array and modifies locally the phase of an incoming beam. Diffracted light from the computer generated phase hologram then forms the desired intensity pattern in the focal plane of an optical system (doublet lens, f=150 mm). The resulting intensity distribution is related to the phase distribution of the beam exiting the SLM by Fourier transform.
Calculation of the required SLM phase pattern (kinoform) has been carried out using an improved version of the Mixed-Region-Amplitude-Freedom (MRAF) algorithm~\cite{pasienski2008high,gaunt2012robust} with angular spectrum propagator. This allows  to simulate numerically the wavefront propagation in the optical system without resorting to paraxial approximation. A region outside the desired ring lattice pattern (noise region) is dedicated to collect unwanted light contributions resulting from the MRAF algorithm's iterative optimization process. This can be seen in the measured intensity pattern in Fig.~\ref{Rings} as concentric, periodic structures surrounding the ring-lattice and can be filtered out by an aperture.

\begin{figure}[tbp]
\centering
\includegraphics[width=78mm]{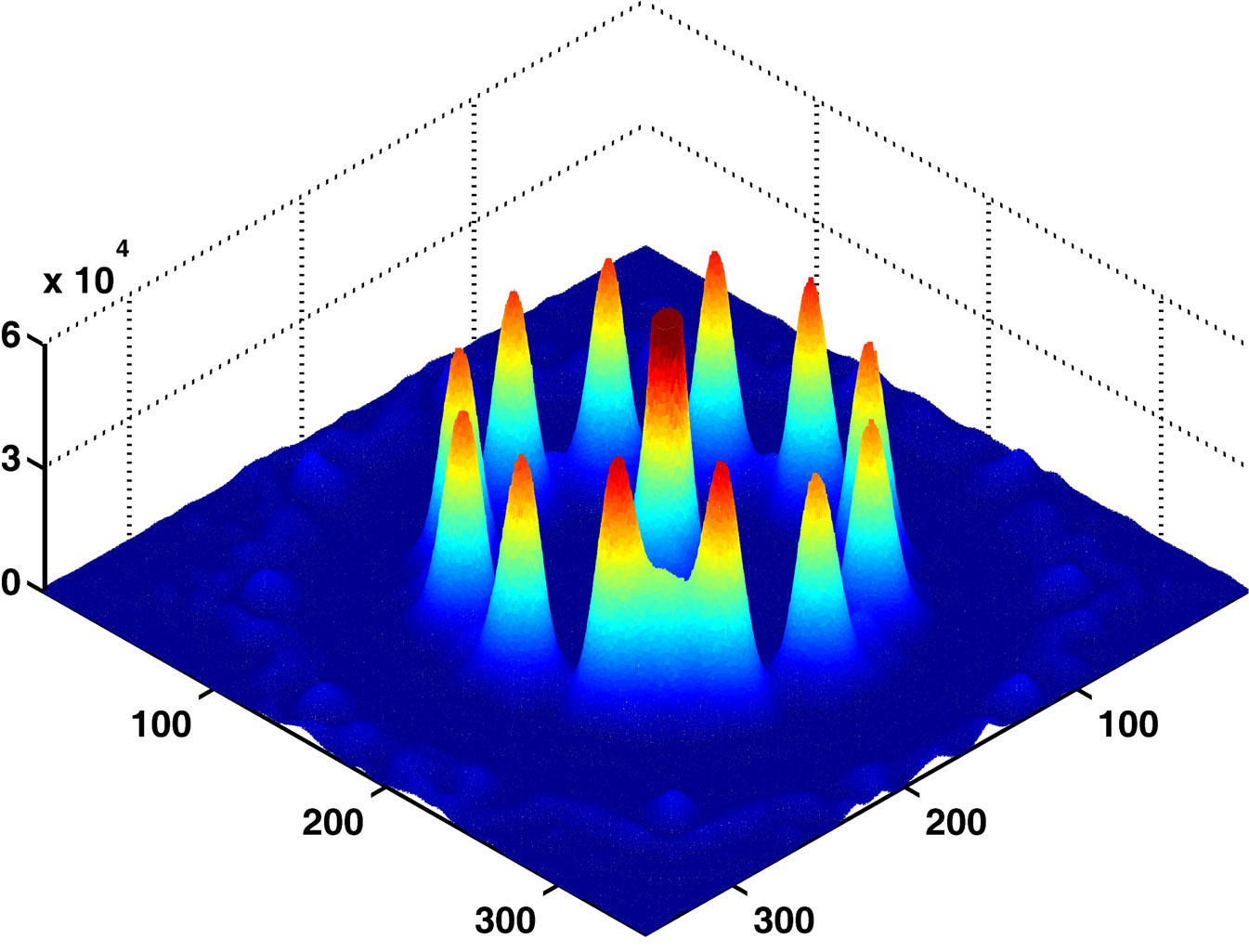}
\caption{Experimental realization of a ring-lattice potential with an adjustable weak link.
	Measured intensity distribution with an azimuthal lattice spacing of 28 $\mathsf{\mu m}$ and a ring radius of 88 $\mathsf{\mu m}$. The central peak is the residual zero-order diffraction.  The size of the structure is scalable and a lower limit is imposed by the diffraction limit of the focusing optics. Reprinted with permission from L. Amico, D. Aghamalyan, F. Auksztol, H. Crepaz, R. Dumke, and L. C. Kwek, Sci. Rep. 4, 4298 (2014), under a Creative Commons 3.0 Unported License.}
\label{exp-onering}
\end{figure}

The ring-lattice potential shown in Fig.~\ref{Rings} and Fig.~\ref{exp-onering} can be readily scaled down from a radius of $\mathrm{\sim 90\; \mu m}$ to $\mathrm{5-10\; \mu m}$ by using a 50x microscope objective with NA=0.42 numerical aperture (Mitutoyo 50x NIR M-Plan APO) as the focusing optics for the SLM beam and with $\mathrm{\lambda_{2}=830\,nm}$ light, suitable for trapping Rubidium atoms. Accounting for the limited reflectivity and diffraction efficiency of the SLM, scattering into the noise region and losses in the optical system only about 5\% of the laser light contributes to the optical trapping potential. However this is not a limiting factor for small ring-lattice sizes in the tenth of micrometer range as discussed here where $\sim 50$ mW laser power is sufficient to produce well depths of several $\mathrm{E_{rec}}$. The generated structures are sufficiently smooth, with a measured intensity variation of 4.5\% rms, to sustain persistent flow-states~\cite{ramanathan2011superflow}.
The barrier height can be dynamically modified at a rate up to 50 ms per step, with an upper limit imposed by the frame update rate of the SLM LCD panel (60 Hz).
\paragraph{Experimental realization  of the ring-lattice potential with three weak links}
\label{platform}
We produce the optical potential using a spacial light modulator (Holey Photonics AG, PLUTO-NIR II), SLM . A collimated Gaussian beam, of 8 mm diameter, is reflected from the SLM's surface forming an image through a $f=200$ mm lens. The light is then split into the two sides of our system, with 10\% of the light in the ``monitoring" arm, and 90\% into the ``trapping" arm used to create a red-detuned dipole trapping potential for a gas of $Rb^{87}$ atoms. A Ti:Saph laser (Coherent MBR-110)  produces a 1W, 828 nm beam, which is spatially filtered and collimated, before reflection on the SLM. To produce the trapping potential the SLM's kinoform is imaged through a 4\textit{f} lens system reducing the beam size to 3 mm diameter and focused through a 50X microscope objective with a 4 mm focal distance and a numerical aperture of NA = 0.42 (Mitutoyo 50X NIR M-Plan APO). The monitoring arm of the system creates an image of the potential through a 10X infinity-corrected microscope objective focused on a CCD camera (PointGrey FL3-GE-13S2M-C). The CCD camera views, therefore, an enlarged image of the optical potential.


\begin{figure}[htb]
    \centerline{\includegraphics[width=0.5\textwidth]{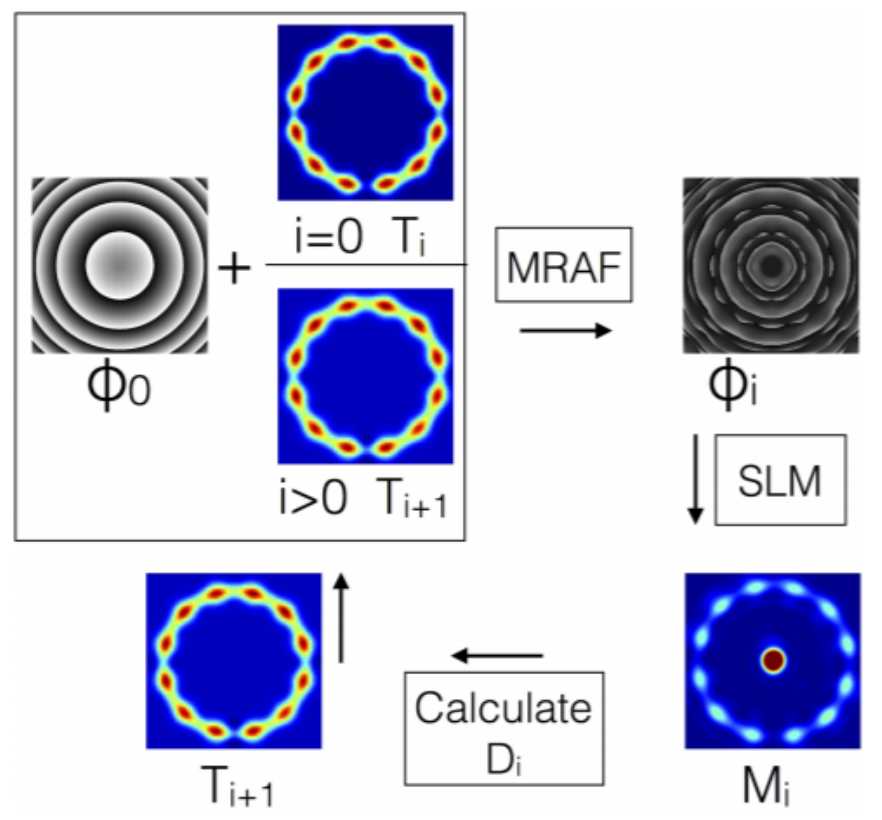}}
    \caption{\label{fig:algor} Our feedback algorithm. Starting at the top left the initial phase and target are used in the MRAF code. This generates the phase guess, $\phi_{i}$, which is uploaded to the SLM and an image captured by the CCD camera, $M_{i}$. This is used to calculate the discrepancy between the image and the original target, and a new target $T_{i+1}$ is created. The loop then repeats. Reprinted with permission from D.  Aghamalyan, N. Nguyen, F. Auksztol, K. Gan, M. M. Valado, P. Condylis, L.-C. Kwek, R. Dumke, and L. Amico, New J. Phys. 18, 075013 (2016), under a Creative Commons Attribution 3.0 License.}
\end{figure}

\begin{figure}[htb]
    \centerline{\includegraphics[width=0.45\textwidth]{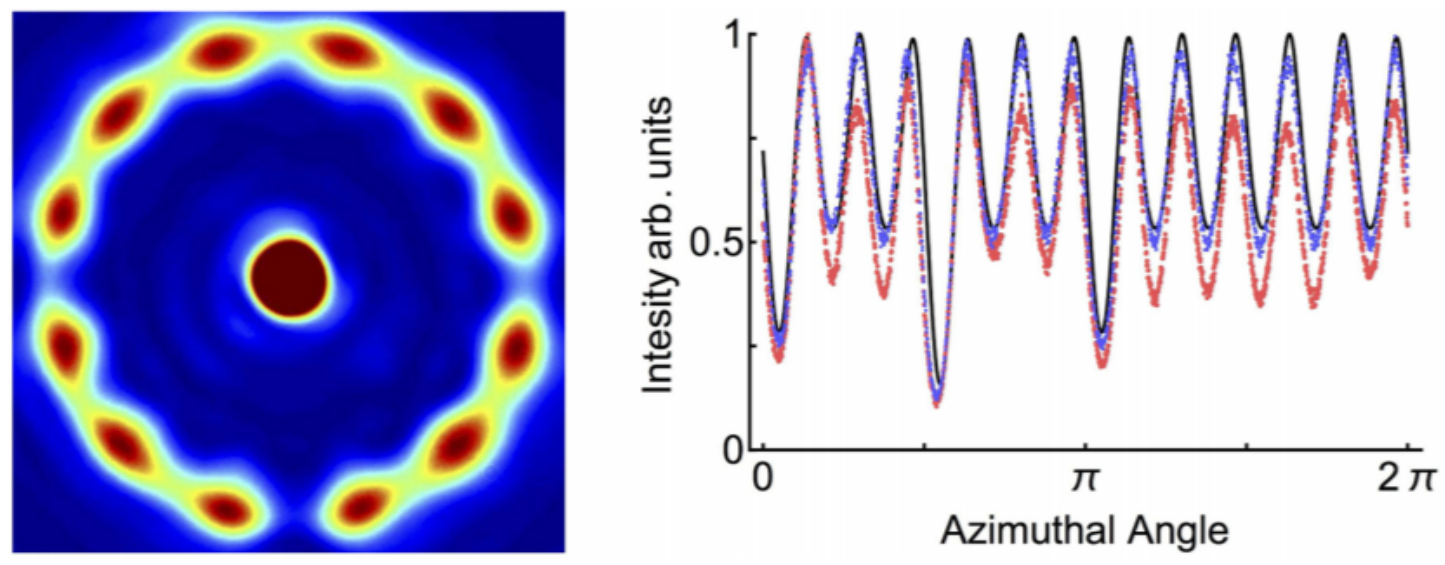}}
    \caption{\label{fig:profile} Left: Final image of the ring lattice after completion of the feedback algorithm. Right: Azimuthal Profile. The solid line plots the target profile.  This is compared to the result after the 1st and 5th iteration of the feedback algorithm (red and blue lines respectively). Reprinted with permission from D.  Aghamalyan, N. Nguyen, F. Auksztol, K. Gan, M. M. Valado, P. Condylis, L.-C. Kwek, R. Dumke, and L. Amico, New J. Phys. 18, 075013 (2016), under a Creative Commons Attribution 3.0 License.}
\end{figure}

To increase the accuracy of the output potential we use the computationally generated kinoform and produce an image of the optical potential in the monitoring arm of our system, and use this as a further source of feedback to the MRAF algorithm. Our method is broadly similar to Bruce \textit{et al} ~\cite{bruce2015feedback}, however it is specialised for producing ring-lattices. Fig.~\ref{fig:algor} shows a flow chart of our improved algorithm. In the first step, the target image, $T_{i}$, and the initial phase, $\phi_0$, is loaded as an input to the MRAF code. This runs for 20 iterations (this was found to be sufficient to get good convergence in most cases) and outputs a phase kinoform, $\phi_{i}$. The kinoform is now applied to the SLM and an image recorded on the camera in the monitoring arm of our system, $M_{i}$. The discrepancy, $D_{i}$, between the original target and the measurement is calculated and used to form an updated target $T_{i+1}$. Here our algorithm differs from \cite{bruce2015feedback} as we take the discrepancy to be $D_i=-(M_i^2+T_0^2)/2T_0$. Also, we do not take into account the whole image, the discrepancy is calculated by comparing the maxima and minima around the azimuthal, 1D, profile of the lattice to the target profile. The targets maxima and minima are then adjusted with $T_{i+1}=T_i+\alpha D_i$, where $\alpha$ is a problem specific feedback gain and $i$ the iteration number. The process now repeats with, $\phi_0$ and $T_{i+1}$, as the inputs to the MRAF code. The feedback gain, $\alpha$, is set to be 0.3 to ensure a quick convergence and this process iterates 30 times. At this point the algorithm is complete and the best image from the set M is selected that minimises the discrepancy below 2$\%$. With this method we produce the ring-lattice potential shown in Fig.~\ref{fig:profile} (left), that on the trapping side of our apparatus creates a scaled-down lattice with radius of 5-10 $\mu$m with more than sufficient power to trap ultra-cold atoms. On the right of Fig.~\ref{fig:profile}, the azimuthal profile around the ring lattice is shown. The red curve indicates the profile on the first iteration of the feedback loop. After 5 iterations (blue curve), the algorithm has converged significantly towards the original target (solid line).

\subsection{Setup for adjustable ring-ring coupling}

In this section  proposal for experimental realization of ring lattices with tunable distance between the rings is suggested by utilizing SLM technique~\cite{pasienski2008high,gaunt2012robust}.
With a SLM arbitrary optical potentials can be produced in a controlled way only in a $2d$-plane -- the focal plane of the Fourier transform lens -- making it challenging to extend and up-scale this scheme to 3d trap arrangements.
The experiment, however, showed (see Fig.\ref{lattice_axial}) that axially the ring structure potential remains almost undisturbed by a translation along the beam propagation axis of $\mathrm{\Delta z\, =\, \pm 2.2\cdot R}$, where $R$ denotes the ring-lattice radius.  {The ring-lattice radius is only weakly affected by an axial shift along z and scales with $\Delta R/R=0.0097\cdot z,$ where $z$ is normalized to the ring-lattice radius.}
For larger axial shifts from the focal plane the quality of the optical potential diminishes gradually. {Based on our measurements this would allow implementation of ring-lattice stacks with more than $10$ rings in a vertical arrangement, assuming a stack separation comparable to the spacing between two adjacent lattice sites.
Propagation invariant beams may allow a potentially large number of rings to be vertically arranged\cite{arnold2012extending}.} \newline
%
\begin{figure}[tbp]
\centering
\includegraphics[scale = 0.3]{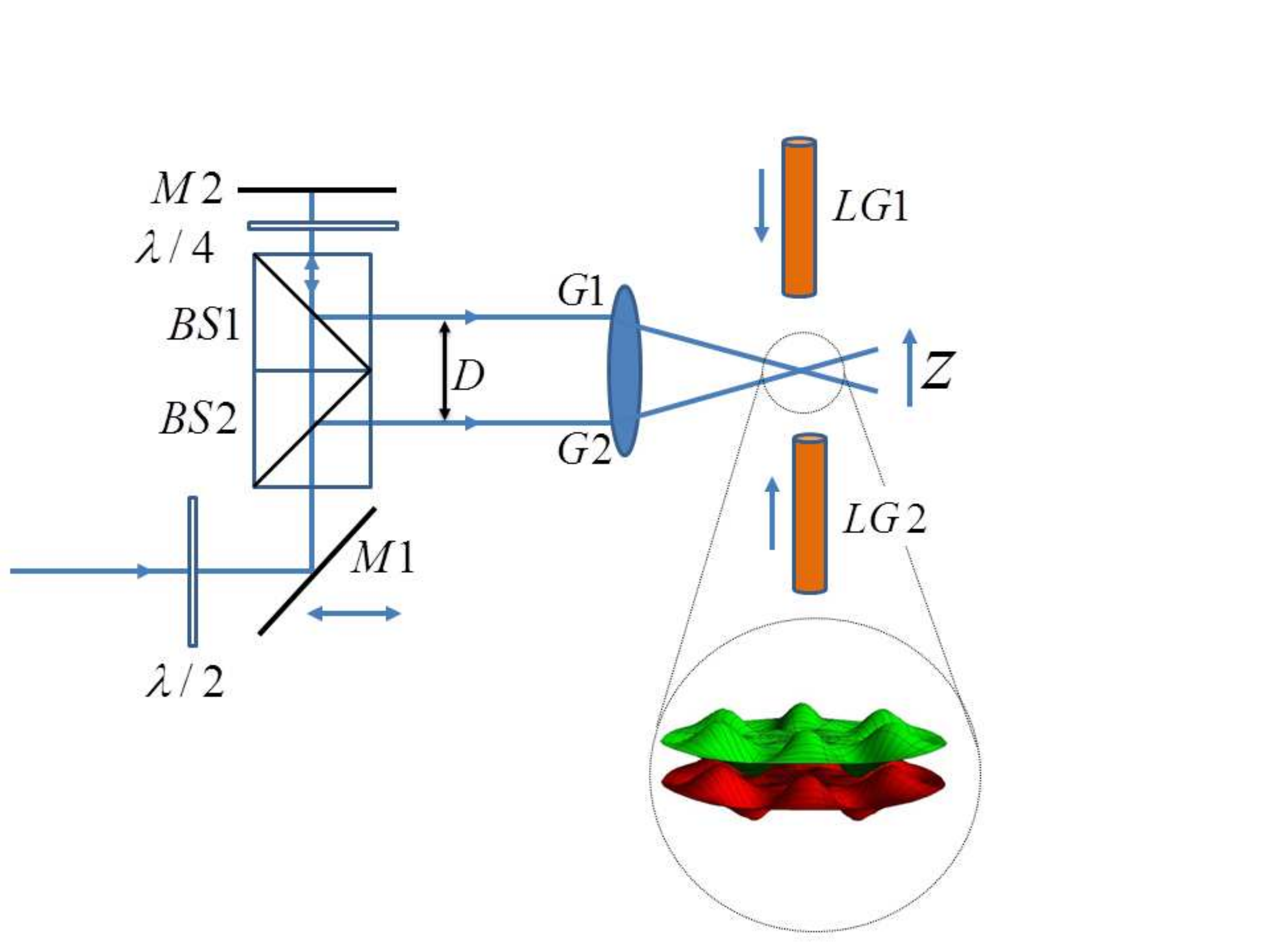}
\caption{Proposed setup for the ring-ring coupling. Two Gaussian laser beams of wavelength $\lambda$ and distance $D$, pass through a lens and interfere in the focal plane ($f$ is the focal length). The distance $D$ can be easily controlled by moving the mirrors. The distance between the fringes is a function of $1/D$ \cite{li2008real}. The resulting Gaussian laser beam with wave vector ${k_{G}={2 \pi D}/{(\lambda f)}} $, then, interferes with two counter propagating Laguerre-Gauss laser beams of amplitude $E_0$. The inset shows the ring lattice potentials separated by $d=\lambda {f}/{D}$. Here $l=6$ and $p=0$. Reprinted with permission from D. Aghamalyan, L. Amico, and L. C. Kwek, Phys. Rev. A 88, 063627 (2013). Copyright 2013 American Physical Society.
}
\label{double-ring potential}
\end{figure}
Besides making the inter-ring dynamics strictly one dimensional, the lattice confinement provides  the route to  the inter-rings coupling.

To allow controlled tunnelling between neighbouring lattice along the stack, the distance between the ring potentials needs to be adjustable in the optical wavelength regime (the schematics in Fig.\ref{double-ring potential} can be employed). A trade-off between  high tunnelling rates (a necessity for fast gate operations)  and an efficient  read out and addressability  of individual stack sites, needs to be analysed. Increasing the lattice stack separation  after the tunnelling interaction has occurred well above the diffraction limit while keeping the atoms confined, optical detection and addressing of individual rings becomes possible.

This arrangement produces equal, adjustable ring-ring spacing between individual vertical lattice sites and can therefore not readily be used to couple two two-ring qubits to perform two-qubit quantum-gates. The SLM method, however, can be extended to produce two ring-lattices in the same horizontal plane, separated by a distance larger than the ring diameter. The separation between these two adjacent rings can then be programmatically adjusted by updating the kinoform to allow tunnelling by mode overlap\cite{anderson2003atomic}. Combined with the adjustable vertical lattice (shown in Fig.\ref{double-ring potential}) this would allow, in principle, two-ring qubit stacks to be circumferential tunnel-coupled to form two-qubit gates
\begin{figure}[tbp]
\centering
{\includegraphics [width=78mm]{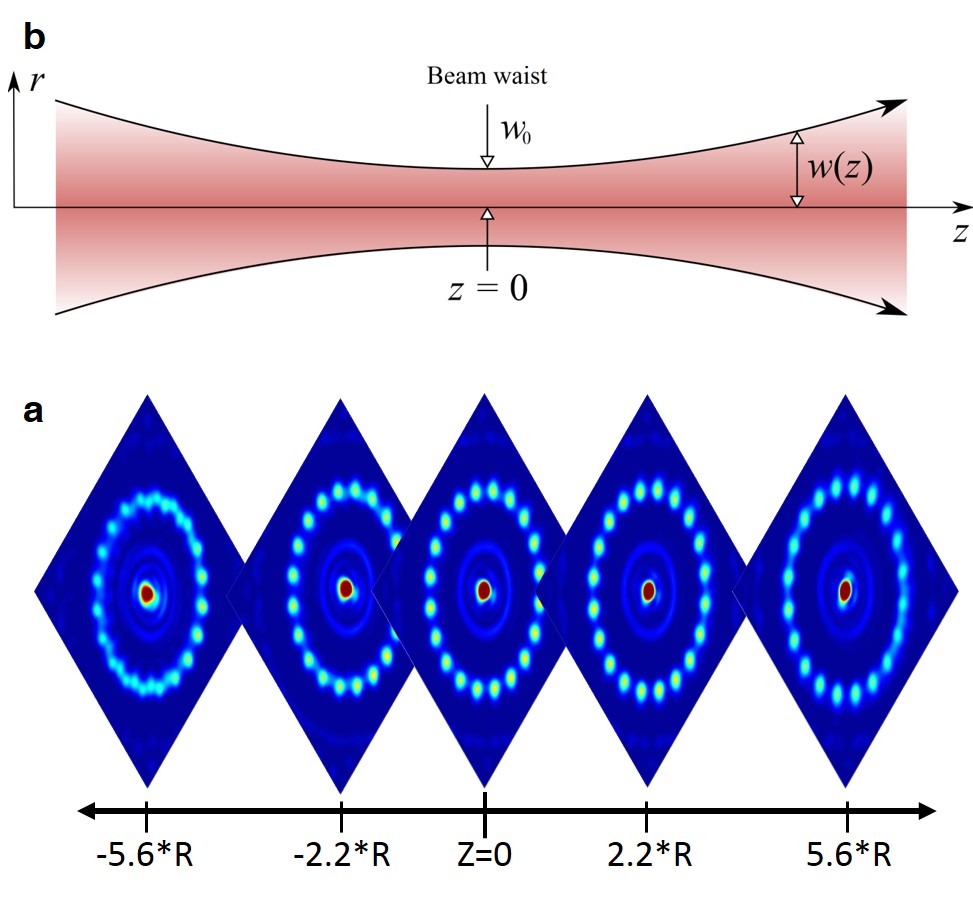} }
\caption{Effect of an axial translation on the ring lattice potential.
	~a) Ring lattice intensity distribution measured at various positions along the beam propagation axis around the focal plane (Z=0). Note that the initial beam, phase modified by the SLM, is not Gaussian any more. The optical potential remains undisturbed by a translation of 2.2 times the ring-lattice radius centred around the focal plane (Z=0). Here R designates the ring-lattice radius of 87.5 $\mathrm{\mu m}$. ~b) This is in contrast to a Gaussian laser beam which exhibits a marked dependence on the axial shift from the focal plane where the beam waist $\omega(z)$ scales with $\sqrt{1+(z/z_{0})^{2}}$ and Rayleigh range $z_{0}$. Reprinted with permission from L. Amico, D. Aghamalyan, F. Auksztol, H. Crepaz, R. Dumke, and L. C. Kwek, Sci. Rep. 4, 4298 (2014), under a Creative Commons 3.0 Unported License. }
\label{lattice_axial}
\end{figure}

\subsection{First experimental demonstration of the interference of atomtronic currents}
A key ingredient to realize the atomtronic qubit is the interference of currents that make up the qubit dynamics. Controlling and interfering currents in-situ of a cold atom circuit is a challenging task, as it requires high coherence and control in the system.
Recently,  interference of currents in an AQUID has been realized experimentally for the first time, which is reviewed in this section~\cite{ryu2020quantum}. 
The interference can be revealed by a periodic modulation of the critical current with applied flux in an atomtronic ring, interrupted by two weak links. Note that the interference has been achieved for a dilute Bose-Einstein condensate which can be described within the mean-field limit as a classical wave equation. While entanglement that is critical for atomic qubits cannot be demonstrated within this experiment, it is nonetheless a first step towards establishing the ingredients for atomic qubits based on superposition of currents.

\begin{figure*}[htbp]
\centering
\includegraphics[width=0.8\textwidth]{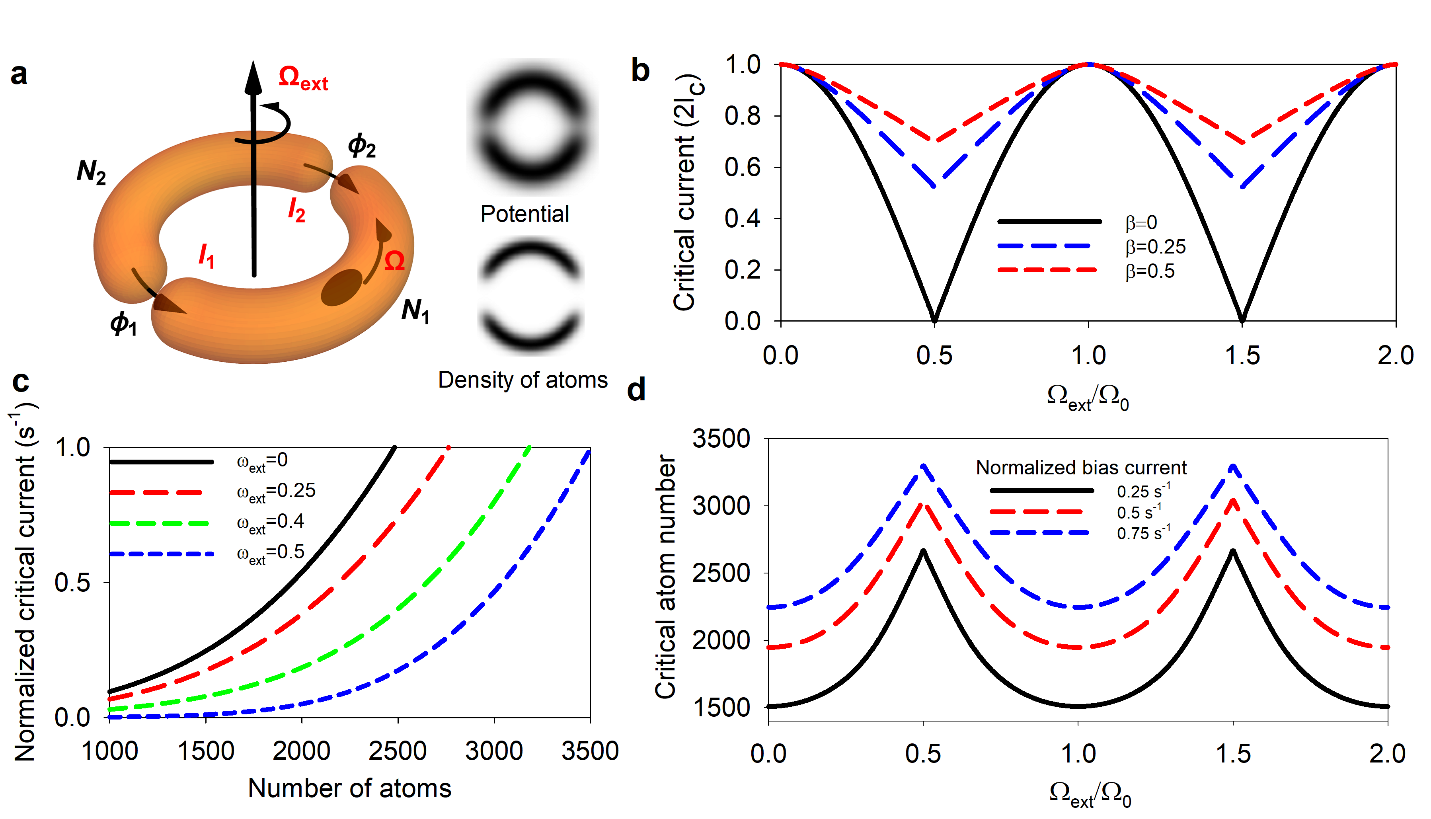}
\caption{Calculation of the periodic modulation of the critical current. \idg{a} Schematic of a double junction atomtronic SQUID. The atomtronic SQUID was created by scanning a single 834 nm laser beam  with 1.7 $\mu$m waist and the barrier full width at half maximum (FWHM) was 2.1 $\mu$m. $\Omega_\text{ext}$ is the rotation rate of the atomtronic SQUID and $\Omega$ is the rotation rate of atoms. $\phi_1$ and $\phi_2$ are the phase differences across the Josephson junctions, $I_1$ and $I_2$ are Josephson junction currents, and $N_1$ and $N_2$ are numbers of atoms in each half. Arrows represent the movement of the junctions. The calculated potential of the atomtronic SQUID and the density of atoms are shown for the radius of 3.85 $\mu$m. \idg{b} Critical current as a function of $\Omega_\text{ext} /\Omega_0$  calculated for different values of $\beta_\text{atom}$. \idg{c} Normalized critical currents ($2I_c /N$ ) where $I_c$  is the critical current and $N$ is the total number of atoms as a function of the number of atoms with different $\omega_\text{ext}$ for the atomtronic SQUID with 3.85 $\mu$m radius. $\beta_\text{atom}$  varies with the number of atoms and the critical current. For each number of atoms, $\beta_\text{atom}$  was calculated to find the variation of the normalized critical current.   \idg{d} Modulation of the critical atom number as a function of $\Omega_\text{ext} /\Omega_0$  for three different normalized bias currents with the 3.85 $\mu$m radius atomtronic SQUID. Reprinted with permission from C. Ryu, E. C. Samson, and M. G. Boshier, Nat. Commun. 11, 3338 (2020), under a Creative Commons Attribution 4.0 International License.}
\label{Fig1-qubit}
\end{figure*}

The periodic modulation of the critical current can be understood by calculating the total current within a model of the atomtronic SQUID based on quantum phase-controlled Josephson junction currents and a toroidal trap geometry (Fig.\ref{Fig1-qubit}a). The total current is the result of  interference of the two Josephson junction currents, given by
\begin{equation}
I_1=\frac{1}{2} (I_t +I_j  )=I_c \sin(\phi_1)              
\end{equation}
\begin{equation}
I_2=\frac{1}{2} (I_t -I_j  )=I_c \sin(\phi_2)              
\end{equation}
where $I_c$ is the critical current of atoms, $I_t$ is the total current, and $I_j$  is the circulating current around the atomtronic SQUID. Because of the toroidal geometry and single valuedness of the wavefunction describing the atoms, the phases should satisfy 
$\phi_1-\phi_2+2\pi \omega=2\pi n$
where $\omega=\Omega/\Omega_0$, with $\Omega$ being the rotation rate of atoms, and n is an integer. The rotation rate of the atoms can be shown to be
\begin{equation}
\omega=\omega_\text{ext}+\beta_\text{atom}\frac{I_j}{I_c}
\end{equation}                                    
where $\omega_\text{ext}=\Omega_\text{ext}/\Omega_0$, $\Omega_\text{ext}$ is the external rotation rate of the atomtronic SQUID,  $\beta_\text{atom}=\frac{2\pi I_c}{N \Omega_0}$,  and $N$ is the total number of atoms. This equation for the rotation rate of atoms can be derived from the relation between the circulating current and the movement of atoms relative to the Josephson junctions. The parameter $\beta_\text{atom}$  is analogous to the screening parameter in the conventional SQUID and can be thought as proportional to the “inductance” which induces the deviation of the rotation rate of atoms from the imposed external rotation rate of the atomtronic SQUID.
Equations (1-4) are equivalent to those of a DC SQUID~\cite{clarke2006squid}, reflecting the fact that the fundamental underlying physics of a double junction atomtronic SQUID and a DC SQUID is the same. In the limit of $\beta_\text{atom}=0$ (for example, when $I_c\approx0$ with much higher barrier height), we can analytically calculate the total currents
$I_t=2I_c  \cos(\pi \omega_\text{ext})\sin(\phi_1-\pi \omega_\text{ext})$.
Thus, the critical current is $|2I_c  \cos(\pi \omega_\text{ext}|$, which establishes a clear modulation of the critical currents by tuning $\omega_\text{ext}$ with a period of $\Omega_0$. With finite $\beta_\text{atom}$ , we can numerically calculate the critical current, and the periodic modulation amplitude decreases with the increasing $\beta_\text{atom}$, as can be seen in Fig.\ref{Fig1-qubit}b.
By using the calculated modulation in Fig.\ref{Fig1-qubit}b, the expected periodic modulation of the critical current in an atomtronic SQUID can be calculated with the Gross-Pitaevskii equation (GPE) in 2D. Fig.\ref{Fig1-qubit}c shows the normalized critical current, which is the critical current of atoms normalized to the number of atoms $\frac{2I_c}{N}$, as a function of the number of atoms for the different rotation rates of the atomtronic SQUID. For a fixed number of atoms, the normalized critical current shown in Fig.\ref{Fig1-qubit}c modulates with rotation rate. However, it is very difficult to experimentally observe this modulation because of the strong dependence of the normalized critical current on the number of atoms and the difficulty in producing a BEC with the same number of atoms consistently. Instead of a fixed number of atoms, a fixed normalized bias current was used, generated by moving Josephson junctions with a fixed speed. When the rotation rate changes, the critical atom number-which is the number of atoms at the transition from DC to AC Josephson effect with the chosen normalized bias current-modulates periodically, as shown in the GPE calculation of Fig.\ref{Fig1-qubit}d. 

\begin{figure*}[htbp]
\centering
\includegraphics[width=0.8\textwidth]{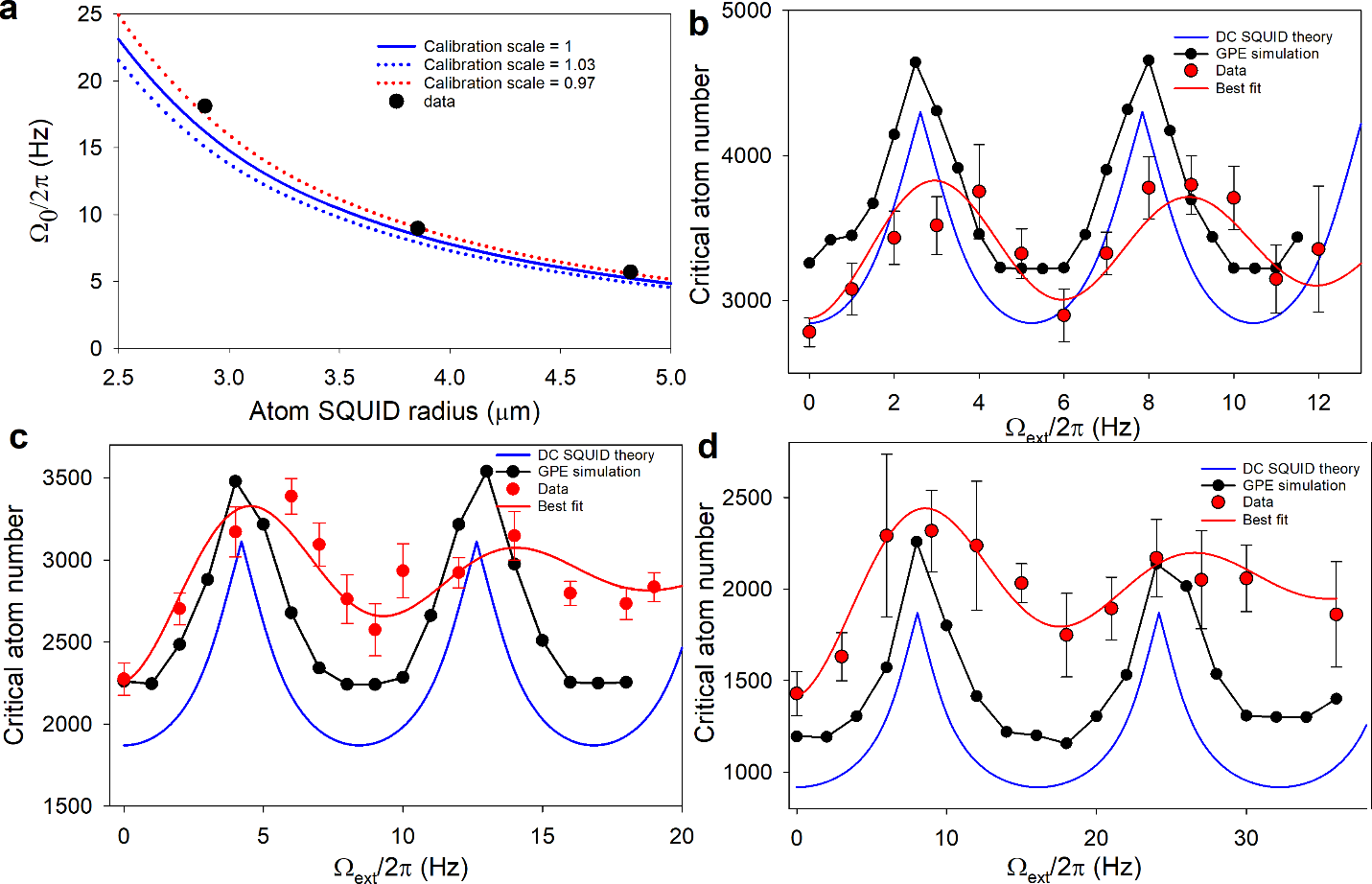}
\caption{Comparison between experiment and theory. \idg{a} Comparison of the measured and calculated values of $\Omega_0/(2\pi)$ (calculation done using GPE). The three curves correspond to calibration scales of the atomtronic SQUID radius. \idg{b,c,d} Critical atom number as a function of the rotation rates obtained with GPE simulation and DC SQUID theory, along with the measured data and the best fit. For \idg{b} the radius is 4.82 $\mu$m, for \idg{c} the radius is 3.85 $\mu$m and for \idg{d} the radius is 2.891 $\mu$m. Reprinted with permission from C. Ryu, E. C. Samson, and M. G. Boshier, Nat. Commun. 11, 3338 (2020), under a Creative Commons Attribution 4.0 International License.}
\label{Fig4}
\end{figure*}

The theoretical prediction of the modulation of the critical current (measured using the critical atom number) are plotted in Fig.\ref{Fig4}. The experimental values clearly show the characteristic modulations of the critical current with flux, revealing the interference of currents in the AQUID.

\subsection{Concluding remarks and outlook}
In this chapter, we have introduced atomtronic qubits constructed with neutral atomic currents flowing   in ring-shaped optical lattice potentials interrupted by few weak links, which give rise to the Atomtronics Quantum Interference Device (AQUID). The effective quantum dynamics of the system is proved to be that one of a two-level system.  The spatial scale of the rings radii would be in the range of 5 to 20 microns. The ring-ring interaction can be realized with the  physical system of two Bose-Einstein condensates, flowing in ring-shaped optical potentials, and mutually interacting through tunnel coupling. Clearly, such systems may be relevant for quantum computation purposes, which was demonstrated further by showing how single and two qubit gates can be obtained in the setup.

The initialization of the qubit can be accomplished, for example, by imparting rotation through light-induced torque from Laguerre-Gauss (LG) beams carrying an optical angular momentum. Stacks of $n \sim 10$ homogeneous ring lattices with tunable distance and stacks of AQUIDs have been realized experimentally (in the lab coordinated by R. Dumke) with Spatial Light Modulators (SLM). Such configurations are realized by making use of the cylindrical symmetry of Laguerre-Gaussian beams and exploiting the flexibility (in terms of generating light fields of different spatial shapes) provided by the SLM devices. Stack of qubits can be realized following very similar protocols.
Indeed, similar goals were carried out by realizing the AQUID with homogeneous condensates (i.e. without lattice modulation)~\cite{wright2013driving,eckel2014interferometric,eckel2014hysteresis,ryu2013experimental,ryu2020quantum,ramanathan2011superflow}. We remark that the lattice confinement brings important added values with respect to that realization. First of all, the gap between the two levels of the qubit displays a more feasible dependence with the number of atoms in the system compared with the case of homogeneous rings with a delta barrier. This is ultimately due to the fact that the barrier can be localized on a lattice spacing spatial scale~\cite{nunnenkamp2011superposition}; thereby the k-mixing-that is the key feature to have a well defined two level system-is not suppressed (as, in contrast, happens for homogeneous condensates with a realistic barrier.  As a second positive feature, the lattice provides a platform for qubit-qubit interaction. These two features, we believe, could ultimately facilitate the exploitation of the device in future atomtronic integrated circuits.

We also reviewed  the construction of a flux qubit employing a ring condensate trapped in a regular lattice potential except for three specific lattice points with a reduced tunneling amplitude. The three weak links solution  was originally suggested in quantum electronics to facilitate the function of the system as a qubit~\cite{mooij1999josephson}. We apply a similar logic leading  to fluxonium from the rf-SQUID:   the continuous quantum fluid, in our system, is replaced by a chain of junctions connecting  the different weak links.  We believe that  the additional lattice helps in adjusting the persistent current flowing through the system. The three weak links architecture, indeed, realizes  a two-level effective dynamics in a considerably enlarged parameter space.
The qubit dynamics can be read-out via time-of-flight measurements. A spiral pattern emerges when the expanding atomic ring with a specific current is interfered with a reference condensate. 
The noise in the time-of-flight images is a hallmark of the entanglement present in the current, allowing to characterize the atomic qubit.
With these methods, the type of current (entangled vs non-entangled), the magnitude and direction can be read out. The depletion at the weak-link can be used to determine the state of the qubit as well. 
This opens up a way to experimentally characterize atomic qubits in the lab.

Recent experiments have demonstrated the interference of currents in atomic SQUIDs for the first time. Oscillations in the critical current with applied flux are a clear hallmark of interference of atomic currents. This result has been achieved in a Bose-Einstein condensate in the dilute limit, such that it can be described within a mean-field description. While entanglement as a key ingredient for atomic qubits has not been demonstrated, this result nonetheless opens up the path to create atomic qubits via superposition of currents and observe their macroscopic entanglement.

Decoherence, of course, is an important issue for our proposal that remains to be studied. We comment, however, that measurements of the decay dynamics of a rotating condensate in an optical ring trap show that the quantized flow states have remarkably long lifetimes, of the order of tens of seconds even for high angular momentum ($l = 10$)~\cite{moulder2012quantized}. Phase slips (the dominant mechanism of decoherence), condensate fragmentation and collective excitations which would ultimately destroy the topologically protected quantum state are found to be strongly suppressed below a critical flow velocity. Although atom loss in the rotating condensate does not destroy the state, it can lead to a slow decrease in the robustness of the superfluid where the occurrence of phase slips becomes more likely. We believe that the decoherence rates could be controlled within the current experimental know- how of the field. The Atomtronics’ positive trend crucially relies on the recent progress achieved in the optics microfabrication field. Thereby, central issues, of the cold atoms system, like scalability, reconfigurability, and stability can be feasibly addressed. In many current and envisaged investigations, there is a need to push for further miniaturization of the circuits. The current lower limit is generically imposed by the diffraction limit of the employed optics. Going to the sub-micron scale, although challenging, might be accessible in the near future. At this spatial scales, mesoscopic quantum effects could be traced out. The scalability of multiple-ring structures will be certainly fostered by tailoring optical potentials beyond the Laguerre-Gauss type(f.i. employing Bessel-Gauss laser beams). A central issue for Atomtronics integrated circuits, is the minimization of the operating time on the circuit and the communication among different circuital parts (i.e. AQUID-AQUID communication). Currently, typical time rates are in the millisecond range, but a thorough analysis of the parameters controlling time rates is still missing.


\section{TRANSPORT AND DISSIPATION IN ULTRACOLD FERMI GASES}
\vspace*{-0.5cm}
\label{TransportFermi}
\par\noindent\rule{\columnwidth}{0.4pt}
{\bf{\small{J.P. Brantut, F. Chevy, M. Lebrat, F. Scazza, S. Stringari}}}
\par\noindent\rule{\columnwidth}{0.4pt}



Atomtronics is based on the flow of quantum gases in circuits or devices. It therefore provides a natural framework in which transport and dissipation, two fundamental dynamical processes, can be observed, studied and controlled. These processes are of fundamental interest in the entire field of many-body physics: first, because they involve not only equilibrium or ground state properties but chiefly that of excitations, they are intrinsically difficult to calculate from first principles. Second, for the same reason, they are very sensitive investigation tools for experimentalists. Third, they underly most of the functionalities of solid-state based quantum devices. 

\begin{center}
\begin{table*}[tp]
\caption{Comparison between cold Fermi gases and electrons in solids}
\vspace*{5pt}
\begin{center}
{\setstretch{1.15}
\begin{tabular}{c|c|c}
 & \,\,Cold Fermi gases\,\, & Electrons in solids\\
 \hline
Interactions     & Contact, tunable & \,\,Coulomb, with density-dependent screening\,\, \\
Internal states & Hyperfine states & Spin \\
Structure shaping\,\, & Light-induced & Gating, crystal growth \\
Energy scales & $E_F\sim 1\,\mu\mathrm{K}$ & $10\,\mathrm{K} < E_F < 10^4\,\mathrm{K}$\\
Density scales & $\sim 10^{12} \,\mathrm{cm}^{-3}$ &  $10^{10}$ (Semiconductors) - $10^{22}\, \mathrm{cm}^{-3}$ (Metals) \\
\end{tabular}}
\end{center}
\label{table1}
\end{table*} 
\end{center}
%

Fermionic quantum gases provide the most direct connection between atomtronics and solid-state electronics. The obvious analogy between the transport of fermionic atoms in light-imprinted structures and that of electrons in condensed matter systems suggests that atomtronics systems could be used as quantum simulators for their electronic counterparts, for which ab-initio modeling is very challenging \cite{cirac2012goals}. The vastly different scales of cold atoms, presented in Table~\ref{table1}, as well as the specific control tools make them especially promising in this perspective.

While electronic systems benefit from their ability to reach low relative temperatures, and from more than a century of development of methods and control protocols for currents and voltages, cold atomic atomtronic systems reach for previously uncharted parameter regimes: they can reach very high relative temperatures without encountering phonons or other disturbances, they offer full control and imaging of the spin degrees of freedom without the need for ferromagnetic materials, and have a rich internal structure that can be leveraged as an extra 'synthetic' dimension \cite{salerno2019quantized}. 
Of particular interest is the possibility of cold Fermi gases to operate in the strongly interacting regime, close to a Feshbach resonance, opening the possibility to realize quantum devices from strongly-correlated matter bypassing the outstanding challenges encountered in solid state systems. 
In this regime, the system is described by the so-called BEC-BCS crossover which interpolates between weakly attractive fermions described by BCS (Bardeen-Cooper-Schrieffer) theory, and a Bose-Einstein Condensate (BEC) of strongly bound molecules \cite{ketterle2008making,giorgini2008theory}. The equilibrium properties of gases in this regime have been extensively investigated in the last decade, and several key thermodynamic properties such as the ground state energy, critical temperature or pairing gap are now known with high precision \cite{zwerger2011bcs,zwierlein2016thermodynamics}. 

The recent years have seen a growing number of experiments exploring the dynamics of fermionic gases in this strongly interacting regime. New systems mixing bosonic and fermionic superfluids provides renewed opportunities to study superfluid flow \cite{ferrier2014mixture,abad2014counter,delehaye2015critical}. The development of two-terminal systems for cold atoms in particular provides the simplest device-like geometry \cite{krinner2017two}. For a long, low dimensional channel, this has allowed for the measurement of particle \cite{brantut2012conduction}, spin \cite{krinner2016mapping} and heat conductances \cite{brantut2013thermoelectric} as well as off-diagonal transport coefficients such as spin-drag or thermopower. For short, planar junctions it realizes a tunnel connection, for which superfluidity yields the celebrated Josephson effect \cite{valtolina2015josephson,burchianti2018connecting,kwon2020strongly,xhani2020critical}. 

This contribution presents some of the most recent development of ultracold Fermi gases in the atomtronics context. In section \ref{sec:SF}, transport phenomena in superfluid Fermi gases are discussed, first in the perspective of the Landau criteria, then in the case of Josephson junctions. In section \ref{sec:Meso}, we describe transport of Fermi gases in mesoscopic channels. In section \ref{sec:SpinDrag}, the physics of the fast spin drag in normal Fermi gases is presented.


%
%

%

\vspace{-1mm}
\subsection{Superfluid transport with Fermi gases}\label{sec:SF}
\subsubsection{Fermionic superfluidity and critical velocity}\label{sec:SFcounterflow}

The first microscopic theory of dissipation in superfluids was proposed by Landau who predicted the existence of critical velocity below which an object in motion in a superfluid feels no drag \cite{landau1941theory}. Landau's original argument was based on constraints imposed by energy and momentum conservation when elementary excitations are shed in the superfluid. In this limit, the critical velocity is given by the minimum value of $\omega(k)/k$, where $\omega(k)$ is the dispersion relation of low-energy modes of the system. For a concave dispersion relation, this is the slope of $\omega(k)$ at the origin and the critical velocity is therefore simply the sound velocity. 

In fermionic systems the excitation spectrum is composed of both bosonic collective modes (the so-called Bogoliubov-Anderson modes) corresponding to phonons and fermionic quasi-particles associated with broken Cooper pairs \cite{combescot2006collective}.  These two sectors lead to different predictions for the critical velocity when interactions are varied across the BEC-BCS crossover. In the BEC regime, where pairs are tightly bound, phonons set the critical velocity, as in a traditional atomic Bose-Einstein condensate. On the contrary, on the BCS side of the resonance, the Cooper pairs are loosely bound and are easily broken by a moving object. In this regime the critical velocity is $v_c\simeq \Delta/p_F$, where $\Delta$ is the excitation gap and $p_F$ is the Fermi momentum. The existence of these two excitation branches leads to a maximum of Landau critical velocity close to the unitary limit that was observed experimentally  by stirring an optical potential in the cloud \cite{miller2007critical,weimer2014critical}.

Recent experiments on atomic mixtures have raised the question of the onset of dissipation in two counterflowing superfluids \cite{ferrier2014mixture,delehaye2015critical,yao2016observation}. Experiments on dual Bose/Fermi superfluids revealed the existence of a critical velocity which was later on interpreted as an extension of Landau's seminal argument similar to parametric down-conversion in quantum optics. In this scenario, the relative motion of the two superfluids can excite {\em pairs} of excitations in the superfluids \cite{castin2014landau,abad2014counter,zheng2014quasiparticle,yukalov2004stratification}.  This modifies the expression of the critical velocity which is equal to the sum of the sound velocities of the two superfluids when phonons limit superfluidity, a prediction that agrees with experimental measurements \cite{delehaye2015critical} performed on mixtures of $^6$Li and $^7$Li.

Let's conclude this subsection by stressing some of the hypotheses underlying Landau's scenario. Firstly, as mentioned earlier, the identification of Landau critical velocity in the phonon sector with sound velocity assumes that the dispersion relation is convex. Although this is true for bosons in free space, this is no longer the case for fermions, for which the coupling with the broken-pair particle-hole continuum bends the dispersion relation downwards \cite{kurkjian2016concavity}. Likewise, the presence of a transverse trapping in experiments leads to a reduction of the critical velocity due to an inversion of the concavity of the dispersion relation at large momenta, a feature first pointed out in weakly interacting Bose-Einstein condensates \cite{fedichev2001critical,tozzo2003bogoliubov} and recently generalized to arbitrary hydrodynamic superfluids \cite{crepin2016hydrodynamic}. Second, Landau's argument assumes that the velocity of the moving disturbance is constant while in experiments the motion of the disturbing potential is usually oscillatory to account for the finite size of the system. By analogy with an accelerated electric charge that radiates electromagnetic wave at an arbitrarily small velocity, Landau critical velocity is suppressed for accelerated disturbances \cite{jin2019hydrodynamic}. Finally, as initially proposed by Feynman and Onsager \cite{onsager1949statistical,feynman1955application} topological defects, such as quantized vortices,  are responsible for the onset of dissipation for stronger disturbances \cite{park2018critical}.

\begin{figure}[b]
\centering
{\includegraphics[width=0.35 \textwidth]{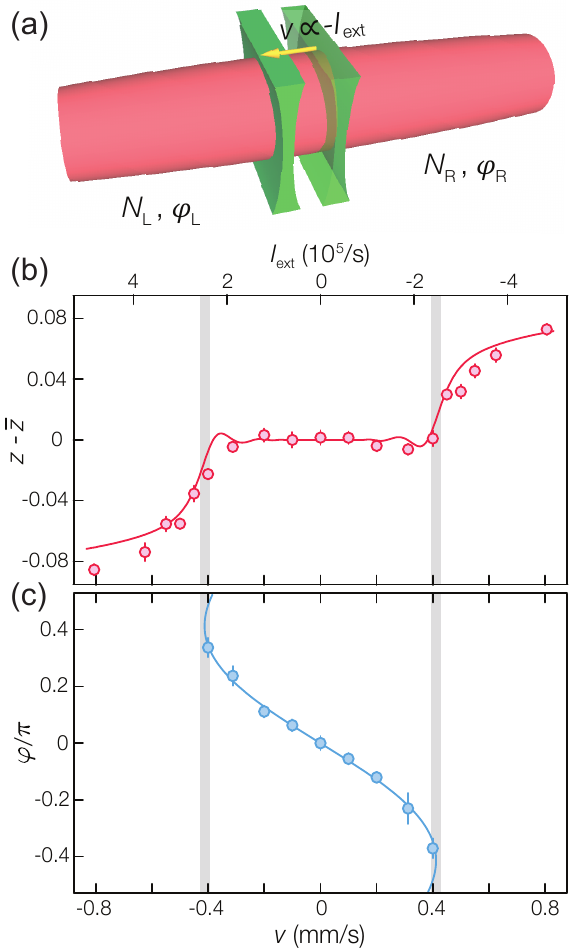}\vspace{-2mm}}
	\caption{\small{Realization of a current-biased Josephson junction between ultracold fermionic superfluids. (\textbf{a}) Two superfluid reservoirs (L, left; R, right) of $^6$Li fermion pairs are weakly coupled through a thin optical repulsive barrier created using a DMD. An external current $I_\mathrm{ext}$ is imposed by translating the tunnelling barrier at a constant velocity $v$. Pair transport is tracked by recording the relative imbalance $z = (N_R - N_L)/(N_R + N_L)$ through \textit{in-situ} absorption imaging, while the order-parameter relative phase $\phi$ is revealed through matter-wave interference. (\textbf{b}) Experimental current-imbalance characteristic, and (\textbf{c}) current-phase relation $I(\varphi)$ for a crossover Fermi gas on the BEC side of the Feshbach resonance. The solid line denotes the fit to a resistively-shunted Josephson junction circuit model, while the shaded vertical lines indicate the extracted $I_c$. 
	Reprinted with permission from W. J.~Kwon, G.~Del Pace, R.~Panza, M.~Inguscio, W.~Zwerger, M.~Zaccanti, F.~Scazza, and G.~Roati, \textit{Science} \textbf{369}, 84–88 (2020). Copyright 2020 AAAS.
	}}
\label{Fig1}
\end{figure}

\begin{figure*}[ht]
\centering
{\includegraphics[width=135mm]{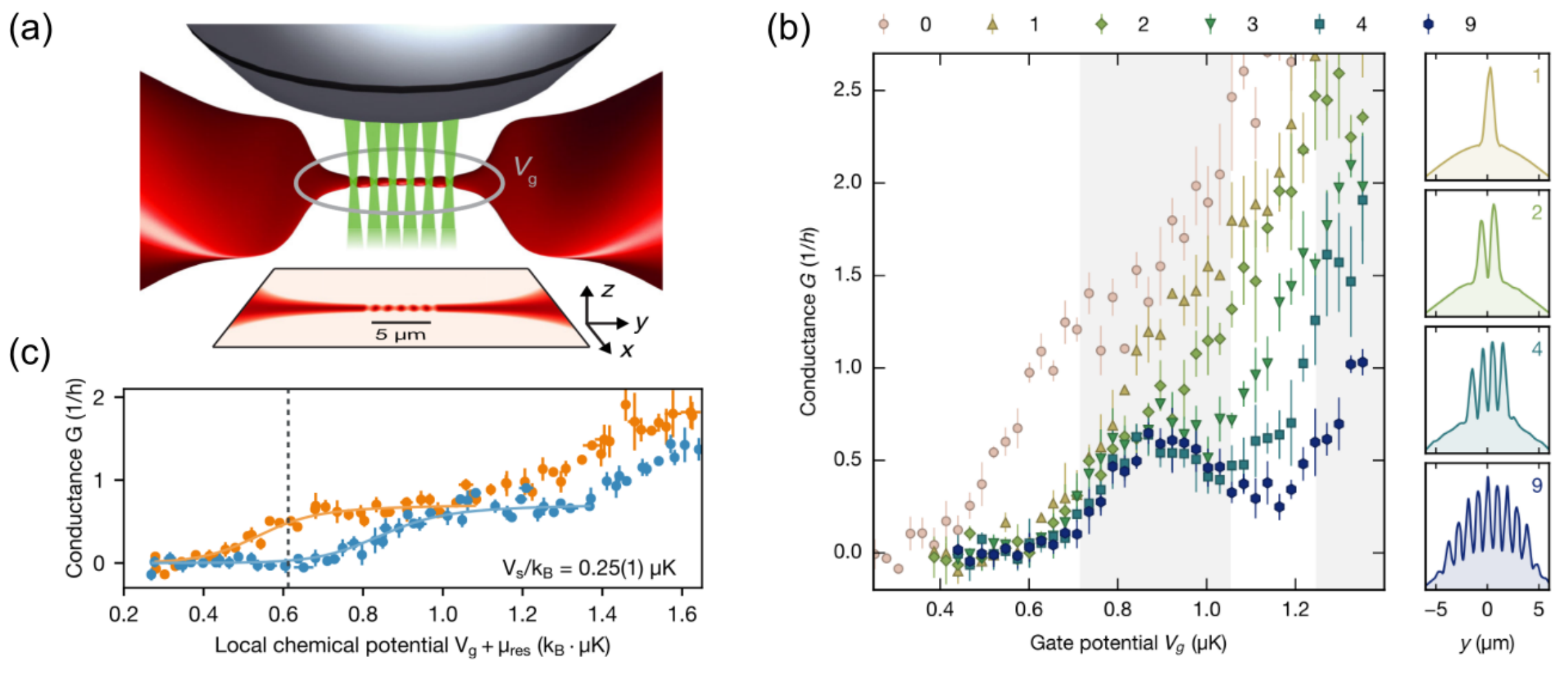}\vspace{-2mm}}
	\caption{\small{Studying mesoscopic transport with ultracold fermions.}
    (\textbf{a}) A degenerate Fermi gas of lithium-6 atoms is shaped using repulsive light potentials into a one-dimensional channel smoothly connected to two macroscopic reservoirs. Spatial light modulation techniques combined with high-resolution optics allow to imprint additional structures to the 1D channel, such as a lattice of variable length.
    (\textbf{b}) For weak interactions, the current through the lattice is proportional to applied bias and particle transport is captured by a linear conductance coefficient. Conductance shows a local minimum as a function of the overall chemical potential, indicating the emergence of a band gap when approaching the infinite lattice limit.
    (\textbf{c}) Spin-dependent quantized conductance in the presence of a near-resonant obstacle focused on the 1D channel, realizing the cold-atom equivalent of a spin filter. 
    Panels (\textbf{a}) and (\textbf{b}) reprinted with permission from M.~Lebrat, P.~Grišins, D.~Husmann, S.~Häusler, L.~Corman, T.~Giamarchi, J.-P.~Brantut, and T.~Esslinger, \textit{Phys. Rev. X} \textbf{8}, 011053 (2018). Copyright 2018 American Physical Society. Panel (\textbf{c}) reprinted with permission from M.~Lebrat, S.~Häusler, P.~Fabritius, D.~Husmann, L.~Corman, and T.~Esslinger, \textit{Phys. Rev. Lett.} \textbf{123}, 193605 (2019). Copyright 2019 American Physical Society.
    }
\label{Fig2}
\end{figure*}

\subsubsection{Josephson currents}\label{sec:Josephson}

The Josephson effect represents a quintessential manifestation of macroscopic quantum phase coherence, stemming from spontaneous symmetry breaking in superfluid states. 
A so-called Josephson junction is typically created by weakly coupling two superfluid order parameters through a thin insulating barrier. In the solid state, this is achieved by separating two superconductors with a nanometer-sized insulating layer. Josephson first predicted that a dissipationless supercurrent $I_s = I_c \sin(\varphi)$ should flow across a tunnel junction in the absence of an applied voltage, associated with the coherent tunnelling of Cooper pairs and sustained only by the relative phase $\varphi$ between the two order parameters. 
The maximum value $I_c$ of the supercurrent is coined the Josephson critical current, and it is directly related to the strength of the tunnel coupling between the two order parameters within the insulating barrier. The measurement of $I_c$ provides a powerful probing tool offering fundamental insights into the microscopic properties of the involved superfluid states, and their robustness against dissipation. For example, for BCS superconductors $I_c$ is directly related to the order-parameter amplitude, i.e. the gap $|\Delta|$, by the Ambegaokar-Baratoff relation. For applied currents above $I_c$, the junction enters a resistive regime, where a non-zero stationary conductance arises from dissipative excitation processes and a finite electro-chemical potential response is generated across the junction.

Experimental studies with atomic superfluids have so far mainly targeted coherent transport in BECs with various geometries and optically engineered weak links \cite{albiez2005direct,levy2007ac,leblanc2011dynamics,ramanathan2011superflow,jendrzejewski2014resistive,ryu2013experimental,eckel2014hysteresis,eckel2014interferometric}. On the other hand, the study of supercurrents between weakly coupled superfluid Fermi gases is of high relevance both from the fundamental and the practical point of view \cite{stadler2012observing,husmann2015connecting,valtolina2015josephson}, since transport therein is crucially influenced and complicated by strong inter-particle interactions and their interplay with fermionic statistics. Only recently, dc Josephson supercurrents have been observed in strongly interacting Fermi gases close to a Feshbach resonance \cite{kwon2020strongly}. Reminiscent of the behavior of the Landau critical velocity across the BCS-BEC crossover \cite{ketterle2008making}, the Josephson critical current was found to exhibit a pronounced maximum around unitarity, resulting from opposite variations of the chemical potential and the pair condensate fraction, the latter playing the role of the order-parameter amplitude throughout the crossover \cite{kwon2020strongly}. 
First experimental investigations of the Josephson effect in quasi-two-dimensional fermionic condensates have also been reported recently, providing information on the connection between condensation and the Berezinskii–Kosterlitz–Thouless superfluid transition \cite{Luick2020}. 
Moreover, experiments showed the breakdown of coherent Josephson transport to be accompanied by the nucleation of topological defects, generated above critical flows by the barrier constriction and subsequently emitted into the superfluid bulk \cite{burchianti2018connecting,xhani2020critical}.
More efforts will be necessary to shed light on the precise mechanisms underlying dissipation in Josephson junctions between crossover Fermi superfluids, and on the interplay of bosonic and fermionic excitation mechanisms, corresponding to the Bogoliubov-Anderson and pair-breaking excitation branches observed for an obstacle moving through the superfluid \cite{combescot2006collective,miller2007critical,weimer2014critical,park2018critical}. 
Such explorations will be essential for our understanding of dissipative transport in highly correlated fermionic systems, and for extending the applications of the Josephson effect to emerging atomtronic devices.

\subsection{Fermionic transport in mesoscopic channels}\label{sec:Meso}

Mesoscopic devices refer to small-size systems whose transport properties are influenced by the quantum nature of matter.
For example, the conductance of a narrow constriction becomes quantized when its width is comparable to the de Broglie wavelength of the particles traversing it. Initially demonstrated with electrons in semiconducting nanostructures \cite{vanwees1988quantized}, mesoscopic transport can be naturally extended to fermionic atoms.

As quantum gases have to be particularly shielded from environmental perturbations they are intrinsically closed systems, which is both a blessing and a complication to study mesoscopic transport phenomena.
On the one hand, the relaxation of thermodynamical quantities involved in transport such as momentum or spin mostly depend on interparticle interactions, which can be tuned for instance via Feshbach resonances.
On the other hand, real-life transport measurements with electrons imply connecting macroscopic leads acting as particle and heat baths to a smaller system of interest, usually treated as a grand canonical ensemble.
With quantum gases, such a paradigm for transport requires partitioning the isolated system into a mesoscopic conductor and two or more macroscopic reservoirs that thermalize fast enough compared to the transport timescales to be considered in thermodynamical equilibrium.
Cold-atom realizations close to such multi-terminal setups include single and multiple Josephson junction arrays of trapped BECs \cite{albiez2005direct,labouvie2015negative}, weak links in ring traps \cite{wright2013driving,eckel2014hysteresis} and planar junctions between two fermionic superfluids \cite{valtolina2015josephson,burchianti2018connecting,kwon2020strongly}.

By reducing the dimensionality of the mesoscopic region, the atomic equivalent of quantum point contacts has been realized, displaying quantized conductance \cite{krinner2015observation}.
Starting from this two-terminal configuration, more complex structures can be engineered by projecting arbitrary light potentials via holographic techniques
\cite{haeusler2017scanning}.
Recently, this technique allowed to investigate the insulating properties of a mesoscopic lattice imprinted site by site within a quantum wire \cite{lebrat2018band}, visible as a suppression of conductance at Fermi energies located in the lattice band gap. 
Strikingly, this insulating behavior persists as attractive interactions are increased to the point where reservoirs become superfluid.
The robustness of the fermionic character of transport can be attributed to the existence of a Luther-Emery liquid \cite{luther1974backward}, a strongly correlated phase of matter distinctive of the 1D character of the channel.
In a more recent set of experiments, optical control in atomic point contacts was extended to spin by using light tuned close to atomic resonance to create local effective Zeeman shifts.
This leads to the realization of an ideal spin filter with cold atoms, one of the most fundamental spintronic devices \cite{lebrat2019quantized}.
In the presence of weak interactions, near-resonant light scattering can be entirely accounted for by including losses in a Landauer-B\"uttiker model \cite{corman2019quantized}.
Such progress towards spin-dependent transport opens avenues for exploring the transport dynamics of strongly correlated heterostructures, where novel nonequilibrium spin and heat transport \cite{bauer2010spin,bergeret2018colloquium} and exotic phases of matter \cite{beenakker2013search} could be observed.

\subsection{Fast spin drag in normal Fermi gases}\label{sec:SpinDrag}

Spin drag is a ubiquitous concept in many branches of physics. It is usually associated with interaction effects which affect the Euler equation for the spin current. Spin drag can be of collisional nature, giving rise to spin diffusion since collisions do not conserve the spin current, or of collisionless nature, being at the origin of non-dissipative dynamics \cite{duine2009spin,sommer2011universal,goulko2013spin,trotzky2015observation,valtolina2017exploring,fava2018observation,nichols2019spin,enss2019universal}. 
Experiments on transverse spin diffusion \cite{trotzky2015observation} (Leggett-Rice effect) in an ultracold Fermi gas along the BCS-BEC crossover have allowed for the determination of the relevant combination $G_0-G_1/3$ of the spin parameters of Landau theory of Fermi liquids. In particular, the parameter $G_1$ accounts for the strength of spin-current interactions. 
Spin drag can also be due to the modification of the equation of continuity in the spin channel, caused by interactions and yielding a violation of the corresponding f-sum rule.  An example of such spin drag behavior of collisionless nature (hereafter called ``fast spin drag'') takes place in the Andreev-Bashkin effect, caused by quantum fluctuations in a mixture of two interacting superfluids \cite{fil2005nondissipative,nespolo2018andreev}. This effect is very tiny and difficult to observe in dilute Bose gases, unless one considers one-dimensional configurations \cite{parisi2018spin} or quantum gases in an optical lattice \cite{nespolo2018andreev}. 
In the following, we will discuss some consequences of fast spin drag in a normal (non-superfluid) interacting mixture of two Fermi gases, where the effect can be sizable and hopefully measurable. Other non-trivial examples examples of fast spin drag concern the dynamical behavior of coherently coupled Bose-Einstein condensed mixtures. 

To investigate the phenomenon of fast spin drag, it is convenient to consider an external perturbation of the form $H_{pert} = -\lambda f({\bf r}) \,\Theta(t)$ applied to the system, where $\Theta(t)$ is the usual Heaviside step function (equal to $0$ for $t<0$  and $1$ for $t>0$) and the function $f({\bf r})$ characterizes the nature of the perturbation, while $\lambda$ is its strength. 
For example, in an ultracold atomic gas a convenient choice is $f({\bf r})=x$, corresponding to a boost generated by an optical potential. If the perturbation is equally applied to both components of the mixture, the velocity acquired by the system is given, for short times, by $v_x= \lambda t/m$, where $m$ is the atomic mass, and we have set $\hbar=1$. A more interesting scenario occurs when the perturbation is applied in a selective way only to one component (hereafter called component $1$). In this case, the velocities acquired by the two components will be different, and for short times they can be easily calculated starting from the many-body wave function of the system which, in the presence of a fast perturbation, takes the form  $\Psi(t)= \exp[i\lambda tx_1 -iHt]\Psi_0$  with $x_1$ the center-of-mass operator relative to the first component. The velocities acquired by the two components are then given by 
\begin{align}
\textcolor{black}{v_1} &\textcolor{black}{=\frac{d}{dt}\langle x_1\rangle = 2\lambda t \langle [x_1,[H,x_1]]\rangle/N= \lambda t (m_1^n+m_1^s)/N\,,}\label{v1aaa}\\
\textcolor{black}{v_2} &\textcolor{black}{=\frac{d}{dt}\langle x_2\rangle = 2\lambda t \langle [x_1,[H,x_2]]\rangle/N = \lambda t (m_1^n-m_1^s)/N\,,}\label{v2aaa}
\end{align}
where $N$ is the total number of particles and we have expressed the double commutators as proper combinations of the energy-weighted moments  $m_1^n=\int d\omega \,\omega \,S(x_1+x_2, \omega)$ and $m_1^s=\int d\omega\, \omega\, S(x_1-x_2, \omega)$  of the dynamic structure factors relative, respectively, to the {\it in-phase} ($x_1+x_2$) and {\it out-of-phase} ($x_1-x_2$) operators. 

In the absence of current interactions, only the component $1$ will feel the external kick. In fact in this case the commutator $[H,x_2]$ commutes with $x_1$, being uniquely fixed by the kinetic energy term in the Hamiltonian, and consequently $v_2=0$ and $m_1^n=m_1^s$.  
While the {\it in-phase} energy-weighted moment $m_1^n$ is given by the model-independent f-sum rule $m_1^n=N/2m$, as a consequence of Galilean invariance, the {\it out-of-phase} moment is instead sensitive to the presence of spin-current interactions \cite{aldrich1976zero}, and in uniform matter it takes the value 
\begin{equation}
\textcolor{black}{m_1^s= N\frac{1}{2m^*}(1+G_1/3)}
\label{Aldrich}
\end{equation}
where $m^*=m(1+F_1/3)$ is the effective mass of quasi-particles, fixed by the Landau's parameter $F_1$ accounting for density-current interactions and determining the low-temperature behavior of the specific heat   \cite{pines1966theory}, while $G_1$ is the spin-current interaction parameter.  
Result (\ref{Aldrich}) reflects the fact that particle-hole excitations, properly accounted for by Landau's theory of Fermi liquids, do not exhaust the f-sum rule, multipair excitations playing a crucial role in providing the remaining difference. The spin energy-weighted sum rule (\ref{Aldrich}) was actually employed \cite{dalfovo1989sum} to estimate the average multipair excitation energy in liquid $^3$He.

The violation of the f-sum rule in the spin channel is responsible for the spin drag effect in a normal Fermi liquid, which, according to Eqs.~(\ref{v1aaa})-(\ref{v2aaa}), takes the form
\begin{equation}
\textcolor{black}{\frac{v_2}{v_1} = \frac{(F_1-G_1)/3}{2+ (F_1+G_1)/3} \; ,}
\label{Fermi}
\end{equation}
revealing that the drag effect ($v_2\ne 0$) vanishes only if the Landau parameters $F_1$ and $G_1$ are equal. 
In a dilute Fermi gas the values of the Landau's parameters are available in both three \cite{lifshitz1960statistical,recati2011spin} and two dimensions \cite{engelbrecht1992landau}, using second-order perturbation theory. For example in 3D one has:
\begin{equation}
\textcolor{black}{F_1= \frac{8}{5\pi^2}(7 \ln2 -1)(k_Fa)^2 \; \; \; ; \; \; G_1= -\frac{8}{5\pi^2}(2+ \ln2)(k_Fa)^2}\\[1mm]
\label{FG}
\end{equation}
showing explicitly that fast spin drag is quadratic in the dimensionless parameter $k_Fa$, where $a$ is the $s$-wave scattering length and $k_F$ is the Fermi wave vector. Similarly to the Andreev-Bashkin effect in interacting superfluids \cite{fil2005nondissipative,nespolo2018andreev}, the fast spin drag exhibited by a normal Fermi gas has consequently a typical beyond-mean-field nature. 

The applicability of Landau theory of Fermi liquids, yielding result (\ref{Fermi}) for the fast spin drag, holds for temperatures much smaller than the Fermi temperature $T_F$. At the same time, the temperatures should be higher than the critical temperature for superfluidity. These conditions can be well satisfied experimentally in the BCS regime of small and negative scattering lengths. For larger values of $k_F|a|$, when the system approaches the unitary regime, its applicability is instead questionable because the superfluid critical temperature is of the order of the Fermi temperature. The experimental determination of the fast spin drag effect along the BEC-BCS crossover would then complement the measurements of the Leggett-Rice effect\cite{trotzky2015observation}, providing a crucial test of Landau's theory and stimulating further theoretical work on spin transport phenomena.

\subsection{Concluding remarks and outlook}\label{sec:Outlook-fermion}
The investigation of transport and dissipation in Fermi gases has only started recently, and many new directions are already emerging. The available light-shaping techniques allow in principle for complex geometries to be investigated \cite{henderson2009experimental,gaunt2012robust}. A particularly appealing situation is the ring trap, which has been successfully explored for weakly interacting bosons \cite{ryu2007observation}. Complex geometries are accessible using the concept of synthetic dimensions \cite{celi2014synthetic}, where multi-terminal geometries are naturally arising from two physical terminals \cite{salerno2019quantized}. Transport of correlated fermions in optical lattices has started recently in bulk lattice systems with promising results on the quantum simulation of the Hubbard model \cite{nichols2019spin,brown2019bad,anderson2019conductivity}.   

The intrinsically low energy scales also implies that currents are weak: translated into electronic scales, the typical currents of fermionic particles in a single mode conductor amounts to fractions of femto-Amperes. Reaching a signal-to-noise ratio comparable with that achieved in the condensed matter context, which would allow for a direct validation of a quantum simulation approach to transport, calls thus for new methods of detection. A practical route is the combination of existing transport systems with single atom sensitive methods that have been demonstrated already, such as fluorescence based counting \cite{neuhauser1980localized,bucker2009single,hume2013accurate} or quantum-gas microscopy \cite{gross2015microscopy}. These methods provide ultimate sensitivity, but still suffer from the technical effects of sample-to-sample preparation noise that scales unfavorably with the number of particles. Ultimately, the limit to signal-to-noise is set by measurement back-action. Achieving this limit would then allow for the reconstruction of the full counting statistics of transport process. Several schemes interfacing atoms with photons in a cavity allow in principle to achieve this limit either in the lattice context \cite{laflamme2017continuous} or for the two-terminal configuration \cite{uchino2018universal}, and experiments directed at implementing such methods have already started. 


The physics of complex atomtronics devices featuring Fermi gases with strong interactions opens many possibilities, also of interest in the condensed matter community at large. An overarching goal is the manipulation of topological superfluids \cite{bernevig2013topological}, such as $p$-wave superfluids or Kitaev chains \cite{kitaev2001unpaired}, which would provide an avenue to study topologically protection of quantum information, thus guiding the development of topological quantum computers.  




%
\section{TRANSPORT IN BOSONIC CIRCUITS}
\vspace*{-0.5cm}
\label{TransportBose}
\par\noindent\rule{\columnwidth}{0.4pt}
{\bf{\small{T. Haug, R. Dumke,  L.-C. Kwek, W. von Klitzing, L. Amico}}}
\par\noindent\rule{\columnwidth}{0.4pt}
%



Atomtronics opens up a new approach to study fundamental problems of transport of quantum matter in various settings with widely different light-controlled atomic circuits \cite{dumke2016roadmap}. Of particular interest is transport generated by attaching a circuit to reservoirs, that induce a directed current through the system.
Transport between atomic reservoirs has been studied to realize fundamental condensed matter systems \cite{brantut2012conduction,krinner2015observation,husmann2015connecting,krinner2017two}. Furthermore,  precise control of the light potentials allow to transport bosonic fluids at hypersonic speeds in ring circuits \cite{pandey2019hypersonic,guo2020supersonic} and in a coherence preserving manner \cite{loiko2014coherent}. Often these basic atomic circuits can be understood by using a simplified lumped element model, that relies only on a few elements \cite{gauthier2019atomtronic}. From there, larger circuits composed of many basic circuits could be constructed to realize large scale atomtronic networks. To this end, there is a considerable interest in understanding the transport through basic circuit elements. 

Recent studies investigated the transport and dynamics in other circuits like rings and Y-junctions \cite{haug2019aharonov,haug2019andreev,haug2019topological,haug2018mesoscopic,navez2016matter,haug2020quantum}.
These systems have been well studied in electronic setups. Surprising differences arise with bosonic atoms instead of fermionic electrons: Andreev reflections, known from superconductor-metal interfaces, can also occur at the interface of two  bosonic condensates: If the density wave excitation in a one-dimensional condensate is transmitted from the first to second condensate, a hole (an excitation with negative amplitude) is reflected back into the first condensate\cite{tokuno2008dynamics,daley2008andreev,zapata2009andreev,watabe2008reflection,haug2019andreev}. 
For ring circuits, Aharonov-Bohm oscillations occur in the current for electronic systems when an a magnetic field is applied to the ring \cite{webb1985observation}. For bosonic rings, this Aharonov-Bohm effect does not occur \cite{haug2019aharonov,tokuno2008dynamics}. As a first step to observe this effect in cold atoms, a recent experiment demonstrated non-reciprocal transport mediated by artificial magnetic fields in closed loops~\cite{gou2020tunable}.
Transport can also be achieved by driving the system parameter in time. Here, topological pumping has been shown to be a robust way to generate transport~\cite{thouless1982quantized,thouless1983quantization}. Here, the circuit parameters are driven periodically in time such that a directed transport arises, which is protected by topological features of the system \cite{lohse2016thouless,haug2019topological_correlations,tangpanitanon2016topological}. For ring systems with applied flux, the transmission becomes fractional in atom number, and highly entangled states can be generated in the process~\cite{haug2019topological}. This promises important applications in quantum-enhanced sensing and quantum information.

\begin{figure}[htbp]
	\centering
	\subfigure{\includegraphics[width=0.48\textwidth]{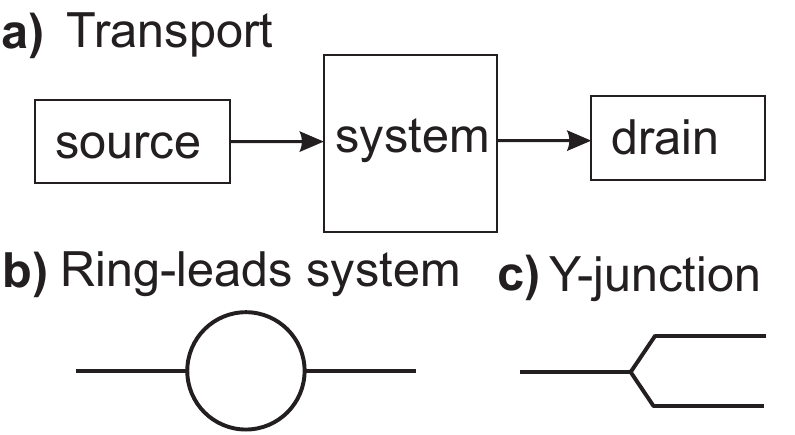}}
	\caption{\idg{a} General transport setup, composed of a source, system and drain. Atoms flow from source, via the system to drain. The current flowing through the system is the quantity of interest, that reveals fundamental features of the system. Specific examples of this kind of setup studied here are a \idg{b} ring-leads system or a \idg{c} Y-junction. }
	\label{Sketch}
\end{figure}

First, we discuss recent advances in matter-waveguides allow to transport cold atoms over long distances (see Sec.\ref{sec:Guides}). Then, we review the transport in two elemental Atomtronic circuits: A ring attached to leads (see Fig.\ref{CentralFigure}, Fig.\ref{currentStatGraph} and Fig.\ref{RingFermionBosonComparison}, and a Y-junction (see Fig.\ref{YDensitysmall} and Fig.\ref{YDensityFermionBosonComp}). A sketch of the systems is shown in Fig.\ref{Sketch}b,c). We investigate different limits: Atoms prepared in a non-equilibrium initial state, with all atoms on one side of the system. Secondly, density wave excitations that propagate through a system filled with atoms. Finally, we also shortly mention topological pumping of atoms in atomtronic circuits (see Fig.\ref{NOON}).

\subsection{\label{sec:Guides}Matterwave guides}
A perfect waveguide allows the guided wave to travel undisturbed over any distance.
In practice, there are always imperfections such as absorption and spatial variations of the guiding potential.
For matterwave guides based on electro-magnetic potentials, absorption plays no role.
In most cases the shapes of the guiding potentials are defined either by a physical structure such as wires in the case of magnetic potentials or by light-fields in the case of dipole traps.  
Examples include, imperfections in the shapes of the wire  \cite{trebbia2007roughness}, the grain size of the copper \cite{jones2003cold} and for the dipole finite amplitude control  \cite{bell2016bose}, diffraction and speckles.
There are a number of solutions to reduce the impact of the imperfections of magnetic waveguides, such as improved manufacturing techniques and periodic current reversal \cite{trebbia2007roughness}. 
Optical techniques employ feedback to image the potential using cold atoms and then correct the imperfections in a feedback loop \cite{bell2016bose}.
Nevertheless, since some level of imperfection in the magnetic wire structure or in the dipole imaging system is unavoidable, waveguides created by artificial structures will always have a certain degree of roughness.

The effect of these imperfections is characterized by the spatial wavelength $\lambda$ and  amplitude $a(k)$  of the modulation of the waveguide potential: A tighter bend will have a stronger effect than a very smooth one. 
For optical traps this strength can be calculated directly by estimating the level of control one has over the optical potential, e.g. by imaging the speckles or by estimating the noise level in the feedback to the shape of the waveguide.
In the case of magnetic waveguides, the variation of the potential can be imaging the break-up of a Bose-Einstein condensate, which is brought close to the wires.
An absolute scale can be established from the resulting images via chemical potential of the BEC.

Increasing the distance between the atoms and the current-carrying conductor decreasing the transverse trapping frequency and reduces the roughness of the waveguide.
For distances $(d)$ from the waveguide much larger than the characteristic wavelength $(\lambda)$, this reduction $(K)$ in roughness can be estimated as a function of the spatial frequency $(k=2\pi/\lambda)$ as \cite{jones2003cold}: 
\begin{equation}
\label{equ:spatial-attenuation}
K(d,\lambda)=(k d)^{-1/2}e^{-k d}
\end{equation}
By increasing $d$ it was possible to observe interference fringes between two condensates after propagating them on a magnetic atomchip waveguide for up to 120\,$\mu$m, albeit reducing the transverse trapping frequency from the kHz level down to $\omega_\perp=2\pi \cdot120$\,Hz \cite{wang2005atom}.
If no propagation is required, even spatially modulated waveguides can exhibit robust coherence \cite{zhou2016robust}.

Very smooth optical dipole matterwave guides can be achieved by weakly focussing a laser beam and taking care to avoid laser speckles.
If  the imaging system is Fourier limited then it cannot produce any structure smaller than the focus, resulting in a perfectly smooth waveguide. 
By the same token, however, no structure other than a simple linear waveguide can be produced by this method. 

A different approach has been recently demonstrated, where the shape of a ring-shaped waveguide is defined by modifying a simple DC quadrupole field using only homogeneous audio-frequency and radio-frequency fields\cite{pandey2019hypersonic}.
In the so-called Time-Averaged Adiabatic Potentials, the radial and vertical confinement is limited to a ring and the  maximum spatial azimuthal frequency that can be addressed is $\phi=4 \pi$.
Since the field generating magnets only have to produce homogeneous and quadrupole fields, they can be far away. 
Eq.(\ref{equ:spatial-attenuation}) predicts a reduction of the field modulations down to a factor $10^{-138} $ of their strength at the magnets, thus practically eliminating them.
This has made it possible to propagate Bose-Einstein Condensates over distances of more than 10\,cm without causing any additional heating.

\begin{figure}[htbp]
		\centering
	\includegraphics[width=0.49\textwidth]{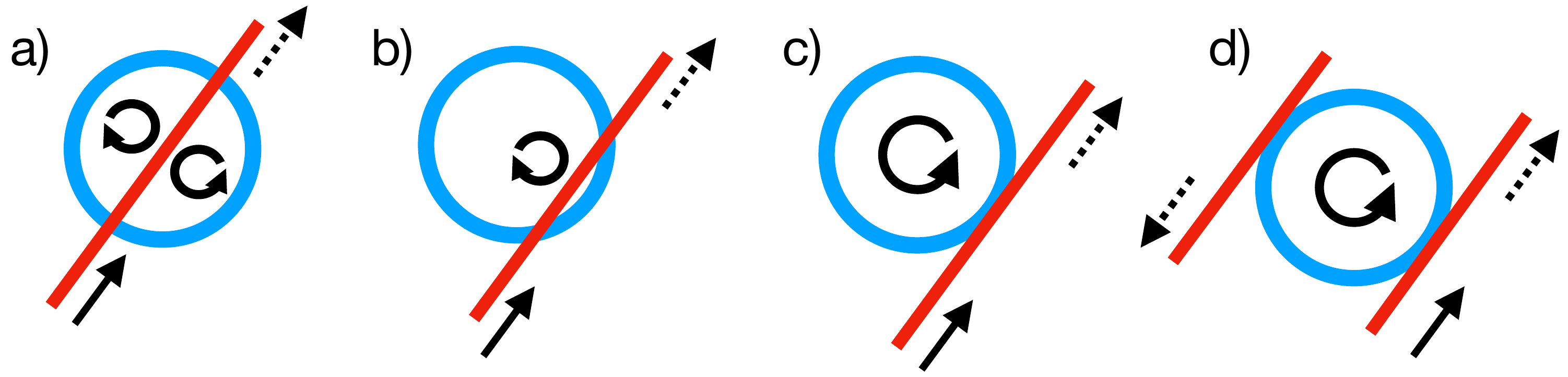}
	\caption{Possible configurations for coupling of a TAAP ring (blue) to an optical guide potential (red). In all cases the atoms are in a magnetically or rotational sensitive state in the TAAP potential, which is then coupled to a magnetically or rotational non-selective state by tunnel coupling or a spatially selective microwave transition. a) a) symmetric ring-lead system, b) asymmetric ring-lead system. c) tangential configuration, where dipole is coupled to only one direction of rotation in the ring. d) TAAP ring coupler between two dipole wave guides.  Note that in a) and b) the coupling on the two sides of the ring can be individually tuned.}
	\label{fig:TAAP-ring-couplers}
\end{figure}

A very interesting perspective is to combine the TAAP rings with optical potentials. 
The standard way to load atoms into the TAAP ring is currently to transfer them from an optical dipole trap \cite{navez2016matter, pandey2019hypersonic}.
Using radio-frequency or microwaves it would be possible to create a beam splitter between the ring and the optical potential. This couples the magnetically or rotational sensitive state in the ring to a magnetically and rotational non-sensitive one in the optical guide. 
Possible configurations would include, for example, a (a-)symmetric ring-lead system (Fig.\,\ref{fig:TAAP-ring-couplers}a and \ref{fig:TAAP-ring-couplers}b), a dipole guide coupled tangentially to the ring (Fig.\,\ref{fig:TAAP-ring-couplers}c). Since the diameter of TAAP rings and therefore their resonant angular momentum is easily tuned they  could act as a velocity-selective resonator. The waveguide could be used to read-out the rotational state of the ring \cite{safaei2019monitoring}
Finally, as shown in Fig.\,\ref{fig:TAAP-ring-couplers}d one could use a TAAP ring to couple two dipole waveguides to each other in a velocity selective fashion, much like wavelength selective multiplexing using tuneable whispering gallery resonators~\cite{boriskina2003tuning,klitzing2001tunable}. 

\subsection{Ring-leads system}\label{Ring}
A widely studied system within mesoscopic physics are rings attached to leads, with an applied magnetic field~\cite{buttiker1984quantum,webb1985observation}. 
This system features the Aharonov-Bohm effect, where currents through the ring can be modulated with an applied magnetic field.
While extensively studied for fermions such as electrons, it is not well understood for bosonic type of systems. Atomtronic setups allow for study of these type of systems in a controlled way for the first time. A theoretical study has been performed in Ref.\cite{haug2019andreev,haug2019andreev} which is reviewed in this section.
Atomtronic setups for transport can be modeled using the Bose-Hubbard model~\cite{bloch2008many}. Here, circuits of ring-leads or Y-junctions as seen in Fig.\ref{Sketch} are modeled as a lattice system. For example for the ring-leads system, the three individual components (source lead, ring and drain) are each modeled as one-dimensional chains with nearest-neighbor tunneling interactions. The different components are then coupled together via tunneling at specific lattice sites.
A ring with an even number of lattice sites $L$ coupled to two leads (see Fig.\ref{Sketch}b) is given by the Hamiltonian ${\mathcal{H}=\mathcal{H}_\text{r}+\mathcal{H}_\text{l}}$.
The ring part of the Hamiltonian is given by
\begin{equation}
\mathcal{H}_\text{r}=-\sum_{j=0}^{L-1}\left(J\expU{i2\pi\Phi/L}\cn{a}{j}\an{a}{j+1} + \text{H.C.}\right)+\frac{U}{2}\sum_{j=0}^{L-1}\nn{}{j}(\nn{}{j}-1)\;,
\end{equation}
where $\an{a}{j}$ and $\cn{a}{j}$ are the annihilation and creation operator at site $j$, $\nn{}{j}=\cn{a}{j}\an{a}{j}$ is the particle number operator, $J$ is the intra-ring hopping, $U$ is the on-site interaction between particles and $\Phi$ is the total flux through the ring. Periodic boundary conditions are applied: $\cn{a}{L}=\cn{a}{0}$.
The two leads dubbed source (S) and drain (D) consist of a single site each, which are coupled symmetrically at opposite sites to the ring with coupling strength $K$. In  both of them, local potential energy and  on-site interaction are set to zero as the leads are considered to be large with low atom density. The lead Hamiltonian is ${\mathcal{H}_\text{l}=-K(\cn{a}{S}\an{a}{0} +\cn{a}{D}\an{a}{L/2} +\text{H.C.})}$,
where $\cn{a}{\text{S}}$ and $\cn{a}{\text{D}}$ are the creation operators of source and drain respectively. 
		
\begin{figure}[htbp]
	\centering
	\includegraphics[width=0.49\textwidth]{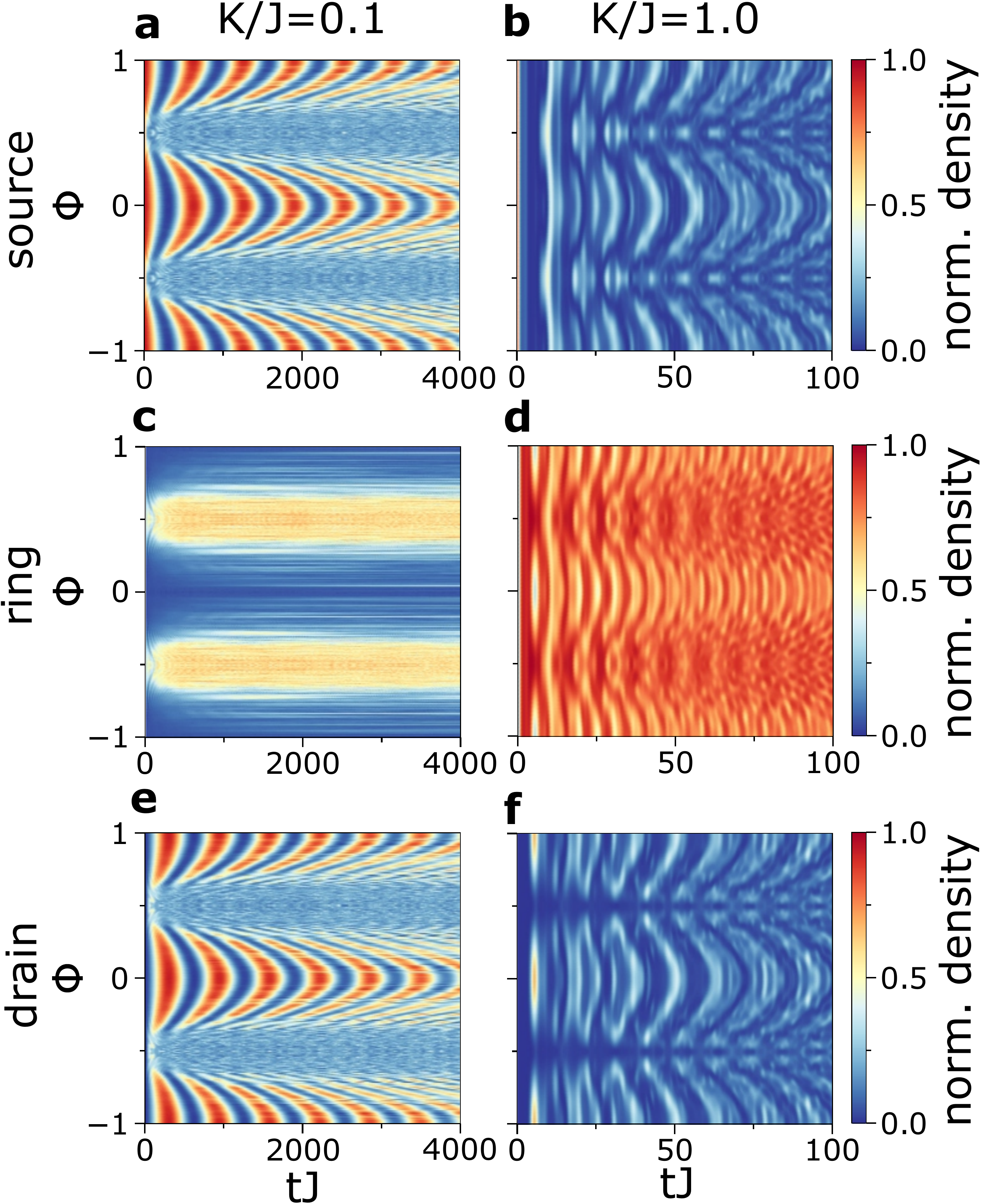}
	\caption{{\bfseries Time evolution of density} in source \idg{a,b}, ring \idg{c,d} and drain \idg{e,f} plotted against flux $\Phi$. \idg{a,c,e} weak ring-lead coupling ${K/J=0.1}$ (on-site interaction ${U/J=5}$). \idg{b,d,f} strong ring-lead coupling ${K/J=1}$ (${U/J=0.2}$). Time is indicated $tJ$ in units of inter-ring tunneling parameter $J$. The number of ring sites is ${L=14}$ with $N_\text{p}=4$ particles initially in the source. The density in the ring is ${n_\text{ring}=1-n_\text{source}-n_\text{drain}}$. Reprinted with permission from T. Haug, H. Heimonen, R. Dumke, L.-C. Kwek, and L. Amico, Phys. Rev. A 100, 041601(R) (2019). Copyright 2019 American Physical Society.}
	\label{CentralFigure}
\end{figure}

The dynamics of transport within this system in various settings is discussed in \cite{haug2019aharonov,haug2019andreev}, which we now review.
A way to probe transport was studied in the case where source and drain consist of only one lattice site each.
Here, the atoms are initially prepared in the source, with ring and drain being empty of atoms. During the time evolution, the atoms flow out of the source lead, and propagate via the ring to the drain.  In the weak-coupling regime ${K/J\ll1}$, the lead-ring tunneling is slow compared to the dynamics inside the ring (see Fig.\ref{CentralFigure}a,c,e). In this regime, the condensate mostly populates the drain and source, leaving the ring nearly empty. As a result, the scattering due to on-site interaction $U$ has a negligible influence on the dynamics.  
With increasing flux $\Phi$ the oscillation becomes faster and the ring populates, resulting in increased scattering and washed-out density oscillations.
In the strong-coupling regime ${K/J\approx1}$, the  lead-ring  and the intra-ring dynamics are characterized by  the same frequency  and cannot be treated separately. Here, a superposition of many oscillation frequencies appears, and after a short time the condensate is evenly spread both in leads and ring (Fig.\ref{CentralFigure}b,d,f). The density in the ring is large and scattering affects the dynamics by washing out the oscillations. Close to ${\Phi=0.5}$, the oscillations slow down, especially for weak interaction, due to destructive interference~\cite{valiente2008two}.

\begin{figure}[htbp]
	\centering
	\subfigure{\includegraphics[width=0.49\textwidth]{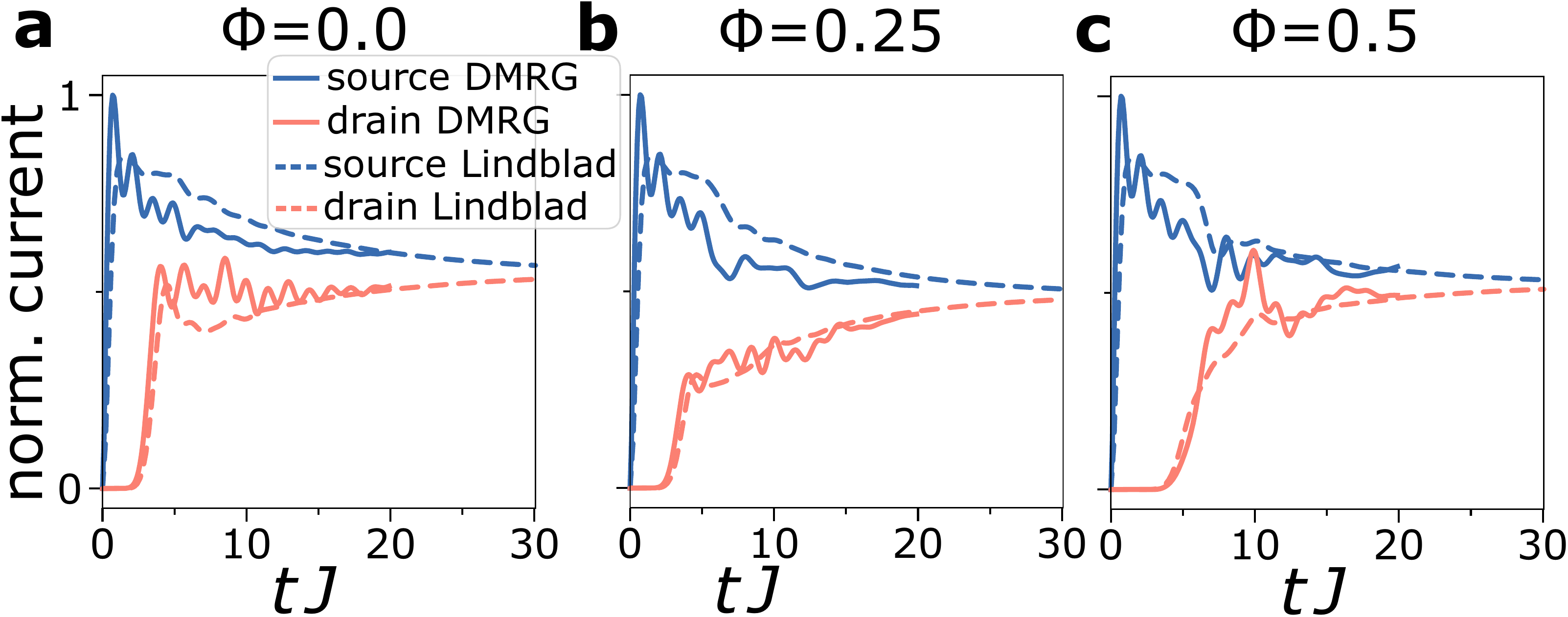}}
	\caption{{\bfseries Current through the Aharonov-Bohm ring} \idg{a-c} Evolution of source and drain current towards the steady state with DMRG (solid line) and Lindblad formalism (dashed) for hard-core bosons, ${K=1}$ and ${L_\text{R}=10}$. For DMRG, both reservoirs and ring are solved with Schr\"odinger equation as a closed system. Source and drain are modeled as chains of hard-core bosons with equal length ${L_\text{S}=L_\text{D}=30}$. Initially, the source is prepared at half-filling ($N_\text{p}=15$) in its ground state (ring and drain are empty) decoupled from the ring (${K(t=0)=0}$). For ${t>0}$ the coupling is suddenly switched on (${K(t>0)=J}$). This setting creates highly non-equilibrium dynamics. Due to numerical limitations, we analyse the short-time dynamics. For the open system, the reservoirs obey Pauli-principle with ${r=0.65}$ and ${\Gamma=1.5}$. Reprinted with permission from T. Haug, R. Dumke, L.-C. Kwek, and L. Amico, Quantum Sci. Technol. 4, 045001 (2019). Copyright 2019 IOP Publishing Ltd. Reprinted with permission from T. Haug, H. Heimonen, R. Dumke, L.-C. Kwek, and L. Amico, Phys. Rev. A 100, 041601(R) (2019). Copyright 2019 American Physical Society.
	} 
	\label{currentStatGraph}
\end{figure}

The authors studied also the dynamics for the case where the source and drain leads consists of many sites, probing the regime of many atoms in an extended system. Here, the large source lead is filled with atoms, and then suddenly coupled to the ring to generate the dynamics.
The authors model this in two ways: In the first case, they solve the full dynamics of leads and ring using DMRG~\cite{white2004real,itensor}. Then, they study an approximate method, where the leads are approximated as a large bath and are traced out. The resulting dynamics is described using the Lindblad master equation. 
\begin{equation*}
\pdif{\rho}{t}=-\frac{i}{\hbar}\left[H,\rho\right]-\frac{1}{2}\sum_m\left\{\cn{L}{m}\an{L}{m},\rho\right\}+\sum_m\an{L}{m}\rho\cn{L}{m}
\end{equation*}
for the reduced density matrix \cite{breuer2002theory} within the Born-Markov approximation with
${L_1=\sqrt{\Gamma}\cn{a}{S}}$, ${ L_2=\sqrt{r\Gamma}\an{a}{S}}$, and ${L_3=\sqrt{ \Gamma}\an{a}{D}}$ ($r$ characterizes the back-tunneling into the source reservoir). 
Then, the steady state of the density matrix is solved ${\pdif{\rho_\text{{SS}}}{t}=0}$~\cite{guo2017dissipatively}. The operator for the current is ${j=-iK(\cn{a}{\text{S}}\an{a}{0}-\cn{a}{0}\an{a}{\text{S}})}$ and its expectation value is ${\avg{j}=\text{Tr}(j\rho_\text{SS})}$. 
In Fig.\ref{currentStatGraph}. a) -c), the authors compare the open system Lindblad approach with a full simulation of both ring and reservoirs using DMRG \cite{white2004real,itensor}. Both methods yield similar results, with the Lindblad approximation smoothing out the oscillation found in DMRG. This shows that leads modeled as Markovian bath without memory is sufficient to describe the dynamics. 
Further, they investigate the evolution of the current towards the steady-state. They find that the initial dynamics depends on the flux, which is a sign of the Aharonov-Bohm effect being initially present. However, the steady-state reached after long times is nearly independent of flux, demonstrating the absence of the Aharonov-Bohm effect in the steady state for interacting bosons.

\subsection{Y-junctions}\label{YExc}
Y-junction is a system consisting of three one-dimensional chains, which are coupled together at a single point (see Fig.\ref{Sketch}c). They have been of wide interest in mesoscopic physics for electronic type systems~\cite{imry2002introduction}. For cold atoms, such systems have been proposed and realized experimentally \cite{kevrekidis2003guidance,andersen2018physical,ryu2015integrated,birkl2001atom,dumke2002interferometer,kreutzmann2004coherence}.
In Ref.\cite{haug2019andreev}, the authors studied theoretically the dynamics of density wave excitations that are transmitted and reflected in a cold atom Y-junction. They are interested in how the system evolves for varying the coupling strength of the Y-junction. They find characteristic regimes of transmission and reflections.

The Hamiltonian for the Y-junction is ${\mathcal{H}_\text{S}+\mathcal{H}_\text{D}+\mathcal{H}_\text{I}}$, with the source lead Hamiltonian (analogue for the two drain leads) 
\begin{equation}\label{HamiltonSource}
\mathcal{H}_\text{S}=-\sum_{j=1}^{L_\text{S}-1}\left(J\cn{s}{j}\an{s}{j+1} + \text{H.C.}\right)+\sum_{j=1}^{L_\text{S}}\frac{U}{2}\nn{s}{j}(\nn{s}{j}-1)\;,
\end{equation}
where $\an{s}{j}$  and $\cn{s}{j}$  are the annihilation and creation operator at site $j$ in the source lead, ${\nn{s}{j}=\cn{s}{j}\an{s}{j}}$ is the particle number operator of the source, $J$ is the intra-lead hopping, $L_\text{S}$ the number of source lead sites and $U$ is the on-site interaction between particles. All units are rescaled in terms of the hopping term $J$. The Hamiltonian $\mathcal{H}_\text{D}$ for the two drain leads have similar Hamiltonians, where one replaces the index $s$ with respective $d$ (for first drain) and $f$ (second drain), and defines the drain length $L_\text{D}$. 
The coupling Hamiltonian between the source lead and the two drain leads is
\begin{equation}
\mathcal{H}_\text{I}=-K\cn{s}{1}\left(\an{d}{1} + \an{f}{1}\right)+ \text{H.C.}\;,
\end{equation}
where $K$ is the coupling strength between source and drain leads. 
The current through the Y-junction is defined as
\begin{equation}\label{Ycurrent}
j_\text{Y}=-iK\cn{s}{0}\an{d}{0} + \text{H.C.} \;,
\end{equation}

To study the propagation of a density excitation through the setup, the authors prepare the system in the ground state of the full Hamiltonian with initially a small local potential offset in the lead Hamiltonian. This will create a localized density bump in the source lead.
Then, they add the following Hamiltonian for the offset potential to the source Hamiltonian
\begin{equation}
\mathcal{H}_P=-\epsilon_\text{D}\sum_{j=1}^{L_\text{S}}\exp\left(-\frac{(j-d)^2}{2\sigma^2}\right)\nn{s}{j}\;,
\end{equation}
where $d$ is the distance of the initial excitation to the junction.
At the start of the time evolution the offset potential is instantaneously switched off. The density bump will propagate as a density excitation in both positive and negative direction, where here only the forward direction is regarded.

The authors develop a method to calculate the amount of transmitted and reflected density waves. They calculate the total density of the incoming wave by taking the first $a$ sites of the source lead at a specific time $t_\text{in}$ when the density waves has entered this region, and subtracting from it the density at time ${t=0}$ before the wave has entered the region
\begin{equation}\label{EqInc}
N_\text{inc}=\sum_{i\in {a\text{ sites of source}}}\left[n_i(t_\text{in})-n_i(0)\right]\;. 
\end{equation}
Here, $n_i(t)$ is the expectation value of the density at the $i$-th site of the system at time $t$.
The transmission coefficient is found by dividing the change in atom number in the drain density by the total density of the incoming wave
\begin{equation}\label{EqTrans}
T=\frac{\sum_{i\in\text{drain}}\left[n_i(t)-n_i(0)\right]}{N_\text{inc}}
\end{equation}
and the reflection coefficient as 
\begin{equation}\label{EqRef}
R=1-T\;.
\end{equation} 

\begin{figure}[htbp]
	\centering
	\subfigimgraised[width=0.24\textwidth]{a}{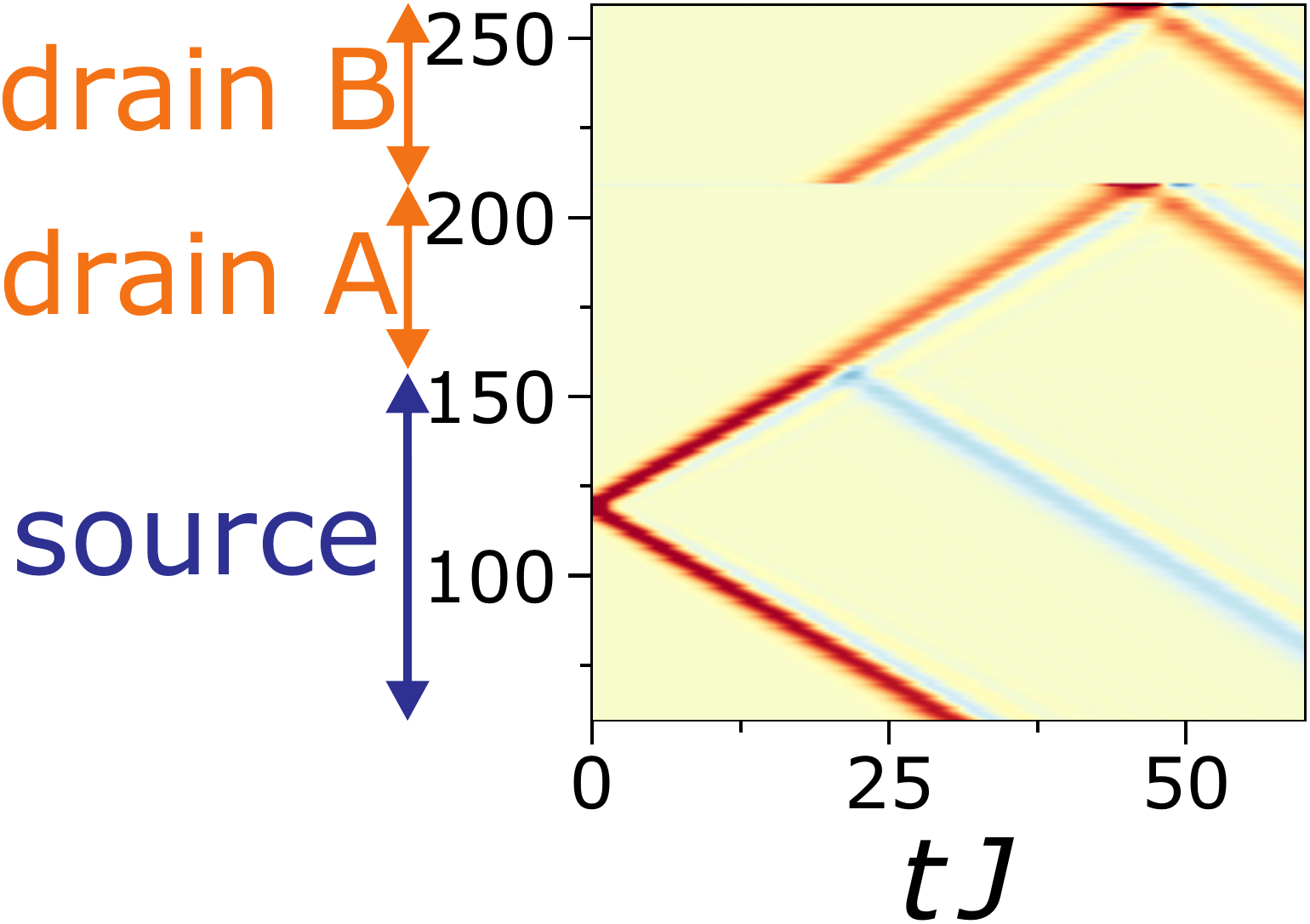}\hfill
	\subfigimg[width=0.24\textwidth]{b}{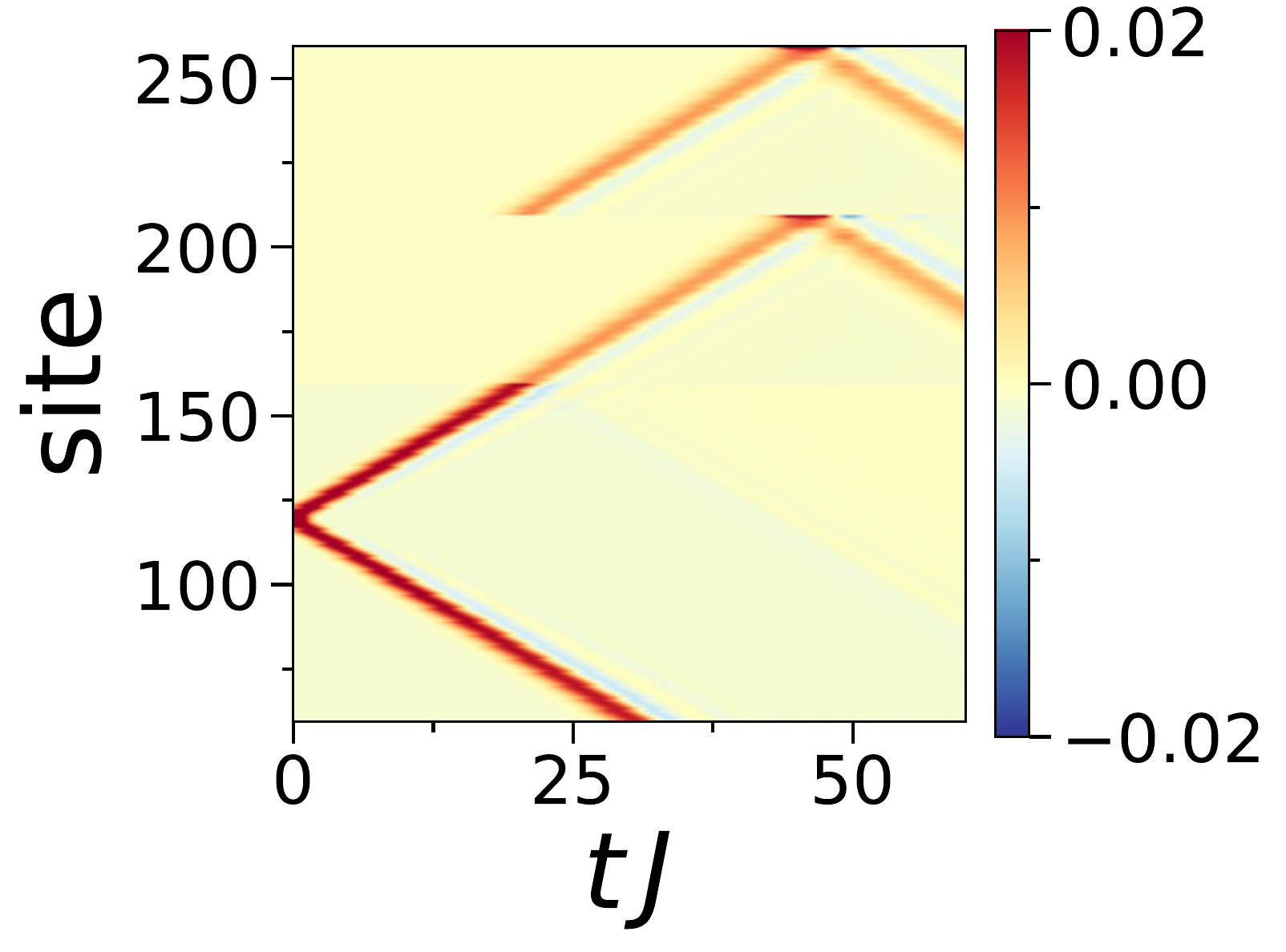}
	\subfigimg[width=0.24\textwidth]{c}{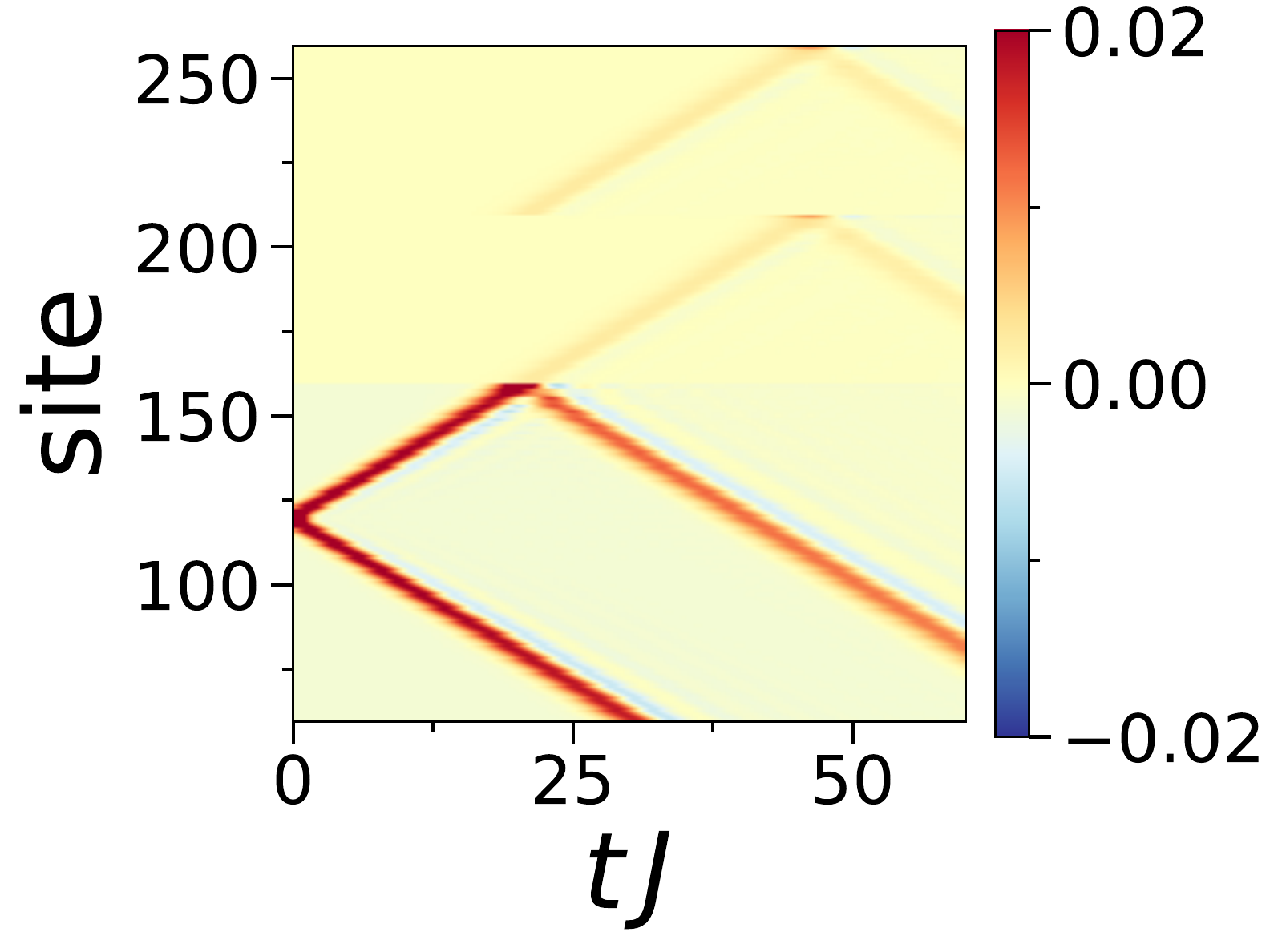}\hfill
	\subfigimg[width=0.24\textwidth]{d}{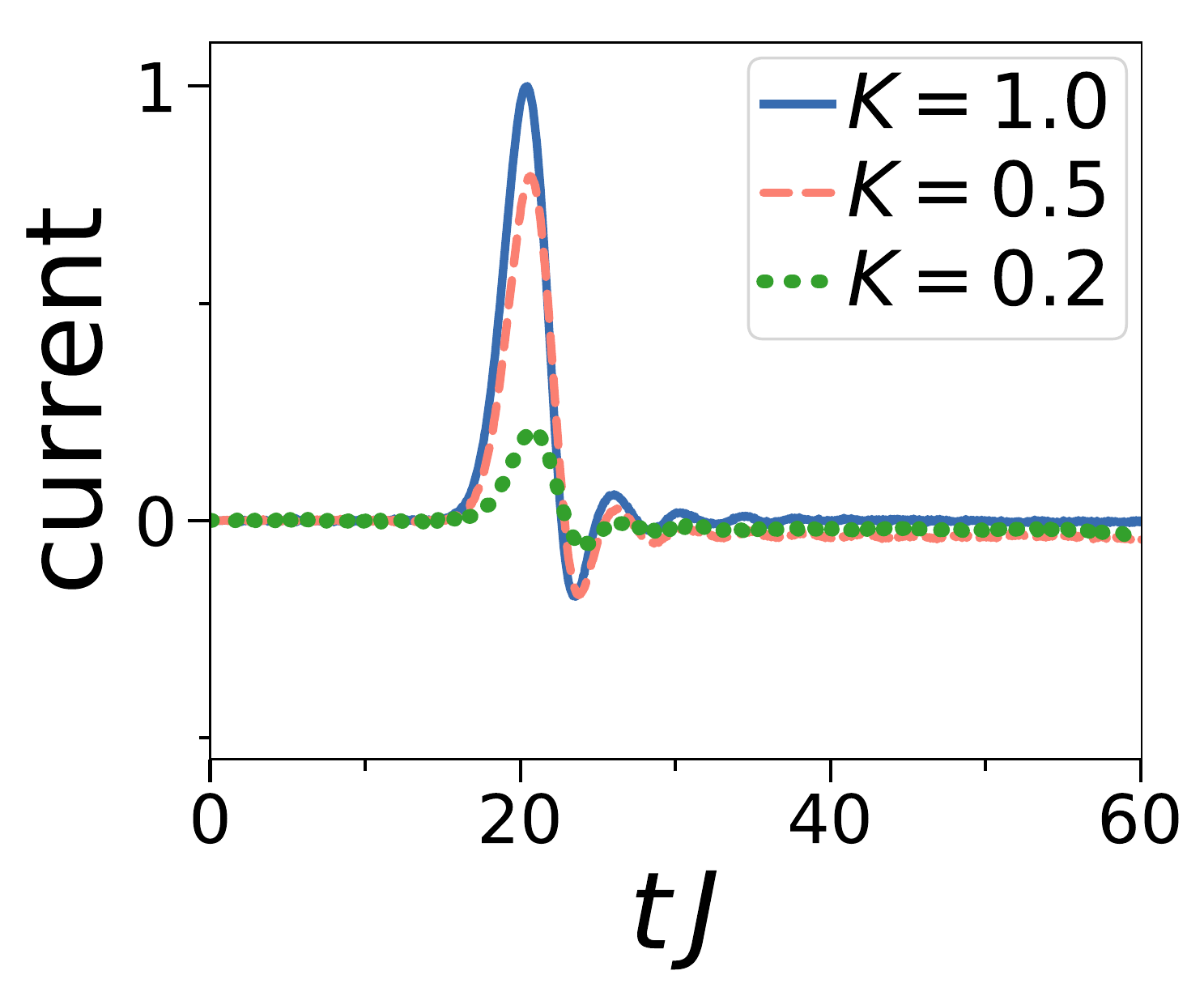}\\
	\subfigimg[width=0.32\textwidth]{e}{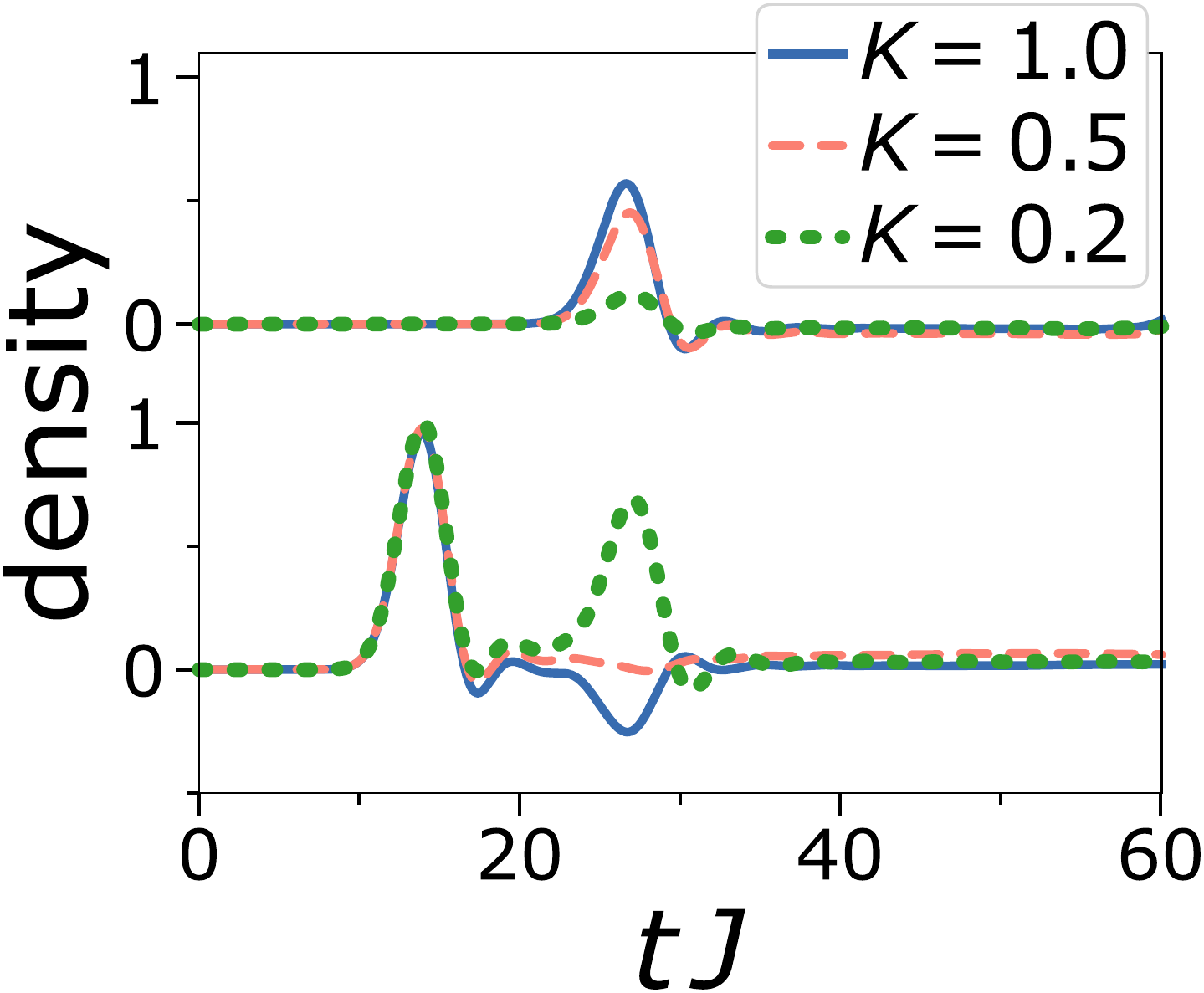}
	\caption{Propagation of small excitation in a Y-junction for the hard-core boson model. The source lead has length ${L_\text{S}=160}$, the drain lead each ${L_\text{D}=50}$, the particle number ${N=130}$, initial distance of the excitation to the junction ${d=40}$ and ${\epsilon_\text{D}=0.3J}$. The source lead is from site 1 to 160, the first drain lead from 160 to 210, and the second one from site 210 to 260. The coupling at the junction (site 160) is \idg{a} ${K=1J}$, \idg{b} ${K=0.5J}$, \idg{c} ${K=0.2J}$. \idg{d} Current through the junction (Eq.\ref{Ycurrent}) in time. \idg{e} \densdescr{170}{175}{145}{150}  For ${K=1J}$ (solid) we observe a negative reflection (Andreev-like), ${K=0.5J}$ (dashed) nearly no reflection, ${K=0.2J}$ (dots) a large positive reflection amplitude. 		Reprinted with permission from T. Haug, R. Dumke, L.-C. Kwek, and L. Amico, Quantum Sci. Technol. 4, 045001 (2019). Copyright 2019 IOP Publishing Ltd. }
	\label{YDensitysmall}
\end{figure}

	\begin{table}
	\begin{tabular}{l|*{3}{c}}
		& ${K=1}$ & ${K=0.5}$ & ${K=0.2}$ \\
		\hline
		transmission & $1.332$ & $0.947$ & $0.207$   \\
		reflection   & $-0.332$ & $0.053$ & $0.793$
	\end{tabular}
\caption{	The table below shows the transmission and reflection coefficients, calculated at ${t=31/J}$ with Eq.\ref{EqInc}-\ref{EqRef} (${t_\text{in}=15}$, ${a=30}$).}
	\end{table}	

The limiting cases of infinitely strong on-site interaction with hard-core bosons is presented in Fig.\ref{YDensitysmall}.
In Fig.\ref{YDensitysmall}a-c, the authors study the propagation for different values of lead coupling $K$. In the source lead an initial excitation is prepared as a small patch of increased density. At ${t=0}$, the potential offset is quenched, and the excitation starts moving in forward and backward direction. The forward moving part of the wave propagates from the source through the junction to the two drain leads.  At the junction between the chains (site 160 in the graph) the wave is both transmitted and reflected. 
For the reflection amplitude, three characteristic reflection regimes are found, which are controlled by the junction coupling $K$.

First, look at the reflection peak as seen in Fig.\ref{YDensitysmall}e) at time ${t J=27}$. 
In the strong coupling regime ${K=1J}$, a negative (Andreev-like) reflection amplitude peak is found. 
For the intermediate coupling regime ${K\approx0.5}$ the back reflection amplitude is very small, and the reflected wave consists of a small, first positive and then negative part, of nearly equal weight.
Finally, for the weak coupling regime with $K$ small, a large positive back-reflection with small transmission is found.
In the table below Fig.\ref{YDensitysmall}, the total transmitted and reflected density at time ${t=31/J}$ is calculated using Eq.\ref{EqInc}-\ref{EqRef}. This gives the transmission and reflection coefficient of the density wave packet. For strong coupling ${K=1J}$ with Andreev reflections, the transmission coefficient is nearly $T\approx 4/3$, which corresponds to the theoretical value predicted for a Y-junction in the limit of weak interaction, within the Gross-Pitaevskii equation~\cite{haug2019andreev}. In this regime, the transmission is larger than the initial density wave, owing to the  the negative reflection $R\approx -1/3$.
Similar dynamics is also found also for finite interaction $U$ within the Bose-Hubbard model~\cite{haug2019andreev}.

\subsection{Differences between fermions and hard-core boson}
Bosons and fermions differ fundamentally in their particle exchange relations: The bosonic many-body wavefunction is symmetric, while fermions are anti-symmetric under exchange of two particles. As a result of these properties, the Pauli principle is enforced for fermions: at a single site only zero or one fermion can exist, while non-interacting bosons do not have this restriction. However, for strongly interacting bosons in the hard-core limit, only one hard-core bosons can be at a single site, mimicking the Pauli principle, while maintaining a symmetric many-body wavefunction. In one dimension, hard-core bosons and fermions are equivalent and a mapping between fermions and hard-core bosons exists, however  this is not the case beyond one-dimensional systems.
The effect of this feature on transport has been studied numerically in detail in Ref.\cite{haug2019andreev}.

\begin{figure}[htbp]
	\centering
	\subfigimg[width=0.33\textwidth]{}{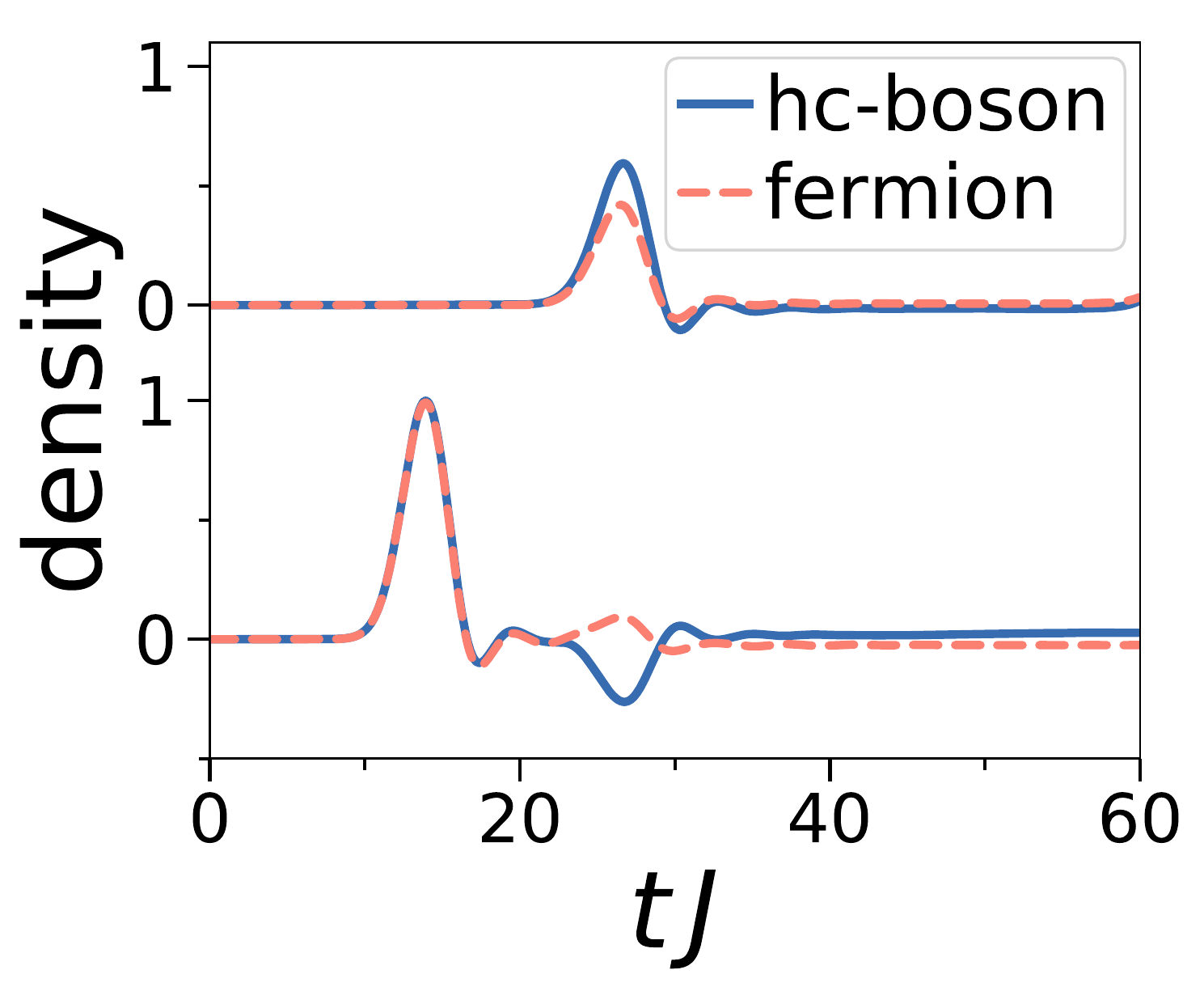}
	\caption{Comparison between hard-core bosons and spinless fermions for the transmission and reflection of a small density excitation in a Y-junction. Bosons show clear negative Andreev-reflection (solid and dashed curve in center bottom), in contrast to fermions. The source lead has length ${L_\text{S}=160}$, the drain lead each ${L_\text{D}=50}$, the particle number ${N=130}$, ${d=40}$, ${K=J}$, initial half-filling and ${\epsilon_\text{D}=0.3J}$. \densdescr{170}{175}{145}{150}.  The transmission coefficients for hard-core bosons is $T=1.332$, while for fermions $T=1.061$ (calculated at ${t=31/J}$ with Eq.\ref{EqInc}-\ref{EqRef}, ${t_\text{in}=15}$, ${a=30}$). Reprinted with permission from T. Haug, R. Dumke, L.-C. Kwek, and L. Amico, Quantum Sci. Technol. 4, 045001 (2019). Copyright 2019 IOP Publishing Ltd.
	}
	\label{YDensityFermionBosonComp}
\end{figure}

In a Y-junction with the same setup as in Sec.\ref{YExc}, fermions and hard-core bosons show fundamentally different types of reflection behavior. Fig.\ref{YDensityFermionBosonComp} shows the density wave for transmission and reflection for both types of particles. Hard-core bosons show a clear Andreev-reflection with negative reflection, while spinless fermions do not. 

\begin{figure}[htbp]
	\centering
	\subfigimg[width=0.33\textwidth]{}{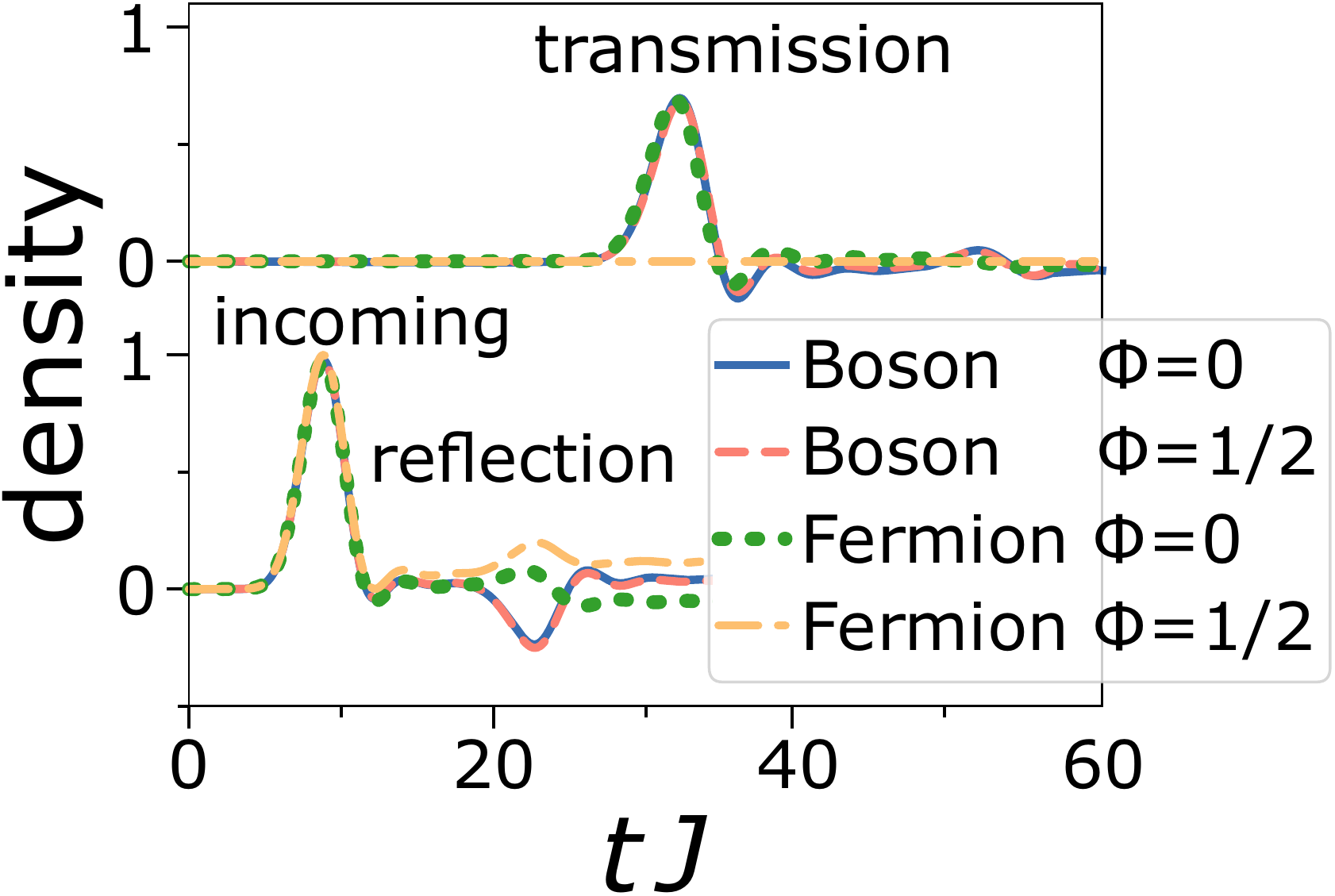}\hfill
	\caption{Propagation of a small density excitation in a ring-lead system for hard-core bosons and spinless fermions for zero and half-flux. Fermion transmission is flux-dependent, while hard-core bosons are flux independent. The source and drain lead has length ${L_\text{S}=L_\text{D}=80}$ and the ring ${L_\text{R}=40}$, the particle number ${N=100}$, strong coupling with ${K=J}$ and ${\epsilon_\text{D}=0.3J}$. \densdescr{130}{135}{65}{70}  Curves stay constant in area shadowed by legend. Reprinted with permission from T. Haug, R. Dumke, L.-C. Kwek, and L. Amico, Quantum Sci. Technol. 4, 045001 (2019). Copyright 2019 IOP Publishing Ltd.
	}
	\label{RingFermionBosonComparison}
\end{figure}

Similar differences arise in the ring-lead system. In a half-filled system, a density wave is excited similar to procedure detailed earlier introduced in Sec.\ref{YExc}.
The reflected and transmitted density wave for zero and half-flux is shown in Fig.\ref{RingFermionBosonComparison}.
First, the authors study the properties of the reflected density wave.
For zero flux, the reflected density wave is different for fermions and bosons. For hard-core bosons, the same characteristic Andreev-like negative reflection peak as seen in the strongly coupled Y-junction. 
However, the reflection dynamics for hard-core bosons is flux independent. 
Now, observe the transmission (upper curve). Spinless fermions  become flux dependent. Here, the fermionic density waves are transmitted for zero flux, while at half-flux zero transmission is observed due to Aharonov-Bohm interference. 
However, for hard-core bosons, the transmission is unaffected by flux. This is again demonstrating the absence of the Aharonov-Bohm effect for bosons.
In short, density excitations for fermions show the Aharonov-Bohm effect, while for interacting bosons the Aharonov-Bohm effect is absent.

\subsection{Entangled state generation with topological pumping in ring circuits}
Directed transport can be also engineered by periodically driving a system. Here, topological pumping, pioneered by Thouless \cite{thouless1982quantized,thouless1983quantization}, can transport excitations, with the added feature that the transport is protected against noise and imperfections due to the topological properties of the system. This has been successfully demonstrated with cold atoms~\cite{lohse2016thouless,takahashi2016}.
Topological pumping is realized by adiabatically driving the parameters of Hamiltonians with topological features periodically in time. This idea can be extended to interacting many-body systems \cite{tangpanitanon2016topological,haug2019topological_correlations}.
This type of driving can be extended to transport atoms through ring-lead circuits as has been shown in Ref.\cite{haug2019topological}, with a similar setup  as introduced for  the ring-lead circuit (see Sec.\ref{Ring}). 
To enable pumping, a time-dependent and spatially varying local potential is added to the system, which id modulated periodically. If engineered correctly, it pumps atoms from the source lead, through the ring, into the drain. For the exact details on the procedure, refer to \cite{haug2019topological}.

\begin{figure}[htbp]
	\centering
	\subfigimg[width=0.4\textwidth]{}{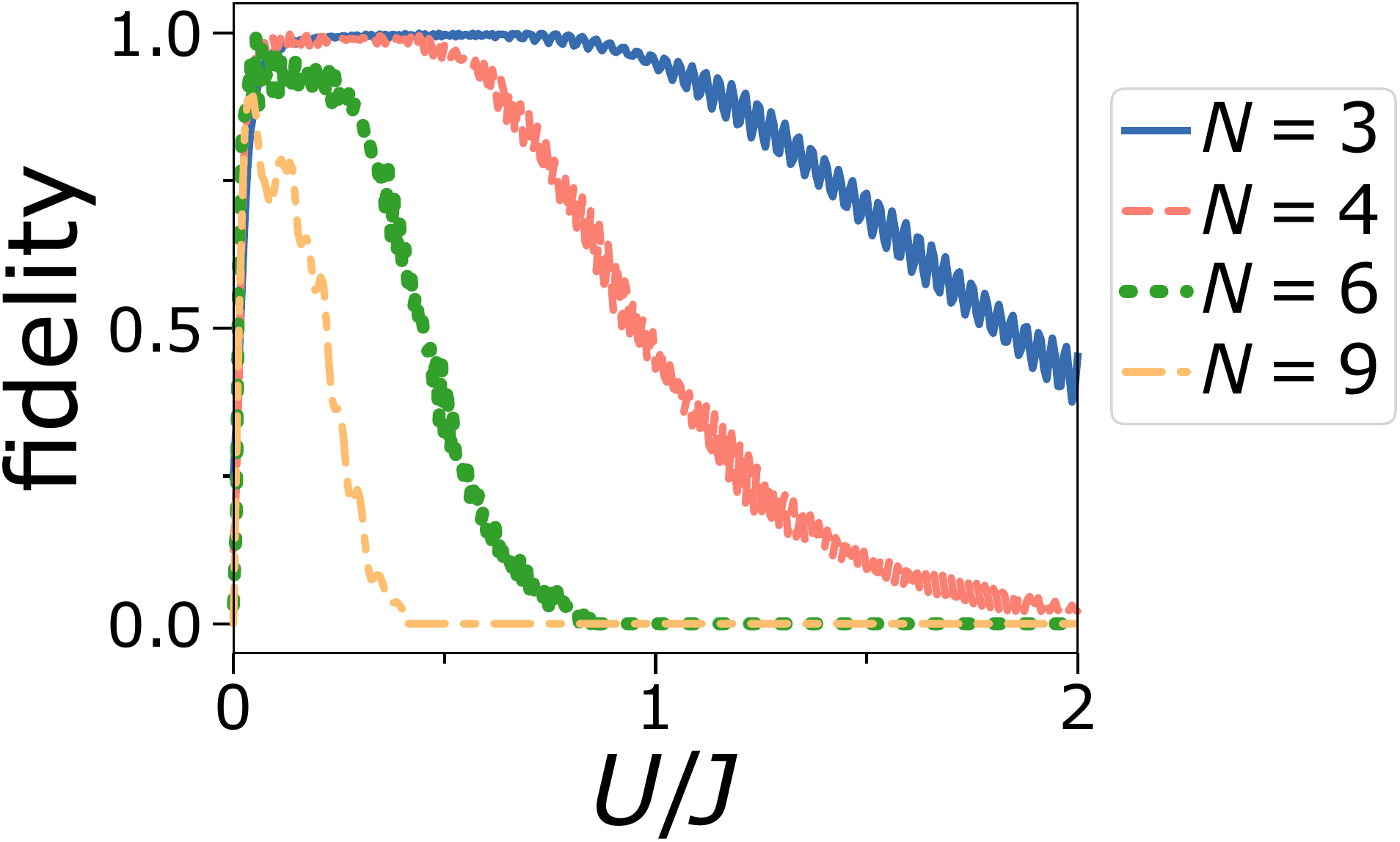}\hfill
	\caption{Numerical simulation of fidelity of creating a NOON-like entangled state by pumping $N$ particles through a simplified ring-lead system of three lattice sites (see text). $N$ particles are placed initially in the source lead, then pumping is switched on, which transport particles into the ring. Then, fidelity of creating a NOON state (particles are in an entangled state of being in either of the two paths of the ring). Fidelity is plotted against interaction $U$ in units of inter-site hopping $J$. Reprinted with permission from T. Haug, R. Dumke, L.-C. Kwek, and L. Amico, Commun. Phys. 2, 127 (2019), under a Creative Commons Attribution 4.0 International License. }
	\label{NOON}
\end{figure}

Here, we will shortly review a key result of this study:
Applying topological pumping to ring-lead systems can be used to create highly entangled quantum states. $N$ atoms are initially placed in the source lead. The pumping is switched on, and the pumping transfers particles into the ring. At the junction of source lead and ring, two possible directions along the ring open up: Either going along the top or bottom path of the ring. Here, the Bose-Hubbard interaction term $U$ can lead to the creation of NOON-like superposition states, where $N$ atoms either go along either of the two paths (${\ket{\Psi_\text{NOON}}=\frac{1}{\sqrt{2}}\ket{0}\otimes(\ket{N0}+\ket{0N})}$). 
This concept can be exemplified by a simplified three site system, consisting of a single source lead site, which is coupled two other sites which represents a part of the ring.
The fidelity ${F=\abs{\braket{\Psi_\text{NOON}}{\Psi}}^2}$ of the creation of the NOON like entangled state is plotted in Fig.\ref{NOON}. For a particular set of parameters, NOON states of up to 6 particles with nearly unit fidelity can be created. For more particles or higher interaction the fidelity decreases due to a exponential suppression of the energy gap. 
This setup is a powerful method to prepare and study highly entangled states of cold atoms.

\subsection{Concluding remarks and outlook}
Transport in quantum many-body systems is a fundamental problem important for quantum information and condensed matter physics.
Cold atoms can be used to simulate these transport problems in novel regimes, that are difficult to realize within other setup.
With recent technologies, transport with nearly no heating can be achieved by extremely smooth atom waveguides. Further, new regimes of transport can be studied by attaching reservoirs of atoms to the system that induce a current through the system. Studying the current can reveal properties of the system that are hard to extract otherwise.

TAAP rings and engineered waveguides made of light allow construction of atomic circuits that can transport cold atoms in various configurations.
Here, we reviewed the properties of two particular atomic circuits that are enabled by this technology, namely ring-lead systems and Y-junctions~\cite{haug2019aharonov,haug2019andreev}. The current through these circuits holds some surprises: Bosonic Y-junctions show Andreev-reflections, known from fermionic superconductor-metal interfaces. By tuning the coupling of the Y-junction, the type of reflection can be tuned, between regular (positive) and negative Andreev reflections. These Andreev reflection are well known from electronic-superconducting interfaces, as such it is surprising to observe them in a bosonic system as well.
For transport through ring-lead systems, the current for interacting bosons is independent of flux and the Aharonov-Bohm effect is absent. This is in stark contrast to fermionic systems, which are highly flux dependent as has been shown in electronic systems. This difference allows one to study the fundamental difference between fermions and bosons due to their anti-symmetric and symmetric many-body wavefunction in a transport experiment, simply by studying the current of the system.  
Finally, by changing the circuit potential in time, topological pumping can be realized to transport atoms in a robust fashion and create highly entangled states of NOON-type. These states could become very useful for quantum-enhanced sensing as the sensitivity of NOON-states increases linearly with the number of entangled particles. For example, the NOON states could be  applied to sense rotation. While the atoms are pumped through the ring, they pick up a phase that is proportional to the rotation affecting the ring times the number of atoms.

In a very interesting future direction transport through non-standard type Hubbard models could be investigated. These types of Hubbard models feature  higher order tunneling and interactions terms that create novel effects and phases. These types of Hamiltonians can nowadays be realized within cold atoms experiments~\cite{dutta2015non}.

The proposed setups can be realized in state-of-the art experiments with both bosonic and fermionic cold atoms. The confinement for the atoms in the form of the circuits can be constructed using DMDs or other light-based potential painting techniques, allowing for arbitrary potential shapes and even time-dependent modulation of the potential. While engineering more complex potentials and driving protocols may be a time-consuming task for humans, new machine learning methods could help to engineer improved potentials and cold atom dynamics automatically without human intervention~\cite{haug2019engineering}.


%


\section{ARTIFICIAL QUANTUM MATTER IN LADDER GEOMETRIES}
\label{Ladder}
\vspace*{-0.5cm}
\par\noindent\rule{\columnwidth}{0.4pt}
{\bf{\small{V. Ahufinger,  R. Citro, S. De Palo, A. Minguzzi,  J. Mompart, E. Orignac,  N. Victorin}}}
\par\noindent\rule{\columnwidth}{0.4pt}



The Fractional Quantum Hall Effect\cite{stormer1999nobel,glattli2002high} is a striking
example of the interplay of
interaction and topology in condensed matter physics. It is
characterized by many fascinating properties such as a precise
quantization of the Hall resistance depending only on fundamental
constants, excitations carrying fractional charges with anyonic
statistics, and dissipationless chiral edge modes. While the effect
has been initially observed with fermions, bosonic analogues have been
proposed by Regnault and Jolicoeur in rotating clouds of ultracold
atoms\cite{regnault2003quantum}.
Recently, the realization in experiments of artificial gauge fields\cite{ruseckas2005non-abelian,lin2011spin-orbit-coupled,galitski2013spin-orbit,dalibard2011colloquium}
has opened another route for observing Quantum Hall phases with
ultracold atoms.
As a first step towards the realization of Quantum Hall phases with
ultracold atoms it is interesting to consider the so called ladder
systems\cite{kardar1986josephson-junction,orignac2001meissner,cha2011two,dhar2012bose-hubbard,dhar2013chiral,tokuno2014ground,uchino2015population-imbalance,petrescu2013bosonic,petrescu2015chiral,petrescu2016precursor,piraud2014quantum,piraud2014vortex,kolley2015strongly,greschner2015spontaneous,greschner2016symmetry-broken,richaud2017quantum,barbarino2016synthetic,taddia2017topological,strinati2017laughlin-like,strinati2018spin-gap,strinati2019pretopological,didio2015persisting,orignac2016incommensurate,orignac2017vortex,citro2018quantum}, \emph{i.e.} two dimensional systems that are of finite size
along one of the dimensions.
Such deceptively simple system is already sensitive to the effect of
the applied flux and can exhibit analogues of the Quantum Hall
phase\cite{petrescu2015chiral,petrescu2016precursor,strinati2017laughlin-like}. Moreover, it shows a wealth of phases, emerging from interplay
of rung and leg tunnel, interactions, artificial gauge field,
filling\cite{dhar2012bose-hubbard,dhar2013chiral,petrescu2015chiral,orignac2016incommensurate,orignac2017vortex}. For bosonic atoms, in low flux, an analogue of the Meissner
phase is obtained\cite{kardar1986josephson-junction,orignac2001meissner}. At high flux, a quasi-long range ordered vortex
phase is
formed\cite{kardar1986josephson-junction,orignac2001meissner}. Interleg
interactions can stabilize an atomic density wave at intermediate
flux\cite{orignac2017vortex,citro2018quantum}. 
For a flux commensurate with the density the analogue of QHE is found\cite{petrescu2015chiral,petrescu2016precursor,strinati2017laughlin-like}. At a different
commensuration between flux and density, an  incommensuration driven
by interchain hopping is obtained\cite{didio2015persisting}.
Furthermore, a variant of the ladder in form of a diamond chain has topological properties  \cite{pelegri2019topological} and allows to simulate quantum magnetism \cite{pinheiro2013XYZ,pelegri2019quantum}.
%






\subsection{A boson ring ladder at weak interactions}
\label{sec:ladder-mean-field}
\paragraph*{Model:} We consider $N$ bosons occupying  two coupled one-dimensional concentric lattice rings subjected to two artificial gauge fields ad organized on a planar geometry. The stacked geometry has also been thoroughly studied~\cite{amico2014superfluid,haug2018mesoscopic}. This system could be experimentally realized \textit{e.~g.}  using  dressed potentials~\cite{perrin2017trapping}, or   Laguerre-Gauss beams~\cite{wright2000toroidal}.
The Hamiltonian reads
\begin{align}
&\hat{H}=\hat{H}_0+\hat{H}_{int}=\nonumber\\ &- J\!\sum_{l=1,p=1,2}^{N_s} \!\!\!\left(b^{\dagger}_{l,p}b_{l+1,p}e^{i\Phi_p} + b^{\dagger}_{l+1,p}b_{l,p}e^{-i\Phi_p}\right)\nonumber \\ &-\!\frac{\Omega}{2}\sum_{l=1}^{N_s}\left(b_{l,1}^{\dagger}b_{l,2}+b_{l,2}^{\dagger}b_{l,1}\right)+\frac{U}{2}\!\!\!\sum_{l=1,p=1,2}^{N_s}b^{\dagger}_{l,p}b^{\dagger}_{l,p}b_{l,p}b_{l,p}
\label{eq1}
\end{align}
where $b_{l,p}$ are the bosonic field operators for the $p$-th ring,  $l$ indicates the site position on each ring made of $N_s$ sites, $J$ is  the tunneling amplitude along each ring, threaded by the fluxes  $\Phi_{1,2}$  respectively  and {$\Omega$} is the inter-ring tunneling amplitude. In the non-interacting regime $U=0$, this model is readily diagonalized, yielding a two-band excitation spectrum
 \begin{align}
E_{\pm}(k)=-&2J\cos(\lambda/2)\cos(k-\Phi)\nonumber\\ \pm &\sqrt{(\Omega/2)^2+(2J)^2\sin(\lambda/2)^2\sin(k-\Phi)^2}.
\label{eq:epm}
 \end{align}
 where we have set {$\lambda = \Phi_1-\Phi_2$} and average flux $\Phi = (\Phi_1+\Phi_2)/2$. Depending on the ratio $\Omega/J$ and on $\lambda$, the lowest band of the excitation spectrum has either one or two minima centered at $k=\Phi$. The ground state of the Bose gas is  a Bose-Einstein condensate occupying the minima of such excitation spectrum. In the case of a single minimum the ground state corresponds to the Meissner phase and in the case of a coherent superposition of the occupancy of the two minima, the ground state is in the  vortex phase. The Meissner phase is characterized by vanishing transverse current and homogeneous density profile. The vortex phase has  non-zero transverse current and density modulations along the ring. The vortex to Meissner phase transition has been experimentally observed in~\cite{atala2014observation}. The chiral current on the ring, \textit{i.~e.} the difference of longitudinal currents among the two rings, is characterized by subsequent jumps each time a vortex enters into the system.
\begin{figure}[h!]
\includegraphics[width=9cm]{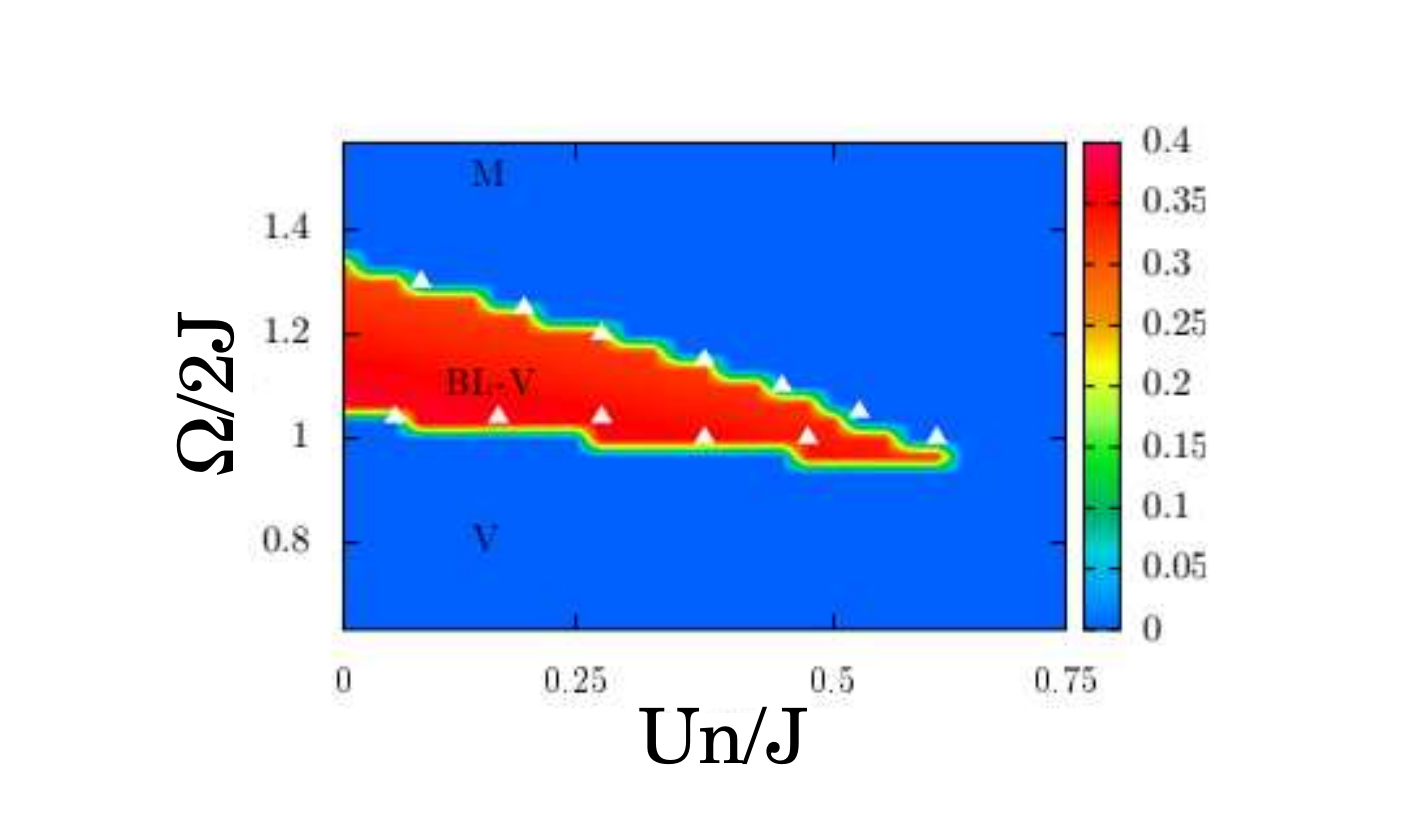}
\includegraphics[width=9cm]{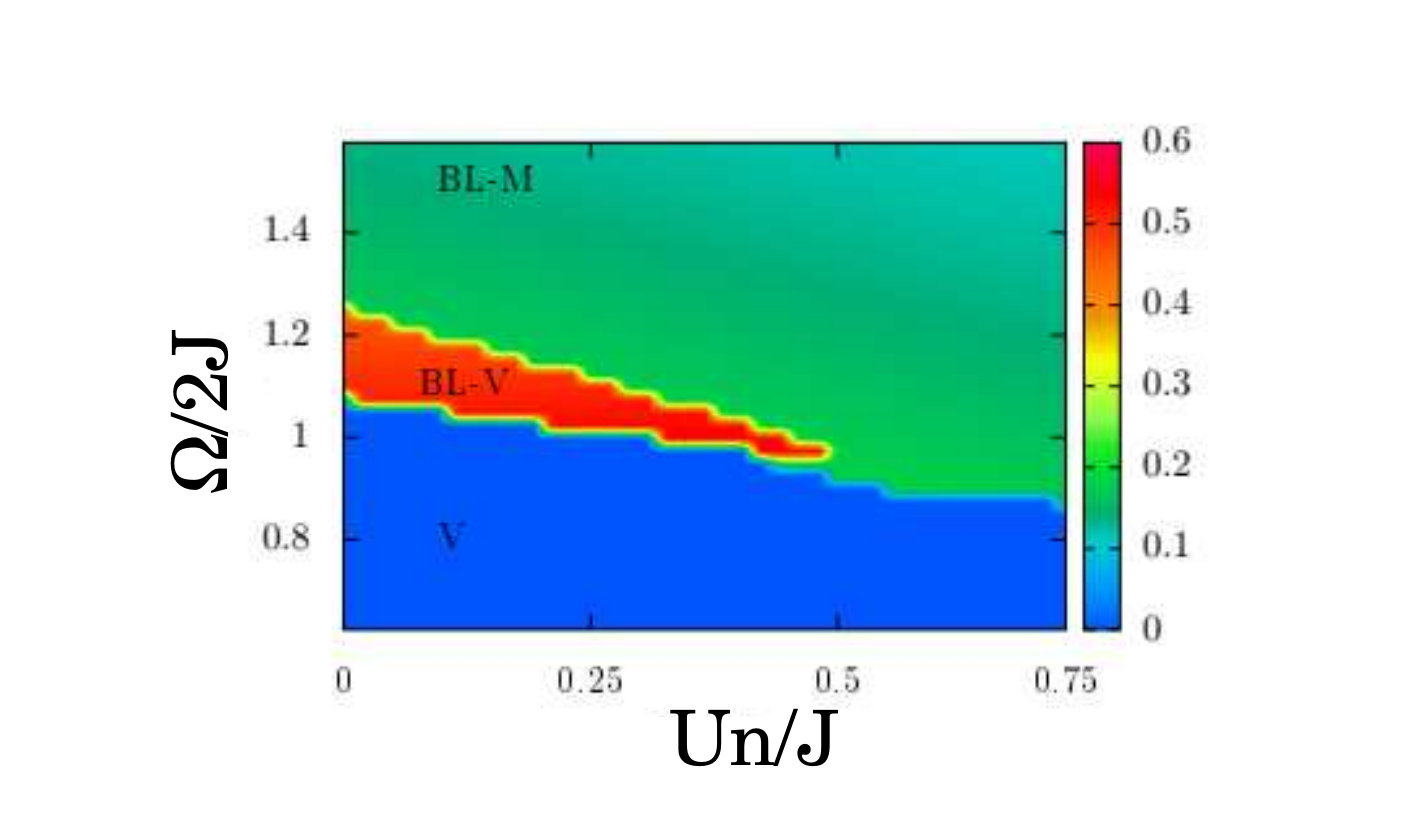}
\caption{\label{fig:victorin2018} (Color online) Color map of the imbalance among particle numbers in each ring,  in the ($\Omega/J$,$UN/JN_s$) plane, for (upper panel) $\lambda=\pi/2$, $\Phi=6\pi/N_s$ and $N_s=20$, (lower panel) $\lambda=\pi/2$, $\Phi=\pi/N_s$ and $N_s=20$ The letters indicate the parameter regimes where we find a biased-ladder phase  (BL-V) where the single-particle spectrum has a double minimum, a Meissner phase (M), a vortex phase (V) and a  biased-ladder phase (BL-M) where the single-particle spectrum has a single minimum. White triangles represent the frontiers between biased-ladder phase the two other phase, namely vortex phase and Meissner phase as calculated with the variational Ansatz including finite size effect.}
\end{figure}

 \paragraph*{Ground state of weakly interacting ring}
 We assume large occupancy of the lattice sites and weak interactions $U/J\ll 1$. In this regime, we describe the system by the mean-field approximation. Setting  $\Psi_{l,p}(t)= \langle b_{l,p}(t)\rangle$ the condensate wave-function, we solve the coupled discrete non-linear Schroedinger equations (DNLSE)
\begin{eqnarray}
\label{dnlse}
i\partial_t \Psi_{l,1}(t)&=& -J\Psi_{l+1,1}(t)e^{i(\Phi+\lambda/2)}-J\Psi_{l-1,1}(t)e^{-i(\Phi+\lambda/2)}\nonumber\\&-&K\Psi_{l,2}(t)+U|\Psi_{l,1}(t)|^2\Psi_{l,1}(t)\\
i\partial_t \Psi_{l,2}(t)&=& -J\Psi_{l+1,2}(t)e^{i(\Phi-\lambda/2)}-J\Psi_{l-1,2}(t)e^{-i(\Phi-\lambda/2)}\nonumber\\&-&K\Psi_{l,1}(t)+U|\Psi_{l,2}(t)|^2\Psi_{l,2}(t)
\end{eqnarray}
 {We have also assumed that the interaction energy is smaller than the bandgap such that the single-band approximation for each ring latice holds.} 
At varying  $U N/J N_s$ and $\Omega/J$, the ground state displays  three phases \cite{victorin2018bosonic}: the vortex (V) and Meissner (M) phases found in the non-interacting regime, as well as the biased-ladder phase, characterized by imbalanced density populations among the two rings and uniform density profile. We denote the latter  (BL-V) or (BL-M) depending whether  for the same values of $\lambda$ and $\Omega/J$ the corresponding non-interacting spectrum has one or two minima.
{Notice that the BL phases are only found at weak interactions, while they are disrupted as the interactions increase \cite{buser2020interacting}. For this reason they are not found in the DMRG calculations of Sec.\ref{sec:strong-interactions}.}

\begin{figure}[h!]
\includegraphics[width=6cm]{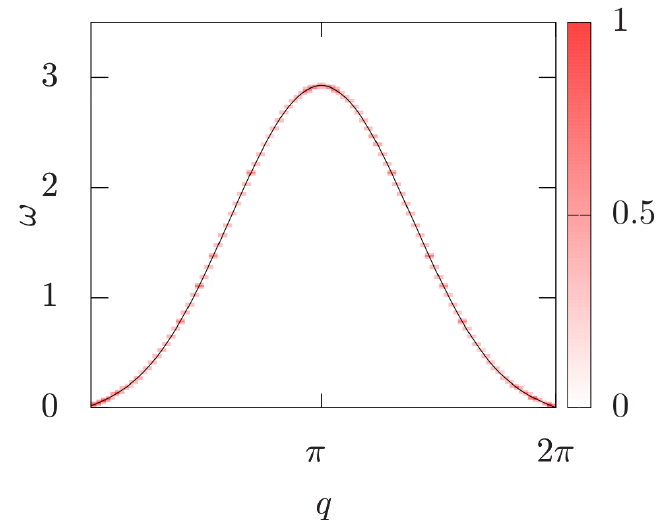}
\includegraphics[width=6cm]{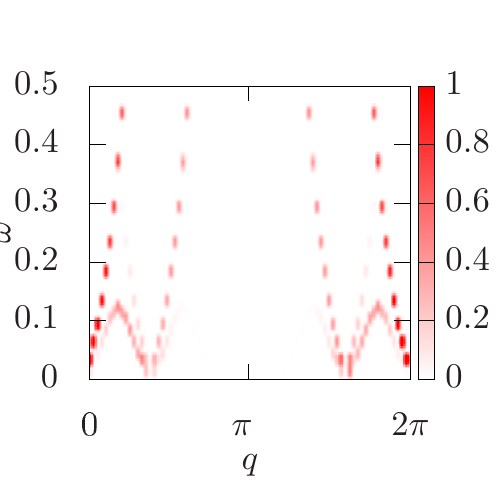}
\caption{\label{fig:victorin2019} (Color online) Dynamic structure factor in the frequency-wavevector plane   in the frequency-wavevector plane (color map, $q$ in units of $1/a$ with $a$ lattice spacing and $\omega$ in units of $J$) Upper panel: in the Meissner phase, for $Un/J=0.2$, $\lambda=\pi/2$, $K/J=3$. Lower panel: in the vortex phase  for $\Omega/J=1.6$  $\lambda=\pi/2$, $Un/J=0.2$. For both panels  $N_s=80$.}
\end{figure}

\paragraph*{Excitation spectrum of weakly interacting ring}
We next present our results for the excitation spectrum of the weakly interacting Bose gas on a ring lattice \cite{victorin2019excitation}.
Withing the Bogoliubov approximation, we set  $\hat{a}_{l,p}=\Psi^{(0)}_{l,p}+\delta\hat{a}_{l,p}$, where $\Psi^{(0)}_{l,p}$ is the ground state solution with chemical potential $\mu$.
we find the  excitation spectrum using the  expansion of the fluctuation operator $\delta \hat{a}_{l,p}$ in normal modes with energy $\omega_\nu$, according to
\begin{align}
\delta \hat{b}_{l,p} =\sum_{\nu} h_{\nu,l}^{(p)}\hat{\gamma}_{\nu}-Q_{\nu,l}^{*(p)}\hat{\gamma}_{\nu}^{\dagger},
\label{Transfo}
\end{align}
The solution Bogoliubov-de Gennes eigenvalue equations for the mode amplitudes  $h_{\nu,l}^{(p)},$ and  $Q_{\nu,l}^{(p)}$ yields the excitation spectrum. We use both eigenvalues and eigenvectors of the Bogoliubov equations to compute the dynamic structure factor
\begin{align}
S_{p,p'}(q,\omega)=\sum_{s\neq 0}|\langle s |\hat \rho_{q}^{(p,p')}|0\rangle|^2\delta(\omega-\omega_s)
\end{align}
As an example, we show the results for the excitation spectrum in the Meissner and in the  vortex phase.  The excitation spectrum is strongly dependent on the phase of the underlying ground state. In the Meissner phase a single Goldstone mode is found, and corresponds to the $U(1)$ symmetry breaking associated to the formation of a Bose-Einstein condensate. In the vortex phase, two Goldstone modes are observed. Indeed in the vortex phase, a second symmetry is broken, \textit{i.~e.} the discrete translational symmetry . We have shown that this is associated to the emergence of supersolidity of the gas, and is corroborated by the calculation of the static structure factor, displaying a well-defined peak, and the first-order correlation function, demonstrating phase coherence.

\subsection{The boson ladder at strong interaction}
\label{sec:strong-interactions}
{Let's consider a spin-1/2 bosons with spin-orbit interaction model\cite{orignac2016incommensurate}, where $\Omega$
is the transverse magnetic field, $\lambda$  the spin-orbit
coupling, $U_{\uparrow\uparrow}=U_{\downarrow\downarrow}=U$ the
repulsion between bosons of identical spins, $U_{\downarrow\uparrow}=U_\perp$ the
interaction between bosons of opposite spins.
 Its Hamiltonian is }
\cite{barbarino2016synthetic,strinati2017laughlin-like}:
\begin{eqnarray}
  \label{eq:full-lattice-ham}
  H=-J \sum_{j,\sigma} (b^\dagger_{j,\sigma} e^{i \lambda \sigma}
  b_{j+1,\sigma} +b^\dagger_{j+1,\sigma}  e^{-i \lambda \sigma}
  b_{j,\sigma}) \nonumber \\
  +\frac \Omega 2 \sum_{j,\alpha,\beta} b^\dagger_{j,\alpha}
  (\sigma^x)_{\alpha\beta} b_{j,\beta}
  + \sum_{j,\alpha,\beta} U_{\alpha \beta} n_{j\alpha} n_{j\beta},
\end{eqnarray}
where {$\sigma=\pm 1/2$} or $\sigma=\uparrow,\downarrow$  is the spin index\cite{celi2014synthetic,saito2017devil,livi2016synthetic}, $j$ the site index, $n_{j\alpha} = b^\dagger_{j\alpha}
b_{j\alpha}$
The Hamiltonian~(\ref{eq:full-lattice-ham}) is mapped onto the Hamiltonian of the two-leg ladder in flux (\ref{eq1}) by making $p=\frac 3 2 -\sigma$, $\Phi_p=(-)^{p-1}\frac \lambda 2$, $U_\perp=0$.  

The low-energy effective theory for the
Hamiltonian~(\ref{eq:full-lattice-ham}), treating $\Omega$ and $U_\perp$ as
perturbations, can be obtained by using Haldane's bosonization of interacting
bosons.\cite{haldane1981effective}
Introducing\cite{haldane1981effective} the fields $\phi_{\alpha}(x)$ and
$\Pi_\alpha(x)$ satisfying canonical commutation relations
$[\phi_\alpha(x),\Pi_\beta(y)]=i\delta(x-y)$ as well as the dual $\theta_\alpha(x) =\pi \int^x dy
\Pi_\alpha(y)$  of
$\phi_\alpha(x)$,  and after introducing the respective combinations of operators {$\phi_{c,s}=(\phi_\uparrow \pm \phi_\downarrow)/\sqrt{2} $},
we can represent the low-energy Hamiltonian as $H=H_c+H_s$, where
\begin{eqnarray}
  \label{eq:bosonized-c}
  H_c=\int \frac{dx}{2\pi} \left[u_c K_c (\pi \Pi_c)^2 +\frac {u_c}
    {K_c} (\partial_x \phi_c)^2 \right]
\end{eqnarray}
 describes the total density fluctuations for incommensurate filling when umklapp terms are irrelevant, and
\begin{eqnarray}
\nonumber
&&H_s = \int \frac{dx}{2\pi} \left[u_s K_s \left(\pi \Pi_s +\frac{\lambda}
{a \sqrt{2}} \right)^2 +\frac {u_s}
{K_s} (\partial_x \phi_s)^2 \right] \\
&&- 2 \Omega A_0^2 \int dx \cos
\sqrt{2} \theta_s + \frac{U_\perp a B_1^2} 2 \int dx \cos \sqrt{8} \phi_s
\label{eq:bosonized}
\end{eqnarray}
describes the antisymmetric density fluctuations. In
Eq.~(\ref{eq:bosonized}) and ~(\ref{eq:bosonized-c}),  $u_s$ and $u_c$ are respectively  the
velocity of antisymmetric and total density excitations, $A_0$ and $B_1$ are non universal coefficients\cite{giamarchi2004quantum}
while
$K_s$  and $K_c$ are  the corresponding Tomonaga-Luttinger (TL)
exponents\cite{citro2018quantum}. They can be expressed as a function of the velocity of
excitations $u$, and  Tomonaga-Luttinger
liquid exponent $K$ of the isolated chain\cite{citro2018quantum}.

For an isolated chain of hard core bosons,  we have $u=2J \sin (\pi
\rho^{0}_\sigma)$ and  $K=1$.
The phase diagram of the Hamiltonian can be determined by looking at physical observables  such as the rung and leg current, momentum distribution and correlation functions. Physical observables can be all represented in bosonization.
The rung current, or the flow of bosons from the  upper leg to
the lower leg, is:
\begin{eqnarray}
\nonumber
  J_{\perp}(j)&=&-i \Omega (b^\dagger_{j,\uparrow} b_{j_\downarrow} -
  b^\dagger_{j,\downarrow} b_{j_\uparrow}). \\
&=& 2 \Omega A_0^2 \sin \sqrt{2} \theta_s +\ldots
\end{eqnarray}
The chiral current, {\it i.e.} the difference between the currents of
upper and lower leg, is defined as
\begin{eqnarray}
\label{eq:spin-current}
J_\parallel(j,\lambda) &=& -iJ \sum_{\sigma}\sigma (b^\dagger_{j,\sigma} e^{i \lambda \sigma}
b_{j+1,\sigma} - b^\dagger_{j+1,\sigma}  e^{-i \lambda  \sigma}
b_{j,\sigma}), \\
&=& \frac{u_s K_s}{\pi\sqrt{2}} \left(\partial_x \theta_s +\frac{\lambda}{a \sqrt{2}} \right).
\end{eqnarray}
The  density difference between the chains $S_j^z=n_{j\uparrow}
-n_{j\downarrow}$, is written in bosonization as:
\begin{eqnarray}
  S_j^z=-\frac{\sqrt{2}}{\pi} \partial_x \phi_s -2 B_1 \sin (\sqrt{2}
  \phi_c -\pi \rho x) \sin \sqrt{2} \phi_s,
\end{eqnarray}
while the density of particles per rung is:
\begin{eqnarray}
  n_j=-\frac{\sqrt{2}}{\pi} \partial_x \phi_c -2 B_1 \cos (\sqrt{2}
  \phi_c -\pi \rho x) \cos \sqrt{2} \phi_s.
\end{eqnarray}

\begin{figure}[h]
\begin{center}
\includegraphics[trim = 12mm 5mm 5mm 5mm,width=82.5mm]{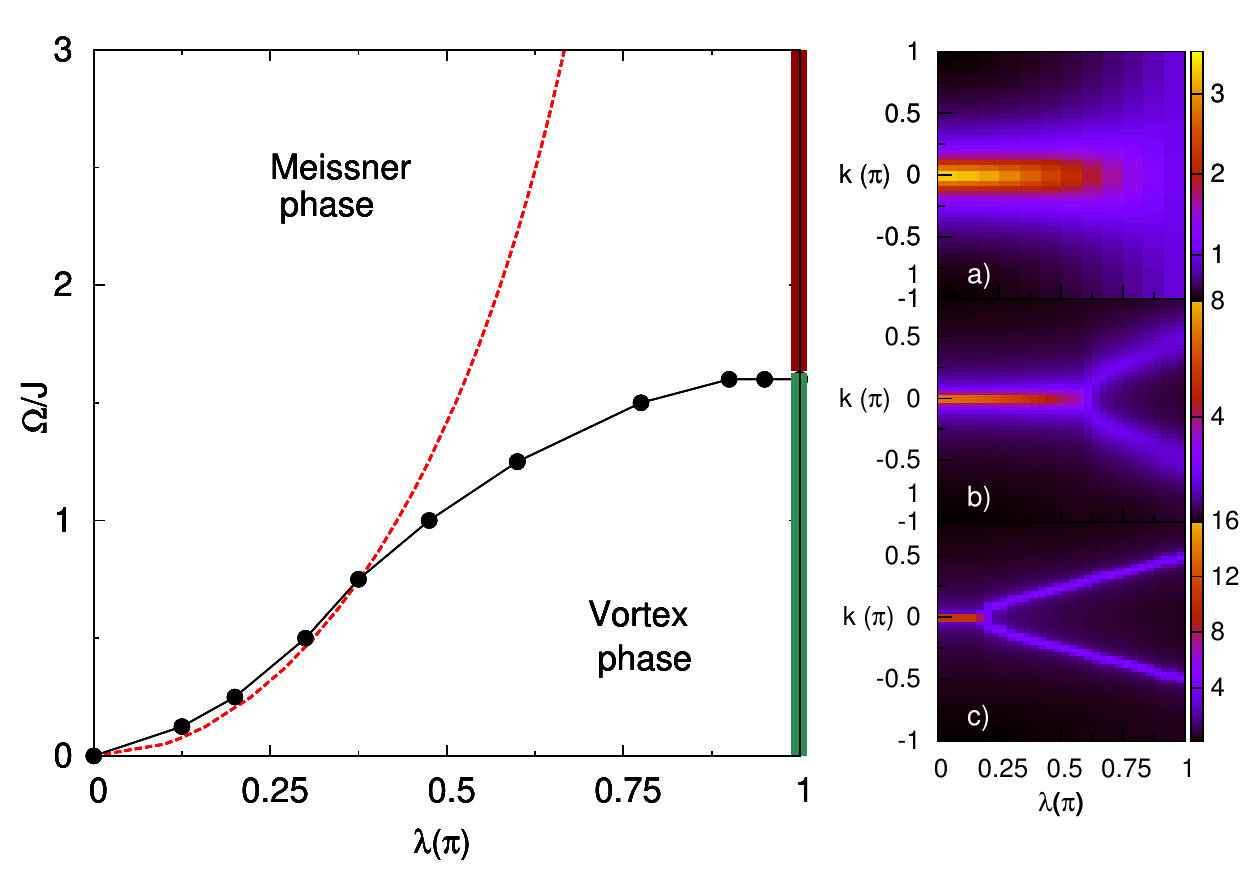}
\end{center}
\caption{Phase diagram for {}hard-core bosons on the two-leg ladder and $U_\perp=0$ as
a function of flux per plaquette $\lambda$ and the interchain
hopping $\Omega$. The boundaries between the Meissner and the Vortex phase are shown by the
red and black solid lines respectively for the non-interacting  and the interacting case.
At $\lambda=\pi$ the thick solid green-line shows the occurrence of the second incommensuration. 
The three insets in the right panel show intensity plots of $n(k,\lambda)$ for the
three different values of $\Omega/J=0.25,1.25$ and $2$. In panels b) and c) the system enters the Vortex
phase for $\lambda > \lambda_c$ : the single peak at $k=0$ splits in two maxima symmetric
around $k=0$ at $\pm q(\lambda)$. At large $\Omega/J=2$, panel a), the system stays always in
Meissner phase and in the vicinity of $\lambda=\pi$, $n(k)$
becomes independent of $k$ indicating the formation of a fully
localized state (thick solid dark-red line).}
\label{fig:hd_ladder}
\end{figure}

When $\Omega\ne 0$, $U_\perp=0$, and $\lambda\rightarrow 0$, the antisymmetric
modes Hamiltonian Eq.~(\ref{eq:bosonized}) reduces to a quantum
sine-Gordon Hamiltonian.
For $K_s>1/4$, the spectrum of $H_s$ is gapped and the system is in the so-called Meissner state\cite{kardar1986josephson-junction,orignac2001meissner} characterized
by $\langle \theta_s\rangle =0$.
In such state, the chiral current increases linearly with the applied flux at
small $\lambda$, while the average rung current $\langle J_\perp
\rangle =0$  and its correlations $\langle  J_{\perp}(j)
J_{\perp}(0)\rangle$ decay exponentially with distance.
The transition from the Meissner to the Vortex phase is signaled by the splitting of the momentum distribution $n(k)$ from $k=0$ to a finite $Q=\sqrt{\lambda^2-\lambda_c^2}$ that depends on the spin-orbit interaction $\lambda$. For this reason the transition falls into the universality class of the commensurate-incommensurate transitions (C-IC). The rung current correlation function develops two symmetric peaks and the spin static structure is linear at low momentum.
The phase diagram for a hard-core bosonic ladder at $n=1$, obtained using density renormalization group (DMRG)
technique\cite{white1993density-matrix,schollwock2005density-matrix}, is shown in Fig.~\ref{fig:hd_ladder} together with the
momentum distribution $n(k)$ as a function of lambda across the C-IC transition.
Compared to the non-interacting case, the phase diagram as a function of $\Omega/J$ and $\lambda$ shows an enlargement of the Meissner phase and its persistence above a certain value of $\Omega/J$.
Above a certain value of $\lambda$ a second incommensuration appears in the rung current correlation functions and the static structure factor. Such incommensuration is associated to the appearance of an extra peak in the rung current correlation function at wavevectors $P=\sqrt{\lambda^2+p(\Omega^2)}$ and $\pi\pm P$, with $p(\Omega)$ a function of the interchain tunneling. For $\lambda=\pi$ the correlation functions show a tendency to a localized regime.

\begin{figure}[h]
\begin{center}
\includegraphics[trim = 12mm 5mm 5mm 5mm,width=82.5mm]{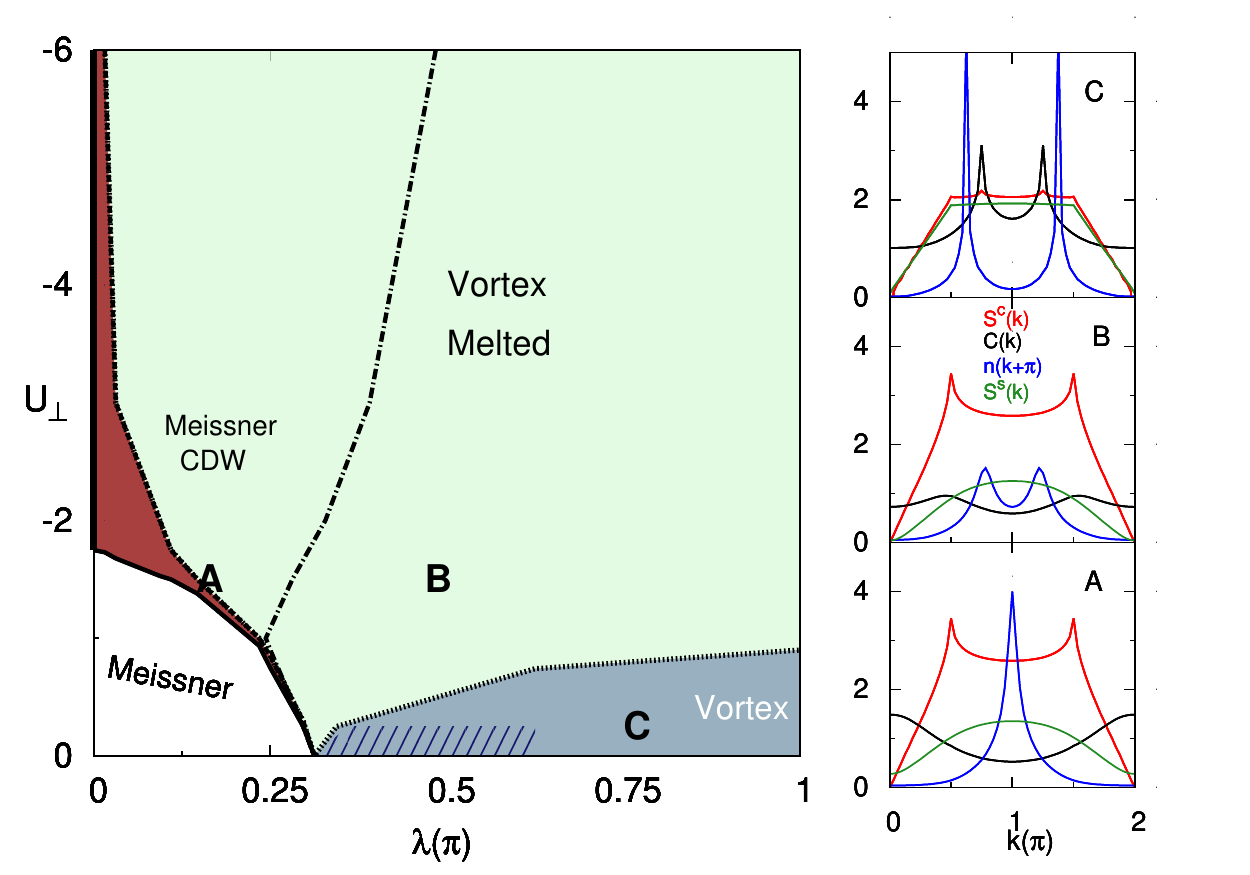}
\end{center}
\caption{(Color online) Phase diagram at $n=0.5$ for a fixed value of interchain
hopping $\Omega/J=0.5$  as a function of the applied flux $\lambda$ and as function of
the strength of the interchain interaction $U_\perp$.
The dashed black line is the boundary between the Meissner phase and the Meissner-CDW phase (dark-red region).
In the Melted vortex phase (light-green region) the dot-dashed line indicates the Lifshitz point.
At large $\lambda$ the Vortex phase is re-established ( light-blue region
under the dotted black line). In the shaded blue area second-incommensuration occurs.
The three insets in the right panel show the behavior of $n(k)$ (blue solid lines), the spin static structure factor $S^s(k)$ 
(solid dark-green line), the charge static structure factor $S^c(k)$ (solid red line) and the rung current-rung current correlation function $C(k)$ (solid black lines), for the three points { $A=(\lambda=0.146\pi, U_\perp=-1.5), B=(\lambda=0.468\pi, U_\perp=-1.5)$ and $C=(\lambda=0.75\pi, U_\perp=-0.25)$ shown in the phase diagram, respectively for the CDW-Meissner, Melted Vortex and Vortex phase.}}
\label{hd_ladder_U}
\end{figure}

With $\Omega\ne 0,U_\perp \ne 0$ the C-IC transition is replaced by a Meissner-to-incommensurate charge density wave (ICDW) which falls into the Ising 
universality class, followed by a melting of the Vortex phase at large enough $\lambda$, going towards a BKT transition when entering the Vortex phase\cite{citro2018quantum}. The melting of the Vortex phase is signaled by the Lorentzian shape of the momentum distribution peaks preceded by a Lifshitz point\cite{hornreich1975critical}.
In the phase diagram, see Fig.~\ref{hd_ladder_U} for the case of a hard-core bosonic ladder in the presence of an attractive
interaction $U_\perp$, obtained using DMRG simulations, it is possible to trace these features.
{As
we increase the interaction strength (panel A in Fig. \ref{hd_ladder_U}) the
charge structure factor develops peaks at $k =\pi/2$ and $k =3\pi/2$ and $S_s(k)$ has an almost quadratic behavior at small wave vectors. The quadratic behavior indicates that spin excitations remain
gapped, while the presence of peaks at $k = \pi/2,3\pi/2$ in $S_c(k)$ is the signature of a zigzag charge density wave. Going to panel B, at increasing $\lambda$, the momentum distribution develops two broad maxima indicating a vortex melted phase, while the rung current correlation function starts to form two bumps. Increasing still $\lambda$, in panel C both the momentum distribution, as well the rung-current correlation
function $C(k)$, develop two separate peaks\cite{orignac2001meissner}, that show negligible
size effects, indicating the presence of an incommensuration. On the other side both the charge and spin structure factors have a linear behavior at small $k$ indicating a gapless phase.}

\subsection{Ultracold atoms carrying orbital angular momentum  {(OAM)} in a diamond chain}

\begin{figure}[t!]
\centering
\includegraphics[width=\linewidth]{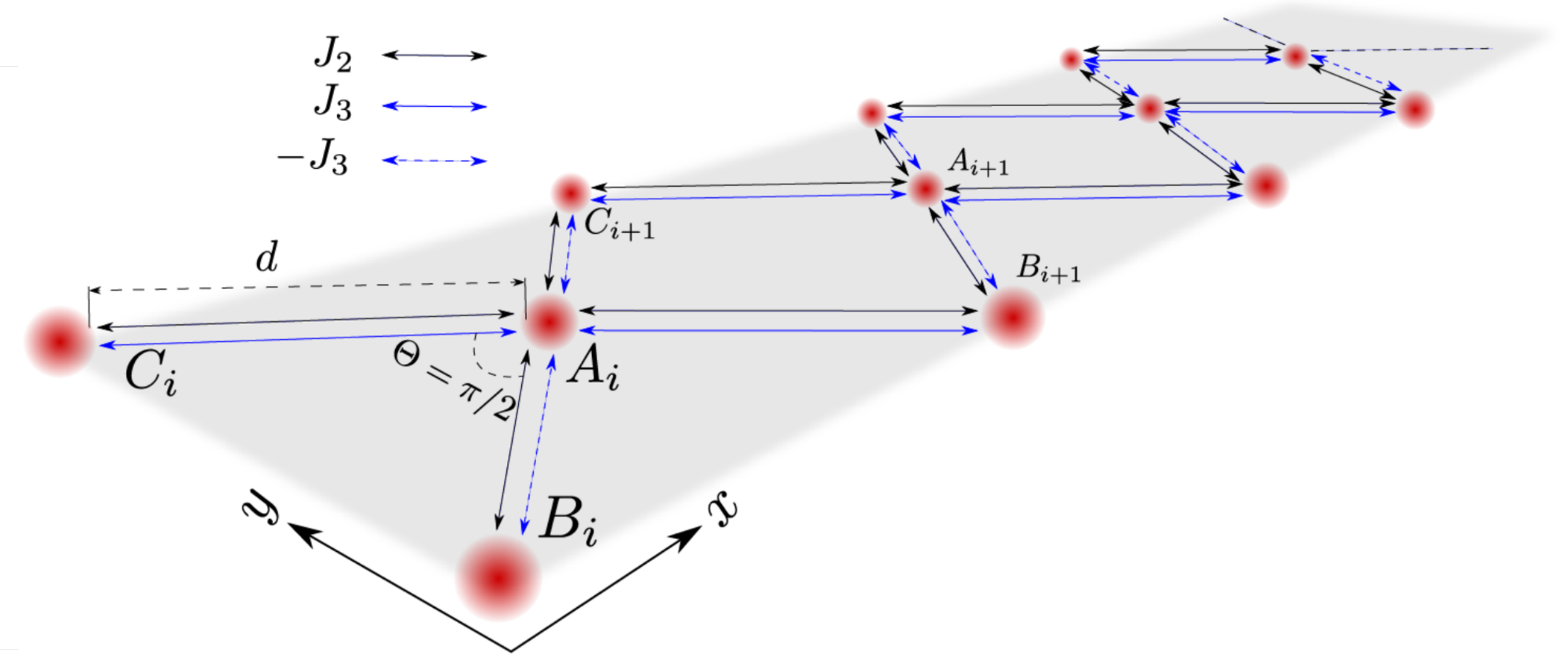}
\caption{{Schematic representation of the considered diamond chain.}}
\label{PhySystem_2}
\end{figure}

\paragraph*{Topological edge states and Aharonov-Bohm caging}
We consider a ladder with a diamond-chain shape with a unit cell formed by three cylindrically symmetric potentials of radial frequency $\omega$, and forming a triangle with central angle $\Theta$ and nearest-neighbor separation $d$ {(see Fig.~\ref{PhySystem_2})}. Non-interacting ultracold atoms of mass $m$ that may occupy the two degenerate OAM $l=1$ states with positive or negative circulation localized at each site are loaded into the ladder. {Such a system could be experimentally implemented, for instance, by exciting the atoms to the $p$-band of a conventional optical lattice \cite{li2016physics,kiely2016shaken,kock2016orbital,wirth2011evidence} or by optically transferring OAM \cite{franke-arnold2017orbital} to atoms confined to an arrangement of ring-shaped potentials, which can be created by a variety of techniques, as discussed in Section III.} Three independent tunneling amplitudes \cite{polo2016geometrically} exist in the system: $J_1$, which corresponds to the self-coupling at each site between the two OAM states with different circulations, and $J_2$ and $J_3$, which correspond to the cross-coupling tunneling amplitudes between OAM states in different sites with equal or different circulations, respectively. For $\Theta=\pi/2$, $J_1$ and $J_3$ acquire a relative phase of $\pi$ along one of the diagonals of the chain and, due to destructive interference between neighboring sites, the self-coupling vanishes everywhere except for the sites at the left edge. The model possesses inversion and chiral symmetry but, due  to  the  two-fold degeneracy, Zak's phases \cite{zak1989berry} are ill-defined. Thus, a series of exact mappings are required to fully characterize topologically the system. In addition, the model here obtained corresponds to a square-root topological insulator \cite{arkinstall2017topological,kremer2018non-quantized}, i.e., the quantized values of the Zak's phases are recovered after taking the square of the bulk Hamiltonian.

\begin{figure}[t!]
\centering
\includegraphics[width=0.7\linewidth]{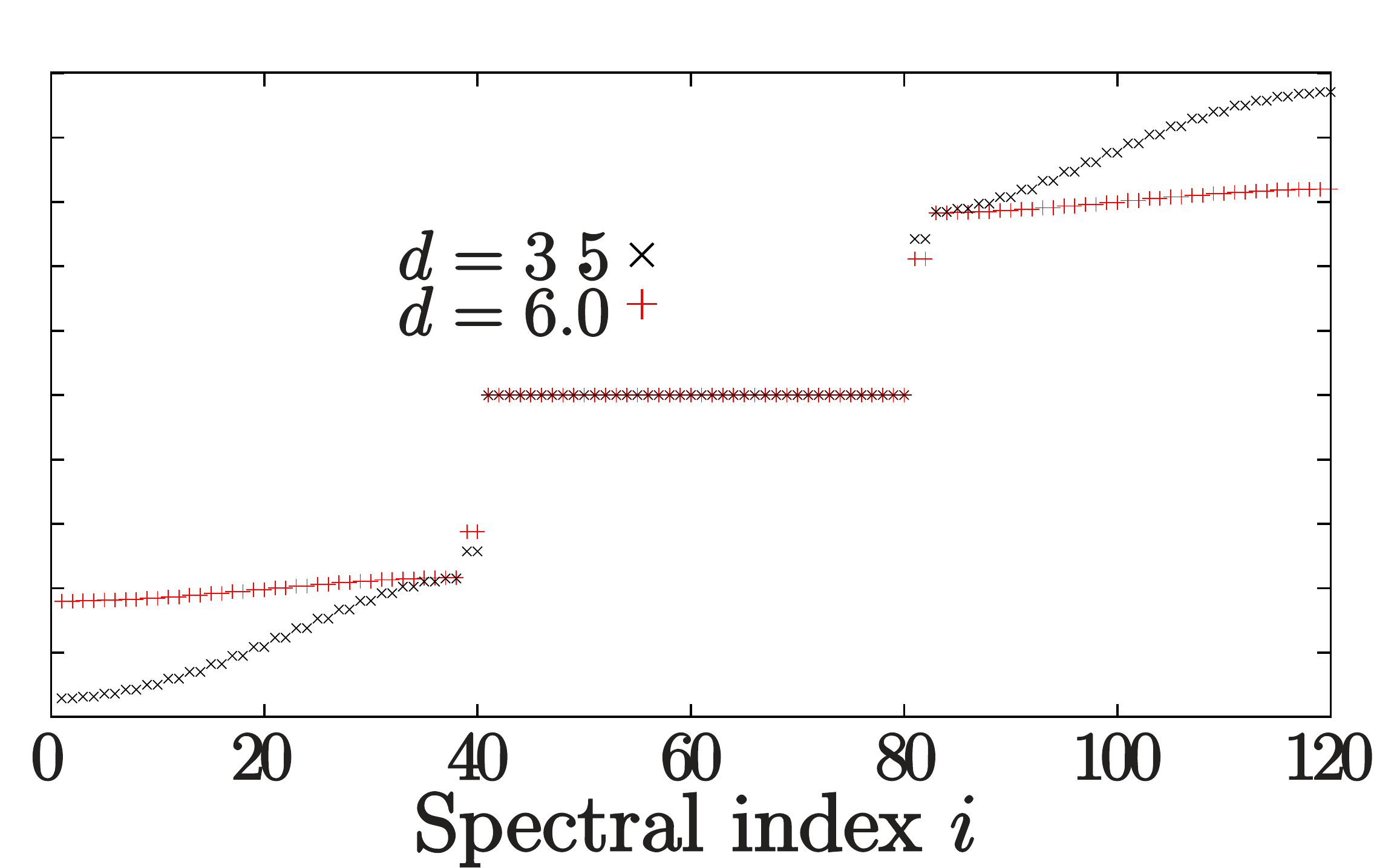}
\caption{ Exact diagonalization spectra of a diamond chain of $N_c=20$ unit cells for $d=3.5\sigma$, corresponding to $J_3/J_2=1.67$ (black solid line) and $d=6\sigma$ corresponding to $J_3/J_2=1.13$ (red dotted line), where $\sigma=\sqrt{\hbar/(m\omega)}$}
\label{PhySystem}
\end{figure}

Under periodic boundary conditions, the diagonalization of the bulk Hamiltonian yields six energy bands in three degenerate pairs and a gap appears in the spectrum. In the $J_2=J_3$ limit, all bands become flat. Exact diagonalization, in the case of open boundary conditions, shows the presence of four in-gap states localized at the right edge of the chain, which persist as long as the energy gap is open (see Fig.~\ref{PhySystem}).

\begin{figure}[t!]
\centering
\includegraphics[width=\linewidth]{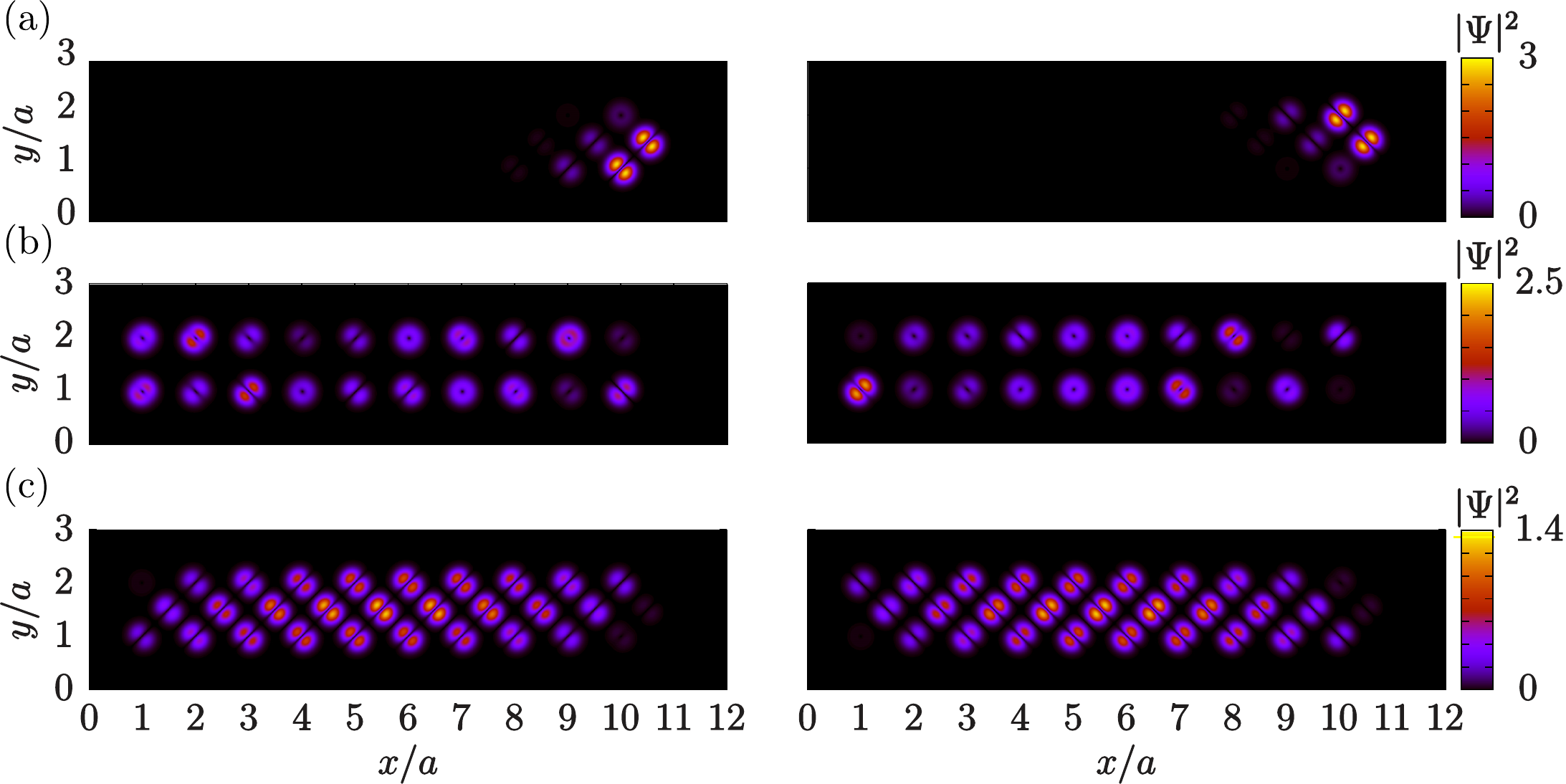}
\caption{Density profiles of numerically obtained eigenstates for a diamond chain of $N_c=10$ units cells and inter-site separation $d=6\sigma$, corresponding to $J_3/J_2=1.13$. (a) Two degenerate edge states. (b) Two states of the flat band. (c) The two degenerate ground states of the system. }
\label{densityplots}
\end{figure}

We perform first a rotation into a basis of symmetric and antisymmetric states, which decouples the diamond chain with six states per unit cell into two independent and identical diamond subchains with three states per unit cell. This explains the two-fold degeneracy of the spectrum and the presence of gaps in the band structure. A second basis rotation maps each of the diamond subchains into a modified Su-Schrieffer-Heeger (SSH) model \cite{su1979solitons} with an extra dangling state per unit cell, which allows to understand the existence of in-gap edge states localized at the right edge of the chain (Fig.~\ref{densityplots}(a)), the zero-energy flat band states without population in the central sites (Fig.~\ref{densityplots}(b)) and the flattening of the bands in the $J_2=J_3$ limit. Fig.~\ref{densityplots}(c) shows the two degenerate ground states of the system. The decoupled subchains do not have inversion symmetry, so that the Zak's phase can yield non-quantized values. Thus, a third mapping to recover inversion symmetry has been introduced \citep{pelegri2019diamond} obtaining a diamond chain with alternating tunneling amplitudes topologically characterized in \cite{marques2018topological, marques2017generalization}. A striking feature of the topology of this model, directly carried over to the original OAM $l=1$ model, is that there is no topological transition across the gap closing point, as can be seen by fixing either $J_2$ or $J_3$ and varying the other across zero. 

Finally, we have also demonstrated that the system can exhibit Aharonov--Bohm caging in the $J_2=J_3$ limit since, in this limit, the states involving the central site of a unit cell can be expressed in terms of flat-band states that occupy solely the four sites surrounding it. Thus, an initial state prepared in an arbitrary superposition of the central sites states will oscillate coherently to its four neighboring sites with a frequency given by the absolute value of the energies of the top/bottom flat-band states without leaving the cage formed by two consecutive unit cells. 
\paragraph*{Simulating quantum magnetism with strongly interacting ultracold bosons}
Up to here, we have neglected interactions among the ultracold atoms. However, as discussed in \cite{pinheiro2013XYZ,pelegri2019quantum}, strongly interacting ultracold bosons loaded into OAM states of lattices of side-coupled cylindrically symmetric traps, e.g., a quasi-one-dimensional ladder of ring potentials or a diamond chain, can realize a variety of spin 1/2 models, including the XYZ Heisenberg model with or without external fields. In \cite{pelegri2019quantum}, we have focused on the Mott insulator regime at unit filling, where each trap is occupied by a single boson and a direct mapping between the degree of freedom corresponding to the two opposite circulations $\pm l$ of the OAM states to a spin 1/2 can be performed. Thus, by tuning the relative phases in the tunneling amplitudes, which depend on the relative orientation between the traps, the system can be used to simulate different spin 1/2 models of quantum magnetism. To this aim, we have computed first, by means of second-order perturbation theory, the explicit dependence of the effective tunneling couplings on the relative angle between the traps. Then, we have discussed for which particular geometries the XYZ Heisenberg model with uniform or staggered external fields could be obtained.
As an example, for a quasi-one dimensional ladder of ring potentials with central angle tuned to $\Theta^l=(2s+1)\pi/(2l)$ with $s\in \mathbb{N} $, single spin flips mediated by  interactions do not take place and only isotropic two-spin flips occur. In this situation, the effective Hamiltonian of the system becomes a XYZ Heisenberg Hamiltonian without external field \cite{pelegri2019quantum}:
\begin{equation}
H^l_{\rm eff}=\sum_{j=1}^N
J_{xx}^l\sigma_j^x\sigma_{j+1}^x+
J_{yy}^l\sigma_j^y\sigma_{j+1}^y+
J_{zz}^l\sigma_j^z\sigma_{j+1}^z
\end{equation}
where $J_{xx}^l=-((J_2^l)^2+(J_3^l)^2)/(2U)$, 
$J_{yy}^l=-((J_2^l)^2-(J_3^l)^2)/(2U)$, and $J_{zz}^l=-3((J_2^l)^2-(J_3^l)^2)/(2U)$, being $J_2^l$ and $J_3^l$ the cross-coupling tunneling amplitudes between states of OAM $l$ possessing equal and different circulations, respectively, while $U$ is the non-linear interaction parameter. 

Worth to highlight, besides engineering different  spin 1/2 models by tuning the geometry of the lattice, the system also allows to adjust the relative strength between the effective couplings by changing the radius of the ring traps and their separation. In fact, we have shown that this additional parameter of control can be exploited in realistic experimental set-ups to explore distinct phases of the XYZ model without external field. Moreover, we have analyzed the effect of experimental imperfections, such as the influence on the tunneling phases of the presence of small fluctuations in the relative angle between the traps. Regarding the physical implementation of the proposal, we have discussed several
possibilities to realize a lattice of ring potentials with a tunable geometry and have analyzed single-site addressing techniques that could allow to retrieve the state of each
individual spin. Finally, we have also investigated the collisional stability of the system and concluded that the anharmonic energy spacing between OAM states introduced by the ring geometry allows extending the lifetime of the Mott state.


\subsection{Concluding remarks and outlook}

The examples detailed in this chapter show that the ring geometry allows both to study the phase diagram  and the main features of the excitation spectrum of the infinite ladder to large accuracy as well as to highlight interesting parity and commensurability effects typical of finite rings. Furthermore, the ring geometry allows for new probes of the various phases \textit{e.~g.}  by the measurement of persistent currents or via  spiral interferometry. It also displays Josephson modes.

In outlook, one should develop suitable theoretical methods to describe the crossover from the weak-interaction and large occupancy regime down to the strongly 
correlated regime reached at large interactions and small filling \cite{victorin2019nonclassical}. 
To make contact with a real experimental situation it is necessary to investigate how much the signatures of these phases are robust against the
finite temperature effects together with the possibility of having long-ranged interactions between the atoms.
Also, the experimental realization of ring ladders seems close to reach and would provide a benchmark of atomtronic devices.

{{\it Acknowledgements}
V.A. and J.M. would like to thank Gerard Pelegrí for fruitful discussions. A.M. and N.V. would like to  thank Paolo Pedri and the late Frank Hekking for discussions.
V.A. and J.M. 
acknowledge financial support from the Ministerio de Economía y Competitividad, MINECO, (FIS2017-86530-P), from the Generalitat de Catalunya (SGR2017-1646), and from the European Union Regional Development Fund within the ERDF Operational Program of Catalunya (project QUASICAT/QuantumCat). A.M. acknowledges funding from the ANR SuperRing (Grant No.  ANR-15-CE30-0012).}





\section{QUANTUM-ENHANCED ATOMTRONICS WITH BRIGHT SOLITONS} \label{Solitons}
\label{SolitonAtomtronics}
\vspace*{-0.5cm}
\par\noindent\rule{\columnwidth}{0.4pt}
{\bf{\small{P. Naldesi, J. Polo, S. A. Gardiner, M. Olshanii, A. Minguzzi, L. Amico}}}
\par\noindent\rule{\columnwidth}{0.4pt}



{Quantum coherent states of macroscopic degrees of freedom are hard to achieve, due to decoherence. Attractive bosons are a very special case study, completely different from repulsive bosons.  In this chapter we revise their  properties in the quantum regime,  such as the chemical composition of its ground state, the transmission across a barrier, the excitation spectrum and the response to rotation, showing that they provide a new type of resource for atomtronics applications.}

\subsection{{Scattering properties of attractive bosons against a barrier}}

\begin{figure}[h!]
\centering
\includegraphics[width=\columnwidth]{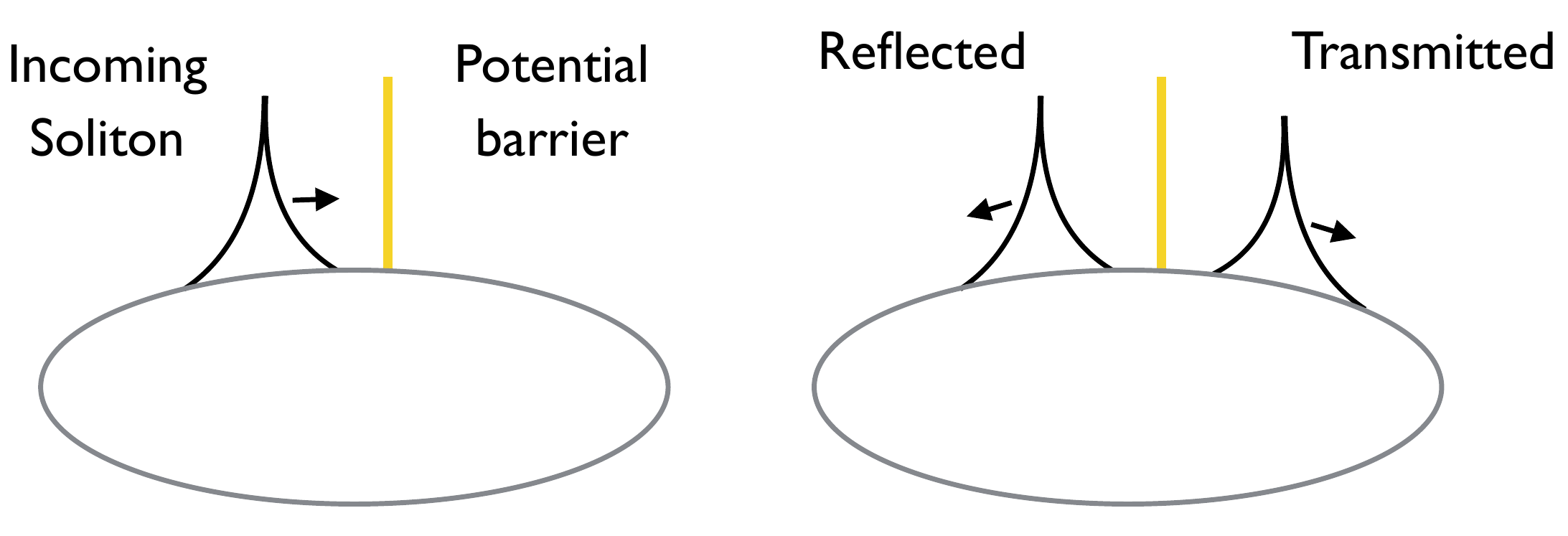}
\caption{Scheme of a soliton scattered from a potential barrier.}
\label{qusup}
\end{figure}

Cold bosonic systems tend to remain in a Bose-condensed state that can be perfectly described by a mean-field theory. However, for attractive condensates, there are points in the space of parameters  where the mean-field theory predicts sudden jumps
{for some macroscopic observables}. 
The relevant example corresponds to a single one-dimensional bosonic soliton\cite{strecker2002formation,khaykovich2002formation,marchant2013controlled} or any other
{many-body bound state}
that is scattered off a barrier {in a typical scattering setup, where a localized wavepacket is prepared and sent towards an obstacle} \cite{pollack2010quantum,boisse2017nonlinear,polo2019traces}. 
{In such scattering events, the nonlinearity of the mean field theory can lead to some ``unsettling'' results. For instance, for}
incident kinetic energies (per particle) below $1/4$ of the magnitude of the soliton chemical potential,
a ``forbidden window''%
{, in the form of a discontinuity, }%
on the axis of the transmission coefficients must emerge \cite{hansen2012scattering} 
(see also \cite{dunjko2015superheated,helm2014splitting}); it appears because in this regime the amount of the incident kinetic energy is insufficient to compensate for the loss of the interaction energy in a 50\%--50\% splitting.
  
As we scan the barrier height from a lower value up, the transmission coefficient increases and at some barrier height, abruptly jumps up \cite{hansen2012scattering}. At the mean-field level, the jump is infinitely sharp. Indeed, a dissociation of the soliton onto the transmitted and reflected parts costs interaction energy, and the incident kinetic energy may not be sufficient to pay for it.

Such a discontinuity is nonphysical. As has been shown in \cite{weiss2009creation,streltsov2009scattering}, the key to ensure the continuity of the transmission coefficient curve is to recognize that at the apparent discontinuity point the condensate becomes fragmented and the transmission events acquire a quantum randomness. This regime will soon be within experimental reach \cite{boisse2017nonlinear}. The good news is  that a highly desirable  Schr\"{o}dinger cat is itself a fragmented state; the bad news is that if the number of occupied one-body orbitals becomes large, the macroscopic coherence becomes unusable.  The paper \cite{weiss2009creation} suggests a secure way of suppressing the undesirable fragments: the soliton kinetic energy must be decreased even further, to a point where the \emph{total\/} kinetic energy becomes less than the chemical potential, thus ensuring no relative motion of the constituent atoms, with only ``cold soliton transmitted'' and ``cold soliton reflected'' allowed orbitals as the result. While conceptually elegant, this method of generating a macroscopic coherence requires center-of-mass kinetic energies $N$ times lower than those currently used ($N$ being the number of atoms in the soliton) and scattering regimes where the barrier becomes completely classical from the soliton center-of-mass point of view.  Accordingly reference \cite{weiss2009creation} suggests using extended center-of-mass coherent wavepackets with a non-zero velocity width, where the barrier is used as a classical velocity filter --- see also \cite{wales2019splitting}. However, a private communication \cite{Ofir_two_orbitals}%
{, and by careful inspection of figures 2 and 3 in reference \cite{streltsov2009scattering},}
indicates --- based on numerical evidence --- that even at moderate kinetic energies, there remain only two populated orbitals. If this is indeed the case, then it is clear what these orbitals are: they are nothing else but the state of the condensate just before and just after the mean-field jump in the transmission coefficient. 

Note that even in the favorable two-orbital case, the macroscopic coherence may still remain unusable due to the entanglement between the center-of-mass motion and possible excitations created during the scattering event. Even if these excitations are small at the level of the BEC wavefunction, the difference between the internal states of the transmitted and reflected condensates may still be large due to the orthogonality catastrophe. Nonetheless, the macroscopic coherence may be potentially preserved if a limited number of atoms is used. The upper bound for this number is yet an open question, which will require an intensive numerical study. 

While in the proposal \cite{weiss2009creation,streltsov2009scattering}, the center-of-mass of the incident soliton is assumed to be in a coherent state prior to the splitting, the cooling of a macroscopic variable to that state is difficult \emph{per se}. However in \cite{yurovsky2017dissiociation,marchukov2019quantum} it is shown that in a factor of four quench of the coupling constant, one can create, while at a finite temperature, an exponentially cold quantum state of a relative distance between the centers-of-mass of two solitons\cite{lai1989quantumI,lai1989quantumII,yeang1999pulse,opanchuk2017one}, itself a macroscopic variable. The theoretical estimates \cite{marchukov2019quantum} show that under realistic experimental conditions, quantum fluctuations of the inter-soliton velocity will lead to an observable inter-soliton separation after a time $$ \tau \approx 4.7\,\text{s} \,\,.$$ 

Further beam-splitting of the inter-soliton distance degree of freedom requires additional study, while its initial coherence is already guaranteed. Classically, the above states would correspond to Gross--Pitaevskii breathers \cite{zakharov1972exact,satsuma1974initial} with 
fluctuating parameters. These have recently been experimentally realized, albeit in the classical regime, as describe in \cite{dicarli2019exitation}.

\subsection{Creation and manipulation of Quantum Solitons}

\subsubsection{Quantum solitons in the Bose-Hubbard model}
Attractive bosons confined in a one dimensional lattice system can be described by the Bose-Hubbard model. {Before moving to atomtronics applications, e.g. a ring lattice, we present here the properties of the ground state and its excitations. For a mesoscopic sample, with limited number of bosons, higher band occupancies are negligible \cite{dutta2015nonstandard} and, even at intermediate and strong attractions the occupancies can be integrated in the model by renormalizing the tunneling and interaction parameters \cite{PhysRevA.94.031601, oosten2003mott, jack2005bosehubbard}.  In this regime, }
the Hamiltonian reads
\begin{equation}
\hat{\mathcal{H}}\!=\!-J\sum_{j\!=\!1}^{L}\left( a_{j}^{\dagger }a_{j+1}+\text{h.c.}%
\right) -\frac{|U|}{2}\sum_{j\!=\!1}^{L}\hat{n}_{j}\left( \hat{n}_{j}-1\right)
\label{BHH-soliton}
\end{equation}%
where the operators $a_{i}^{\dagger }$ obey the canonical commutation relations $[a_{i},a_{j}^{\dagger }]\!=\!\delta _{ij}$, $n_{i}\!=\!a_{i}^{\dagger }a_{i}$ is the number operator at the site $i$; the operators $a_{i}$ and $L$ is the number of sites in the chain. The parameters $J$, $U$ in \eqref{BHH-soliton} are the hopping amplitude and the strength of the on-site interaction, respectively. Periodic boundary conditions are implemented requiring that $a_{1}^{\dagger }a_{L}\!=\!a_{L}^{\dagger }a_{1}$. The lattice is loaded with $N$ bosons. While some exact results are available for $N\!=\!2$ \cite{valiente2008two,boschi2014bound,polo2020exact} and an effictive model can be built to explain the spectrum for $N\!=\!3$ \cite{mattis1986few,valiente2010three}, for larger number of particles the system is not solvable and numerical simulations are necessary \cite{naldesi2019rise,naldesi2019angular}. 
\paragraph*{2-particle sector}
The problem of two attracting bosons on a lattice is exactly solvable {\em \`a la} coordinate Bethe Ansatz by transforming the wave function in the center of mass and relative coordinates. This solution is also valid in the presence of an synthetic gauge field \cite{polo2020exact}. The eigenstates of the system form two bands depending on the nature of relative momentum. For imaginary solutions we have the lowest energy branch composed by $L$ bound-state (solitons), while the real solutions correspond to scattering states which form the second band at higher energy. The energy gap separating the two increases with interactions and it is found that for $U/J \geq 4$ the two bands completely detach for each momenta \cite{polo2020exact}.
\paragraph*{N-particle sector}
For larger number of particles, the BHM \eqref{BHH-soliton} is not solvable by the coordinate Bethe ansatz. The failure results because of finite probabilities that a given site is occupied by \textit{more than two} particles, whose interaction cannot be factorized in 2-body scattering \cite{haldane1980solidification, choy1980some, choy1982failure}. 

\begin{figure}[h!]
\centering
\includegraphics[width=\columnwidth]{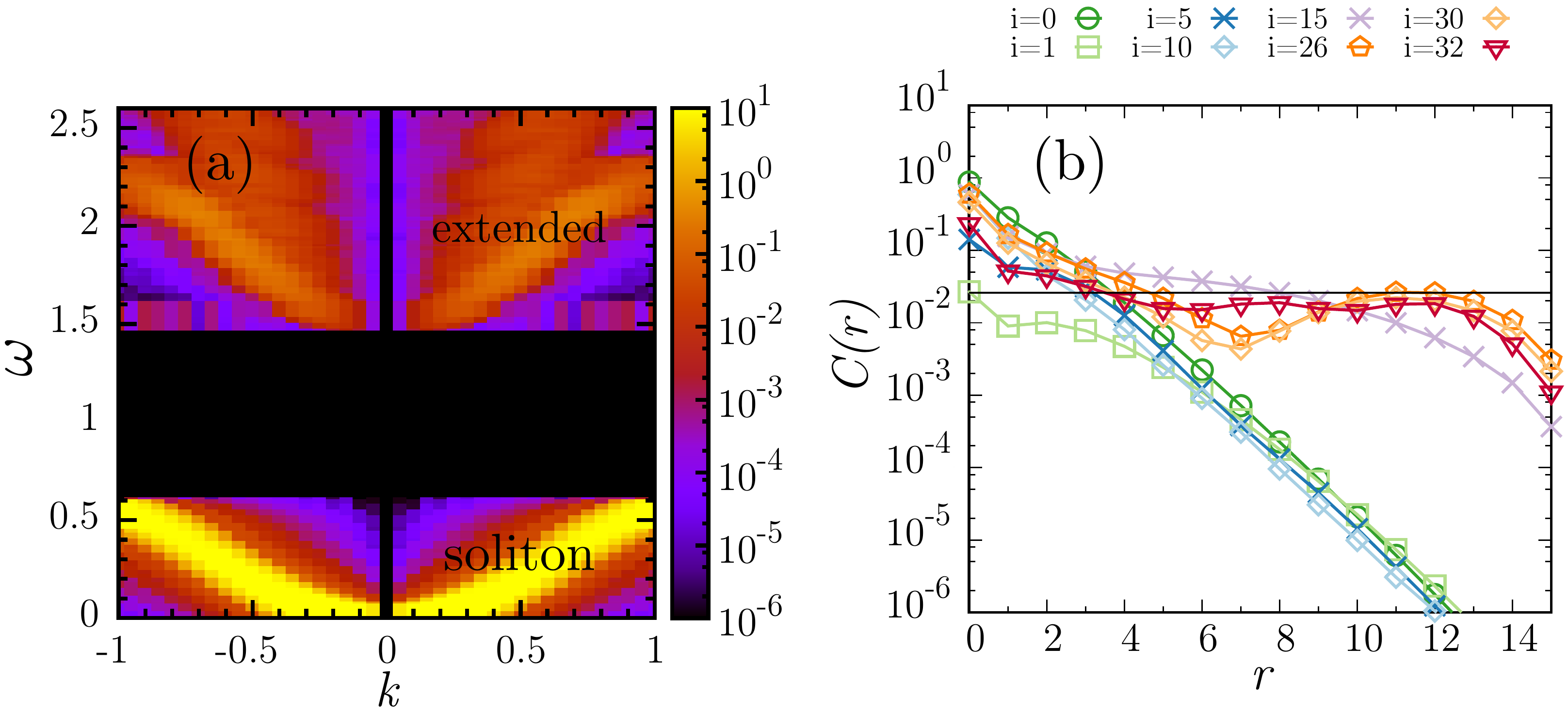}
\caption{Panel (a): dynamical structure factor $S(k,\omega )$ for a chain of $L\!=\!30$ sites. Numerical results for $N\!=\!5$ particles and interactions $U\!=\!1.2$. Panel (b): Density-Density correlation function $C(r)$ for $N\!=\!5$ particles in a chain of $L\!=\!30$ sites and interactions $U\!=\!0.6\!<\!U_{c}$. Correlations are computed over several excited states labelled by i (i-th excited state, i$\!=\!0$ correspond to the ground state).}
\label{Swk}
\end{figure}

Information on the available excitations in the system as a function of their momentum $k$ and energy $\omega$ is provided by the dynamical structure factor $S(k,\omega )$:
\begin{equation}
S(k,\omega )\!=\!\sum_{\alpha \neq 0}\sum_{r}|\left\langle \alpha \right\vert e^{-i k r} \hat{n}_{r}\left\vert 0\right\rangle |^{2}\delta (\omega -\omega _{\alpha }).
\label{strfac}
\end{equation}
where $\hat{n}_{r}$ is the number operator acting on the site $r$, $\ket{0}$ is the ground state and $\alpha$ labels the states with increasing energy (ie $\alpha\!=\!1$ is the first excited state).
The peaks of $S(k,\omega )$ reconstruct the energy bands of the system~ \cite{mattis1986few,valiente2010three} and are shown in \figref{Swk}.
Numerical results show a scenario similar to the two-particle case with a low-energy band that is separated from the rest of the spectrum. The nature of such a band can be analyzed by the study of correlation functions: $C(r)\!=\!\left\langle n_{L/2}\:n_{L/2+r} \right\rangle$.
The numerical analysis shows that the lowest energy band is composed of {many-body} bound states.
In fact all these states are characterized by an exponential decaying of correlations $C(r)\!\sim\!\exp (-r/\xi)$. The correlation length $\xi$ is fixed only by the interactions and decreases with increasing $U$. 
For states belonging to the second branch $C(r)$ approaches, at intermediate distances, a plateau $\sim n_{as}\!=\!(N/L)^{2}$, before dropping down when approaching the walls of the box. We thus can conclude that the higher branch contains extended states. {Notice that, at difference from the continuum case, where a Bethe Ansatz solution is available and one can tell the nature of the state by checking whether the rapidities are real or complex,  in the lattice case there is no exact solution, hence no way to tell whether they are scattering states, $N-1$-body bound states etc. So the dynamical structure factor is very practical to visualize all types of excitations.} Also in this case the bands gap increases with interactions and the critical interaction to have a complete detachment of the bands scales like $U_c \sim 1/N$.
{\paragraph{Soliton stability}}
Finally, we devise a specific dynamical protocol to study the solitons stability and evidence the features of the band structure.

By initially breaking the lattice translational symmetry with an attractive potential $\mathcal{H}_{i}(\mu ,U)\!=\!\mathcal{H}(U)+\mu (U)n_{i_{0}}$, a soliton is pinned in a given site $i_{0}$ of the lattice, and then let it expand by removing the pinning. In this way, while for small $U$ we populate both scattering and bound states, for $U\!>\!U_{c}$ when the gap separates the two bands, mostly bound states are populated.

\begin{figure}[h!]
\centering
\includegraphics[width=\columnwidth]{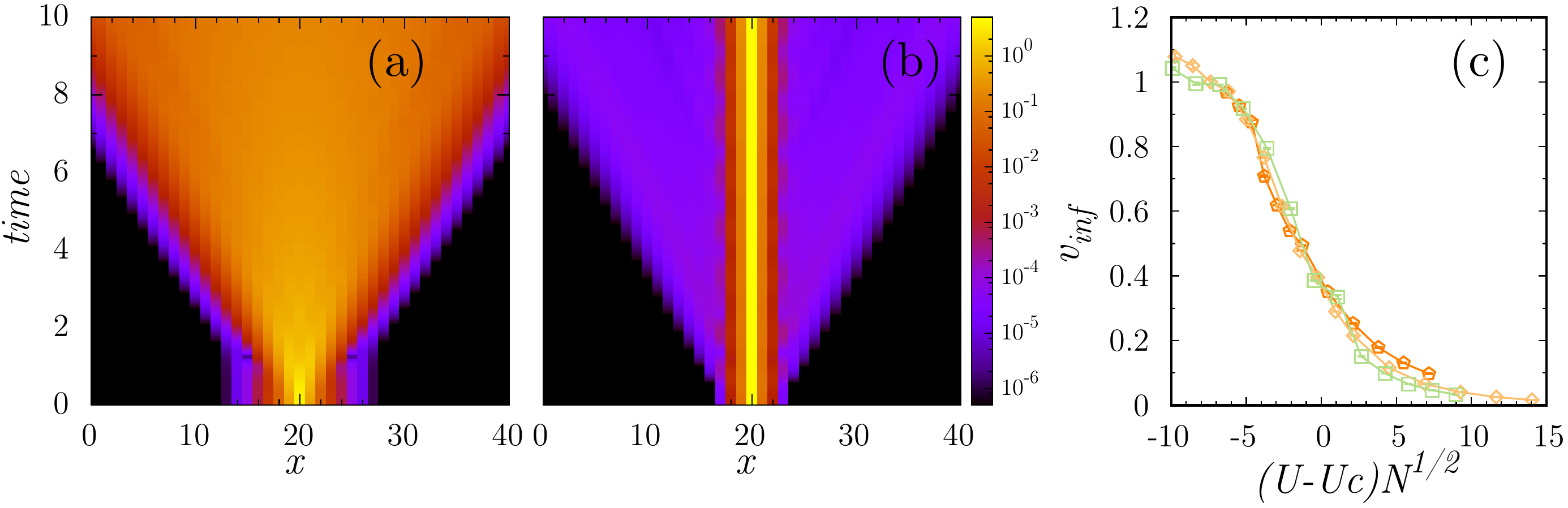}
\caption{Panels (a-b): expansion of a soliton composed by $N\!=\!5$ particles, pinned to the center of a chain with $L\!=\!41$ sites for interactions $U\!=\!0.4$ and $U\!=\!1.8$. Panel (c): asymptotic expansion velocity $v_{\infty }$ as a function of $U\!-\!U_c(N)$ and of $(U\!-\!U_c(N))\sqrt{N}$.}
\label{expansion}
\end{figure}
In Fig.~\ref{expansion} (a-b) we show the expansion dynamics of the density for two cases: $U\!<\!U_{c}$, and $U\!>\!U_{c}$. Increasing the interaction strength, the density profile remains closer and closer to the one of the initial state. Only a small fraction is spreading into the chain leading to a higher stability of the soliton. This phenomenon can be studied more quantitatively by analyzing the expansion velocity: $v(t)\!=\!(d/dt)\sqrt{R^{2}(t)\!-\!R^{2}(0)}$, with $R^{2}(t)\!=\! (1/N)\sum_{i\!=\!1}^{L}n_{i}(t)\left( i\!-\!i_{0}\right)^2$ and its asymptotic value at large times $v_{\infty }$. 
The inspection of $v_{\infty }$ in Fig.~\ref{expansion} (c) further shows the difference between the two regimes. While there is no criticality in the system close to $U_c$, $v_{\infty }$ displays a peculiar scaling behaviour and acts like an order parameters for the system.

\subsubsection{Solitons in rotation}

{As an application to atomtronics, attracting bosons can be used to devise a new type of interferometer, based on superposition of persistent current states.}
The effect of an induced rotation, or more generally of a (syntethic) gauge field, on such a system as been extensively studied \cite{dalibard2011colloquium}.
It's in fact well know that quantum system in a ring geometry displays a staircase response to an applied gauge field of intensity $\Omega$. 
The induced angular momentum increases in quantized steps as a function of $\Omega$ \cite{moulder2012quantized,wright2013threshold} and the amplitude of persistent currents displays periodic oscillations with $\Omega$ \cite{byers1961theoretical,onsager1961magnetic}. The periodicity of such oscillation is completely fixed by the effective flux quantum present in the system, and does not depend on the intensity of particle-particle interactions \cite{beenakker1991nanoelectronics}. 
In the following, without loss of generality, we will refer only to the case of an artificial gauge field induced by a global rotation at angular frequency $\Omega$. Our discussions can be applied to any type of artificial gauge fields.

For strongly correlated one-dimensional bosons with attractive interactions, as we discuss in the following, the nature of flux quantum is non trivial, due to the formation of many-body bound states. 
This feature has dramatic effects on the persistent current 
that oscillates with a periodicity $N$ times smaller than in the standard case corresponding to repulsive interactions. 
Remarkably, the  periodicity depends on interaction, which leads to an extension of  the Leggett theorem.

\paragraph*{Continuous ring}

For a continuous ring, the system can be described through the Bose-gas integrable theory, i.e. the Lieb-Liniger model~\cite{amico2004universality}. 
This is the case when the density $N/L$ of bosons, where $N$ is the particle number and $L\!=\!2\pi R$ is the perimeter of the ring of radius $R$, is small. {A well established limiting procedure allows to link the lattice and continuous models (see eg \cite{amico2004universality,polo2020exact} for a discussion).}
For such systems, exact results are well known~\cite{lieb1963exact}. 
The Lieb-Liniger Hamiltonian in the rotating frame reads:
\begin{equation}
\hat{\mathcal{H}}_{LL}\!=\!\sum_{j\!=\!1}^N\frac{1}{2m}\bigg(p_j-m\Omega R\bigg)^2 + g \sum_{j<l} \delta(x_j-x_l) -E_\Omega,
\label{LL}
\end{equation}
where $m$ and the $p_i$'s are respectively the mass and the momentum of each particle, $L_z\!=\!\sum_{j\!=\!1}^N L_{z,j}$ is the total angular momentum of the $N$ particles, $g$ is the interaction strength and $E_\Omega\!=\!N m \Omega^2 R^2/2$.

The solution of the model dramatically change according to the sign of the interactions.
For repulsive interactions, independently on their strength, the ground state energy $E_{GS}$ is periodic in $\Omega$ with period $\Omega_0\!=\!\hbar/mR^2$. The persistent current in the rotating frame defined as $I_p\!=\!-(\Omega_0/\hbar)\partial E_{GS}/\partial \Omega$ displays a sawtooth behaviour versus $\Omega$ \cite{beenakker1991nanoelectronics}, corresponding to a staircase behaviour of angular momentum $L_z$.

For attractive interactions the scenario changes completely; the ground state is a many-body bound state, {\it i.e.} a 'molecule' made of $N$ bosons, corresponding to the quantum analog of a bright soliton \cite{kanamoto2005symmetry,calabrese2007correlation,naldesi2019rise}.  
The ground state energy for arbitrary $\Omega$ then reads
\begin{equation}
 E_{GS}\!=\! \frac{\hbar^2}{2 M R^2}\left(\ell -N\frac{\Omega}{\Omega_0}\right)^2 - {\frac{N (N^2-1)g^2}{12}},
 \label{egs-ll}
\end{equation}
where the second term accounts for the interaction energy $E_{int}$ and is independent on the rotation frequency. This result clearly shows how, under the effect of the artificial gauge field, attracting bosons effectively behave as a single massive object of mass $M\!=\!Nm$. The energy displays a $1/N$-periodicity as a function of the artificial gauge field, $\Omega$, in units of $\Omega_0$ corresponding to {\it fractionalisation} of angular momentum per particle.

\begin{figure}[h!!!!t!!]
\centering
\includegraphics[width=\columnwidth]{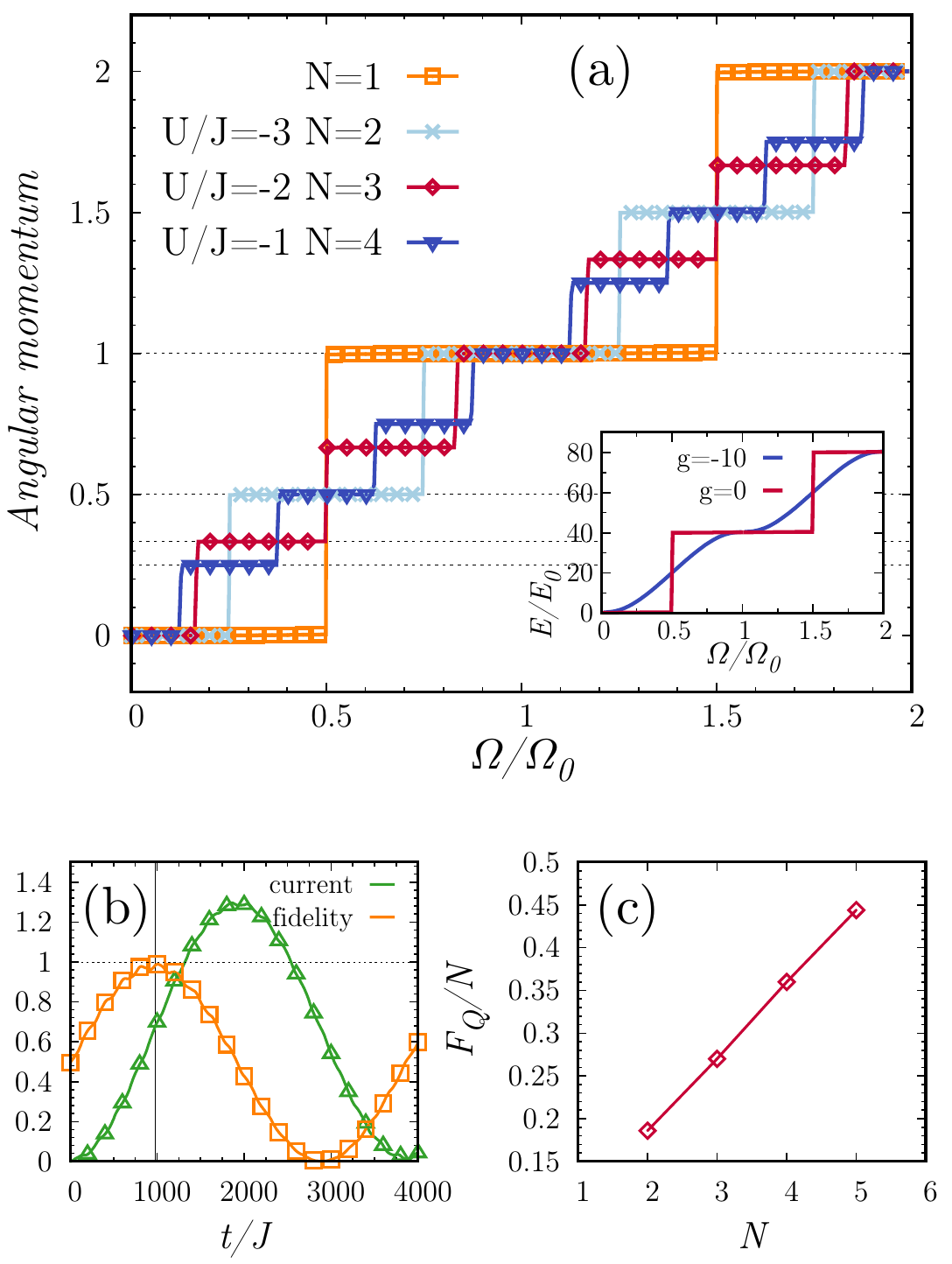}
\caption{Panel (a): average angular momentum per particle  (inset: GP analysis) as a function of the artificial gauge field for different particle number and for particular values of interaction strength.  Panel (b): time dependent current (in units of the hopping constant $J$) following a quench from $\Omega/\Omega_0\!=\!0$ to $\Omega/\Omega_0\!=\!1/2$. Here we set $L=28$, $N=3$, $U/J=-0.51$ and $\Delta_0/J=0.015$. Panel (c): Quantum Fisher information as a function of the particle number showing  the Heinsenberg-limited behaviour $F_Q\propto N^2$.
}
\label{fig:fractional}
\end{figure}

\paragraph*{Lattice ring}
When the density of particles is not small, the lattice effects, that break the integrability of the model, start to be relevant. In this situation the system is well described by the Bose-Hubbard Model (BHM):
\begin{equation}
\hat{\mathcal{H}}_{BH}\!=\! \sum_{j\!=\!1}^{N_s} \frac{U}{2} n_{j}\left( {n}_{j}-1\right) -J\left( e^{-i \tilde{\Omega}}a_{j}^{\dagger }a_{j+1}+\text{h.c.} 
\right) ,
\label{BHH-lattice}
\end{equation}

where $a_{j}$ and $a_{j}^{\dagger }$ are site $j$ annihilation and creation Bose operators and $n_{j}\!=\!a_{j}^{\dagger }a_{j}$ is number operator. The parameters $J$, $U<0$ in \eqref{BHH-lattice} are respectively the hopping amplitude and the strength of the on-site interaction, $N_s$ being the number of sites in the lattice and $\tilde{\Omega}\doteq 2\pi \Omega/(\Omega_0N_s)$ for brevity. 

In the lattice model~(\ref{BHH-lattice}), the center-of-mass and relative coordinates, at any finite interaction, cannot be decouple. This feature has a profound implication on the behaviour of persistent current. As we will discuss below, in contrast with the continuous theory, here the persistent current periodicity does depend on interaction strength. 

In Fig.~\ref{fig:fractional} we show the numerical results angular momentum: also in this case the $1/N$ periodicity in $\Omega/\Omega_0$ of the persistent currents emerges, as well as fractionalization of angular momentum.
While fractionalization always occurs, the $1/N$ periodicity, is affected by the interplay between system size and interaction strength. 

When interactions are sufficiently large the 'size of the many-body bound state', i.e. the decay length of the density-density correlations \cite{naldesi2019rise}, is much smaller than the size of the system.
Upon decreasing the interactions, the size of the many-body bound state increases more and more over the chain and the solitonic nature of the state gets less and less pronounced.
All the observed features are  purely quantum many-body effects tracing back to specific quantum correlations since they completely disappear in a mean-field Gross-Pitaevskii description of the system.

Angular momentum fractionalization and the related persistent current periodicity can be observed with standard time-of-flight (TOF) techniques. Measuring the distributions of the atoms after releasing the trap confinement and turning off interactions we have access to the momentum distribution, defined as
$n(\mathbf{k}) \!=\! |w(\mathbf k)|^2 \sum_{j,l} e^{i \mathbf{k}\cdot(\mathbf{x}_j-\mathbf{x}_l)} \langle a^\dagger_j a_l\rangle$. In fact we find that the mean-square radius of the distribution increases in fractional steps of  for $\Omega/\Omega_0\!=\!\ell/N$ \cite{bretin2004fast}.

\subsubsection{Entangling solitons with different $L_z$}

We finally demonstrate how the scenario above can be harnessed to  create specific entangled states of persistent currents. Such entangled states are characterised by an increased sensitivity to the effective magnetic field that reaches the Heiseberg limit. In the following, we propose a specific dynamical protocol that allows us to create such type of state.

Since the Hamiltonians in Eqs.(\ref{LL}), (\ref{BHH-lattice}) commute with the total angular momentum, dynamically mix entangle states with different angular momentum, the rotational invariance of the system needs to be broken.
The ring is then interrupted with a potential barrier of strength $\Delta_0$ localized in a single lattice site.
Then the artificial gauge field is quenched from $\Omega\!=\!0$ to $\Omega\!=\!\Omega_0/2$. This procedure is capable to dynamically entangle the angular momentum state at $\Omega\!=\!0$, ie $L_z\!=\!0$, with the one  at $\Omega\!=\!\Omega_0$, ie $L_z\!=\!N$ (see  Fig.~\ref{fig:fractional}), yielding    $\ket{\psi}_{NOON} \!=\! \frac{1}{\sqrt{2}} \left( \ket{L_z\!=\!0} + \ket{L_z\!=\!N} \right)$ when the current reaches the half of its maximum value. 

The response of such a state to an external rotation is $\ket{\psi(\phi)} \!=\!e^{i \phi \hat{L}_z/\hbar} \ket{\psi}_{NOON}$, and the quantum Fisher information \cite{caves1994statistical,pezze2009entanglement} $F_\mathcal{Q} \!=\! 4 \left(\bracket{\psi'(\phi)}{\psi'(\phi)} - |\bracket{\psi'(\phi)}{\psi(\phi)}|^2 \right)$, being $\ket{\psi'(\phi)}\!=\! \partial \ket{\psi(\phi)}/\partial \phi$. For our state we find $F_\mathcal{Q}\sim N^2$, ie it reaches the Heisenberg limit - see Fig.~\ref{fig:fractional}. The corresponding sensitivity $\delta \phi$, therefore, is 
\begin{equation}
\delta \phi \ge \frac{1}{(F_\mathcal{Q})^{1/2}} \!=\!\frac{1}{ N} ,
\end{equation}
This shows that entangled states of quantum solitons with different angular momenta lead to a quantum advantage of the sensitivity {for rotation detection}. {Notice  that this type of entangled state is completely different  from a superposition state obtained by splitting a soliton by  a barrier in real space, which could be used eg as a gravimeter. In both cases the main experimental limitation is due to unwanted fluctuations (thermal, technical, etc) and particle losses. The latter are nevertheless expected to play a minor role for the small particle numbers considered in this setup.}

\subsection{Concluding remarks and outlook}
In this chapter, we studied  attractive bosons in the quantum regime. Its ground state is an $N$-body bound state, which on a lattice is protected by a gap with respect to the first branch of excitations, corresponding to scattering states. We have shown that this implies the stability of a soliton initially prepared in a pinning site.  We have also shown that quantum solitons on a ring display an enhanced response to artificial gauge field $\Omega$, with a $1/N$ periodicity as a function of $\Omega/\Omega_0$. This corresponds to fractionalization of angular momentum per particle, intrinsically due to the presence of many-body bound states. Finally, we have identified a protocol to create a non-classical superposition of angular momentum states by a suitable quench of the artificial gauge field, based on  angular momentum fractionalization. The use of quantum coherent macroscopic superposition states in atom interferometry devices can increase considerably the phase sensitivity. The states studied in this chapter can yield an $N$-fold enhancement in sensitivity to rotation in a ring-based gyroscope. In a typical configuration, a localized barrier can split the solitons in two waves propagating in clock-wise and anticlockwise that can ultimately recombine producing interference fringes with a specific npattern.  Controlling the  effects of decoherence, losses, and the identification of the optimal working parameters for bright solitons based  interferometers are important challenges to overcome.





\section{ATOMTRONICS WITH ALKALINE-EARTH-LIKE METAL ATOMS}
\label{alkaline}
\vspace*{-0.5cm}
\par\noindent\rule{\columnwidth}{0.4pt}
{\bf{\small{D. Wilkowski, W.J. Chetcuti, C. Miniatura, L.-C. Kwek, L. Amico}}}
\par\noindent\rule{\columnwidth}{0.4pt}



\subsection{Why alkaline-earth-like metal atoms?}
\label{Why alkaline-earth-like}

\begin{figure}
\includegraphics[width=0.5\textwidth]{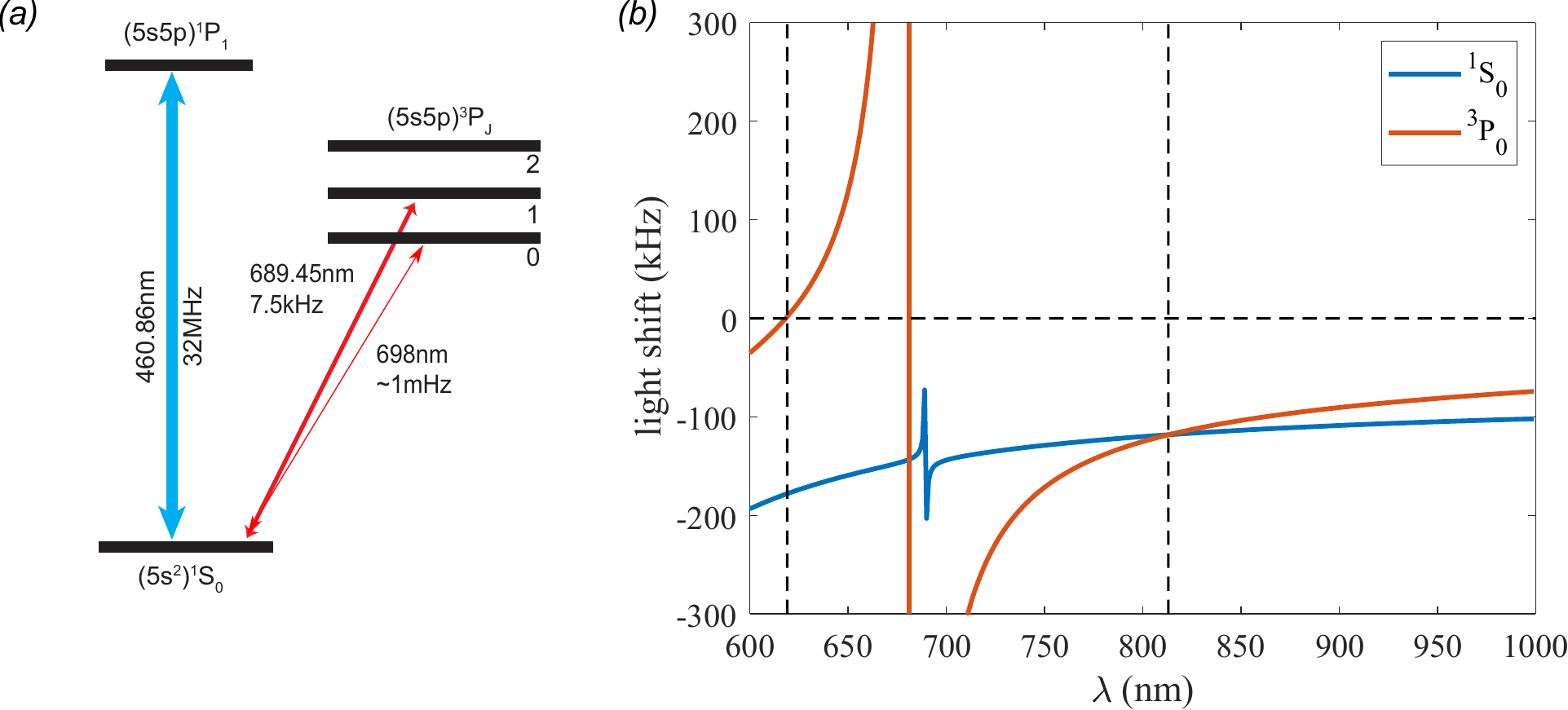}
\caption{(a) Energy levels and transitions of interest for Strontium.
The transitions linewidths and wavelengths are indicated in
the figure. (b) Light shifts of the $^1S_0$ and $^3P_0$ states of Sr as a function of the wavelength of the dipole trap. The excited-state and ground state light shifts are exactly equal at the magic wavelength $813\,$nm. The light shifts are calculated for a laser power of $600\,$mW and a beam waist of $65\,\mu$m.  \label{Sr_Spectrum_Light_Shift}}%
\end{figure}

Over the past two decades, the number of experiments using ultracold alkaline-earth-like metal atoms have considerably increased. Indeed, these atoms have singlet and triplet electronic spectra that offer interesting alternatives over the usual doublet spectrum of the more commonly used alkali metal atoms. For the purpose of illustration, we show the energy levels and transitions of interest for strontium atoms (Sr) in Fig. \ref{Sr_Spectrum_Light_Shift}(a). Laser cooling and magneto-optical traps are achieved using the electric dipole-allowed singlet $^1$S$_0\rightarrow ^1$P$_1$ transition. For heavy elements (Sr, Yb, Hg), the singlet-triplet intercombination line is strong  enough to allow further cooling. For example, for Sr atoms, reaching temperatures at the single photon recoil limit on large atomic ensembles \cite{yang2015high} can be achieved simply with the usual Doppler cooling technique \cite{chalony2011doppler}. In addition, since multiple scattering is limited, large space phase densities can be reached compared to alkali metal atoms \cite{katori1999magneto}. Laser cooling on intercombination lines is then efficient, and provides a favourable starting point to reach quantum degeneracy with evaporative cooling techniques. The latter was obtained for several isotopes of Yb and Sr, such as $^{84}$Sr (0.6$\%$)\cite{stellmer2009bose,escobar2009bose}, $^{86}$Sr (9.9$\%$)\cite{stellmer2010bose}, $^{87}$Sr (7.0$\%$)\cite{tey2010double,desalvo2010degenerate}, $^{174}$Yb (31.8$\%$)\cite{takasu2003spin} , $^{173}$Yb (16.1$\%$)\cite{fukuhara2007degenerate}, $^{176}$Yb (12.8$\%$)\cite{fukuhara2009all}, and also for $^{40}$Ca (96.9$\%$)\cite{kraft2009bose}. The percentage, given in parenthesis, is the relative abundance of the isotope.

Importantly, we note that the spin-singlet ground state is not sensitive, or only weakly so (for nuclear spin of fermionic isotopes), to magnetic fields.
Magnetic trapping is thus excluded as well as the possibility of using magnetic Feshbach resonances to tune the scattering length and, in turn, interactions. One has to rely on optical dipole traps and zero-field interactions to implement evaporative cooling. There have been attempts to control the scattering length by optically dressing the ground state level to some molecular bound states in the excited level \cite{enomot2008optical} but the lifetimes of such dressed states remain too short to be of practical interest \cite{nicholson2015optical}. 

In addition to the electric-dipole transition and the intercombinaison lines, alkaline-earth-like metal atoms possess a clock transition connecting the ground state $^1S_0$ to the long-lived excited state $^3P_0$. Since these states are energetically well separated, their light shifts, induced by a far-off-resonant laser light, can be engineered almost with high control. An illustrative example is given in Fig. \ref{Sr_Spectrum_Light_Shift}(b) for Sr. For instance, in the so-called magic configuration, where light shifts exactly compensate, the transition frequency becomes almost insensitive to the trapping optical field leading to applications in precision frequency and time measurements \cite{takamoto2005optical}. Here atoms are trapped in optical lattices to allow for long interrogation times in a massively parallel configuration, giving the best clock uncertainty to date \cite{bothwell2019jila}. Aside from obvious metrological applications, the clock transitions are also suitable for strongly-correlated many-body phenomena that may be difficult to be addressed elsewhere, such as the Kondo effect and the heavy Fermion manifestation \cite{foss2010probing,riegger2018localized}. These experiments can be performed in an optical lattice with a wavelength corresponding to weak light shifts in the excited state and a stronger confinement in the ground state [$\sim 619\,$nm for Sr as shown in Fig. \ref{Sr_Spectrum_Light_Shift}(b)]. Moreover, the fermionic isotopes (Sr and Yb) possess a nonzero nuclear spin. Thus, a spin exchange interaction between two atoms is present in the cold collision regime; one in the fundamental orbital $^1S_0$ and the other in the excited orbital $^3P_0$ \cite{scazza2014observation}. The resonant character of this Feshbach-type exchange interaction has been shown for Yb\cite{hofer2015observation,pagano2015strongly} and opens the door for quantum simulation of strongly correlated 2-orbital quantum gases\cite{riegger2018localized}. 

The fermionic isotopes, $^{87}$Sr et $^{173}$Yb, have also the interesting property of possessing a nuclear spin larger than $1/2$ ($I=9/2$ and $I=5/2$ respectively) decoupled from the electronic orbital \cite{gorshkov2010two,cazalilla2014ultracold} (beside the above mentioned spin exchange interaction). As a consequence, the many-body Hamiltonian of these systems do not depend on the nuclear spin orientation: They are invariant under the $SU(N)$ symmetry with a dimension $N=2I+1$ much larger than $N=2$ (corresponding to a spin-1/2 fermion), going thus beyond the usual $SU(2)$ symmetry. Numerous theoretical efforts have been pursued to better understand such $SU(N)$ systems, in particular their magnetic \cite{gorshkov2010two} and topological properties, and their quantum phase transitions \cite{capponi2016phases}. On the experimental side, important results have been obtained on the Yb Mott insulator \cite{taie20126,hofrichter2016direct}. Ordered magnetic phases above the N\'eel temperature could be observed \cite{ozawa2018antiferromagnetic} because the entropy per spin component was reduced by the Pomeranchuk effect, relaxing the temperature constraint on the gas \cite{gorshkov2010two,hazzard2012high}. 

The Sr fermionic isotope has also been used to generate artificial gauge fields. An effective spin-orbit coupling, mediated by the clock transition has been studied in a one-dimensional lattice \cite{wall2016synthetic,bromley2018dynamics}. The goal here is to act on the ultracold gas to obtain many-body states of metrological interest. In another work, non-Abelian gauge transformations have been reported using two dark states of a tripod laser scheme\cite{leroux2018non}. This configuration appears to be promising for atomtronics and will be discussed in more detail in the following Section.  

\subsection{Effective Abelian and non-Abelian gauge fields}
\label{Gauge_Field}

\begin{figure}
\includegraphics[width=0.5\textwidth]{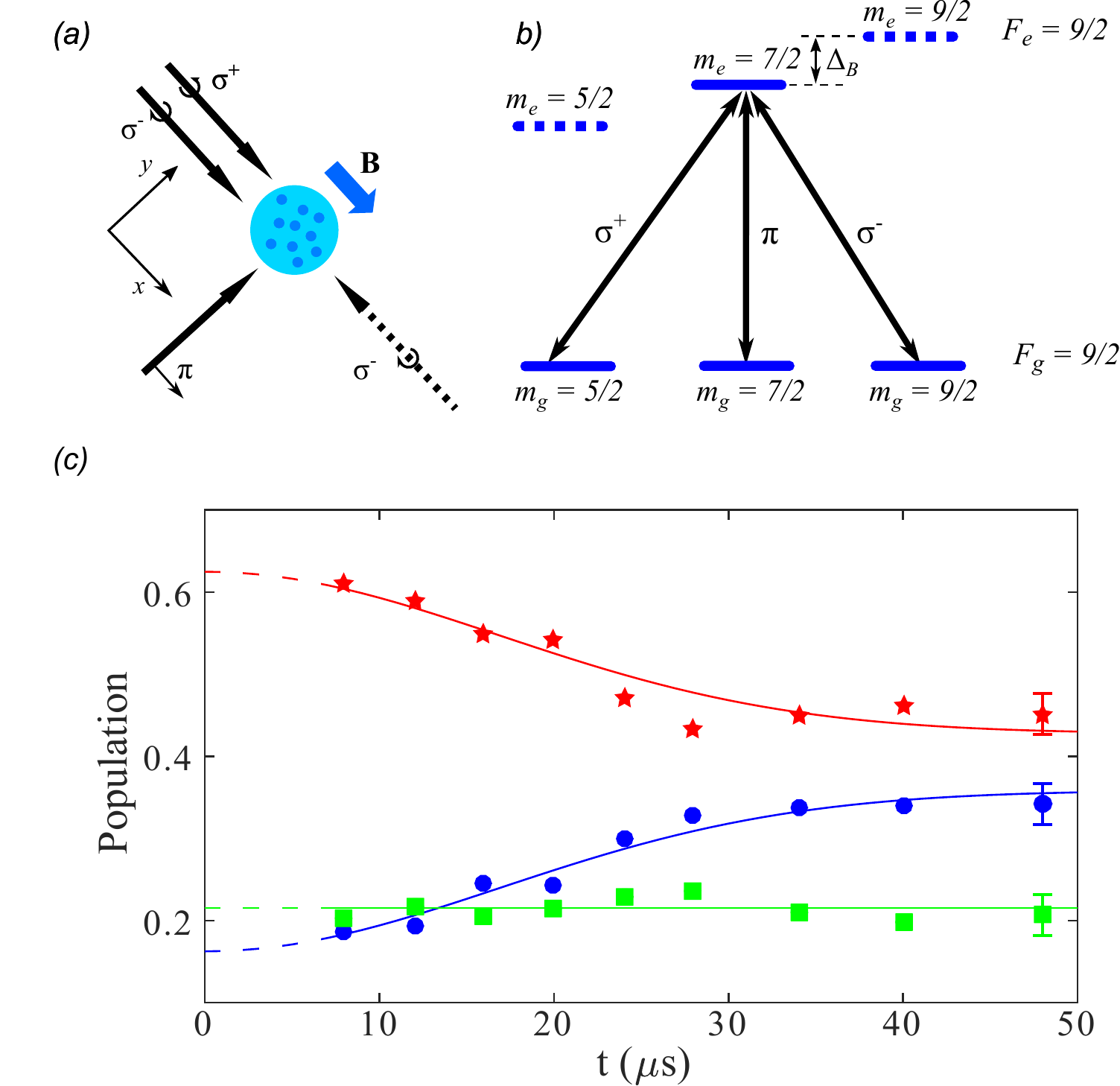}
\caption{(a) Propagation directions of the tripod laser beams (full arrows) and their polarizations along a magnetic field bias ($B =67\,$G). For the non-Abelian gauge field discussed in the text, the $\sigma^-$ laser beam direction is flipped (dashed arrow). (b) Energy levels and experimentally relevant transitions. The magnetic
bias field lifts the degeneracy of the different Zeeman manifolds
and allows to address each transition individually. The Land\'e $g$-factors are
indicated for each hyperfine level. The black arrows connecting the ground state to the excited states correspond to the
tripod beams. (c) Time evolution of the bare-state populations after
the tripod ignition. Red stars, green squares, and blue circles correspond to $|m=9/2\rangle$, $|m=7/2\rangle$, and $|m=5/2\rangle$ states respectively. 
Adapted with permission from F. Leroux, K. Pandey, R. Rehbi, F. Chevy, C. Miniatura, B. Grémaud, and D. Wilkowski, Nat. Commun. 9, 3580 (2018), under a Creative Commons Attribution 4.0 International License.
\label{Gauge_field_fig}}%
\end{figure}

We discuss here the implementation of effective gauge fields for ultracold alkaline-earth metal atoms in the presence of a general atomtronics circuit. We require the spatial scales of the atomtronics circuit to be larger than the laser wavelength used to create the gauge field such that the adiabatic approximation always holds \cite{dalibard2011colloquium}. State differently, the artificial gauge field should act in continuous bulk space and be unaltered by the presence of the atomtronics circuit. This excludes lattice-type structures but one can still implement the gauge field using off-resonant coherent Raman beams as used in the group of I. Spielman \cite{lin2011spin}. Alternatively, $SU(2)$ gauge fields can be generated within the dark-state manifold of a four-level resonant tripod scheme \cite{ruseckas2005non}. Using a double tripod scheme, the symmetry can be further extended to $SU(3)$ \cite{hu2014u}. Since dark states are sensitive to ground state energy fluctuations, this scheme is appropriate for alkaline-earth-like metal fermions which only possess a nuclear spin well protected from their environment (magnetic fields, collisions). 

Recently\cite{leroux2018non}, we implemented a tripod scheme on a cold gas of $^{87}$Sr containing around $10^5$ atoms using the $F_g = 9/2 \rightarrow F_e = 9/2$ intercombination line at 689 nm. The cold sample \cite{yang2015high} is prepared in a crossed optical dipole trap where atoms are optically pumped in the stretched $m = F_g$ magnetic substate and Doppler cooled at a temperature around $0.5\,\mu$K. A magnetic bias field isolates a particular tripod scheme among the excited and ground Zeeman substate manifolds. The three coupling laser beams are set on resonance with their common $|m = 7/2, F_e = 9/2\rangle$ excited state [see Fig. \ref{Gauge_field_fig}(a) $\&$ (b)].


Under the adiabatic approximation, the Hamiltonian describing the quantum state evolution in the dark-state manifold \cite{dalibard2011colloquium} reads 
\begin{equation}
H=\frac{1}{2M}({\bf \hat{p}} \openone-\textbf{A})^2+W,
\label{eq:Scho}
\end{equation}
where ${\bf \hat{p}} = -i \hbar \boldsymbol \nabla$ is the momentum operator, $\openone$ is the identity operator in the internal dark-state subspace, and $M$ the atom mass. With equal and constant Rabi frequencies amplitudes, and for the orientation of our laser beams [see Fig. \ref{Gauge_field_fig}(a)], the vector and scalar potential are \cite{leroux2018non}
\begin{equation}
\textbf{A}=\frac{2\hbar(\textbf{k}_2-\textbf{k}_1)}{3} \, \mathcal{M}, \hspace{1cm} W=-\frac{4E_R}{9} \, \mathcal{M},
\label{eq:A}
\end{equation}
where $E_R=\hbar\omega_R=\hbar^2k^2/(2M)$ is the atomic recoil energy and $\bf k_j$ is the wavevector of laser beam $j=1,2,3$ (with $1\equiv\sigma^+$, $2\equiv\pi$, and $3\equiv\sigma^-$). The matrix $\mathcal{M}$ reads,
\begin{equation}
\mathcal{M} =  \left(
  \begin{array}{cc}
    3/4 & -\sqrt{3}/4 \\
    -\sqrt{3}/4 & 1/4 \\
  \end{array}
\right).
\end{equation}
Since the components of the vector potential commute, this gauge field is Abelian. In Fig. \ref{Gauge_field_fig}(c), we show the result of a ballistic expansion of the cold atomic cloud on the bare state populations. The red stars, green cubes, and blue triangles correspond to the $|m=9/2\rangle$, $|m=7/2\rangle$, and $|m=5/2\rangle$ populations respectively, whereas the curves correspond to the evolution given by the Hamiltonian, Eq. (\ref{eq:Scho}), with thermal averaging \cite{leroux2018non}. The relaxation of the $|m=9/2\rangle$, and $|m=5/2\rangle$ populations is due to the thermal averaging, and the temperature is proportional to the characteristic relaxation time of the system \cite{leroux2018non}. 

Remarkably, since the gauge field is Abelian and homogeneous, the field strength (or Berry curvature), i.e. the curl of the vector potential, is zero. Hence, there is no Lorentz forces acting on the system. This result can be simply understood through a simple physical argument: Mechanical forces here come from photon redistribution among the tripod lasers with different propagation directions. With our laser configuration, such photon exchanges would induce a population change only for state $|m=7/2\rangle$. However, this population remains constant, confirming the absence of light-assisted forces.

The situation becomes more complex if one flips the direction of one of the laser along the $x$-axis. For instance, if the laser $\sigma^-$ is flipped, the gauge potential now reads
\begin{equation}
\textbf{A}=\frac{2\hbar}{3}\left(\textbf{k}_1\mathcal{N}+\textbf{k}_2\mathcal{M}\right),
\label{eq:NA}
\end{equation}
where
\begin{equation}
\mathcal{N} =  \left(
  \begin{array}{cc}
    9/4 & \sqrt{3}/4 \\
    \sqrt{3}/4 & 3/4 \\
  \end{array}
\right).
\end{equation}
The two components of the vector potential do not commute, so the gauge field becomes non-Abelian. In the context of atomtronics, spin precession and related spin-orbit-like coupling can play an important role. For instance, the spin precession leads to spatial oscillation of the wave-packet (as in relativistic {\it Zitterbewegung} effect) whereas the momentum operator still commutes with the Hamiltonian \cite{huo2014solenoidal,beeler2013spin,vaishnav2008observing}. 
 Moreover, it was shown that the characteristic double-well energy dispersion of a spin-orbit coupled system, leads to a Josephson effect in momentum space, with presence of supercurrents \cite{hou2018momentum}. Further applications and potential research objectives are give in the next Section.  
{
\subsection{Persistent Current of SU($\textbf{N}$) Fermions}\label{pcsun}

Atomtronics can provide key contributions to  mesoscopic physics, exploring physical situations that are hard, if not impossible, to explore with standard implementations. One of the purest expressions of mesoscopic behaviour is the persistent current. There have been several studies on the persistent current of bosonic systems in ring-shaped circuits. In this article, these are summarized in Chapters~\ref{Persistent_toroid} and~\ref{DeviceSensors}.   Atomtronic circuits comprised of ultracold fermions are much less explored. In~\cite{chetcuti2020persistent},  the persistent current  of interacting multi-component SU($N$) fermions is studied. The system is modelled by the SU($N$)  Hubbard model~\cite{capponi2016phases} with repulsive interaction; the particles are confined in a ring-shape circuit pierced by an effective magnetic field.  As discussed in the Chapt.\ref{Persistent_toroid}, the  zero temperature  persistent current $I (\phi )$ is defined as
\begin{equation}\label{eq:pcs}
I(\phi ) = -\frac{\partial E_{0}}{\partial \phi}
\end{equation}

By applying a combination of Bethe ansatz \cite{sutherland1968further}\cite{capponi2016phases,sutherland1975model} and numerical analysis, it is  demonstrated how the persistent current displays a specific dependence on the parameters characterizing the physical conditions of the system. A combination of spin correlations, effective magnetic flux and interaction brings about a peculiar phenomenon: spinon creation in the ground state.
\begin{figure}[h!]
\includegraphics[width=0.45\textwidth]{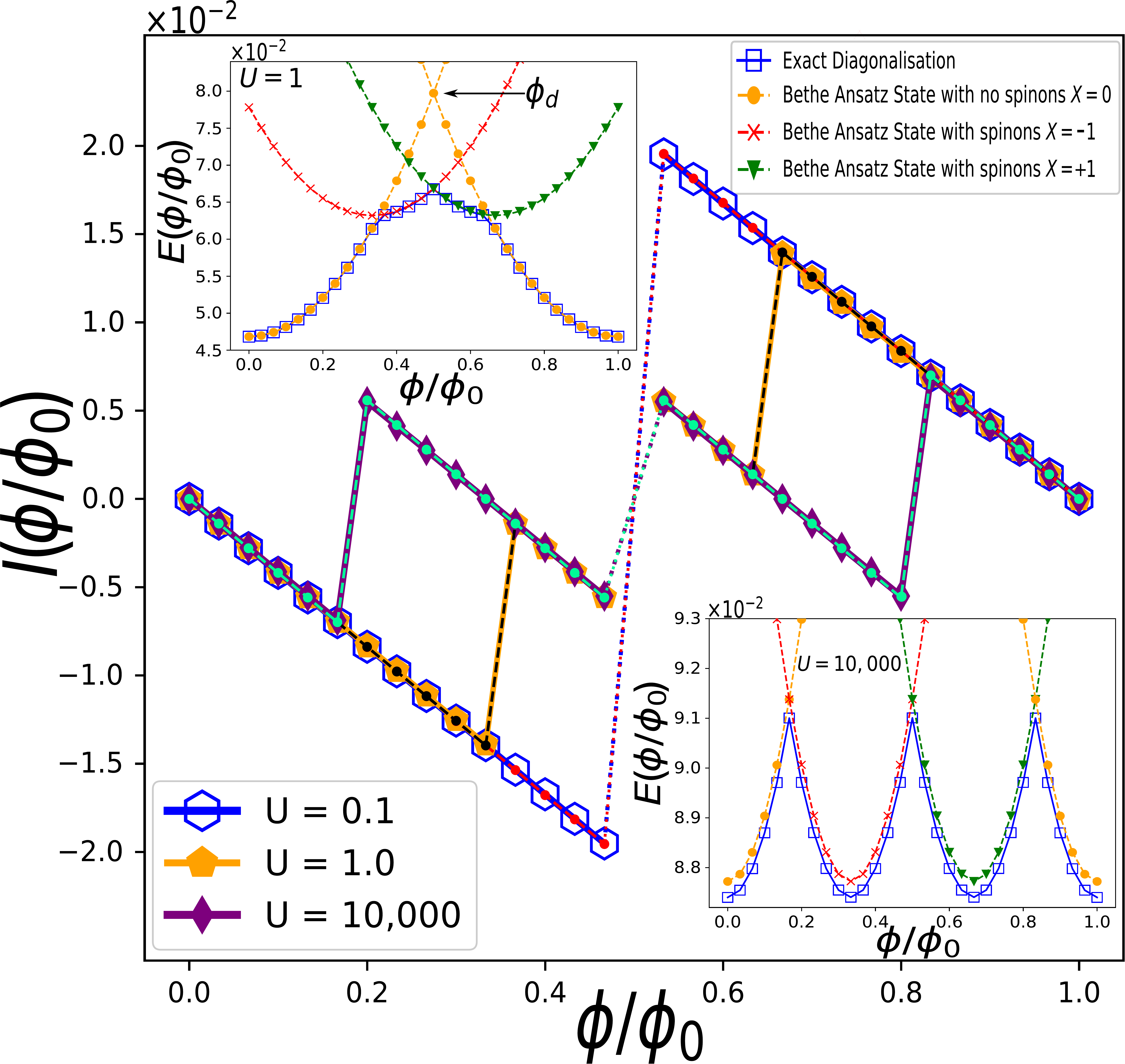}
\caption{Persistent current $I(\phi )$ at incommensurate filling for SU(3) fermions with different interaction strengths U in the dilute filling regime of the Hubbard model. The exact diagonalization $L = 30$, $N_{p} = 3$ is monitored with the Bethe ansatz of the Sutherland- Gaudin-Yang model. The Insets show how the Bethe ansatz energies need to be characterized by spinon quantum numbers in order to be the actual ground state. At $U = 0$, the ground state energy is a periodic sequence of parabolas meeting at degeneracy points $\phi_{d}$ ($\phi_{d} = 1/2$ for the case displayed in the figure). Figure is taken from reference~\cite{chetcuti2020persistent}.
\label{PCS}}%
\end{figure}

Creation of spinons in the ground state leads to a re-definition of the elementary flux quantum $\phi_{0}$, which fixes the periodicity of the current.  From Figure~\ref{PCS}, one can clearly observe how the profile of the persistent current changes with increasing $U$, which reflects the periodic $1/N_{p}$ oscillations in the ground state energy that in the large interaction regime, results in $N_{p}$ parabolic cusps/segments. 

Such fractionalization of the flux observed here is very different from the one typically observed in bosonic system with attractive interactions (see Section~\ref{Solitons}): While the fractionalization in bosonic systems arise from the formation of bound states, for repulsing  fermions the phenomenon is a direct manifestation of   the coupling between the spin and matter degrees of freedom.

Spinon creation in the ground state displays a marked dependence on the number of spin components, highlighting distinguishing features between SU(2) and SU($N$) fermions for $N>2$. In particular, for integer fillings, at variance with their standard two spin component fermions counterpart~\cite{lieb1968absence}, SU($N$) fermions with $N>2$ undergo a Mott quantum phase transition for a finite value of the interaction. Despite its mesoscopic nature, the persistent current is able to detect the onset of the Mott transition marked by a clear finite size scaling.  Furthermore, the presence of a Mott gap suppresses spinon creation in the ground state.  Lastly, a specific SU($N$) parity effect is shown to hold whereby the current is diamagnetic (paramagnetic) in nature for systems comprised of $(2n+1)N$ [$(2n)N$] number of fermions, with $n$ being an integer. This result generalizes a prediction by Onsager, Byers-Yang and Leggett~\cite{byers1961theoretical,onsager1961magnetic,leggett1991granular}. 
}

\subsection{Concluding remarks and outlook}
\label{Outlook}

\begin{figure}
\includegraphics[width=0.5\textwidth]{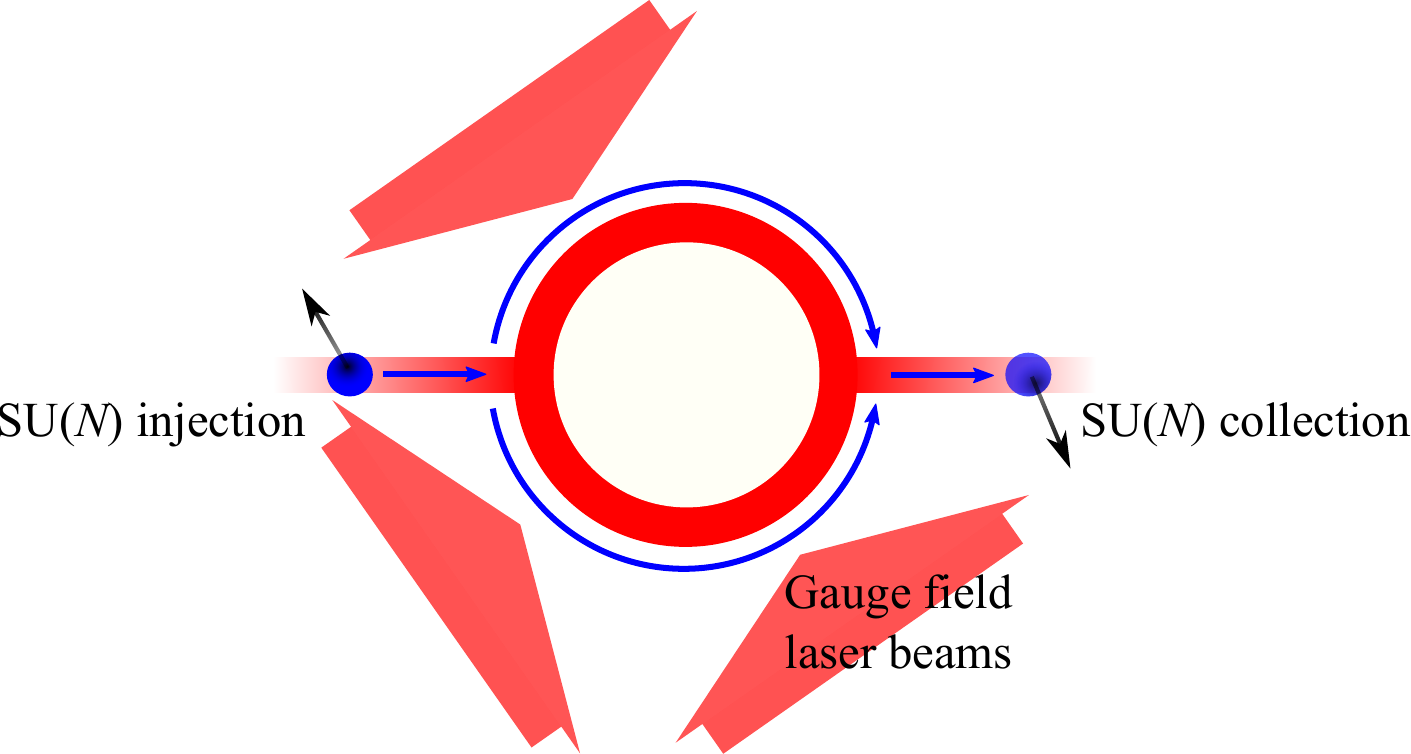}
\caption{Sketch of a (non-Abelian) $SU(N)$ Aharonov-Bohm matter-wave device. The tripod laser beams illuminate the full ring structure. With a single tripod scheme, we have $N=2$. With a double tripod scheme, we have $N=3$.\label{Atomtronics_fig}}%
\end{figure}

In section \ref{Why alkaline-earth-like}, we saw how to cool down alkaline-earth-like metal atoms to degeneracy and how we can take advantage of the fermionic isotopes and clock transition to address $SU(N)$ Hamiltonian and engineer the light shifts between two orbital states almost at will. In Section \ref{Gauge_Field}, we discussed the laser tripod scheme to implement (homogeneous) Abelian and non-Abelian gauge fields in the bulk [see Fig. \ref{Gauge_field_fig}(b)]. In this Section, we discuss several potential research objectives and applications in elementary atomtronics circuit in the presence of $SU(N)$ symmetry and/or gauge fields.

In the simplest instance, one can consider rectilinear and ring-shaped quasi-1D guides. The practical implementation can be done following methods discussed in Section \ref{sculpted_light}.  
Then, one can generate a non-Abelian synthetic gauge field in which the atomic spin will also change its orientation. This way, spin Hall current might be present even if the gauge field is constant and uniform in space \cite{jacob2007cold,huo2014solenoidal,beeler2013spin} (see also Section \ref{Gauge_Field}). This system can be implemented with various gauge structures, for example: Uniform Abelian gauge field, spin-orbit configurations (in uniform non-Abelian gauge field), and synthetic magnetic fields such as a uniform or a monopole configuration. All these configurations can be realized using the tripod scheme developed for $SU(2)$ systems. A natural extension of the tripod scheme can also be used for exploring the $SU(3)$ symmetry \cite{hu2014u}. 
One can fabricate Non-Abelian Aharonov-Bohm matter-wave interferometers operating with a $SU(N)$ fermionic fluid. To this end, one shall attach source and drain leads to the ring-shaped optical potential to inject and collect the quantum gas (See Fig. \ref{Atomtronics_fig}). Alternatively, the wires can be suppressed and the out-of-equilibrium dynamic of the ultracold gas can be investigated in the ring only. While the bosonic case has been largely studied (See  sections \ref{ringCondensates}, \ref{Persistent_toroid} and \ref{TransportBose}
 and reference therein), the interacting fermionic case and the role of its spin internal structure remains largely unexplored. A  number of theoretical questions need to be tackled to understand the dynamic of the system. For example: effects of the quasi-1D geometry,  the interplay between charge and spin degrees of freedom at mesoscopic scales and the role of finite temperatures, impact of quantum statistics (Pauli Blocking). Interestingly, persistent current states shall exist if pairing and superfluidity can occur and becomes superfluid and reaches the antiferromagnetic regime at strong repulsive interactions (Tonks regime). 
 In the out-of-equilibrium regime, one can study the response to a quench in an isolated system. The integrable regimes could be explored by employing the machinery  discussed in   \ref{NonEqDyn}, \ref{DynProt}. One can also add localized barriers interrupting the ring (see \ref{ringCondensates},\ref{PhaseSlips}, \ref{DeviceSensors}, \ref{MQD}).  Such a scheme would provide the implementation of an AQUID operating with a fermionic quantum fluid under a non-Abelian synthetic gauge field.

Based on the fermionic nature of the alkaline-earth atoms, it would be  interesting to transpose the standard electronic and/or spintronics circuits to atomtronics circuits operating with neutral-atom fermionic species with enhanced control and flexibility. 

The {\it primum mobile} for  circuits with flowing SU($N$) matter  was theoretically analysed recently\cite{chetcuti2020persistent}.  It would be interesting to study configurations for atomic SQUIDs exploiting the $SU(N)$ features. For 'SU($N$) atomtronics,    it would be also interesting to generalize the Datta-Das Transistor (DDT), the  fundamental building block of spintronics circuits\cite{datta1990electronic}, to ultracold gas system. Major steps, toward that goal, were done theoretically \cite{vaishnav2008spin}, and experimentally, on Rb BEC \cite{beeler2013spin}, and recently on strontium ultracold gas using a tripod scheme \cite{Madasu20}. Among the possible specific added values of the latter implementation, a fermionic atomic DDT operating  with an ultracold alkaline-earth-like gas, can be extended to gauge field with higher symmetry ($SU(3)$ at least), which can be generated using a double tripod laser scheme \cite{hu2014u}.
$SU(N)$ fermionic systems have triggered a great interest to explore their magnetic properties both theoretically \cite{gorshkov2010two} and experimentally \cite{ozawa2018antiferromagnetic}. One can exploit atomtronics circuits to probe $SU(N)$ matter. For example, in the spirit of solid-state physics I-V characteristics, one could define a new route for the diagnostic of the different many-body quantum regimes in terms of the current flowing through the $SU(N)$ system. Specifically, one could focus on fermionic systems that realize the $SU(N)$ Heisenberg or Hubbard models in the rectilinear/ring-shaped potentials attached to source and drain leads. In these structures, the transport coefficients can be derived by monitoring the densities in the source and drain leads. One can consider investigating transport in the $SU(N)$ Kondo impurity model. The effect of disorder, Anderson localization and many-body localization \cite{kramer1993localization,asada2002anderson,su2018role,abanin2019colloquium} could be explored with fermionic atomtronics using $SU(2)$ and $SU(3)$ spin-orbit coupling.  A similar logic could be employed to study the BCS-BEC crossover\cite{strinati2018bcs}.

\section{MANIPULATING RYDBERG ATOMS}\label{Rydberg}
\vspace*{-0.5cm}
\par\noindent\rule{\columnwidth}{0.4pt}
{\bf{\small{W. Li and  O. Morsch}}}
\par\noindent\rule{\columnwidth}{0.4pt}



Atoms excited to high-lying energy states (with principal quantum number $n$ larger than $\approx 15$) are known as Rydberg atoms \cite{gallagher1994rydberg}. They have considerably longer lifetimes than atoms in low-lying excited states, and much  larger (by several orders of mangnitude, with strong scaling with $n$) electric polarizability as well as dipole and van der Waals interactions. Rydberg atoms have been studied for several decades, with renewed interest sparked by the invention of  laser cooling, which made more accurate studies possible, and also due to the advent of quantum computation and quantum simulation, for which Rydberg atoms are a promising building block \cite{saffman2010quantum}. Generally, the combination of controllability, strong interactions and long coherence times make Rydberg atoms promising candidates for the realization of future quantum information technologies. In a broader context of quantum technologies, Rydberg atoms have also been explored for sensitive detection of electric fields and towards quantum state transfer between microwave and optical domains.

For the purposes of atomtronics, Rydberg atoms are an interesting system to study in regard to the propagation of excitations in disordered or ordered arrays. In fact, many transport properties, both in the quantum and semiclassical regimes, can be studied using Rydberg excitations. While Rydberg atoms have not been used for atomtronics applications (as understood in this review) so far, they might represent a valuable addition to the atomtronics toolbox in the future. In this spirit, the present chapter presents a few recent results on percolation phenomena studied in a gas of ultra-cold Rydberg atoms as well as on microwave control of Rydberg atoms.

\subsection{Driven-dissipative Rydberg systems}
An important aspect of transport phenomena is the interplay between an external drive and the natural dissipation of the system, which has been investigated by several groups in recent years  \cite{helmrich2018uncovering,helmrich2020signatures, whitlock2019diffusive,ding2020phase,letscher2017bistability}. In samples of ultra-cold Rydberg atoms (with temperatures around $T\approx 120\,\mathrm{\mu K}$, so on the timescales of typical experiments atomic motion can be neglected) we can study this interplay by driving a transition between the ground state of the atom (87-rubidium in our case) and a high-lying Rydberg state with $n\approx 70-80$. In our experiments in Pisa we use $S$ states (zero angular momentum), for which the van der Waals interaction is repulsive. This interaction leads to two distinct many-body effects. For resonant driving, it prevents the excitation of more than one Rydberg atom inside the "blockade sphere"; this is known as the dipole blockade \cite{urban2009observation,gaetan2009observation}. On the other hand, for off-resonant driving the van der Waals interaction can lead to the compensation of the detuning if a ground state atom is at a certain "facilitation distance" from a Rydberg atom \cite{simonelli2016seeded,valado2016experimental}. At that distance, the off-resonant driving is shifted into resonance and thus the excitation of the ground state atom is "facilitated". 

It turns out that by adding the natural decay of a Rydberg state due to spontaneous emission (with timescales of a few hundred $\mathrm{\mu s}$) it is possible to realize a paradigmatic model from statistical physics called directed percolation \cite{marcuzzi2015non}, which can be used to study such diverse processes as epidemic spreading, wildfires or the onset of turbulence. This model can characterized by two processes in a spin-1/2 example: offspring production, in which a "spin up" causes a nearby "spin down" to flip its state at a certain rate; and sudden death, in which a "spin up" spontaneously flips down. For our Rydberg system, these two processes can be directly translated into facilitation with rate$\Gamma_\mathrm{fac}$  and spontaneous decay with rate$\Gamma_\mathrm{spon}$ . We note here that both processes are incoherent (in particular, we choose a Rabi frequency for Rydberg excitation that is smaller than the decoherence rate). From statistical physics we know that this directed percolation model exhibits a phase transition between its absorbing state (all spins "down", or all atoms in the ground state) and an active state in which, on average, a macroscopic number of spins are "up" (i.e., atoms are in Rydberg states).  

\begin{figure}
\includegraphics[width=\columnwidth]{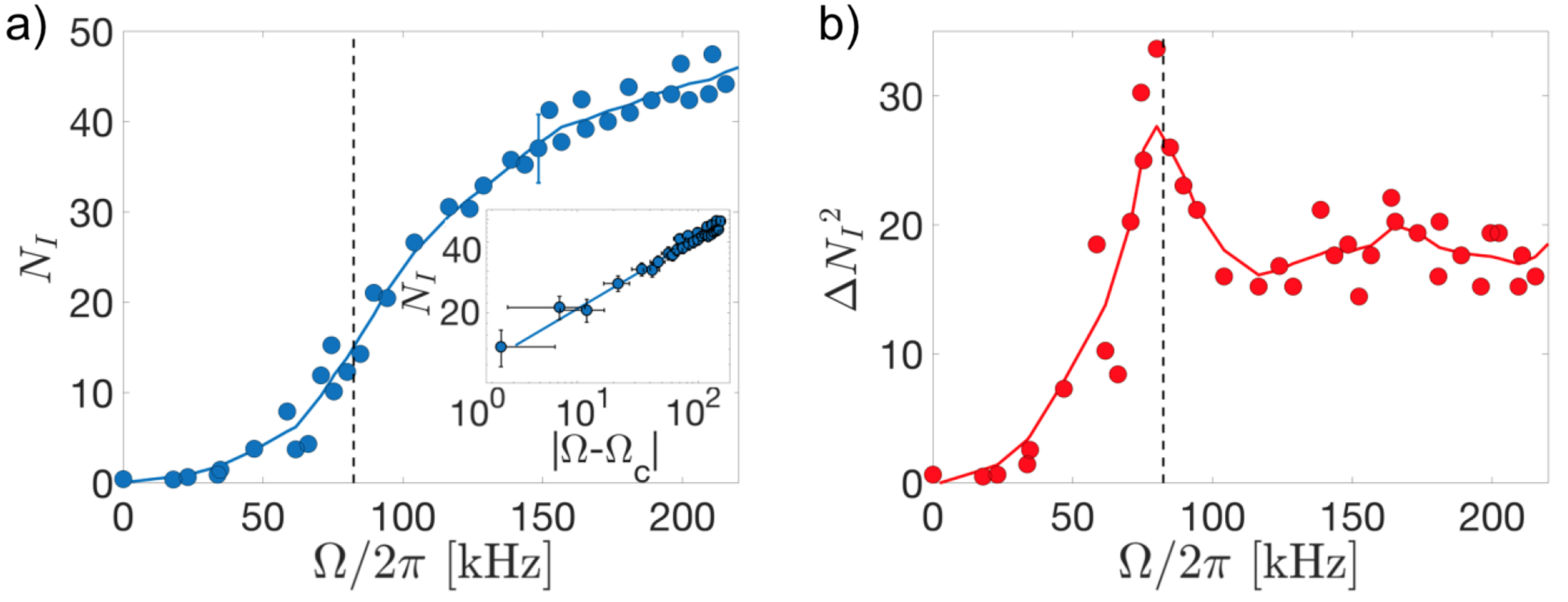}     
\caption{Evidence for an absorbing state phase transition in a Rydberg gas.  a) Number of excitations in the stationary state as a function of $\Omega$ (the solid line is a sliding average to guide the eye). The inset shows a power-law fit around the critical value $\Omega_c$, indicated by the dashed line in the main figure. In b) the peak in the variance plotted as a function of $\Omega$ indicates the critical point. Adapted with permission from R.  Gutiérrez, C. Simonelli, M. Archimi, F. Castellucci, E. Arimondo, D. Ciampini, M. Marcuzzi, I. Lesanovsky, and O. Morsch, Phys. Rev. A96, 041602(R) (2017). Copyright 2017 American Physical Society.}\label{exp_absorbing_results}
\end{figure}

We realized this model using Rb Rydberg atoms in a magneto-optical trap \cite{gutierrez2017experimental}. By varying the Rabi frequency of the off-resonant laser driving (a two-photon excitation via an intermediate $6P$ state was used) we were able to scan the ratio $\Gamma_\mathrm{fac}/\Gamma_\mathrm{spon}$ across the critical value for the absorbing-state phase transition ($\Gamma_\mathrm{fac}$ is related to $\Omega$ via $\Gamma_\mathrm{fac}=(\Omega^2/2\gamma)$ , where $\gamma$ is the decoherence rate). Fig.\ref{exp_absorbing_results} shows the results of those experiments. In order to prepare the system away from the absorbing state with all atoms in the ground state (from which, by definition, the system cannot escape), we initially excited around $30$ Rydberg atoms in the cloud and then allowed the system to evolve under constant driving for $1.5 \,\mathrm{ms}$ before measuring the number of Rydberg excitations by field ionization. The directed percolation phase transition is visible both in the plot of the number of excitations as a function of $\Omega$ (Fig. \ref{exp_absorbing_results} a)) and as a peak in the variance of the number of excitations (Fig. \ref{exp_absorbing_results} b)).

This is one example of a transport/percolation problem implemented using cold Rydberg atoms. In future experiments, this concept can be extended to (partially) coherent driving  \cite{marcuzzi2016absorbing,buchhold2017nonequilibrium} and/or ordered arrays  \cite{turner2017quantum, schlosser2020assembled}, as well to tailored and controllable dissipation.

\subsection{Microwave-optical conversion using Rydberg atoms}

Rydberg atoms feature transitions of very large dipole moments in the microwave frequency range~\cite{gallagher1994rydberg}, which has been utilized for sensitive detection of microwave electric field~\cite{sedlacek2012microwave,simons2016using} and for efficient conversion from microwave to optical photons~\cite{han2018coherent}. In quantum simulation using Rydberg atoms, nearby Rydberg states are commonly encoded as spin states, and their populations and dynamics can be conveniently manipulated with microwave radiation~\cite{orioli2018relaxation,deleseleuc2019observation}.

Here we present a demonstration of coherent microwave-to-optical conversion of classical fields via six-wave mixing in Rydberg atoms. In quantum regime, such coherent conversion is essential for coupling superconducting qubits operating at microwave frequencies to photonic qubits used in quantum communication over long distances~\cite{kimble2008quantum}, and therefore has been intensively pursued in quite a few different physical systems~\cite{lambert2020coherent}.

\begin{figure}
\includegraphics[width=\columnwidth]{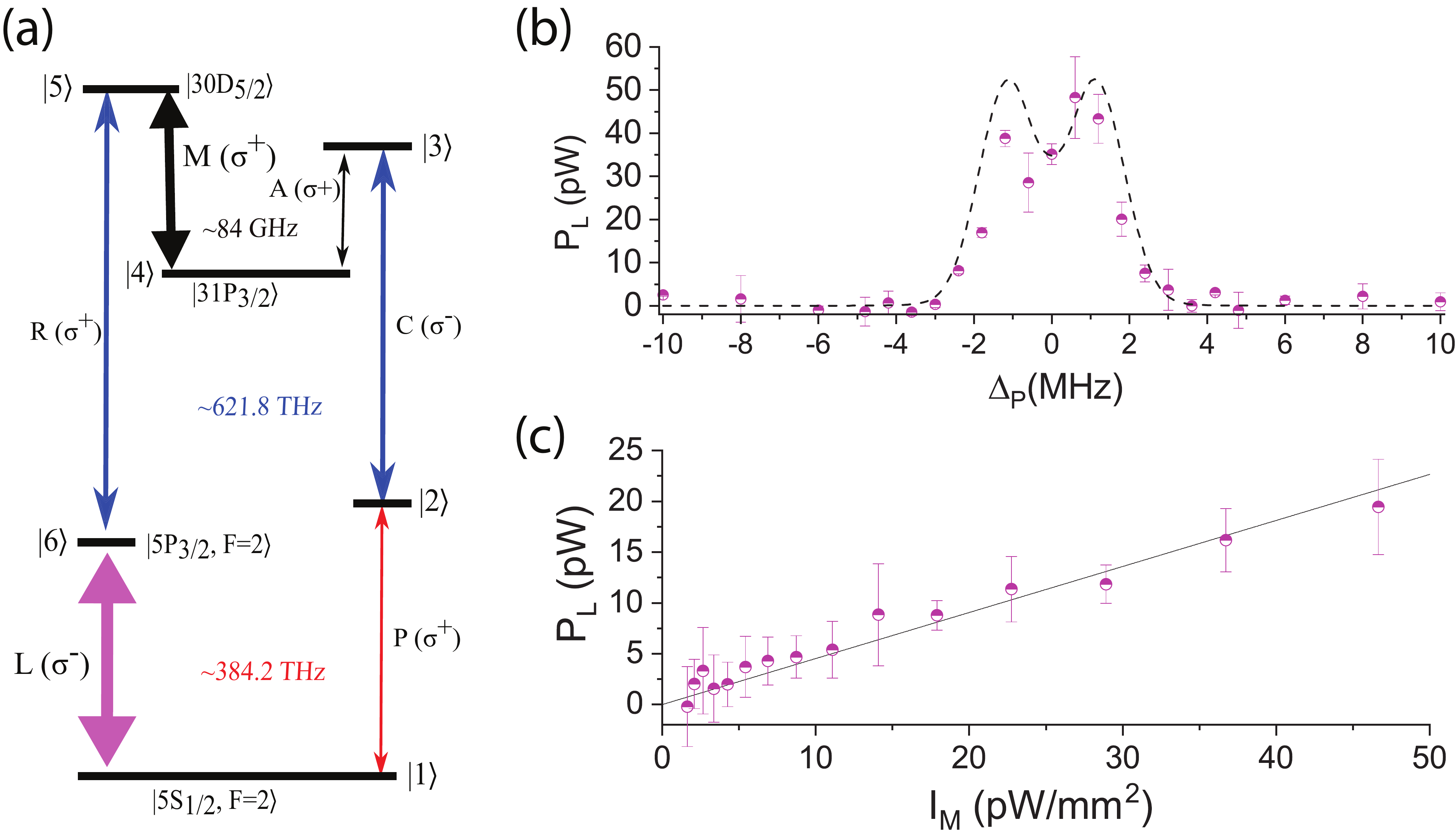}     
\caption{Efficient microwave-to-optical conversion using Rydberg atoms.  (a) Energy level diagram and coupled transitions. The polarization of the fields are indicated inside parentheses. (b) Spectrum of the generated light power $P_L$. The dashed line is a simulated result obtained using Maxwell-Bloch equations. (c) The power of the generated light is plotted versus $I_M$ in the range of 0 to $50\ \textrm{pW/mm}^2$. The solid line is the result of a linear fit. Adapted with permission from T. Vogt, C. Gross, J. Han, S. B. Pal, M. Lam, M. Kiffner, and W. Li, Physical Review A 99, 023832 (2019). Copyright 2019 American Physical Society.}\label{microwave-conversion}
\end{figure}

The principle of our conversion experiment using Rydberg atoms is as follows. A cloud of cold polarized $^{87}$Rb atoms is illuminated by four auxiliary electromagnetic fields P, C, A, R as well as the microwave field M to be converted. By non-linear frequency mixing of the six waves in the atomic medium, the field M is converted into the optical field L. The chosen configuration of energy levels is displayed in Fig.~\ref{microwave-conversion}(a), where the six waves are near-resonant with the atomic transitions shown in the figure with $|1\rangle \equiv |5S_{1/2}, F = 2, m_F = 2\rangle$, $|2\rangle \equiv |5P_{3/2}, F = 3, m_F = 3\rangle$, $|3\rangle \equiv |30D_{3/2}, m_J = 1/2\rangle$, $|4\rangle \equiv |31P_{3/2}, m_J = -1/2\rangle$, $|5\rangle \equiv |30D_{5/2}, m_J = 1/2\rangle$, and $|6\rangle \equiv |5P_{3/2}, F = 2, m_F = 1\rangle$. In the absence of the microwave field M, the system is in the configuration of microwave dressed electromagnetically induced transparency involving Rydberg states (Rydberg EIT), formed by the two optical waves P and C, and the auxiliary microwave field A. Once the M and R fields are added, the coherence induced between the ground state  $|1 \rangle$ and the intermediate state  $|6 \rangle$ triggers the generation of the converted optical field L.

A typical spectrum of the measured power $P_L$ of the generated L field vs. the input P field detuning $\Delta_P$ is shown in Fig.~\ref{microwave-conversion}(b). The conversion is most efficient around $\Delta_P=0$, which is consistent with the non-linearity responsible for the frequency mixing being maximum close to resonance. The behavior $P_L$ for $\Delta_P=0$ is approximately linear as a function of the input intensity $I_M$ of field M, as shown in Fig.~\ref{microwave-conversion}(c). Given $P_L \approx \alpha I_M$, a linear fit to the data yields the photon conversion efficiency of the process to be $\eta=0.051$. This conversion efficiency is seventeen times larger than the one reported in Ref.~\cite{han2018coherent}, and this enhancement is due to an improved experimental configuration, which makes the conversion occur over a longer distance. Our theoretical study shows that by using a carefully selected energy level scheme to minimize the absorption of the input P field when propagating through the conversion medium, a conversion efficiency above 50\% can be reached even with all-resonance six-wave mixing similar to that in Fig.~\ref{microwave-conversion}(a)~\cite{vogt2019efficient}.

This conversion method is an application example from the strong coupling between microwave and Rydberg atoms. To reach near-unit conversion efficiency for quantum state transfer at the single photon level, one may consider implementing stimulated Raman adiabatic passage~\cite{gard2017microwave}, or tuning two of the fields (for example fields C and A) off-resonance to realize an effective two-photon transition in our system~\cite{kiffner2016two,vogt2018microwave}. Besides its potential for quantum state transfer, this method of conversion into optical photons is also promising for the sensitive real-time detection of microwave or THz fields.

\subsection{Concluding remarks and outlook}
While the experiments outlined in this section do not yet make a direct contribution to atomtronics, it is likely that future studies of transport phenomena could make use of the techniques presented here. In the case of Rydberg atoms, Rydberg excitations coupled via dipole-dipole and van der Waals interactions - rather than the atoms themselves - are transported. Seed or source excitations can be injected into a cloud (or ordered array) of atoms at well-defined positions. In particular, using the recently developed patterning techniques based on dipole trap arrays (either using micro-mirror devices, holographic methods [24] or custom-made microlens-arrays [162]) it will be possible to conduct excitation-transport experiments using source-drain con-figurations (exploiting the high spatial resolution for excitation and detection). In this way, ring-shaped circuits or other, more complicated transport topologies could be explored. Coupling of the Rydberg atoms to laser or microwave sources could then be used to further tailor the interaction between the Rydberg atoms or for inducing additional dissipation/dephasing in the system. This will allow one to study the crossover between incoherent hopping and coherent transport. Finally, it is conceivable that studies of Rydberg excitation transport could be combined with “regular” atomtronics, resulting in a hybrid system in which both excitations and matter are transported (either independently or possibly coupled to each other).


%
%

%

{\it Acknowledgments}
W.L. acknowledges the support by the National Research Foundation, Prime Minister's Office, Singapore and the Ministry of Education, Singapore under the Research Centres of Excellence programme.


%



\bigskip

{\bf{Data Availability Statement}}. Throughout this article,  data sharing is not applicable to this article as no new data were created or analyzed in this study. 


\providecommand{\noopsort}[1]{}\providecommand{\singleletter}[1]{#1}

\end{document}